%% file: main.tex
\documentclass[11pt,oneside,a4paper]{article}
\usepackage{amssymb}
\usepackage{amsmath}
\usepackage[titletoc,title]{appendix}
\usepackage[T1]{fontenc}
\usepackage{graphicx}
\usepackage{float}
\usepackage{subfig}
\usepackage{newtxtext,newtxmath}
\usepackage{url}
\usepackage{color}
\usepackage{hyperref}
\usepackage{siunitx}
\usepackage{miller}

\usepackage[style=numeric-comp,sorting=none,backend=biber,firstinits=true]{biblatex} 
\definecolor{blu}{rgb}{0.,0.,1.}
\definecolor{red}{rgb}{1.,0.,0.}
\definecolor{burgundy}{rgb}{0.5, 0.0, 0.13}
\definecolor{crimsonred}{rgb}{0.6, 0.0, 0.0}
\definecolor{persianblue}{rgb}{0.11, 0.22, 0.73}
\definecolor{forestgreen}{rgb}{0.13,0.35,0.13}

\makeatletter
\g@addto@macro\bfseries{\boldmath}
\makeatother

\hypersetup{colorlinks, citecolor=crimsonred, linkcolor=persianblue, urlcolor=crimsonred}

\input{macro.tex}

\begin{document}

\begin{center}
\LARGE{\bf HIKE, High Intensity Kaon Experiments \\ at the CERN SPS}

\vspace{1cm}
\LARGE{\bf Letter of Intent}

\renewcommand{\thefootnote}{\fnsymbol{footnote}}
\vspace{1.0cm}
\LARGE{The HIKE Collaboration\footnote{Address all correspondence to \texttt{hike-eb@cern.ch}}} 
\end{center}
\vspace{2.0cm}
\addtocounter{footnote}{-1}
\renewcommand{\thefootnote}{\arabic{footnote}}

\begin{figure}[H]
\centering
\includegraphics[width=10cm]{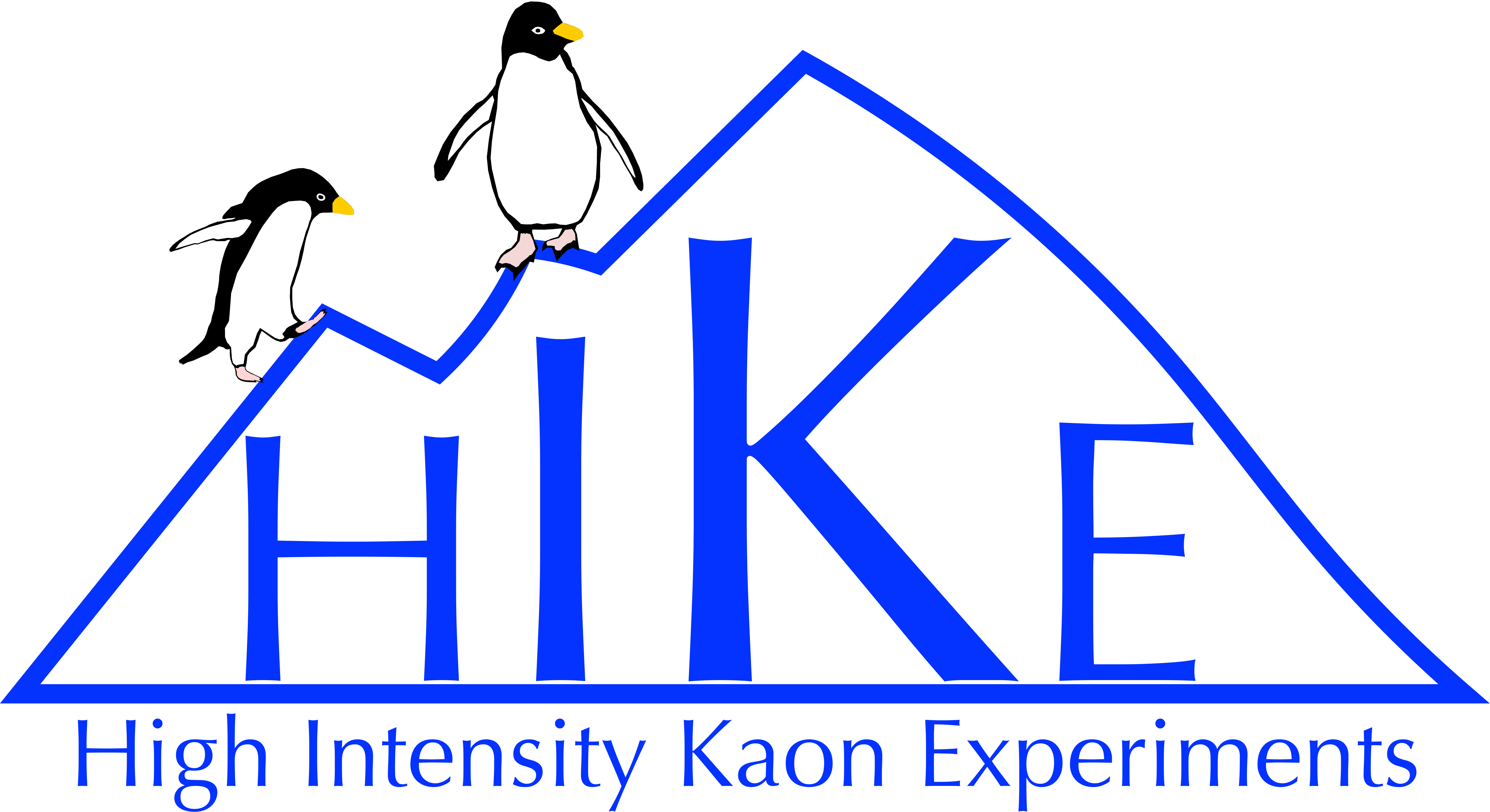}
\label{fig:my_label}
\end{figure}
\vspace{2.0cm}

\begin{abstract}
A timely and long-term programme of kaon decay measurements at a new level of precision is presented,
leveraging the capabilities of the CERN Super Proton Synchrotron (SPS).
The proposed programme is firmly anchored on the experience built up studying kaon decays at the SPS over the past four decades, and includes rare processes, CP violation, dark sectors, symmetry tests and other tests of the Standard Model.
The experimental programme is based on a staged approach involving experiments with charged and neutral kaon beams, as well as operation in beam-dump mode. The various phases will rely on a common infrastructure and set of detectors.
\end{abstract}

\newpage

\input{authors.tex}

\newpage
\setcounter{tocdepth}{2}
\tableofcontents

\newpage
\setcounter{page}{1}
\pagestyle{plain}


\input{goal.tex}

\input{motivation.tex}

\input{physics.tex}

\input{fips.tex}


\input{beam_mm.tex}

\input{phase1.tex}

\input{phase2.tex}

\input{klever.tex}

\input{dump.tex}

\input{detectors.tex}

\input{DAQ.tex}

\input{Computing}

\input{infrastructure.tex}

\input{Conclusions.tex}

\newpage 

\printbibliography[heading=bibintoc] 

\end{document}

%% file: macro.tex
\newcommand{\Fig}[1]{Figure~\ref{#1}}

\newcommand{\Sec}[1]{Section~\ref{#1}}

\newcommand{\Tab}[1]{Table~\ref{#1}}

\ifdefined\qtyproduct
\else
  \ifdefined\NewCommandCopy
    \NewCommandCopy\qtyproduct\SI
  \else
    \NewDocumentCommand\qtyproduct{O{}mm}{\SI[#1]{#2}{#3}}
  \fi
\fi

%% file: authors.tex
\begin{center}
{\Large The HIKE collaboration}
\end{center}
\begin{flushleft}
 E.~Cortina Gil$^{1}$,   
 J.~Jerhot$^{1}$,   
 N.~Lurkin$^{1}$,   
 T.~Numao$^{2}$,   
 B.~Velghe$^{2}$,   
 V.~W.~S.~Wong$^{2}$,   
 D.~Bryman$^{3}$,   
 L.~Bician$^{4}$,   
 Z.~Hives$^{4}$,   
 T.~Husek$^{4}$
% ...
%
\end{flushleft}
%
%
%
%
$^{1}$
Universit\'e Catholique de Louvain, B-1348 Louvain-La-Neuve, Belgium \\
$^{2}$
TRIUMF, Vancouver, British Columbia, V6T 2A3, Canada \\
$^{3}$
University of British Columbia, Vancouver, British Columbia, V6T 1Z4, Canada \\
$^{4}$
Charles University, 116 36 Prague 1, Czech Republic \\
% ...
\clearpage

%% file: goal.tex
\section{Overview of HIKE}


Over the past 75~years, experimental studies of kaon decays have played a singular role in propelling the development of the Standard Model. As in other branches of flavour physics, the continuing experimental interest in the kaon sector derives from the possibility of conducting precision measurements, particularly of suppressed rare processes, that may reveal the effects of new physics with mass-scale sensitivity exceeding that which can be explored directly, e.g., at the LHC or next-generation colliders. Because of the relatively small number of kaon decay modes and the relatively simple final states, combined with the relative ease of producing intense kaon beams, kaon decay experiments are in many ways the quintessential intensity-frontier experiments.

Over the past four decades, the CERN North Area has been the host to a successful series of precision kaon decay experiments. Among the many results obtained by these experiments is the discovery of direct CP violation, widely quoted to be among the major discoveries
made at CERN. 

\textbf{Continuation of high-intensity kaon experiments at CERN, including future upgrades (providing a unique probe into BSM physics complementary to the $B$-sector) has been identified as an essential scientific activity in the 2020 Update of the European Strategy for particle physics, and is strongly supported in the national roadmaps across Europe.}

The High Intensity Kaon Experiments (HIKE) project represents a broad, long-term programme at CERN after LS3, based in the North Area ECN3 experimental hall, covering all the main aspects of rare kaon decays and searches accessible via kaon physics, from ultra-rare kaon decays to precision measurements and searches for new phenomena. 
HIKE is intended to continue the very successful tradition of kaon experiments at CERN in ECN3, the latest of which is the currently operating NA62 experiment. HIKE will profit from a beam intensity increase by a factor between four and six with respect to NA62, and cutting-edge detector technologies.
This will allow HIKE to play a pivotal role in the quest for New Physics (NP) at the sensitivity required by the present experimental limits and theoretical models, in a wide range of possible masses and interaction couplings.
\textbf{The programme will consist of multiple phases, first with charged and then neutral kaon beams, as well as periods in beam dump mode.} The long decay volume and detector characteristics needed for kaon physics make HIKE suitable to search for new feebly-interacting, long-lived particles, providing unique sensitivity to forward processes. The detector design is challenging but at least one technological solution exists for each subsystem, thanks also to the synergy with HL-LHC. The various phases allow for insertion, repositioning or removal of specific detector elements depending on the physics requirements, while the overall setup remains broadly the same. 

\subsection{The scientific goals of HIKE}

The HIKE programme consists of several phases using shared detectors and infrastructure: a charged kaon phase between LS3 and LS4, and neutral kaon experiments commencing after LS4 including a phase
with tracking, and a phase optimised specifically for the $K_L\to\pi^0\nu\bar\nu$ measurement (KLEVER).



\vspace{2mm}
\noindent {\bf Phase 1: a multi-purpose $K^+$ decay experiment}

\noindent The physics of $K^+$ decays will be scrutinised with the highest precision, expanding and improving the physics programme of the NA62 experiment presently running in ECN3. New detectors will replace those of NA62 with the goal of improving the performance and sustaining higher rates.

\newpage
\noindent The principal goals of this phase are:
\begin{itemize}
\vspace{-0.5mm}
\item Measurement of the $K^+\to\pi^+\nu\bar\nu$ branching ratio to 5\% relative precision, matching the SM theoretical uncertainty.
\vspace{-0.5mm}
\item Precision measurements of $K^+\to\pi^+e^+e^-$ and $K^+\to\pi^+\mu^+\mu^-$ decays, and a precision lepton universality test.
\vspace{-0.5mm}
\item Searches for lepton flavour/number violating decays $K^+\to\pi(\pi^0)\mu e$,
$K^+\to\pi^-(\pi^0)\ell^+\ell^+$, $K^+\to\ell_1^-\nu\ell_2^+\ell_2^+$ and $\pi^0\to\mu^\pm e^\mp$ 
at the ${\cal O}(10^{-12})$ sensitivity.
\vspace{-0.5mm}
\item Measurement of the quantity $R_K=\Gamma(K^+\to e^+\nu)/\Gamma(K^+\to\mu^+\nu)$ to 0.1\% precision.
\vspace{-0.5mm}
\item Measurement of the ratios of the branching ratios of the main decay modes $K^+\to\pi^0\ell^+\nu$, $K^+\to\pi^+\pi^0$, $K^+\to\pi^+\pi^+\pi^-$ to a relative precision of a few per mille.
\vspace{-0.5mm}
\item Improvements of existing measurements of rare decays, including $K^+\to\pi^+\gamma\gamma$,
$K^+\to\pi^+\gamma\ell^+\ell^-$,
$K^+\to\pi^+\pi^0\gamma$, $K^+\to\pi^+\pi^0e^+e^-$.
\vspace{-0.5mm}
\item Searches for production of feebly-interacting particles in $K^+$ decays.
\vspace{-0.5mm}
\item Collection of a dataset in the beam-dump mode (with appropriate time sharing with the kaon mode), aiming for a factor~10 improvement in sensitivity to decays of feebly-interacting particles with respect to what NA62 can achieve by LS3. 
\end{itemize}

\noindent {\bf Phase 2: a multi-purpose $K_L$ decay experiment}

\noindent The physics of the $K_L$ decays will be addressed, including $K_L$ decays into charged particles in the final state. This mode of operation will require some modifications of the detector layout with respect to the charged mode. The main objectives of this phase are:
\begin{itemize}
\vspace{-0.5mm}
\item Observation of the ultra-rare decays $K_L\to\pi^0 e^+e^-$ and $K_L\to\pi^0\mu^+\mu^-$, or establishment of stringent upper limits on the branching ratios of these decays at the ${\cal O}(10^{-11})$ level.
\vspace{-0.5mm}
\item Measurement of the $K_L\to\mu^+\mu^-$ decay branching ratio to 1\% relative precision.
\vspace{-0.5mm}
\item Searches for lepton flavour violating decays including $K_L\to(\pi^0)(\pi^0)\mu^\pm e^\mp$ and $K_L\to e^\pm e^\pm\mu^\mp\mu^\mp$ with ${\cal O} (10^{-12})$ sensitivity.
\vspace{-0.5mm}
\item Measurement of the ratios of the branching ratios of the main decay modes $K_L\to\pi^\pm \ell^\mp\nu$,
$K_L\to\pi^+\pi^-(\pi^0)$,
$K_L\to\pi^0\pi^0(\pi^0)$ to a relative precision of a few per mille.
\vspace{-0.5mm}
\item Collection of a further dataset in the beam-dump mode (with appropriate time sharing with the kaon mode), and characterisation of the neutral beam necessary to proceed to the third phase of HIKE.
\end{itemize}


\noindent {\bf Phase 3 (KLEVER): measurement of the $K_L\to\pi^0\nu\bar\nu$ decay}

\noindent The experimental apparatus will be modified to specifically address the $K_L\to\pi^0\nu\bar\nu$ decay, and to measure its branching ratio to a 20\% relative precision. To optimise the measurement, the tracking systems will be removed from the fiducial volume and additional photon veto detectors installed. Although this phase will focus on $K_L\to\pi^0\nu\bar\nu$, the setup will allow study of additional topics:
\begin{itemize}
\vspace{-0.5mm}
\item Searches for additional flavour-changing neutral current decays, such as $K_L\to\pi^0\pi^0\nu\bar\nu$. 
\vspace{-0.5mm}
\item Limits on the production of a long-lived, feebly interacting particle $X$ in $K_L\to\pi^0X$ decays, as a by-product of the measurement of $K_L\to\pi^0\nu\bar\nu$.
\vspace{-0.5mm}
\item Searches for $K_L$ decays to feebly-interacting particles $X$ such as $K_L\to XX$, with $X\to\gamma\gamma$.
\vspace{-0.5mm}
\item Searches for forbidden decays such as $K_L\to\pi^0\gamma$.
\end{itemize}

%% file: motivation.tex
\section{Scientific context}

The Standard Model (SM) of particle physics describes the experimental results obtained so far by experiments with exceptional precision. The culmination of this success was the discovery of the Higgs boson by the ATLAS and CMS collaborations at the LHC in 2012~\cite{ATLAS:2012yve,CMS:2012qbp}. Thus the SM provides a solid foundation for the present understanding of elementary particles and their interactions.
Despite this success, cosmological observations, experimental tensions, and theoretical motivations strongly suggest that the SM is an approximation of a more fundamental theory.
One paradigm assumes that this theory lies above the electroweak (EW) mass scale and manifests itself in terms of new particles with masses well above the Higgs boson mass and having sizeable interactions with the SM fields.
Another paradigm accounts for extensions of the SM that predict particles with masses below the EW scale, feebly interacting with the SM fields.
Such a low-energy NP paradigm is usually referred to as a Feebly Interacting Particles (FIPs) scenario.
Both high-scale BSM and FIPs models provide explanations for the main open questions in modern physics: the nature of dark matter, the baryon asymmetry of the universe, cosmological inflation, the origin of neutrino masses and oscillations, the strong CP problem, the hierarchical structure of the Yukawa couplings, the hierarchy problem, and the cosmological constant.

HIKE intends to study NP in both high-scale BSM and FIPs directions. 
In this respect, HIKE is perfectly aligned with the recommendation of the 2020 Update of the European Strategy for Particle Physics~\cite{EuropeanStrategyGroup:2020pow}
, where a whole chapter is devoted to flavour physics research with the statement: 
``Experimental hints for deviations from SM predictions in flavour processes are one of our best hopes to direct research towards the right energy scale where NP can be found''.

\subsection{Experimental context for high-scale BSM physics}
\label{sec:ExperimentalStudiesOfBSMPhysics}
The experimental techniques to investigate the presence of BSM physics at high mass scales are direct searches for massive particles or processes forbidden by the SM, and precision measurements of observables precisely known within the SM.
The first method is usually referred to as {\it direct searches}; the second as {\it indirect searches}.

Direct searches allow unambiguous identification of NP already in single processes.
After the observation and the study of the Higgs boson, direct searches for BSM particles were a primary goal of the LHC experiments ATLAS and CMS.
Still, the LHC centre-of-mass energy limits the sensitivity of this technique to the TeV scale, or slightly above. Besides, a direct observation alone gives little information on the BSM structure.
The absence of any significant direct observations of NP so far sets strong bounds on several types of BSM models up to the TeV scale.
The bounds are still statistically limited in the TeV region, and the LHC high-luminosity programme will allow them to be extended to the kinematic limits, if no observations emerge in the meantime.
Another type of direct searches focuses on the study of processes forbidden by the SM due to accidental SM symmetries like such as charged lepton flavour (CLF) or lepton number (LN) conservation.
These searches typically involve the study of hadron or lepton decays forbidden by the SM, for example, $\tau\to\mu\mu\mu$ and 
$K_L\to\mu^\pm e^\mp$.
Presently, these searches are statistically limited, and experiments at the high-intensity frontier are planned with increased sensitivity. 

Indirect searches exploit the possibility that NP particles affect low-energy observables virtually, via loop-level corrections to the SM prediction.
The golden modes of indirect searches are theoretically clean observables of rare processes in the SM, i.e. occurring at loop-level at the lowest order, because NP here may favorably compete with SM.
Indirect searches allow sensitivities to NP up to mass scales well above the TeV, but may require measurements of several observables to provide conclusive evidence of NP and to determine its nature.
Given the limits to NP at TeV-scale already set by LHC through direct searches, indirect searches represent a promising tool to further boost BSM searches.
Experimental flavour physics, measurements of Higgs couplings, and measurements in the gauge sectors of the SM ($W$, $Z$, top masses and gauge couplings) are typical examples of indirect searches.
The existing measurements are broadly consistent with the SM and already suggest that NP may occur well beyond the TeV scale.
Nevertheless, tensions between experiments and the SM exist mainly in flavour physics observables, and statistical limitations or the lack of measurements of rare processes demonstrate the need for additional efforts and motivate future directions in experimental particle physics.
Presently, the most significant experimental tensions with the SM are:
\begin{itemize}
\item the violation of lepton flavour universality in $b\to s$ and $b\to c$ 
transitions reported by LHCb and $B$-factories~\cite{LHCb:2014cxe,LHCb:2015wdu,LHCb:2021trn,LHCb:2015svh,LHCb:2017avl,Belle:2019oag,BELLE:2019xld}, with a global significance of about 3--4$\sigma$~\cite{Aebischer:2019mlg,Altmannshofer:2021qrr,Cornella:2021sby,Isidori:2021vtc}; 
\item the violation of CKM unitarity in the first row, known as the  ``Cabibbo angle anomaly'', with a significance of about 2--3$\sigma$~\cite{Cirigliano:2022yyo,Seng:2022wcw,Bryman:2021teu};
\item the inconsistency between the exclusive and inclusive measurements of the CKM matrix element $|V_{cb}|$ at the level above $3\sigma$~\cite{PDG};
\item the inconsistency between the experiment and the SM prediction for the anomalous magnetic dipole moment of the muon at the level of 4.2$\sigma$~\cite{Keshavarzi:2021eqa}.
\end{itemize}

The sensitivity of the indirect searches to BSM physics scales with the size of the datasets, which is particularly important in case of rare processes.
Moving to higher intensities is the way to increase statistics, however this requires improvements of the detector performance to cope with harsher experimental environments.
Upgraded detectors also allow the improvement of the accuracy of already precise measurements, overcoming systematic uncertainties whenever they are the main limiting factor.
The indirect search approach is at the heart of the LHC intensity frontier programme, the motivation for the upgrade of the LHCb experiment and for the continuation of the $B$-factory programme with the Belle~II experiment.


\subsection{Experimental context for feebly-interacting particles}

Direct searches for new low-mass particles can provide experimental insight into the models involving dark sectors and predicting the existence of FIPs~\cite{Beacham:2019nyx,Agrawal:2021dbo}. FIPs can be produced in meson decays or hadronic interactions of protons, and can decay to SM particles.
They are produced in rare processes and are expected to have long lifetimes with respect to typical experimental scales as a result of their feeble couplings to the SM fields.

Experiments may search for FIP production or decays. The first method focuses on the associated production of FIPs with SM particles in meson decays; see Ref.~\cite{Goudzovski:2022vbt} for a review of possible production processes in kaon decays. The meson mass ($\pi$, $K$, $D$, $B$) determines the kinematic region that can be explored.
Some examples are production of heavy neutral leptons in $K^+\to \ell^+N$, dark photons in $\pi^0\to\gamma X$, and dark scalars in $B\to KS$ or $K\to \pi S$.
These kinds of searches can be pursued by flavour experiments, like LHCb and NA62, but also at ATLAS and CMS.
The second method is based on FIP production, mainly in proton-nucleon interactions, and the detection of the subsequent FIP decays.
Forward configurations and long experimental setups are best suited for this type of search, as those of fixed-target experiments, but also LHCb.
The solid angle covered by the detector defines the kinematic range of the FIPs that can be studied. Examples of FIP decays include dark scalars or dark photon decays $S/A^\prime\to\mu^+\mu^-$ and heavy neutral lepton decays $N\to\pi^\pm\mu^\mp$.
Searches for FIPs performed so far have not reported any hints of NP, and only limits in the plane of the coupling versus FIP mass have been set within the framework of various NP models.
Nevertheless, it is worth continuing to explore the parameter space for FIP models.
High beam intensities and experiments with appropriate geometries are necessary for the pursuit of these studies.


\subsection{The role of HIKE}

The HIKE experiment is primarily a flavour physics experiment, with a fixed target configuration appropriate for the study of decays of kaons at high energies, on the order of 100~GeV.
The study of the kaon sector makes HIKE complementary to LHCb and Belle~II:
kaons provide different, and in some cases higher, sensitivity to NP than $B$ and $D$ mesons.
Measurements of kaon observables and correlations among them can help to find NP through the comparison with corresponding SM expectations and investigation of the present experimental inconsistencies with SM. 
More generally, the comparison between the flavour picture emerging from kaons with that from $B$ mesons is a powerful tool to investigate the indirect effects of NP: to shed light on existing experimental anomalies, push the sensitivity to mass scales beyond those attainable by $B$ and $D$ only, and provide insights to the flavour structure of possible NP models.

Several models of BSM physics predict effects on kaon observables, ranging from supersymmetric to non-supersymmetric models, NP involving heavy $Z^\prime$ bosons, vector-like couplings, leptoquarks, and extra dimensions, in both minimally flavour violating (MFV) scenarios or with new sources of flavour violation. A common feature is that bounds on the parameter space of these models coming from direct searches at LHC and $B$ and $D$ physics have only a marginal impact on the possible effects on kaon observables, even more so if the models are non-MFV and non-supersymmetric. In contrast, the main constraints on the effects of NP on kaon observables come from the precisely measured parameters $\varepsilon_K$ and $\Delta M_K$. 
Table~\ref{tab:models} gives a selected list of the most relevant NP models that may affect significantly kaon observables~\cite{Aebischer:2022vky}.

\begin{table}[h]
\centering
\caption{Compendium of new-physics models relevant for kaon processes~\cite{Aebischer:2022vky}.}
\vspace{-2mm}
\begin{tabular}{c|c|c}
\hline
NP scenarios & Process & References \\
\hline
Z-FCNC & $K^+ \rightarrow \pi^+ \nu \bar{\nu}$, $K_L \to \pi^0 \nu \bar{\nu}$, $\varepsilon^\prime / \varepsilon$ & \cite{Buras:2015jaq,Aebischer:2020mkv,Bobeth:2017xry,Endo:2018gdn,Endo:2016tnu} \\
Z$^\prime$ & $K^+\to\pi^+ \nu\bar\nu$, $K_L \to\pi^0\nu\bar\nu$, $\varepsilon^\prime / \varepsilon$, $\Delta M_K$ & \cite{Buras:2012jb,Buras:2015jaq,Aebischer:2020mkv,Buras:2012dp,Aebischer:2019blw} \\
Simplified models & $K_L \to \pi^0 \nu \bar{\nu}$, $\varepsilon^\prime / \varepsilon$ & \cite{Buras:2015yca} \\
LHT & All $K$ decays & \cite{Blanke:2006eb,Buras:2006wk,Blanke:2015wba} \\
331 models & Small effects in $K\to\pi\nu\bar\nu$ & \cite{Buras:2012dp} \\
Vector-like quarks & $K^+ \to \pi^+ \nu \bar{\nu}$, $K_L \to \pi^0 \nu \bar{\nu}$, $\Delta M_K$ & \cite{Bobeth:2016llm} \\
Supersymmetry & $K^+ \to \pi^+ \nu \bar{\nu}$, $K_L \to \pi^0 \nu \bar{\nu}$ & \cite{Buras:2004qb,Isidori:2006qy,Blazek:2014qda,Altmannshofer:2009ne,Tanimoto:2016yfy,Kitahara:2016otd,Endo:2016aws,Crivellin:2017gks,Endo:2017ums} \\
2HDM & $K^+ \to \pi^+ \nu \bar{\nu}$, $K_L \to\pi^0\nu \bar{\nu}$ & \cite{Chen:2018ytc,Chen:2018vog} \\
Universal extra dimensions & $K^+ \to \pi^+ \nu \bar{\nu}$, $K_L \to \pi^0 \nu \bar{\nu}$ & \cite{Buras:2002ej,Buras:2003mk}\\
Randall-Sundrum models & All rare $K$ decays & \cite{Blanke:2008yr,Duling:2009sf,Albrecht:2009xr,Casagrande:2008hr,Bauer:2009cf} \\
Leptoquarks & All rare $K$ decays & \cite{Bobeth:2017ecx,Fajfer:2018bfj,Marzocca:2021miv} \\
SMEFT & Several processes in $K$ system & \cite{Aebischer:2020mkv,Aebischer:2018csl} \\
SU(8) & $K^+ \to \pi^+ \nu \bar{\nu}$, $K_L \to \pi^0 \nu \bar{\nu}$ & \cite{Matsuzaki:2018jui}\\
Diquarks & $K^+\to\pi^+\nu\bar\nu$, $K_L\to\pi^0\nu\bar\nu$, $\varepsilon_K$ & \cite{Chen:2018dfc,Chen:2018stt}\\
Vector-like compositeness & $K^+\to\pi^+\nu\bar\nu$, $K_L\to\pi^0\nu\bar\nu$, $\varepsilon_K$ & \cite{Matsuzaki:2019clv} \\
\hline
\end{tabular}
\label{tab:models}
\end{table}

Presently, the main limitation to the investigation of the above models, and more generally to the quest for NP with kaons, comes from the experimental precision of the measurements of the kaon observables. To a lesser extent, the limitation is due to theoretical uncertainties and the precision of the measurements of the SM parameters, which will eventually be improved in the upcoming years.
The primary goal of HIKE is to improve the accuracy of the measurements, wherever they exist, in order to match and possibly challenge the theory precision, to study and measure for the first time channels not yet observed, and to search with unprecedented sensitivity for kaon decays forbidden by the SM.

A natural by-product of HIKE is the possibility to search for kaon decays to FIPs, a line of inquiry that is already being pursued with success by NA62. In addition, the fixed-target configuration and the high beam intensity requirement make HIKE suitable to search for decays of FIPs, exploring regions mainly below the $D$ mass, but with unprecedented sensitivity.

%% file: physics.tex
\section{Physics with high-intensity kaon beams}


Kaons offer a unique opportunity to investigate BSM physics in a way complementary to $B$ or $D$ mesons, lepton physics, and EDM searches.
The strategy is to perform multiple experimental studies ranging from measurements of branching ratios and form factors of kaon decays to searches for processes forbidden or not existing within the SM.
A global fit to the experimental inputs provides the best sensitivity to NP, although single measurements also put significant constraints. Evidence for NP can also show up as inconsistencies between the observed flavour pattern and that expected from the SM, violation of accidental SM symmetries like lepton flavour universality (LFU), lepton flavour or number (LFV or LNV), and direct production of new low-mass particles in kaon decays.

Because the BSM paradigm assumes NP at high masses, NP is expected to affect kaon processes at short-distance (SD) scales.
The link between kaon-physics observables and BSM physics relies on the SM theoretical knowledge of these observables.
The development of the theoretical understanding of the kaon sector is focused exclusively on calculations of observables relevant to pin down SD contributions from experimental results. 
Examples are: the SM calculation of the $K^+\to\pi^+\nu\bar\nu$ and $K_L\to\pi^0\nu\bar\nu$ decay branching ratios at NNLO leading to a few percent precision on these quantities; 
the recent calculation of the CP violation parameter $\varepsilon^\prime/\varepsilon$;
computations of the branching ratios for the rare decays $K^+\to\pi^+\ell^+\ell^-$, $K_L\to\pi^0\ell^+\ell^-$, $K_{L,S}\to\mu^+\mu^-$.

\vspace{3mm}
\noindent It is convenient to group kaon decays into the following classes.
\begin{itemize}
\item {\bf Flavour-changing neutral currents} (Section~\ref{sec:rare-kaon-decays}). The $K\to\pi\nu\bar\nu$, $K_L\to\pi^0 \ell^+\ell^-$, $K^+\to\pi^+ \ell^+\ell^-$  and $K_L\to\mu^+\mu^-$ decays are suppressed in the SM, occurring at loop-level at the lowest order. The corresponding SD contributions are related to CKM matrix elements and are generally calculated with high precision within the SM. Together with $\varepsilon_K$, rare decays allow the unitary triangle to be over-constrained, independently from $B$ processes.
With the exception of $K\to\pi\nu\bar\nu$ decays, the above decays receive important contributions from long distance (LD) physics.
The theoretical framework used to study these processes involves low-energy effective theories, like Chiral Perturbation Theory (ChPT), which require experimental inputs (chiral parameters) to provide and validate the predictions, as well as lattice QCD.
\newpage
\item {\bf Lepton flavour universality tests} (Section~\ref{sec:leptonic-kaon-decays}). 
The ratios ${\cal B}(K^+\to\pi^+e^+e^-)/{\cal B}(K^+\to\pi^+\mu^+\mu^-)$ and ${\cal B}(K^+\to e^+\nu)/{\cal B}(K^+\to\mu^+\nu)$ are extremely sensitive probes of lepton flavour universality. Another sensitive test is based on the ratio ${\cal B}(K\to\pi e\nu)/{\cal B}(K\to\pi\mu\nu)$, where both neutral and charged kaon decays ($K_L \to \pi^\pm \ell^\mp\nu$ and $K^+ \to \pi^0\ell^+\nu$) can be used.
\item {\bf Lepton flavour and number violating decays} (Section~\ref{sec:lfv}).
Decay modes forbidden by the SM by either lepton flavour or number conservation include $K^+\to\pi^-(\pi^0)\ell_1^+\ell_2^+$, $K^+\to\pi^+\mu^\pm e^\mp$, $K^+\to\ell_1^-\nu\ell_2^+\ell_2^+$, $K_L\to(\pi^0)(\pi^0)\mu^\pm e^\mp$ and $\pi^0\to\mu^\pm e^\mp$. Any experimental signal of these processes would constitute direct evidence for NP. It should be noted that kaon beams represent sources of tagged, Lorentz-boosted $\pi^0$ mesons.
\item {\bf Tests of low-energy QCD} (Section~\ref{sec:physics-low-energy-qcd}).
The abundant kaon decays $K\to\pi\pi$ and $K\to\pi\pi\pi$ occur at the tree level and are dominated by LD physics. The decays are described by ChPT which is the low-energy effective field theory of QCD. Global fits to the measured branching ratios and kaon lifetimes of the $K\to\pi\pi$ and $K\to\pi\pi\pi$ decays play a crucial role in the extraction of the ChPT low-energy constants. Rare and radiative decays including $K\to\pi\gamma\gamma$, $K\to\ell\nu\gamma$, $K\to\pi\ell\nu\gamma$, $K\to\pi\pi\gamma$, $K\to\pi\pi\pi\gamma$,  $K\to\pi\pi e^+e^-$, and $K\to\pi\pi\ell\nu$ are also dominated by LD physics and similarly provide access to the ChPT parameters.

\item {\bf CKM first-row unitarity tests} (Section~\ref{sec:semileptonic-kaon-decays}). Branching ratios and form factors of the tree-level $K^+\to(\pi^0)\ell^+\nu$ and $K_L\to\pi^\pm\ell^\mp\nu$ decays represent the principal experimental inputs for the determination of CKM matrix element $V_{us}$, with input from lattice QCD.
\item {\bf Production of feebly-interacting particles in kaon decays} (Section~\ref{sec:fips}) is predicted by FIP models involving new particles below the kaon mass scale. Examples include production of a heavy neutral lepton ($K^+\to\ell^+N$), or a dark scalar ($K^+\to\pi^+S$).
\end{itemize}
HIKE intends to address experimentally all these processes at a new level of sensitivity, improving the existing experimental picture to fully exploit the complementarity of kaon with respect to heavy-quark physics in the quest for NP.


\subsection{Flavour-changing neutral currents}
\label{sec:rare-kaon-decays}

Kaon decays proceeding at loop level in the SM are denoted ``rare'', and typically have branching ratios below $10^{-7}$. They provide a tool independent from $B$ and $D$ physics to test the flavour structure of the SM by constraining the unitary triangle via loop-level observables.

\begin{figure}
\includegraphics[width=1.0\textwidth]{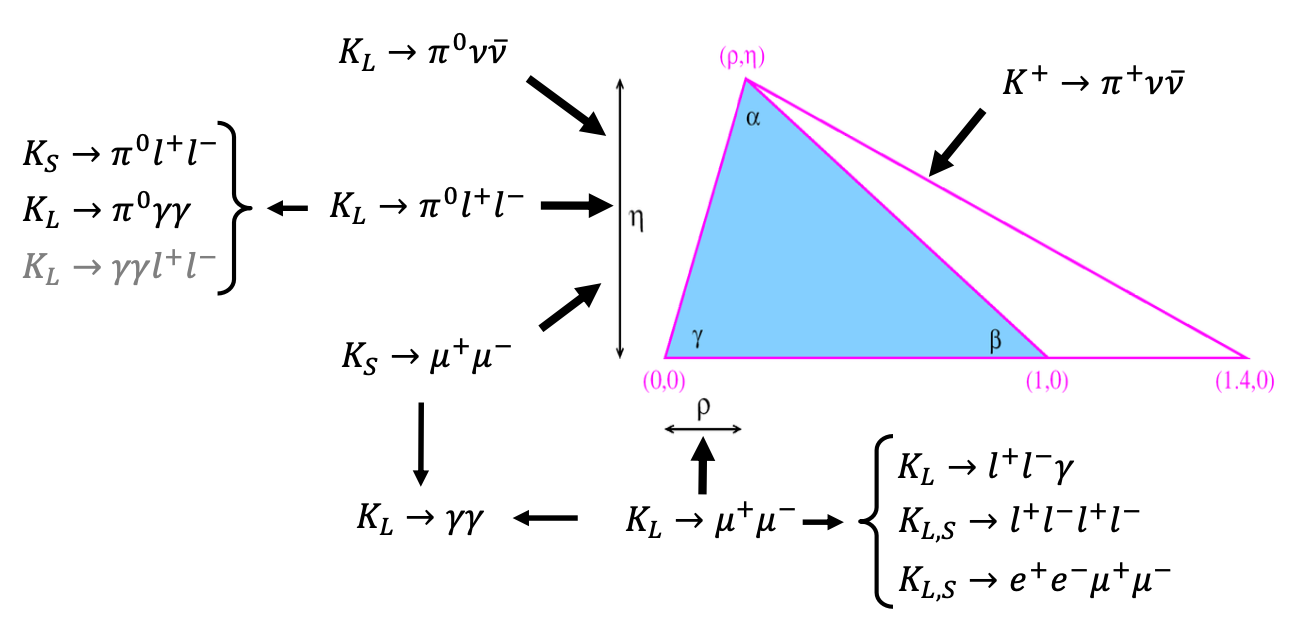}
\vspace{-6mm}
\caption{Relation between kaon rare decay modes and the parameters $\rho$ and $\eta$ of the unitary triangle (UT).
The direct link between decay modes and the UT indicates short distance terms dependent on  $\rho$ or $\eta$ contributing to the corresponding decay amplitudes.
Decays not directly connected to the UT are relevant to interpret the experimental results of the decay modes to which they are related.}
\label{fig:ckmk}
\end{figure}

The link between rare kaon decays and flavour physics in the $\rho-\eta$ plane is schematically shown in Fig.~\ref{fig:ckmk}. 
CP-violating short distance (SD) physics contributes to the $K_L\to\pi^0\nu\bar\nu$, $K_L\to\pi^0e^+e^-$, $K_L\to\pi^0\mu^+\mu^-$, and $K_{S}\to\mu^+\mu^-$ decays.
The amplitudes of these decays thus depend on the height, $\eta$, of the unitary triangle.
The SD physics contributing to the decay $K_L\to\mu^+\mu^-$ is CP-conserving, and its amplitude depends on the base, $\rho$, of the unitary triangle.
The amplitude of the rare decay $K^+\to\pi^+\nu\bar{\nu}$ contains terms depending on both $\eta$ and $\rho$.

The $K_L\to\pi^0e^+e^-$, $K_L\to\pi^0\mu^+\mu^-$ and $K_L\to\mu^+\mu^-$ decays are LD dominated.
The extraction of their SD components from experimental data proceeds through the study of ancillary decay modes, listed in Fig.~\ref{fig:ckmk} in association with the rare decays mentioned above.
In contrast, LD contributions are sub-dominant to the amplitudes of the $K^+\to\pi^+\nu\bar\nu$ decay, and negligible for the $K_L\to\pi^0\nu\bar\nu$ decay. As a matter of fact, these two decay modes belong to the theoretically cleanest probes of the flavour structure of the SM among all the kaon and $B$ meson decays.


\subsubsection{$K\to\pi\nu\bar\nu$ decays}
\label{sec:PhysicsKpinunu}

The $K\to\pi\nu\bar\nu$ decays involve $s\to d$ quark transitions. They proceed through box and penguin diagrams, which are SD dominated because of the leading contribution of the virtual $t$ quark exchange.
The quadratic GIM-mechanism and the $t\to d$ Cabibbo suppression make place these processes among the rarest meson decays in the SM.
Their decay amplitudes can be parameterised in terms of the precisely measured $K^+\to\pi^0e^+\nu$ decay, allowing theoretical computation free from hadronic uncertainties.
The SM predictions of the  $K^+\to\pi^+\nu\bar\nu$ and $K_L\to\pi^0\nu\bar\nu$ branching ratios can be written in terms of CKM parameters as~\cite{Buras:2015qea}
\begin{align}
{\cal B}(K^+\to\pi^+\nu\bar{\nu})&=(8.39\pm0.30)\times10^{-11}\left(\frac{|V_{cb}|}{0.0407}\right)^{2.8}\left(\frac{\gamma}{73.2^\circ}\right)^{0.74},
\label{eq:pnnsm_Kp}
\\
{\cal B}(K_L\to\pi^0\nu\bar{\nu})&=(3.36\pm0.05)\times10^{-11}\left(\frac{|V_{ub}|}{3.88\times10^{-3}}\right)^2\left(\frac{|V_{cb}|}{0.0407}\right)^2\left(\frac{\sin{\gamma}}{\sin{73.2^\circ}}\right)^2.
\label{eq:pnnsm_KL}
\end{align}
Here $V_{cb}$ and $V_{ub}$ are elements of the CKM matrix, and $\gamma$ is the angle of the unitary triangle defined as $\arg[(-V_{ud}V_{ub}^*)/(V_{cd}V_{cb}^*)]$.
The theoretical uncertainty depends on the NNLO approximation of the computation of the Feynmann diagrams, on the radiative corrections, and on the estimates of the long distance contributions due to the exchange of the $u,d,c$ quarks.
The prediction for the neutral mode prediction is more precise than that for the charged mode because the neutral amplitude is purely imaginary, allowing cancellation of the long distance corrections.
In either case, the main contribution to the  uncertainties for the branching ratio calculation comes from the precision of the measurements of the CKM matrix elements, known as the parametric uncertainty, and could be as large as 9\% if the current measurements of $V_{cb}$, $V_{ub}$ and $\gamma$ from the CKM fit are considered.
A suitable combination of external parameters nearly independent of new physics contributions allows the reduction of the parametric uncertainty, reducing the overall relative uncertainties in the branching ratios to 5\%~\cite{Buras:2021nns}.
Other recent calculations reach a similar quantitative conclusion~\cite{Brod:2021hsj}.
A global fit to the CKM unitary triangle, instead, allows 
the parametric uncertainty to be factored out, and offers a powerful tool to compare the SM flavour structure arising from kaon and $B$ physics.
Equations~\ref{eq:pnnsm_Kp} and~\ref{eq:pnnsm_KL} also indicate that precise measurements of $K\to\pi\nu\bar\nu$ decays would shed light on CKM parameters historically measured with $B$ mesons and suffering from long-standing experimental tensions like $|V_{cb}|$.

The extreme SM suppression makes these decays particularly sensitive to NP. From a model-independent point of view, the $K\to\pi\nu\bar\nu$ decays probe NP at the highest mass scales~\cite{Buras:2015qea}, of the order of hundreds of TeV.
Existing experimental constraints on NP affect the $K\to\pi\nu\bar\nu$ branching ratio weakly.
Model-dependent scenarios predict sizeable deviations of the branching ratios from the SM, as well as correlations between the branching ratios for the  charged and neutral modes, depending on the model~\cite{Chen:2018ytc,Bobeth:2017ecx,Bobeth:2016llm,Endo:2016tnu,Endo:2017ums,Crivellin:2017gks,Blanke:2015wba,Bordone:2017lsy} (see also Table~\ref{tab:models}).

The NA62 experiment at CERN has observed $20$ candidate events for the decay $K^+\to\pi^+\nu\bar\nu$ with 7~expected background events and 10~expected SM signal events~\cite{NA62:2021zjw}, leading to the measurement  $\mathcal{B}(K^{+}\rightarrow\pi^{+}\nu\bar{\nu})=(10.6^{+4.0}_{-3.4}|_{\text{stat}}\pm0.9_{\text{syst}})\times10^{-11}$ at $68\%$ confidence level. 
This represents the most precise measurement to date of this process, providing first evidence for its existence and falsifying the background-only hypothesis with $3.4\sigma$ significance.
The experiment is currently taking data in Run~2 (2021--LS3) with the aim of reaching a ${\cal O}(10\%)$ uncertainty measurement, and has demonstrated its ability to sustain nominal beam intensity. The NA62 experiment has therefore shown that the decay-in-flight technique works well and is scalable to larger data samples. 

Beyond the $\mathcal{B}(K^{+}\to\pi^+\nu\bar\nu)$ measurement, it is important to establish if the decay has a purely vector nature as expected within the SM. In BSM scenarios, an additional scalar contribution to the decay is predicted~\cite{Deppisch:2020oyx,Crosas:2022quq}, leading to possibly LNV/LFV $K^+\to\pi^+\nu\nu$ contributions, and since the neutrinos are not detected, a $\nu\bar\nu$ pair cannot be directly distinguished from a $\nu\nu$ pair, and the measurement of the branching ratio alone is insufficient to understand the nature of the process. 
Because of the suppression from the small neutrino mass, there is a negligible interference between vector and scalar contributions, and the experimentally measured branching ratio is given by~\cite{Deppisch:2020oyx, Aebischer:2022vky}:
\begin{equation}
\mathcal{B}(K^{+}\rightarrow\pi^{+}\nu \overset{\textbf{\fontsize{2pt}{2pt}\selectfont(---)}}{\nu} ) = \mathcal{B}(K^{+}\rightarrow\pi^{+}\nu\bar{\nu})_{\rm SM} + \sum_{i \leq j}^{3} \mathcal{B}(K^{+}\rightarrow\pi^{+}\nu_{i}\nu_{j})_{\rm LNV}.
\end{equation}
To fully test the SM prediction of a purely vector, lepton-number-conserving $K^{+}\rightarrow\pi^{+}\nu\bar{\nu}$ decay, the momentum-transfer spectrum must be studied to rule out the additional scalar BSM $K^{+}\rightarrow\pi^{+}\nu\nu$ contribution to the overall experimentally measured rate. Such a measurement requires a much larger dataset, as offered within the HIKE programme.

The current upper limit on the branching ratio of the $K_L\to\pi^0\nu\bar\nu$ decay is $3\times10^{-9}$ at 90\% CL, set by the KOTO experiment at J-PARC~\cite{KOTO:2018dsc}. The KOTO experiment is currently taking data with the goal of reaching ${\cal O}(10^{-11})$ sensitivity in the next 5~years. In the longer term, an ambitious upgrade, KOTO Step-2, is planned to begin construction after 2025 in a proposed extension of the Hadron Experimental Facility, with the goal of measuring the branching ratio to within $\sim$20\% in about three years of data taking~\cite{NA62KLEVER:2022nea}.


\subsubsection{$K_L\to\pi^0\ell^+\ell^-$ decays}
\label{sec:KLpi0ll}

The ultra-rare $K_L\to\pi^0\ell^+\ell^-$ decays represent another set of theoretically clean ``golden modes'' in kaon physics, second only to $K\to\pi\nu\bar\nu$ in terms of their importance, and allowing for the direct exploration of new physics contributions in $s\to d\ell\ell$ transitions (to be compared to $b\to s\ell\ell$ transitions). In the SM, they are dominated by the indirect (mixing-induced) CP violation and its interference with the direct CP-violating contribution. The decay rates can be enhanced significantly in the presence of large new CP-violating phases, in a correlated way with the effects in $K_L\to\pi^0\nu\bar\nu$ and $\varepsilon^\prime/\varepsilon$~\cite{Aebischer:2022vky}. The SM expectation for the decay branching ratios is~\cite{Isidori:2004rb,Friot:2004yr,Mescia:2006jd}
\begin{eqnarray}
{\cal B}(K_L\to\pi^0 e^+e^-) & = & 3.54^{+0.98}_{-0.85}~ \left(1.56^{+0.62}_{-0.49}\right)\times 10^{-11}, \nonumber\\
{\cal B}(K_L\to\pi^0\mu^+\mu^-) & = & 1.41^{+0.28}_{-0.26}~ \left(0.95^{+0.22}_{-0.21}\right)\times 10^{-11}, \nonumber
\end{eqnarray}
with the two sets of values corresponding to constructive (destructive) interference between direct and indirect CP-violating contributions.

Experimentally, the most stringent upper limits (at 90\% CL) of the decay branching ratios have been obtained by the KTeV experiment~\cite{KTeV:2003sls,KTEV:2000ngj}:
\begin{displaymath}
{\cal B}(K_L\to\pi^0 e^+e^-) < 28\times 10^{-11}, \quad {\cal B}(K_L\to\pi^0 \mu^+\mu^-) < 38\times 10^{-11}.
\end{displaymath}
The irreducible $K^+\to\gamma\gamma\ell^+\ell^-$ background, whose rate vastly exceeds that of the signal decays $K_L\to\pi^0\ell^+\ell^-$, represents a significant problem in achieving experimental sensitivity at the SM level, especially for $K_L\to\pi^0 e^+e^-$~\cite{Greenlee:1990qy}.


\subsubsection{$K^+\to\pi^+\ell^+\ell^-$ decays}
\label{sec:Kplus_pill}

The amplitudes of the $K^+\to\pi^+\ell^+\ell^-$ decays ($\ell=e,\mu$) are LD-dominated and their branching ratios are $\mathcal{O}(10^{-7})$.
The differential decay amplitude of these decays with respect to the di-lepton invariant mass depends on two form factor parameters, denoted $a_+$ and $b_+$, that lepton universality predicts to be independent from the flavour of the leptons.
Differences between $a_+$ and $b_+$ between the electron and muon channels can be correlated to LUV effects that can explain the possible observed lepton anomalies in $B$ physics~\cite{Crivellin:2016vjc,DAmbrosio:2018ytt}.

The NA62 experiment has recently reported a new measurement of the form factors of the $K^+\to\pi^+\mu^+\mu^-$ decay using data collected in 2017--2018~\cite{NA62:2022qes}.
The excellent kinematic resolution allowed a nearly background-free selection of $3\times10^4$ decays.
The decay $K^+\to\pi^+\pi^+\pi^-$ was used for normalisation, and the parameter values $a_+ = -0.575 \pm 0.013$ and $b_+ = -0.722 \pm 0.043$ were extracted from a fit to the di-lepton mass spectrum.
This improves significantly on the previous measurements in the muon channel while also consistent within uncertainty with measurements of the same parameters performed both in the muon and electron channels~\cite{E865:1999ker,NA482:2009pfe,e865:1999kah,NA482:2010zrc}.
The NA62 experiment is expected to improve the precision of the LFU test in the next years, addressing also the measurement of the $K^+\to\pi^+e^+e^-$ form factors. 
However a significant increase in precision is required to perform a LFU test with comparable sensitivity to NP models as in current results from $B$ physics~\cite{Crivellin:2016vjc,DAmbrosio:2022kvb}. HIKE will offer a significantly larger data sample, further reducing statistical uncertainties, which remain a large fraction of the total uncertainty in the most recent NA62 measurement~\cite{NA62:2022qes}. 


\subsubsection{$K_L\to\mu^+\mu^-$ decays}
The $K_L\to\mu^+\mu^-$ is a rare and helicity suppressed decay.
Its branching ratio is driven by LD contributions, with a sub-dominant CP-conserving SD component~\cite{Buras:1997fb}.
The SM prediction exhibits large uncertainties due to the sign ambiguity of the interference between the LD and SD terms contributing to the decay amplitude.
The sign ambiguity leads to two SM predictions:
\begin{equation}
\text{[LD$+$]:}~~\left(6.82^{+0.77}_{-0.24}\pm0.04\right)\times10^{-9},~~~\text{[LD$-$]:}~~\left(8.04^{+1.66}_{-0.97}\pm0.04\right)\times10^{-9}.
\end{equation}
The first uncertainty comes from the calculation of the LD terms, dominates the overall precision, and makes the uncertainty quite asymmetrical.
This asymmetry is reflected in the computation of the $K_L\to\mu^+\mu^-$ rate with the inclusion of NP.
This decay has been studied experimentally and the measured branching ratio is $(6.84\pm0.11)\times10^{-9}$~\cite{E871:2000wvm}.
The sign ambiguity and the LD uncertainties prevent a clean theoretical interpretation of this result. Theoretical efforts are foreseen in the next few years to resolve the sign ambiguity through an appropriate matching between LD and SD contributions and to reduce the LD uncertainty by at least a factor of two. 
This could allow a sizeable reduction of the uncertainty of the SM prediction, opening the possibility to exploit the high sensitivity of this decay to BSM physics models.

The $K_S\to\mu^+\mu^-$ decay provides a complementary, theoretically clean observable~\cite{DAmbrosio:2017klp,Dery:2021mct}. While the total $K\to\mu^+\mu^-$ rate is dominated by long-distance (LD) physics, the measurement (in a dedicated high-intensity $K_S$ experiment) of CP violation in the interference of mixing and decay via the time-dependent rate enables the extraction of the purely CP odd short-distance amplitude, which is predicted within the SM with an ${\cal O}(1\%)$ uncertainty.

The study of the $K_S\to\mu^+\mu^-$ decay is the main goal of the kaon physics programme of the LHCb experiment. The high production rate of kaons in proton-proton collisions at 13~TeV partly compensates the small LHCb acceptance for kaon decays due to the long lifetime. LHCb has recently reported an upper limit on the $K_S\to\mu^+\mu^-$ branching fraction using the combined Run 1+2 dataset, ${\cal B}(K_S\to\mu^+\mu^-) < 2.1\times 10^{-10}$ 90\% CL~\cite{LHCb:2020ycd}, to be compared to the SM prediction, ${\cal B}_{\rm SM}(K_S\to\mu^+\mu^-) = (5.18\pm1.50)\times 10^{-12}$~\cite{Ecker:1991ru, DAmbrosio:2017klp}. The ultimate expected LHCb sensitivity for this decay is close to the SM branching ratio and is expected to be limited by the statistical uncertainty on the background subtraction and the signal yield~\cite{AlvesJunior:2018ldo}.

%


\subsubsection{Global sensitivity to New Physics}
\label{sec:global_fit}

\begin{figure}
\begin{center}
\resizebox{1.0\textwidth}{!}{\includegraphics{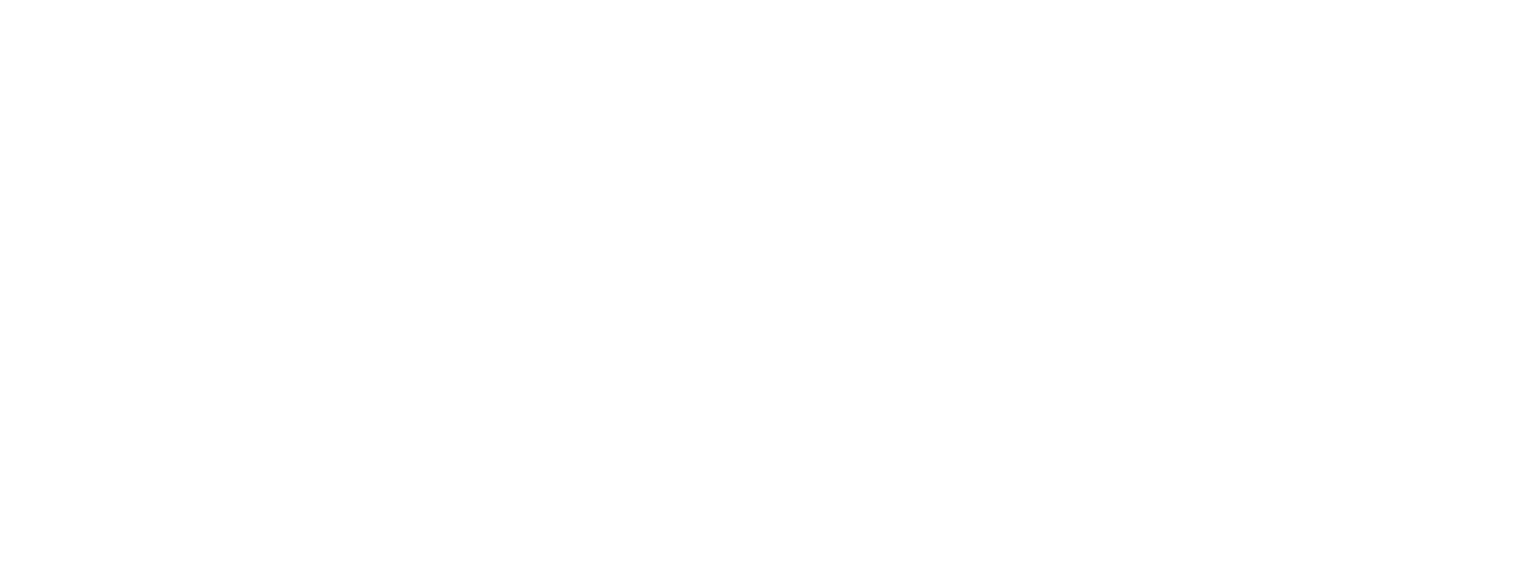}}
\vspace{-11mm}
\end{center}
\caption{Global fit to current data (purple, labelled as ``current fit'' in the legend) and the results of two projections discussed in Ref.~\cite{DAmbrosio:2022kvb}: A, light-red, where central values of observables currently only constrained with upper bounds are set to the SM predictions; B, dark-red, where central values for all of the observables are best-fit projections from existing data. The SM and best-fit point with the current data are indicated with black and purple crosses, respectively. The two figures represent the two possible signs of the long-distance contribution to the $K_L\to\mu^+\mu^-$ process; the sign ambiguity may be resolved with future data.}
\label{fig:cfitk}
\end{figure}

The rare decays discussed above offer the possibility to look for NP using a global fit technique. As an example, Fig.~\ref{fig:cfitk} shows the result of this technique using a low energy effective theory approach~\cite{DAmbrosio:2022kvb}. 
In this approximation, the physics is described in terms of an effective Hamiltonian, which is an expansion series of fermion current operators.
The coefficient of this expansion is the Wilson coefficient $C_k$, where $k$ is an index running over the expansion terms.
The assumption is that NP modifies $C_k$ with respect to the SM values.
The Wilson coefficients are then fitted using experimental values of observables from rare kaon decays as input variables.
The results shown in Fig.~\ref{fig:cfitk} assume NP scenarios compatible with the LFUV anomalies observed in $B$ decays.

\subsection{Lepton flavour universality tests}
\label{sec:leptonic-kaon-decays}

Lepton flavour universality (LFU) is a cornerstone of the SM postulating that the lepton coupling to gauge bosons is independent of lepton type (``flavour''), in contrast with the flavour-dependence of quark interactions. The origin of the observed LFU, which is not associated with any known symmetry and is thus not a fundamental conservation law, is a major question in modern physics.
Recently observed tensions between the SM predictions and experimental results such as the measurement of the anomalous magnetic moment of the muon~\cite{Muong-2:2021ojo}, the Cabibbo angle anomaly~\cite{Coutinho:2019aiy}, and the $B$-meson anomalies~\cite{LHCb:2017avl,LHCb:2019hip,LHCb:2021trn},
suggest possible LFU violation and motivate further the quest for improved experimental LFU measurements.

The kaon sector offers opportunities for precision lepton flavour universality tests by measuring the ratio ${\cal B}(K^+\to\pi^+e^+e^-)/{\cal B}(K^+\to\pi^+\mu^+\mu^-)$ as discussed in the framework of a global fit in a low-energy effective theory approach in Sections~\ref{sec:Kplus_pill} and~\ref{sec:global_fit}, as well as ${\cal B}(K^+\to e^+\nu)/{\cal B}(K^+\to\mu^+\nu)$ and ${\cal B}(K\to\pi e\nu)/{\cal B}(K\to\pi\mu\nu)$, exploiting cancellation of hadronic effects~\cite{Bryman:2021teu,Crivellin:2016vjc}.

The ratio of purely leptonic decay rates of the charged kaon $R_K = \Gamma(K^+ \to e^+ \nu)/\Gamma(K^+ \to \mu^+ \nu)$ is strongly suppressed in the SM by conservation of angular momentum and an extremely sensitive probe of~LFU.
Its SM expectation $R^{\rm SM}_{K} = (2.477 \pm 0.001) \times 10^{-5}$~\cite{Cirigliano:2007xi} is known to excellent (0.4\textperthousand) precision and the currently most precise (4\textperthousand) experimental result is $R_K = (2.488 \pm 0.007_{\rm stat} \pm 0.007_{\rm syst}) \times 10^{-5}$~\cite{NA62:2012lny}. The ratio $R_K$ is highly sensitive to the possible violation of LFU naturally arising in new physics scenarios involving sterile neutrinos~\cite{Abada:2012mc, Bryman:2019bjg}, leptoquarks~\cite{Crivellin:2021egp}, massive gauge bosons~\cite{Capdevila:2020rrl}, or an extended Higgs sector~\cite{Girrbach:2012km, Fonseca:2012kr}.
Variations of $R_K$ up to a few per mille from its SM expectation are predicted by these models, without contradicting any present experimental constraints.

An analogous LFU test is based on the ratio of semileptonic kaon decay rates $R_K^{(\pi)} = \Gamma(K \to \pi e \nu)/\Gamma(K \to \pi \mu \nu)$, where both neutral and charged kaon decays ($K_L \to \pi^\pm \ell^\mp\nu$ and $K^+ \to \pi^0\ell^+\nu$) can be used.
For a given neutral or charged initial state kaon, the Fermi constant,
$V_{us}$, short-distance radiative corrections, and the hadronic form factor at zero momentum transfer cancel out when taking the ratio $R_K^{(\pi)}$~\cite{Bryman:2021teu}. Therefore, in the SM this ratio is entirely determined by phase space factors and long-distance radiative corrections~\cite{Cirigliano:2001mk, Cirigliano:2004pv, Cirigliano:2008wn, Seng:2021boy,Seng:2022wcw}.


%
%



\subsection{Lepton flavour and lepton number violation}
\label{sec:lfv}

In the SM, lepton number and lepton flavour are conserved due to accidental global symmetries. Additionally, the SM requires neutrinos to be strictly massless due to the absence of right-handed chiral states. However the discovery of neutrino oscillations has demonstrated non-zero neutrino mass and non-conservation of lepton flavour in the neutrino sector. On the other hand, no evidence has yet been found for lepton flavour violation (LFV) in the charged lepton sector, or lepton number violating (LNV) processes. Any such observation would be a clear indication of NP. 

Mixing of active neutrinos can, in principle, mediate LFV decays in the SM. However the branching ratios are vanishingly small, and therefore any observation of LFV kaon decays would be clear evidence of new physics. Such LFV kaon decays are predicted in BSM scenarios with light pseudoscalar bosons (ALPs)~\cite{Cornella:2019uxs}, a $Z^\prime$ boson~\cite{Landsberg:2004sq,Langacker:2008yv} or leptoquarks~\cite{Pati:1974yy,Bordone:2018nbg}.
As discussed in Section~\ref{sec:ExperimentalStudiesOfBSMPhysics}, several hints of violations of lepton flavour universality (LFUV) have been reported. Models which predict LFUV and explain these potential anomalies in general naturally predict LFV processes~\cite{Glashow:2014iga,Calibbi:2015kma,Greljo:2015mma}. Therefore search for LFV processes are of strong interest.

In BSM scenarios, LNV kaon decays are predicted for example in models with Majorana neutrinos. In the minimal Type-I seesaw model~\cite{Minkowski:1977sc,Gell-Mann:1979vob,Mohapatra:1979ia}, the neutrino is a Majorana fermion with a mass term that violates lepton number by two units.
Non-zero neutrino masses make it possible, in principle, to distinguish experimentally between the possible Dirac and Majorana natures of the neutrino. Strong evidence for the Majorana nature of the neutrino would be provided by the observation of lepton number violating (LNV) processes, including kaon decays.


The recent results of LFV/LNV searches with the NA62 Run~1 (2016--2018) dataset~\cite{NA62:2019eax,NA62:2021zxl,NA62:2022tte} include upper limits on the $K^+\to\pi^-\ell^+_1\ell^+_2$ decay branching ratios at the ${\cal O}(10^{-11})$ level, already leading to stringent constraints on active-sterile mixing angles between Majorana neutrinos. Below the kaon mass, these constraints are competitive with those obtained from neutrinoless double beta decay~\cite{Littenberg:2000fg,Atre:2005eb,Atre:2009rg,Abada:2017jjx}. The NA62 searches are not limited by backgrounds, and indicate that HIKE Phase~1  would make significant improvements in sensitivity to LFV and LNV $K^+$ decays, extending the limits to the ${\cal O}(10^{-12})$ level. On the other hand, the existing limit on the LFV $K_L\to\mu^\pm e^\mp$ decay~\cite{BNL:1998apv}, which represents a stringent constraint on several BSM scenarios, would be improved significantly by the HIKE Phase~2; further details are provided in Section~\ref{sec:phase2:sensitivity}.



\subsection{Tests of low-energy QCD}
\label{sec:physics-low-energy-qcd}

Most kaon decays are governed by physics at long distances.
These decays can be described by chiral perturbation theory (ChPT), which is the low-energy effective field theory of QCD.
The ChPT framework determines the kaon decay amplitudes in terms of the so-called low-energy constants, which are determined using experimental data. Therefore, measurements of kaon decay rates and form factor parameters of various decay channels represent essential tests of the ChPT predictions and crucial inputs to the theory at the same time. A comprehensive overview of kaon decays and their relation to the ChPT  can be found in Ref.~\cite{Cirigliano:2011ny}.

The HIKE dataset will provide a unique opportunity to perform precision measurements of rare and radiative decays including $K\to\pi\gamma\gamma$, $K\to\ell\nu\gamma$, $K\to\pi\ell\nu\gamma$, $K\to\pi\pi\gamma$, $K\to\pi\pi\pi\gamma$,  $K\to\pi\pi e^+e^-$, and $K\to\pi\pi\ell\nu$ of both $K^+$ and $K_L$ mesons.
Possible studies of the abundant decay modes $K\to\pi\pi$ and $K\to\pi\pi\pi$ will also provide important inputs to future ChPT parameter fits.


\subsection{Test of first-row CKM unitarity}
\label{sec:semileptonic-kaon-decays}

Measurements of semileptonic kaon decays $K\to\pi\ell\nu$ provide the principal input for the extraction of the CKM parameter $V_{us}$, while the ratios of (semi)leptonic $K^+$ and $\pi^+$ decay rates are used to extract the ratio $V_{us}/V_{ud}$, with inputs provided from lattice QCD~\cite{Aoki:2021kgd}. Determination of $V_{us}$ from kaon, pion, and $\tau$ decays, combined with $V_{ud}$ measurement from super-allowed beta decays~\cite{Hardy:2020qwl} and neutron decays~\cite{Czarnecki:2018okw,Seng:2020wjq}, gives rise to a $3\sigma$ deficit in first-row CKM unitarity relation known as the Cabibbo angle anomaly; a tension of similar significance is observed between $K\to\ell\nu$ and $K\to\pi\ell\nu$ rates~\cite{Cirigliano:2022yyo,Bryman:2021teu}.
The uncertainty in $V_{us}$ comes in equal parts from the experimental errors and theoretical uncertainties in the ratio of decay constants, $f_K/f_\pi$, and the $K\to\pi\ell\nu$ form factor, $f_+(0)$. Substantial improvements in the lattice QCD calculations of these hadronic factors are expected in the next five years, thanks to decreased lattice spacing and accurate evaluation of
electromagnetic effects~\cite{Carrasco:2015xwa, Boyle:2019rdx,Feng:2021zek,Seng:2020jtz}. Significant progress on the calculation of radiative corrections has been achieved recently, reducing the uncertainties on the long-distance electromagnetic corrections to a negligible level~\cite{Seng:2019lxf,Seng:2021boy,Seng:2021wcf,Seng:2022wcw}.
In order to shed light on the Cabibbo angle anomaly, improved measurements of the principal $K^+$ and $K_{L,S}$ branching ratios are essential (and it should be noted that no $K_L$ decay measurements have been made in the past decade). In particular a precision measurement of the ratio of $K^+\to\pi^0\mu^+\nu$ and $K^+\to\mu^+\nu$ rates is well-motivated~\cite{Cirigliano:2022yyo}. In case the Cabibbo angle anomaly persists, HIKE would be able to make precision measurements of (semi)leptonic $K^+$ and $K_L$ decays based on special datasets collected with minimum-bias triggers at low beam intensity. 

It should be noted that the Belle~II experiment is planning to extract $V_{us}$ at an improved precision from a suite of inclusive and exclusive measurements of $\tau$ decays, including ${\cal B}(\tau^-\to K^-\nu)/{\cal B}(\tau^-\to\pi^-\nu)$, combined with theory improvements~\cite{Belle-II:2022cgf}. This would provide a possibility to cross-check the kaon results.


\subsection{Production of feebly-interacting particles in kaon decays}
\label{sec:fips}

An alternative paradigm to the searches for NP at high energy is the search for feebly-interacting particles (FIPs), which would be the manifestations of weakly-interacting NP potentially at much lower energy scales.
Interactions of FIPs with SM fields are possible through BSM extensions involving low-dimension operators, the so-called portals~\cite{Beacham:2019nyx,Agrawal:2021dbo}, the simplest of which are summarised in Table~\ref{tab:PortalsTable}. For further discussion of the benchmark scenarios, see also Section~\ref{sec:DumpSensitivity}.

\begin{table}[b]
\centering
\caption{Summary of generic portal models with FIPs~\cite{Beacham:2019nyx,Agrawal:2021dbo}.}
\vspace{-3mm}
\begin{tabular}{c|c}
\hline
Portal & FIPs \\ 
\hline
Vector & Dark photon, $A^{\prime}$ \\ 
Scalar & Dark Higgs/scalar, $S$ \\ 
Fermion & Heavy neutral lepton (HNL), $N$ \\ 
Pseudoscalar & Axion/axion-like particle (ALP), $a$ \\ 
\hline
\end{tabular}
\vspace{-3mm}
\label{tab:PortalsTable}
\end{table}

Due to the availability of large datasets and the suppression of the total kaon decay width, kaon decays represent uniquely sensitive probes of light hidden sectors via FIP production searches. The possible search strategies have been reviewed recently~\cite{Goudzovski:2022vbt}, and the following have been identified as the most promising. Searches for the $K\to\pi X_{\rm inv}$ decay by extension of the $K\to\pi\nu\bar\nu$ measurements, where $X_{\rm inv}$ represents an invisible particle, represent a unique probe into the dark-scalar and ALP parameter space. Searches for heavy neutral lepton ($N$) production in $K^+\to\ell^+N$ decays are approaching the seesaw neutrino mass models with ${\cal O}(100~{\rm MeV})$ sterile neutrinos~\cite{Abdullahi:2022jlv}. Searches for resonances in the $K\to\pi\ell^+\ell^-$ and $K\to\pi\gamma\gamma$ decay spectra are complementary to searches at beam-dump experiments for a significant ALP mass range. 
Finally, searches for a leptonic force mediator ($X$) in $K^+\to\mu^+\nu X$ decays can probe a region of parameter space providing an explanation for the muon $g-2$ anomaly~\cite{Krnjaic:2019rsv}.

Experimentally, the NA62 Run~1 dataset has been used to establish the analysis methods for searches for dark-scalar and HNL production in $K^+$ decays, producing world-leading limits on the dark-scalar coupling~\cite{NA62:2021zjw,NA62:2020xlg,NA62:2020pwi} and HNL mixing parameters~\cite{NA62:2020mcv,NA62:2021bji} below the kaon mass. Further details on the experimental aspects and HIKE sensitivity projections are provided in Sections~\ref{sec:KpiXSensitivity} and~\ref{sec:K-HNL-Sensitivity}.

%% file: fips.tex
\section{Physics in the beam-dump mode}
\label{sec:DumpPhysics}

Thanks to the high-intensity beam and high-performance detectors (redundant particle identification capability, highly efficient veto systems and high resolution  measurements of momentum, time, and energy), HIKE can achieve the sensitivities required for competitive searches for production and decay of long-lived light mediators in a variety of new-physics scenarios.

A number of possible Standard Model (SM) extensions aimed at explaining the abundance of dark matter in our universe predict a new ``hidden sector'' with mediator fields or matter fields in the MeV--GeV mass range. The realisations of such scenarios are usually classified in terms of the Lagrangian operators (the so-called ``portals'' listed in Table~\ref{tab:PortalsTable}) connecting SM particles to the new mediators. The related coupling constants are small, thus evading the existing exclusion bounds and justifying the label of ``feebly interacting'' given to these new-physics models~\cite{Beacham:2019nyx,Agrawal:2021dbo}. The mass of the mediator and hidden-sector matter fields and the couplings are free parameters of the models.
The relevant features of the phenomenology of such models are:
\begin{itemize}
\item The mediators can be produced in proton-nucleus interactions through a number of mechanisms, the most important being direct Primakoff production and meson-mediated tertiary production. The specific production mechanism for each mediator type differs in terms of production cross section and momentum-angle spectra of the mediator emitted.
\item At the 400-GeV energy of SPS protons, mediators with momenta above 10~GeV have decay lengths ranging from tens of meters to tens of kilometres in a suitable, interesting interval of the feeble coupling.
\item Due to the feeble interaction with the SM particles, the emitted mediators can reach the decay volume after punching through tens of meters of traversed material before decaying.
\end{itemize}

As demonstrated by the NA62 Run~2 operation~\cite{MinucciICHEP2022}, the intense P42 400~GeV proton beam extracted from the CERN SPS and the HIKE setup can be exploited to search for production and decay of the emitted mediators. For this purpose, HIKE can be operated in the so-called beam-dump mode. 
Considering that HIKE will have the same type of forward geometry as NA62, the HIKE beam-dump operation will be particularly sensitive to specific physics cases in which the hidden sector mediator is produced by Primakoff scattering or light-meson mediated decays. Prominent examples are the vector and some axion-like particle scenarios. We address a number of new physics scenarios in Section~\ref{sec:dump}, tentatively assuming a total integrated statistics of $10^{19}$~POT for HIKE Phase~1, and $5\times 10^{19}$~POT for the complete HIKE programme.

%% file: beam_mm.tex
\section{High-intensity beams}
\label{sec:beam}

As the acronym HIKE suggests, the beam line must provide significantly higher beam intensities than those that are now delivered to the NA62 experiment, on the order of 4 to 6 times of that currently provided, which would correspond to between 1.2 and $2 \times 10^{13}$ protons on the T10 target per SPS extraction. Here, we assume that a typical running year has 200~days with 3000 spills each, including an SPS uptime of about 80\%, and a flat top length of 4.8~seconds.

The proposed HIKE program includes both a charged beam ($K^{+}$) phase and a neutral beam ($K_{L}$) phase. For both parts of the program, protons at much higher intensity have to be transported via the P42 beam line from the T4 target to the kaon production target T10. Conceptually, the $K^+$ beam is proposed to be the same as the present beam line, but consolidation and upgrades are needed to handle the higher beam intensities. Improved beam instrumentation will keep the beam losses better under control. A more precise calibration of the primary beam intensity measurements by the secondary emission monitors at T4 and T10 is important for beam optimisation, and  
for normalisation purposes in dump mode. Efforts in this respect have already been started by the BI and EA groups at CERN. The neutral beamline is an entirely new design and requires minor local modifications to the last section of the P42 line.


\subsection{Beam delivery to the kaon production target}
\label{sec:proton_beam}

The P42 beam line started operation in 1980 and hardly has been modified since then~\cite[Chapter 5]{Banerjee:2774716}. Most of the equipment dates from then and is showing strong effects of ageing. Many issues and suggestions for mitigation have been described in the report from the PBC Conventional Beams working group~\cite{Gatignon:2650989}, which has recently been updated. The first phase of a vast consolidation program for the CERN North Area, titled NACONS, has recently been approved and funded, and the second phase has been prepared~\cite{Kadi:2018, Kadi:2019, Kadi:2021}. This will allow restoration of the equipment and beam infrastructure to a reliable state. Some upgrades are required to cope with the up to 6 times higher beam intensity, including operational requirements from the rest of the North Area complex and with stricter safety and radiation protection requirements compared to the original installation. Many studies have been performed by the SPS Losses and Activation working group (SLAWG)~\cite{Balhan:2668989, Fraser:ipac17-mopik045, Fraser:ipac19-wepmp031}, the Physics Beyond Collider (PBC) Proton Delivery working group~\cite{KoukoviniPlatia:2675225}, the PBC Conventional Beams working group~\cite{Gatignon:2650989}, and more recently, by the PBC ECN3 beam delivery task force~\cite{Brugger:2022}, focusing on the possibility to extract and deliver such high intensities towards TCC8 and ECN3.

\begin{figure}[tb]
\centering
\includegraphics[width=0.62\textwidth]{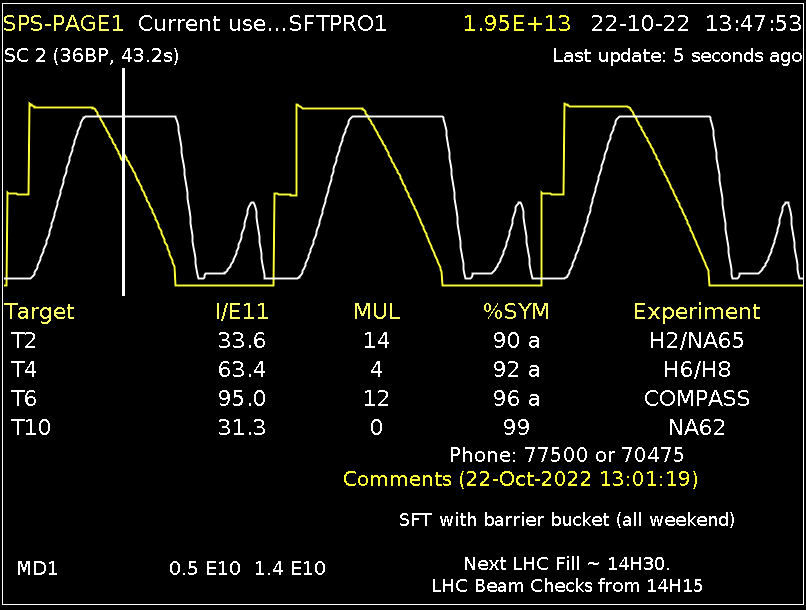}
\caption{The spill configuration used on 22 October 2022, as seen from SPS Page~1.}
\label{fig:screenshot}
\end{figure}

\begin{figure}[h]
\centering
\includegraphics[width=\textwidth,height=8.5cm]{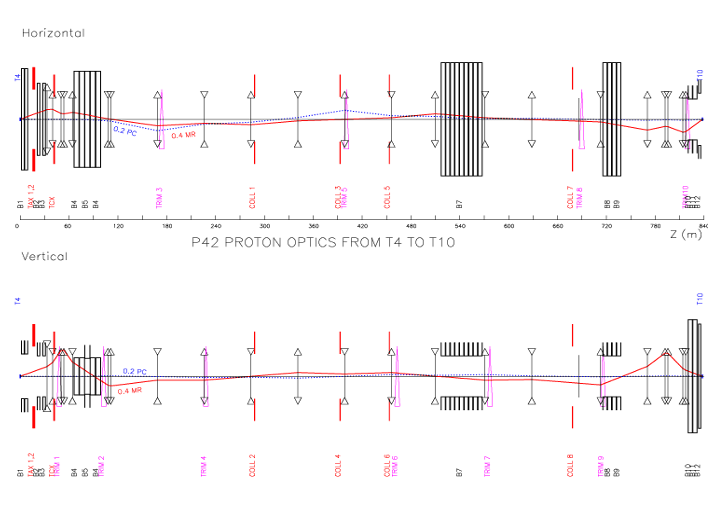}
\vspace{-11mm}
\caption{Beam optics for the P42 primary beam line, delivering protons from the T4 production target to the T10 kaon production target. The red line depicts the local values of the R12 (horizontal) and R34 (vertical) beam transport matrix terms. The dotted blue line shows the local dispersion terms R16 (horizontal) and R36 (vertical).}
\label{fig:P42optics}
\end{figure}

The present optics of the P42 beam is shown in Fig.~\ref{fig:P42optics}. In order to reduce the attenuation of the primary beam in the T4 target, several options are considered. One is to increase the vertical beam size so that most of the beam bypasses the target without interacting. Another option is to run dedicated cycles for ECN3 in which the beam is bumped to pass above or below the target plate. This option could allow in principle even higher intensities than $2 \times 10^{13}$ ppp on the T10 target.
At present, the beam is provided in spills of 4.8~seconds, typically twice per super-cycle of about 40 to 45~seconds during day time, and two equally-spaced spills during 33.6~seconds during nights. One spill every 14.4~seconds is also technically possible and will be the standard cycle adopted during the night from 2023 (Fig.~\ref{fig:screenshot}). 

In recent years, on average 3000~spills per day are delivered, including the typical availability of SPS beams of about 80\%. The figure of merit is the time on flat top per year, which is proportional to the duty cycle and the running time. It is also important that the beam intensity is delivered as uniformly as possible. At the time of the NA62 proposal one assumed an effective spill of 3.3~seconds, but thanks to recent improvements we anticipate that it can be significantly better, ideally better than 4~seconds for a 4.8~second spill. The spill length can be modified to some extent, but it is important to maintain the duty cycle as well as the instantaneous intensity, respecting both the RMS power limitation of the SPS and transfer lines as well as the maximum intensity that the SPS is able to accelerate. The latter implies that the intensity per spill shall be kept proportional to the spill length.

Studies by the SY-STI group in the PBC context suggest that the TAXs in P42 and K12 are already running close to the maximum intensity allowed~\cite{Gatignon:2650989}. An upgrade is being studied at the technical level in the framework of the ECN3 task force. This implies a substantial upgrade of the T10 target-TAX complex to withstand the higher beam power and to be compatible with RP and maintenance requirements. A new target design could be inspired by the concepts of the CNGS target.
The target material might be changed from beryllium to carbon for better handling. However, also more dense target materials can be considered as an option outside of the present baseline, as they could be advantageous in particular for the neutral beamline. The new K12 TAX will 
be based on the concept that is developed for the P42 TAX. TAX holes will be defined to ensure compatibility with the charged beam as well as the neutral beam. It will probably require a different choice of material in the TAX blocks and cooling closer to the beam impact point.
The present interlocks, P0-Survey and Dump Control, protect both the beam line equipment and the NA62 experiment against incorrect magnet currents, T10 target and K12 TAX cooling problems and closed vacuum valves in the K12 beam line. For the higher intensities in the future, this system must be sped up. This is planned within the NACONS program baseline, independently of the proposed high intensity upgrade for ECN3, profiting from a new generation of power converters. On the shorter time scale, some improvements are already under consideration, compatible with the presented HIKE timeline.  
Many studies are on-going concerning radiation dose levels to equipment and on surface, which must remain under control also at the higher intensities. All these issues will be addressed in the forthcoming report from the ECN3 beam delivery task force.

\begin{figure}[h]
\centering
\vspace{-8mm}
\includegraphics[width=\textwidth, trim={3cm 0.1cm 0 3cm},clip]{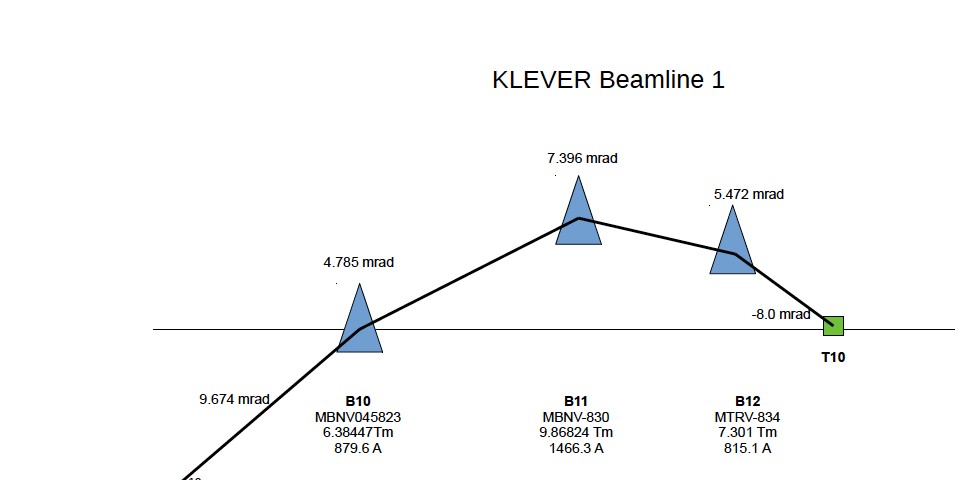}
\caption{Modifications of the T10 target steering dipoles for increasing the production angle to 8~mrad, as required for the KLEVER phase of the HIKE programme.}
\vspace{-10mm}
\label{fig:prodangleklever}
\end{figure}

For a $K_L\to\pi^0\nu\bar\nu$ experiment, the production angle of the kaon beam has to be increased to 8~mrad instead of the zero production angle used for the charged mode. This can be achieved by a minor modification of the P42 beam line, i.e. realigning the three last dipoles as documented in the Conventional Beams Working Group Report~\cite{Gatignon:2650989}, shown in Fig.~\ref{fig:prodangleklever}. 
The high-intensity proton beam will ultimately impinge on the kaon production target, followed by the TAX beam dumps.


\subsection{Charged kaon beamline}

\begin{figure}[h]
\centering
\vspace{-4mm}
\includegraphics[width=0.9\textwidth,height=10cm]{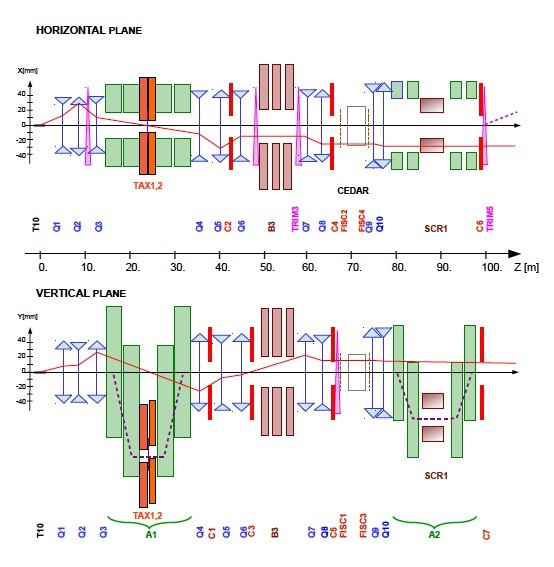}
\vspace{-5mm}
\caption{Beam optics for the K12 beamline. The red line depicts the local values of the R12 (horizontal) and R34 (vertical) beam transport matrix terms. The dotted black line shows the local dispersion terms R16 (horizontal) and R36 (vertical).}
\label{fig:K12optics}
\end{figure}

The HIKE charged kaon beamline will remain conceptually as it is now, with essentially the same beam properties apart from the beam intensity (Fig.~\ref{fig:K12optics}). Consolidation of some equipment is required in the framework of NACONS, such as the K12 machine protection system in synergy with the P42 interlock upgrade. The main upgrades concern the T10 production target and the K12 TAX, including the needed shielding. A pre-study was presented in the Conventional Beams Working Group report~\cite{Gatignon:2650989} and is being followed up further in the working group. 

The K12 beam is a mixed beam containing 6\% of kaons at 75~GeV/$c$. A study has been performed to evaluate if the beam can be enriched by RF separation. RF separation requires a much longer beamline, implying that more kaons decay before the fiducial volume. This decreases not only the kaon flux but also the initial kaon fraction in the beam before enrichment. An RF-separated beam cannot fulfil the HIKE requirements, even with state-of-the-art RF systems~\cite{Appleby:2765940}.

\newpage


\subsection{Neutral kaon beams}
\label{sec:neutral_beam}

\begin{figure}[h]
\centering
\includegraphics[width=0.9\textwidth,height=6.5cm]{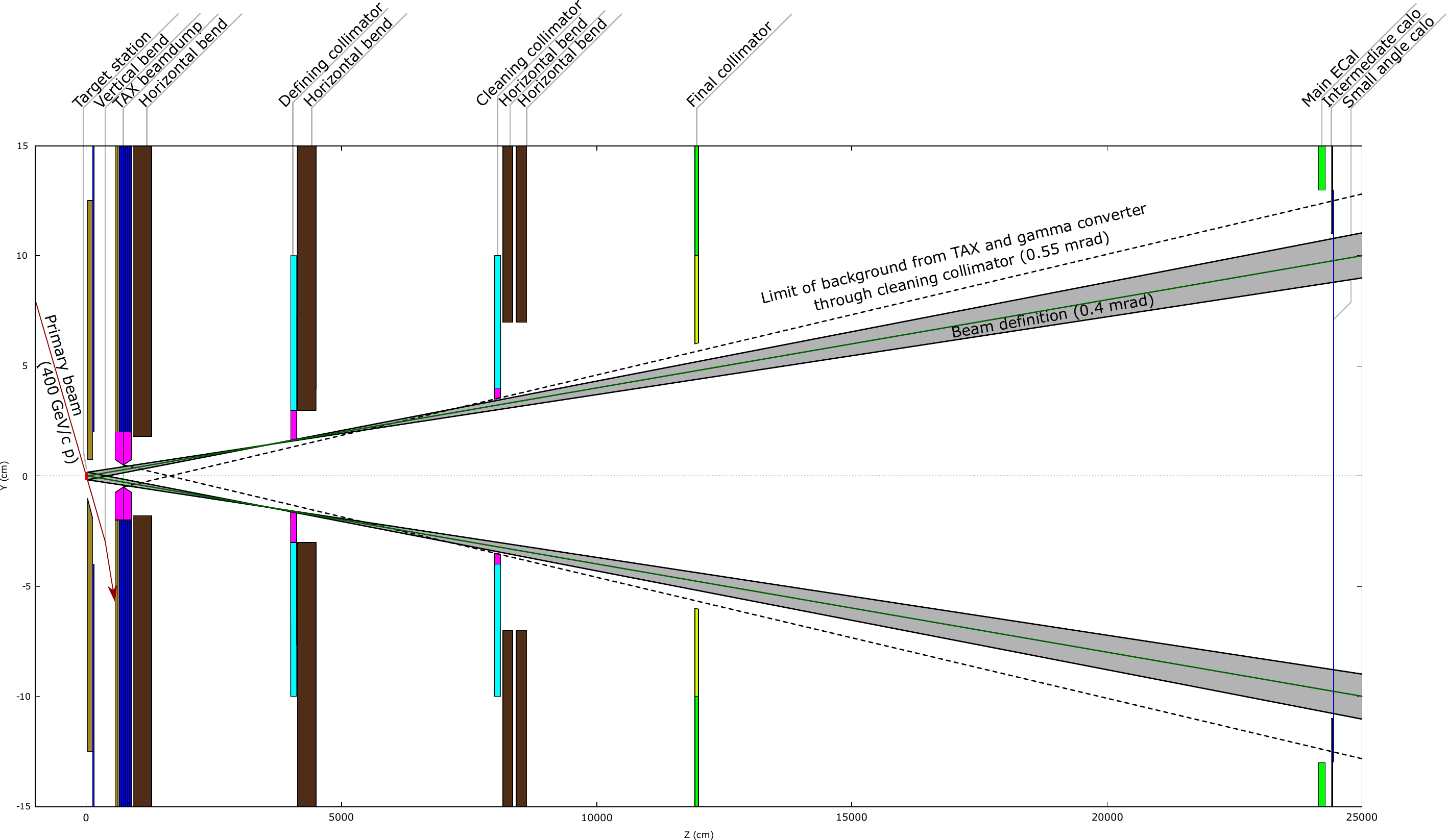}
\vspace{-1mm}
\caption{Baseline design of neutral beamline for HIKE Phases~2 and 3.}
\label{fig:klever_beam}
\vspace{-20mm}
\end{figure}

The original design and simulations for the KLEVER beamline were performed by the Conventional Beams Working Group~\cite{Gatignon:2650989}. The design is based on the experience with the $K_L$ beam lines for NA31 and NA48, following guidelines laid out in Ref.~\cite{Gatignon:2730780}, and is shown schematically in Fig.~\ref{fig:klever_beam}.
It forms the basis for the beamline to be used also for HIKE Phase~2, and will be extended in length for the precise measurement of $K_L\to\pi^0\nu\bar\nu$ (KLEVER).


\newpage

\begin{figure}[h]
\centering
\includegraphics[width=0.5\textwidth]{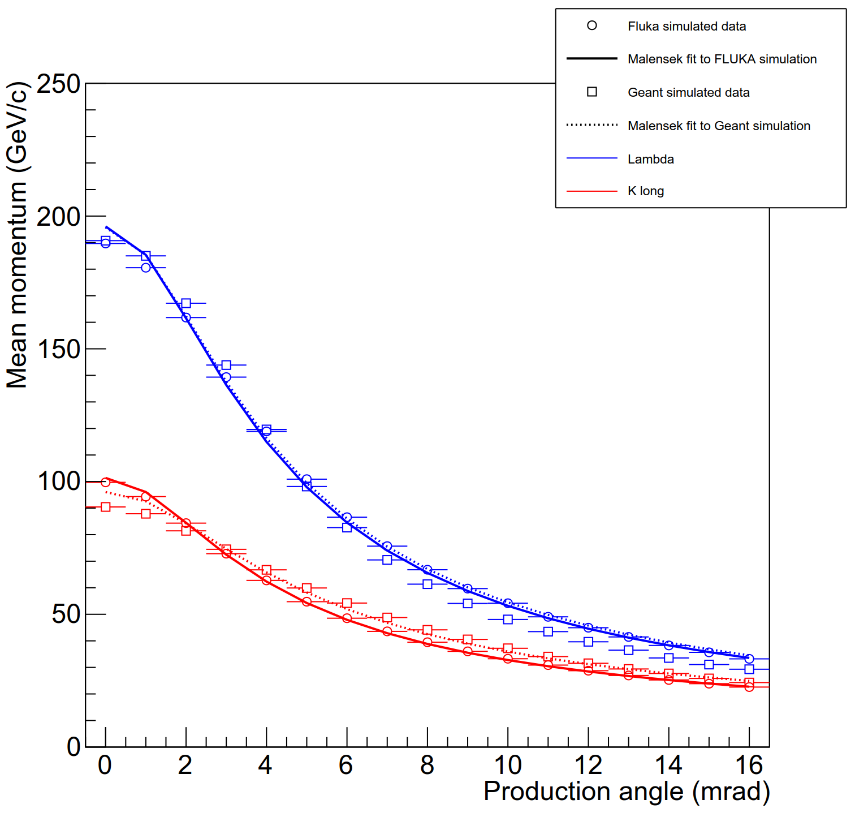}%
\includegraphics[width=0.5\textwidth]{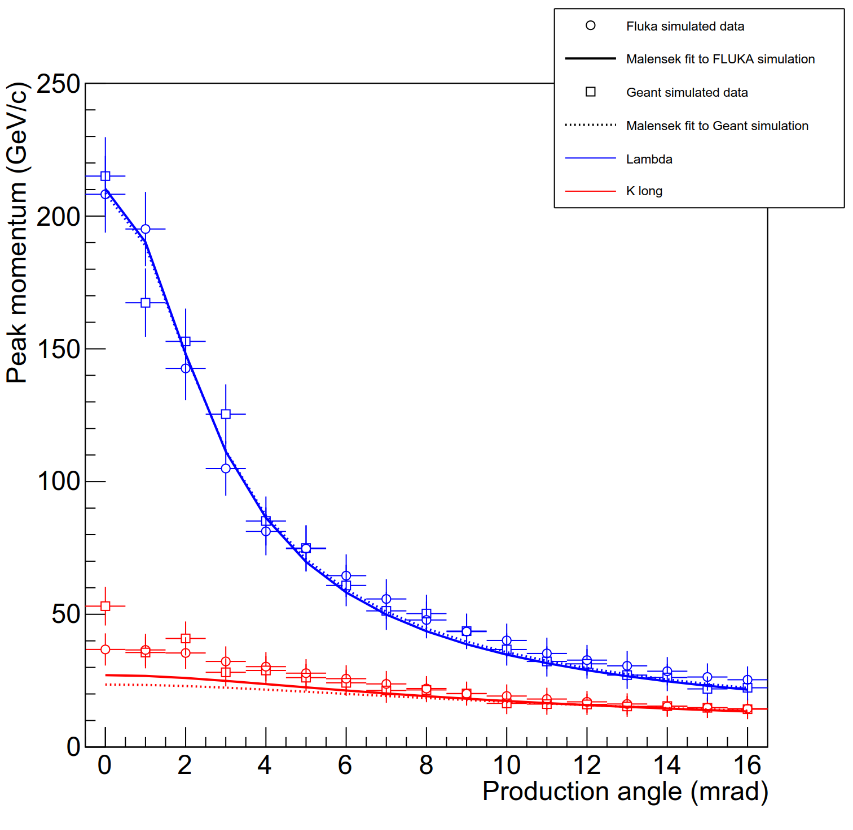}
\vspace{-3mm}
\caption{Mean (left) and peak (right) momentum for $K_L$ (red) and $\Lambda$ (blue) components vs beam production angle, from FLUKA (squares) and Geant4 (circles), with solid (dashed) curves from fits to the form in Ref.~\cite{Malensek:1981em}.}
\label{fig:beam_mom_ang}
\end{figure}
\begin{figure}[p]
\centering
\includegraphics[width=0.5\textwidth]{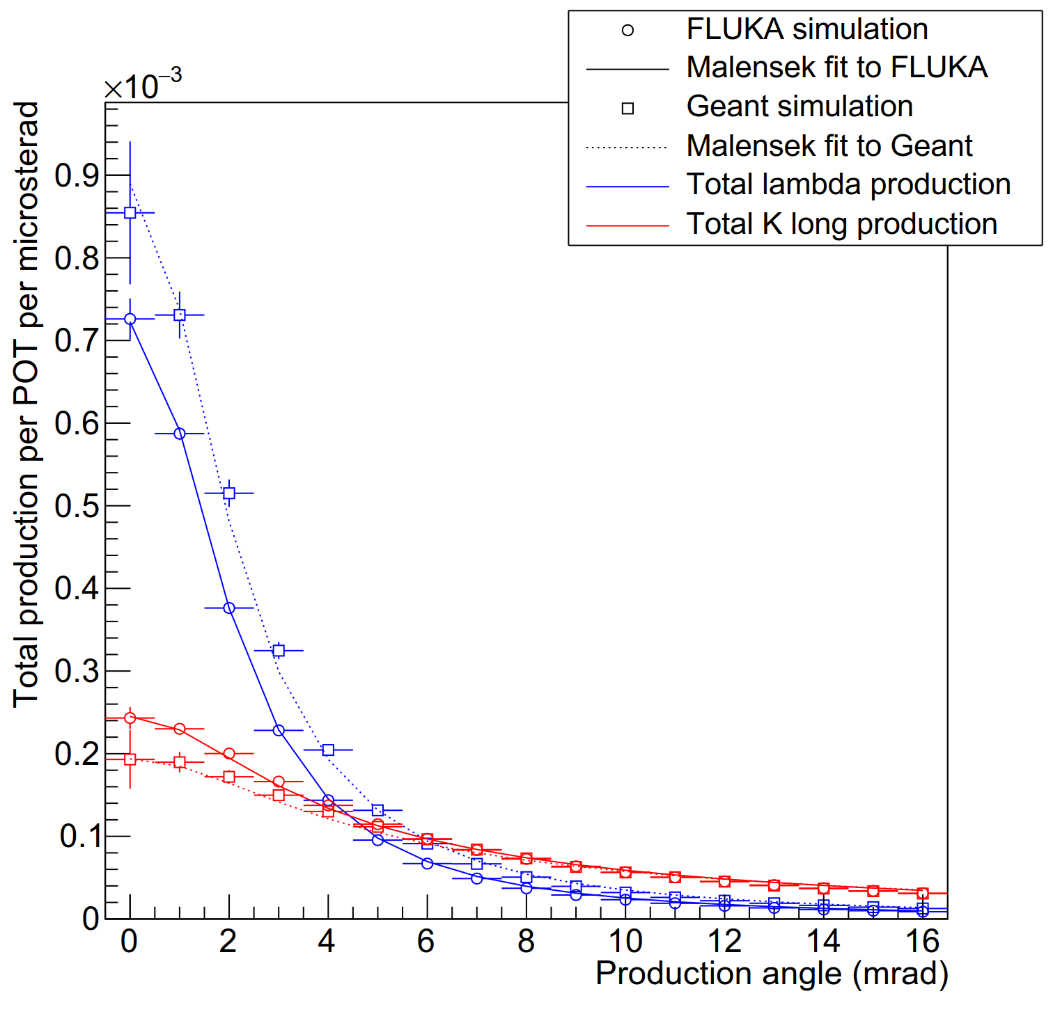}%
\includegraphics[width=0.5\textwidth]{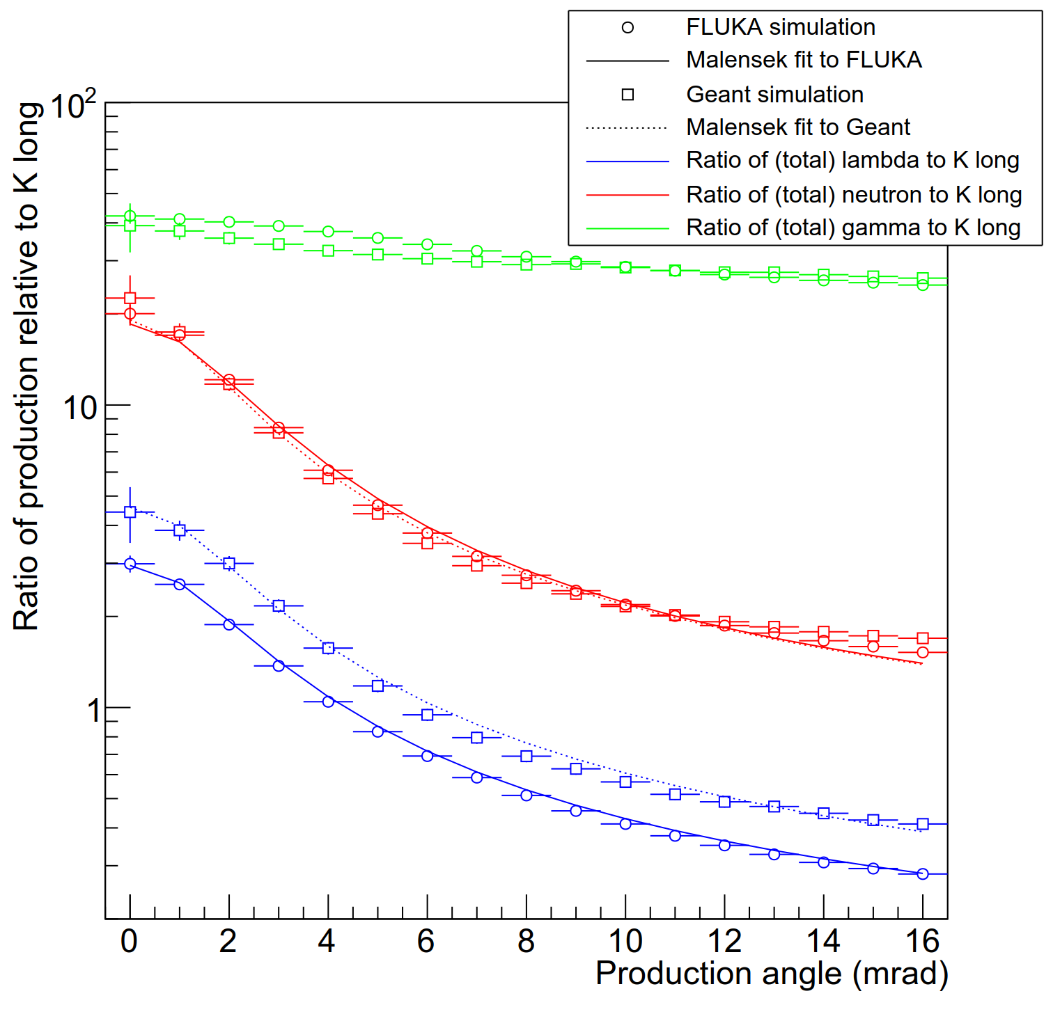}
\vspace{-4mm}
\caption{Left: neutral beam fluxes per POT per $\mu$sr of acceptance for $K_L$ (blue) and $\Lambda$ (red) vs beam production angle. Right: ratios of $\gamma$ (green), $n$ (red), and $\Lambda$ (blue) to $K_L$ fluxes vs beam production angle. FLUKA (Geant4) results are shown with squares (circles), with solid (dashed) curves from fits.}
\label{fig:beam_prod_ang}
\end{figure}
\begin{figure}[p]
\centering
\includegraphics[width=0.5\textwidth]{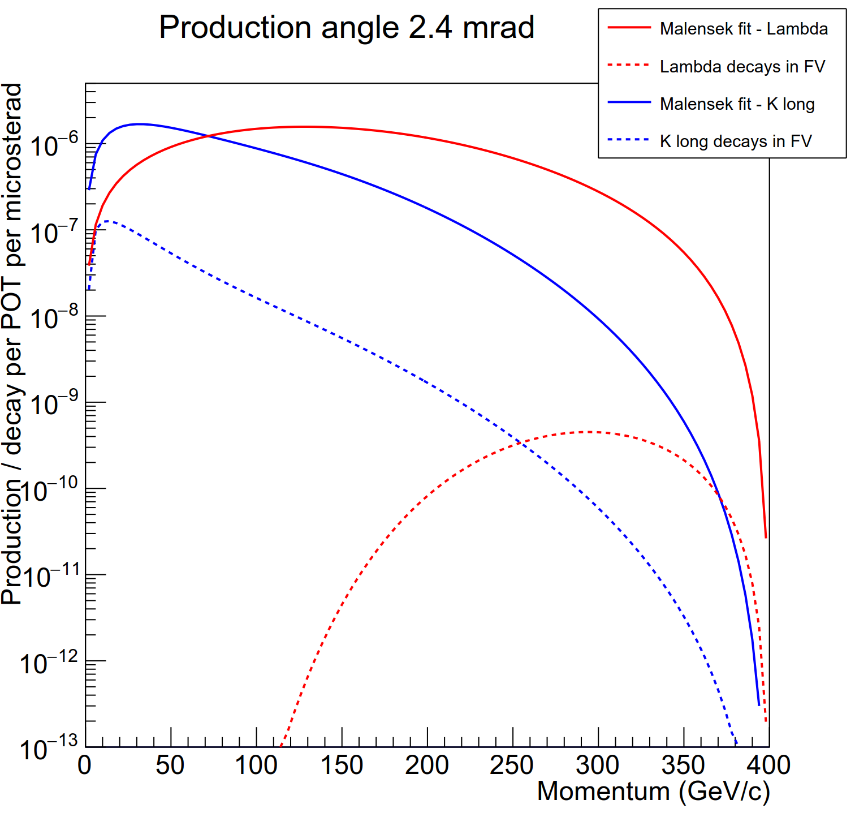}%
\includegraphics[width=0.5\textwidth]{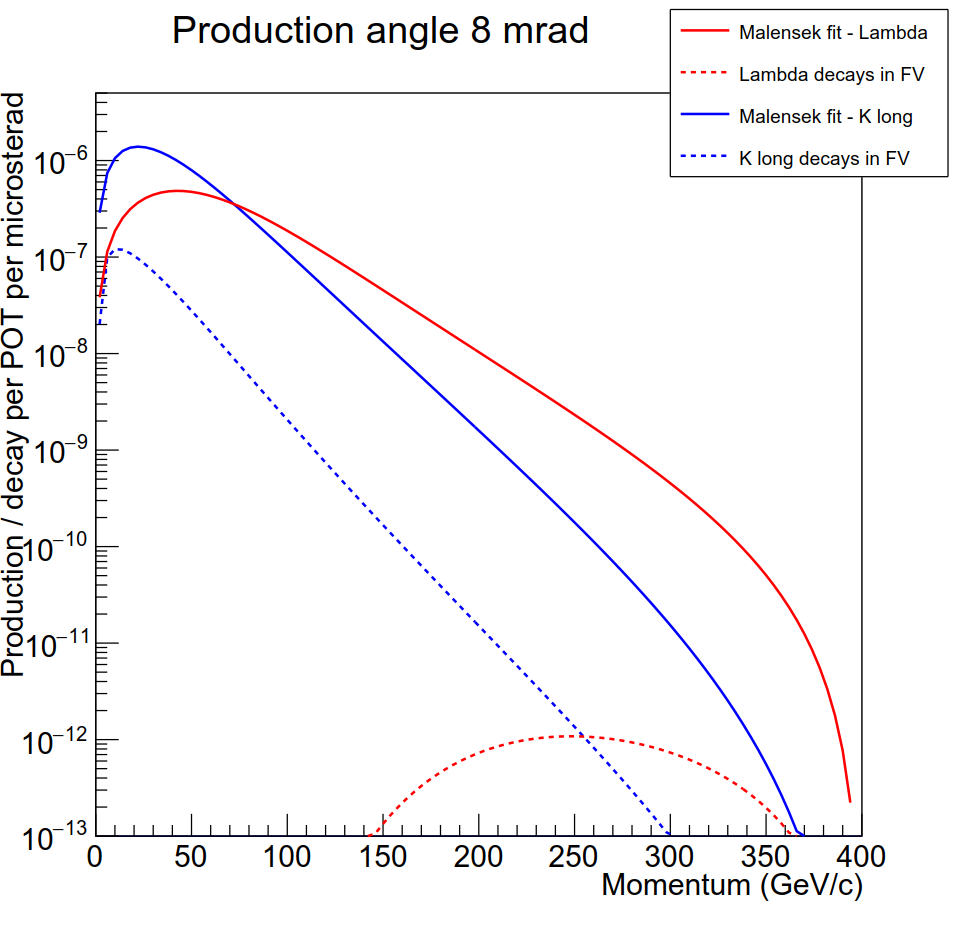}
\vspace{-4mm}
\caption{$K_L$ (blue) and $\Lambda$ (red) beam momentum spectrum (number of particles in beam per POT per $\mu$sr of neutral beam acceptance) for production angles of 2.4~mrad (left) and 8~mrad (right). The solid curves show a parameterisation~\cite{Malensek:1981em} of FLUKA results at production; dashed curves show the momentum spectra for particles decaying in an indicative fiducial volume of the experiment extending from 135~m to 195~m downstream of the target.}
\label{fig:beam_mom}
\end{figure}

The primary proton beam impinges on the T10 target with a downward vertical angle of 8~mrad for KLEVER, and possibly 2.4~mrad for the $K_L$ experiment with tracking. These choices balance several factors for the measurements of $K_L\to\pi^0\nu\bar\nu$ 
and $K_L\to\pi^0\ell^+\ell^-$, respectively.
As seen from Figs.~\ref{fig:beam_mom_ang} and~\ref{fig:beam_prod_ang}, a smaller production angle increases $K_L$ flux and hardens the $K_L$ spectrum, facilitating the task of vetoing background channels with extra photons, such as $K_L\to\pi^0\pi^0$. On the other hand, with harder input momentum spectra, a smaller $K_L$ fraction and a larger $\Lambda$ faction decay inside the fiducial volume. For the $K_L\to\pi^0\nu\bar\nu$ measurement,
$\Lambda\to n\pi^0$ decays constitute a potentially dangerous background. Neutron interactions on residual gas in the vacuum tank and from halo neutrons striking the detector constitute another dangerous background, which is reduced by going to larger angle. The production angle of 8~mrad softens the $\Lambda$ spectrum, so that almost all $\Lambda$ hyperons decay before the start of the fiducial volume. For the $K_L\to\pi^0\nu\bar\nu$ measurement, the reduced $K_{L}$ flux is compensated to some extent by an increase in the fraction of $K_L$ mesons that decay in the fiducial volume.
However, this is not true for $K_L\to\pi^0\ell^+\ell^-$, and moreover, background from $\Lambda$ decays and neutron interactions is not a concern in this case. A production angle as small as possible is therefore preferred for HIKE Phase~2, subject to limitations from the neutron flux in the beam at very small angles. A production angle of 2.4~mrad (the same as for the $K_L$ beam in NA48) is a likely choice. 
The $K_L$ and $\Lambda$ momentum spectra for production angles of 2.4 and 8~mrad are shown in Fig.~\ref{fig:beam_mom}, and the fluxes and other parameters are tabulated in Table~\ref{tab:beam_prod}.


\begin{table}[h]
\centering
\caption{Parameters of $K_L$, $n$, $\Lambda$ production at angles of 2.4 and 8.0~mrad, obtained from the FLUKA simulation of the beamline. An indicative fiducial volume (FV) extending from 135~m to 195~m from the production target is considered.}
\vspace{-2mm}
\begin{tabular}{lcc}
\hline
& 2.4 mrad & 8.0 mrad \\ \hline
Mean $K_L$ momentum (GeV/$c$) & & \\
\hspace{3ex}at production & 79 & 39 \\
\hspace{3ex}in FV & 46 & 26.4 \\
Mean $\Lambda$ momentum (GeV/$c$) & & \\
\hspace{3ex}at production & 151 & 66 \\
\hspace{3ex}in FV & 285 & 251 \\
$K_L$ rate ($10^{-6}$/pot/$\mu$sr) & & \\
\hspace{3ex}at production & 187 & 73 \\
\hspace{3ex}in FV & 6.8 & 4.1  \\
$\Lambda$ rate ($10^{-6}$/pot/$\mu$sr) & & \\
\hspace{3ex}at production & 310 & 37 \\
\hspace{3ex}in FV & 0.053 &  0.000132\\
\hline
\end{tabular}
\label{tab:beam_prod}
\end{table}


\begin{table}[h]
\centering
\caption{Particle fluxes in the neutral beam, obtained from the FLUKA simulation. The rates in MHz/GHz assume a primary beam intensity of $6.7\times10^{12}$ pot/s.}
\vspace{-2mm}
\small
\begin{tabular}{lcccccc}
\hline
\multicolumn{1}{c}{Beam component}  
& \multicolumn{2}{c}{Before converter} 
& \multicolumn{2}{c}{After converter} 
& \multicolumn{2}{c}{After final collimator} \\
& pot$^{-1}$ & GHz & pot$^{-1}$ & MHz & pot$^{-1}$ & MHz \\ \hline
$\gamma$ & & & & & & \\
\hphantom{2em} $E>1$~GeV & $4.2\times10^{-3}$ & 27.9
& $9.1\times10^{-4}$ & 6100 & $2.97\times10^{-5}$ & 198 \\
\hphantom{2em} $E>5$~GeV & $2.19\times10^{-3}$ & 14.6 
& $7.1\times10^{-5}$ & 470 & $7.9\times10^{-6}$ & 53 \\
\hphantom{2em} $E>10$~GeV & $1.55\times10^{-3}$ & 10.3 
& $1.81\times10^{-5}$ & 121 & $3.15\times10^{-6}$ & 21 \\
\hphantom{2em} $E>30$~GeV & $6.3\times10^{-4}$ & 4.2
& $2.06\times10^{-6}$ & 13.7 & $6.2\times10^{-7}$ & 4.1 \\
$n$ & & & & & & \\
\hphantom{2em} $E>1$~GeV & $4.3\times10^{-4}$ & 2.88
& $4.2\times10^{-4}$ & 2820 & $6.7\times10^{-5}$ & 440 \\
$K_L$ & & & & & & \\
\hphantom{2em} $E>1$~GeV & $1.37\times10^{-4}$ & 0.91
& $1.29\times10^{-4}$ & 870 & $2.11\times10^{-5}$ & 140 \\
\hline
\end{tabular}
\label{tab:beam_rates}
\end{table}


The target is immediately followed by a first collimator that stops hadrons outside the beam acceptance before they decay into muons. The non-interacting protons are swept further downward by a strong dipole magnet.
Protons and charged secondaries are dumped in the TAX, which has apertures that allow the beam to pass. A photon converter consisting of 9$X_0$ of high-$Z$ material is positioned at the centre of the TAX between the two modules and reduces the flux of high-energy photons ($E > 5$~GeV) in the neutral beam by two orders of magnitude. A thinner oriented crystal converter is an optional possibility under study. Shower products emerging from the TAX are swept horizontally by a dipole magnet. Three collimators define the beam acceptance in a clean way. The defining collimator is located at $1/3$ of the distance to the final collimator and defines the beam angular acceptance to not more than $\pm$0.4~mrad, which is necessary to measure the transverse momentum of the $\pi^0$ to sufficient precision in order to reject background from $K_L\to\pi^0\pi^0$ decays. A cleaning collimator stops debris from scatterings in the jaws of the defining collimator, and a final collimator stops scattering products from the cleaning collimator. The defining and cleaning collimator are each followed by strong sweeping magnets. The final collimator will be active, and will define the start of the fiducial volume. Sufficient space is available to add more collimation stages, depending on future design iterations within the Conventional Beams Working Group.


\subsubsection{Simulation of the $K_L$ beamline}
\label{sec:kl-beamline-sim}

\begin{figure}[p]
\centering
\includegraphics[width=0.45\textwidth]{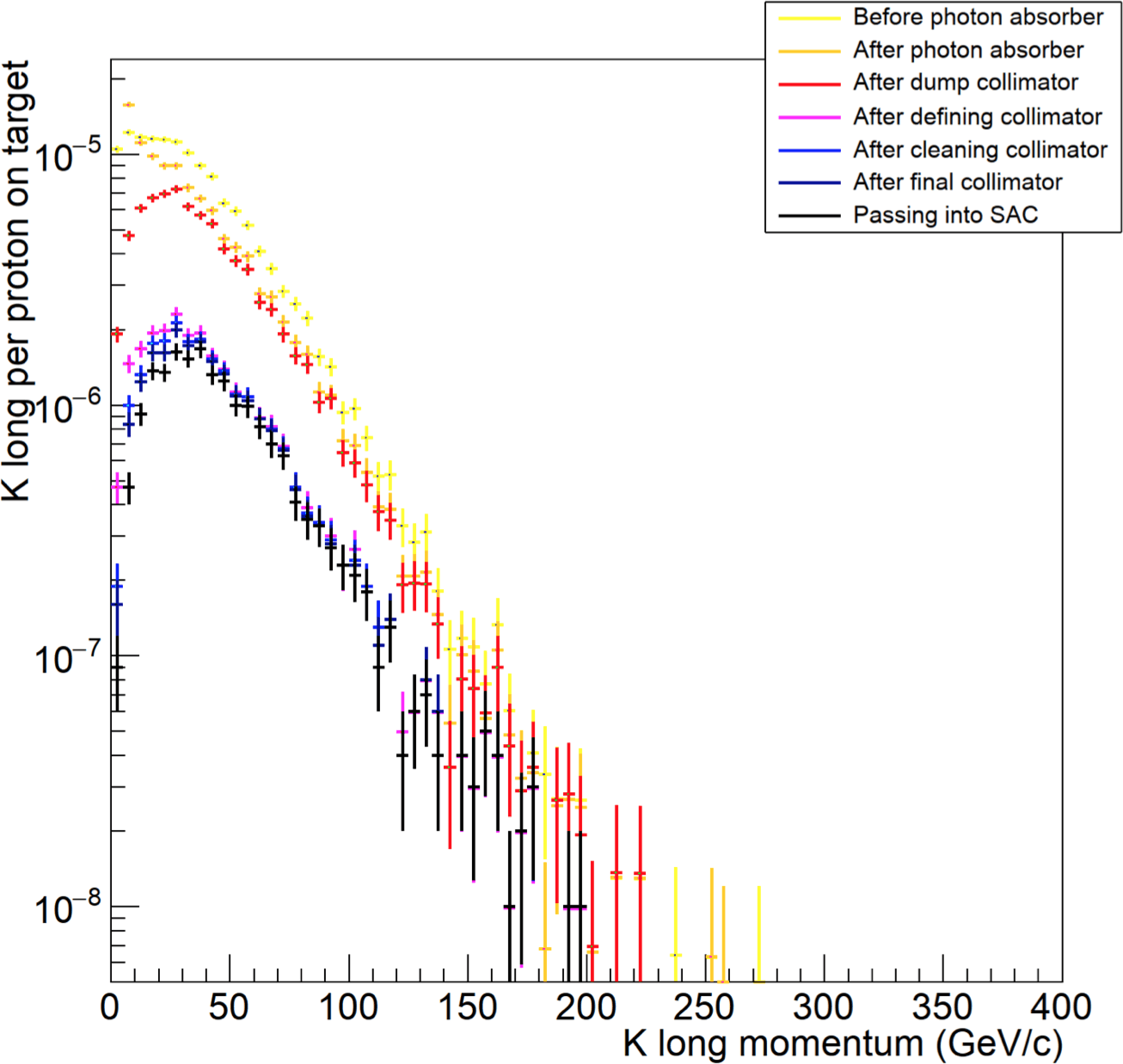}\\
\includegraphics[width=0.45\textwidth]{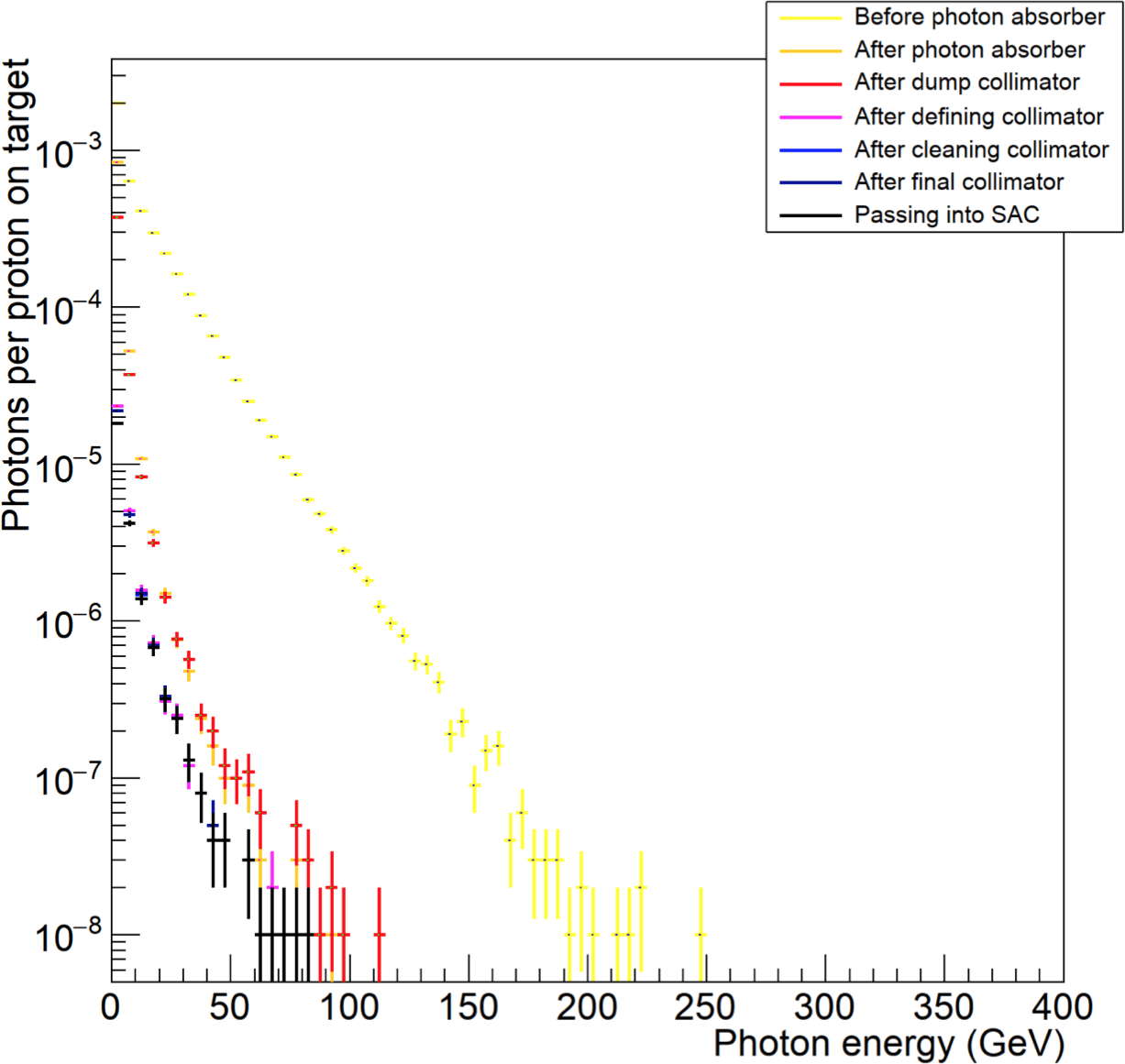}%
\includegraphics[width=0.45\textwidth]{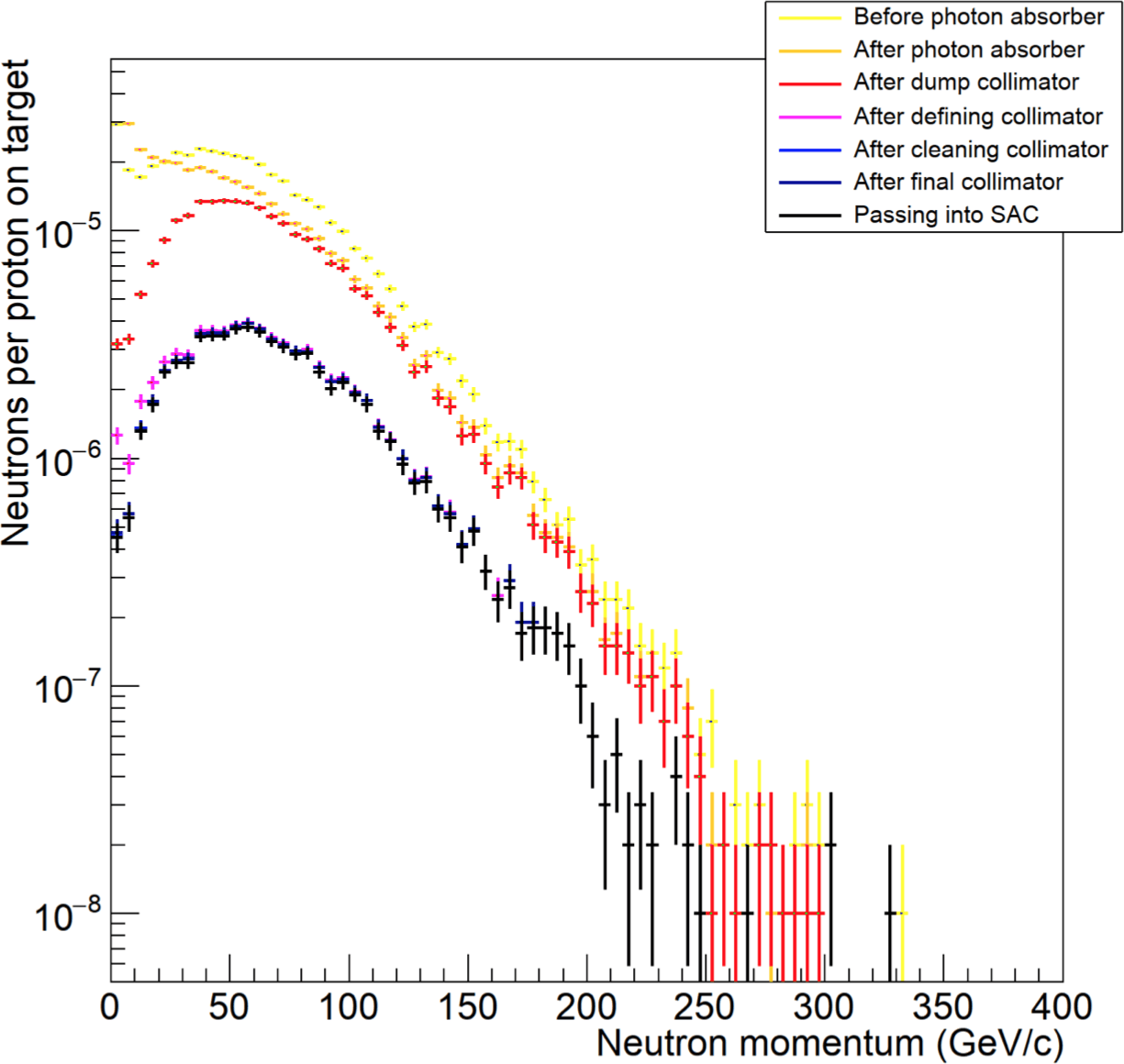}
\vspace{-1mm}
\caption{Momentum distributions for $K_L$ mesons (top), photons (bottom left), and neutrons (bottom right) in the neutral beam at various points along the neutral beamline.}
\vspace{-11mm}
\label{fig:beam_flux}
\end{figure}

A detailed FLUKA simulation of the entire beamline has been developed by the Conventional Beams Working Group. Fig.~\ref{fig:beam_flux} shows the momentum distributions for $K_L$, photons,
and neutrons in the beam at various stages of the beamline simulation: at generation (at the exit from the target), after the converter and each of the collimators, and at the end of the beamline, at the front face of the small-angle calorimeter (SAC). 
The normalisation of these distributions provides the particle fluxes in the neutral beam per incident proton. 
Of particular interest are the photon and neutron fluxes after the final
collimator.
The fluxes of photons, neutrons and $K_L$ in the beam are summarised
in Table~\ref{tab:beam_rates}. For the purposes of the sensitivity estimates for the KLEVER phase, $2.1\times10^{-5}$ $K_L$ mesons enter the detector per proton incident on the target.
The beam rates in the table
assume a primary beam intensity of 
$2\times10^{13}$~ppp with a (pessimistic) effective spill 
length of 3~seconds.
In addition to the 140~MHz of $K_L$ mesons entering the detector, there are 440~MHz of neutrons: the $n/K_L$ ratio of about 3 observed for particle production at 8~mrad is not significantly changed during transport of the 
neutral beam. In addition, there are about 50~MHz of photons with $E>5$~GeV entering the detector, most of which are incident on the SAC at the downstream end of the detector (the
condition $E<5$~GeV corresponds to the SAC threshold for KLEVER).
The FLUKA simulation also contains an idealised representation of the KLEVER experimental setup, for the purposes of evaluating rates on the detectors from beam halo. These rates, with specific reference
to the KLEVER experimental geometry, are discussed in \Sec{sec:klever_rates}.

As seen from Fig.~\ref{fig:beam_flux} and Table~\ref{tab:beam_rates}, the photon converter in the TAX dramatically reduces the photon flux in the beam, especially for high energy photons. This is critical to avoid blinding the SAC at the downstream end of the beamline. 
In the baseline design, the converter is a tungsten prism of 32.9~mm thickness, corresponding to 9.4$X_0$, or 7.3~photon conversion lengths. This thickness is chosen to keep the rate of photons with $E > 5$~GeV below 40~MHz at the entrance to the SAC.
On the other hand, this thickness corresponds to 58\% of a nuclear collision length and 33\% of an interaction length, so that about 35\% of $K_L$ mesons and 40\% of neutrons interact in or are scattered out of the beam by the converter.


\begin{figure}[p]
\centering
\includegraphics[width=0.45\textwidth]{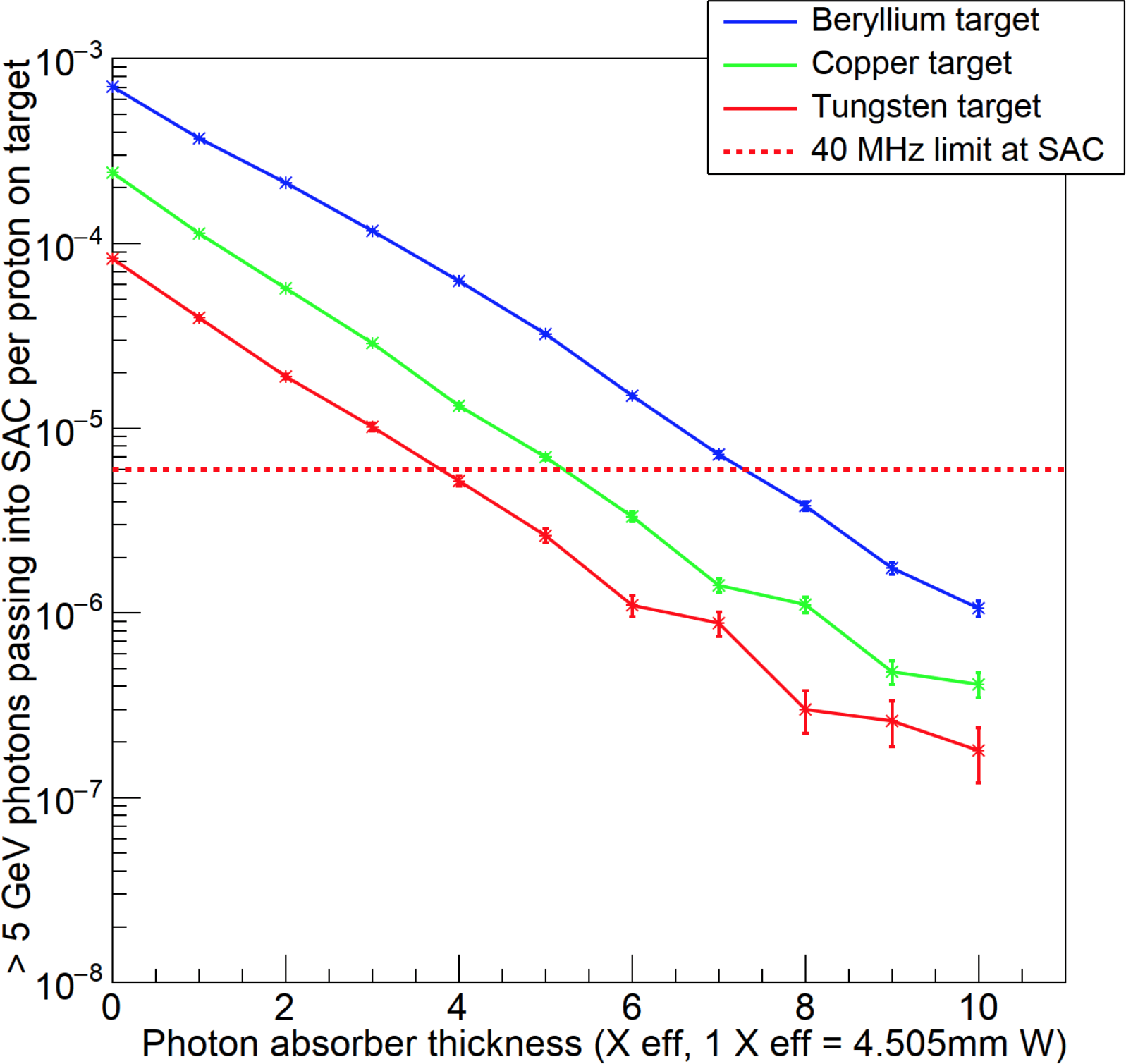}
\vspace{-1mm}
\caption{Photon fluxes (for $E>5$~GeV) at the SAC as functions of photon converter thickness for beryllium, copper, and lead targets. The dotted line indicates the flux corresponding to a rate of 40~MHz, considered to be the maximum tolerable rate for high-energy photons on the SAC.}
\vspace{-5mm}
\label{fig:targ_photons}
\end{figure}


The use of a high-$Z$ material for the target is an interesting alternative to the current baseline. Most of the prompt photons come from $\pi^0$ decays within the target, and a high-$Z$ material causes many of these photons to be converted immediately, while the overall thickness of the target in nuclear interaction lengths is unchanged. No large differences are observed in terms of hadronic production for targets of identical thickness in nuclear interaction lengths~\cite{vanDijk:2018aaa}.
Photon rates at the SAC for beryllium, copper and tungsten targets (each with a nuclear interaction length equivalent to 400~mm Be), determined with the FLUKA simulation, are shown in Fig.~\ref{fig:targ_photons}. The required converter thicknesses for beryllium, copper, and tungsten targets of equivalent thickness are 7.3, 5.2, and 3.8 photon conversion lengths, respectively, resulting an effective increase in the $K_L$ flux into the detector of 15\% (28\%) for a copper (tungsten) target relative to a beryllium target. Changing the target material strongly impacts the target design,
as the energy deposit per unit of volume is much higher in a high-$Z$ target than in beryllium, adding to cost and complexity of the new target design.

Another promising technique to reduce the photon content in the beam is to use a crystal metal photon converter. If the crystal axis is aligned with the incoming photon direction, the coherent effect of the crystal lattice promotes pair production, leading to an effective decrease in the photon conversion length~\cite{Bak:1988bq,Kimball:1985np,Baryshevsky:1989wm}. The effects of coherent interactions increase with photon energy and for decreasing angle of incidence.
A series of exploratory tests was performed with a set of tungsten crystals at the CERN SPS in summer 2018, together with the AXIAL collaboration~\cite{Soldani:2022ekn}. In particular, a commercial quality tungsten crystal of 10-mm thickness was targeted with a tagged photon beam. 
When the \hkl<111> crystal axis was aligned with the beam to within 2.5~mrad, the multiplicity of charged particles was found to be enhanced by a factor 1.6--2.3 for photon energies over the range of 30--100 GeV. Ref.~\cite{Soldani:2022ekn} also reports simulations validated by this result that suggest that a crystal of this type could be used to reduce the thickness of the photon converter by 15--20\% at no cost in effectiveness. 
Such a solution appears to be relatively easy to implement.


\subsubsection{Beam for Phase~2 (a multi-purpose $K_L$ experiment)}

The proposed HIKE Phase~2 is based on the $K_L$ beam, but a detector very similar to the Phase~1 experiment, including a tracking system. This phase of the programme will make use of the neutral beamline described above with a production angle chosen to optimise the statistics for the measurement of $K_L\to\pi^0\ell^+\ell^-$ decays, likely 2.4~mrad. Alternative beamline options can be studied. The full FLUKA simulation will be performed when the choice of production angle is finalised. For the sensitivity estimates in Section~\ref{sec:phase2}, the entries in Table~\ref{tab:beam_rates} are scaled to account for production angle of 2.4~mrad (Table~\ref{tab:beam_prod}).




\subsubsection{Beam for Phase~3 (KLEVER: measurement of $K_L\to \pi^0\nu\bar\nu$ decay)}
\label{sec:neutral_beam_ext}

After the initial neutral beam studies for the KLEVER phase of HIKE were completed, it was discovered that the background from $\Lambda\to\pi^0 n$ decays is more difficult to suppress than previously thought. Therefore, the amount of $\Lambda$ decays in the fiducial region has to be reduced further by several orders of magnitude, whilst preserving as much as possible the kaon flux. 
Various options were considered, such as lengthening the distance between the target and the start of the fiducial volume, reducing the primary proton beam energy (for example, from 400 to 300~GeV), and increasing the production angle (for example, from 8~mrad to 20~mrad).
Lengthening the beam line by 150~m seems to be the best option, and several possibilities are under consideration.
Moving the T10 target further upstream is one of them, but this may have complex radiation protection implications and might require the whole experiment to be installed on a slope, depending on the exact target location.
The more straightforward option might be to prolong the ECN3 cavern by 150~m, implying civil engineering works. Simulations show that this brings the background from $\Lambda$ decays to an acceptable level (\Sec{sec:klever_sens}). The technical details and costs of the different options are under discussion with the Civil Engineering group of CERN. 

An extension of the ECN3 cavern implies that the detector will be installed 150~m downstream of the present NA62 location. 
At the moment, we do not contemplate changing the transverse dimensions of the setup. 
The positions of the beamline elements would change as noted in Table~\ref{tab:ext_layout}. The beam solid angle and hence the $K_L$ flux decreases by a factor of $(0.256/0.400)^2 \approx 0.41$. Since the $p_\perp$ reconstruction at the downstream end of the FV depends on the beam spot size, it is not trivial to recover the sensitivity of the experiment by enlarging its transverse dimensions, and we favour the other approaches discussed in \Sec{sec:klever_sens}. The full FLUKA simulation of the extended configuration will be performed as soon as the design is finalised. For the sensitivity estimates, the rates in Table~\ref{tab:beam_rates} can be scaled to account for the reduction in solid angle from the extension.

In case resources allow to advance the construction of the ECN3 extension, HIKE Phase~2 can be performed with the extended beamline and the detector in the new location. This would eliminate the need for infrastructural work to change beamline configurations between Phase~2 and KLEVER and allow precise information on the beam composition and halo to be obtained during Phase~2, thereby improving the understanding of beam-related backgrounds in KLEVER.

\begin{table}[h]
\centering
\caption{Collimator positions and other parameters for standard and extended KLEVER beamlines.}
\vspace{-2mm}
\begin{tabular}{lcc}
\hline
Configuration & Standard & Extended \\ \hline
Target position [m] & 0 (T10) & 0 (T10) \\
Defining collimator position [m] & 40 & 90 \\
Cleaning collimator position [m] & 80 & 180 \\
Final collimator position [m] & 120 & 270 \\
Beam opening angle [mrad] & 0.400 & 0.256 \\
\hline
\end{tabular}
\vspace{-5mm}
\label{tab:ext_layout}
\end{table}

%% file: phase1.tex
\section{Phase 1: a multi-purpose $K^+$ decay experiment}
\label{sec:phase1}
\vspace{-1mm}

\subsection{Experimental layout}

While the success of the NA62 experiment has proven that its layout is suitable for a precision ${\cal B}(K^+\to \pi^+\nu\bar\nu)$ measurement, new or upgraded detectors will replace those of NA62 with the goal of sustaining secondary-beam rates four times higher than those of NA62, to substantially boost the statistical sensitivity. 
To this end, new technologies will be used, in synergy with the upgrades of the LHC experiments.
While some detectors will be renovated for the start of the HIKE $K^+$ phase (Phase 1), others are already intrinsically fast in terms of detector technologies but will need readout upgrades.
Therefore a staged approach will be used, where renovated or new detectors are inserted as soon as they are needed and ready, while maintaining the general principle that changes must serve the remaining phases of the programme once they are applied.

\begin{figure}[h]
\begin{center}
\resizebox{\textwidth}{!}{\includegraphics{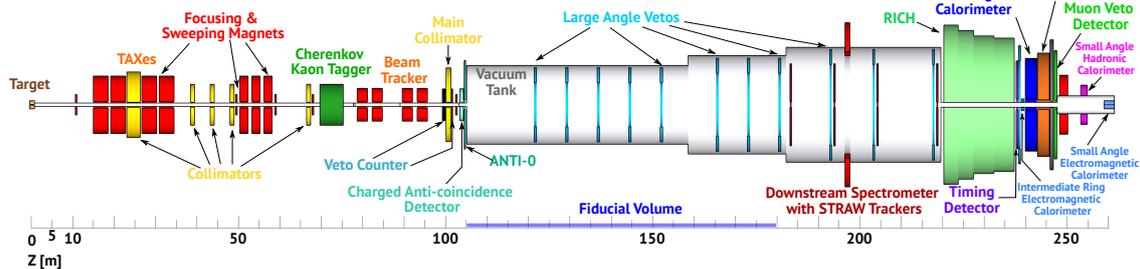}}
\put(-425,107){\color{blue}{\bf\LARGE HIKE Phase~1 ($K^+$)}}
\end{center}
\vspace{-8mm}
\caption{HIKE Phase~1 layout, with an aspect ratio of 1:10.}
\label{fig:phase1-layout}
\vspace{-8mm}
\end{figure}

In the baseline configuration, the $K^+$ beam is produced by $1.2\times10^{13}$ $400~{\rm GeV}/c$ protons/spill at zero angle on target. The mean $K^+$ momentum at the entrance to the decay volume is 75~GeV/$c$. The layout (Fig.~\ref{fig:phase1-layout}) includes a beam tracker, a kaon-identification detector, and a veto counter upstream of the decay volume; a main tracking system with the MNP33 magnet in the decay volume; 
calorimeters, vetoes and particle identification detectors downstream of the decay volume; and a large-angle veto system surrounding the decay volume and part of the region downstream.


\subsection{Physics sensitivity}



\subsubsection{$K^+\to\pi^+\nu\bar\nu$ measurement}
\label{sec:kpnn_sens}

The first phase of HIKE will be dedicated to a detailed measurement of the $K^+\to\pi^+\nu\bar\nu$ decay, with a target of reaching a branching ratio measurement with $\mathcal{O}(5\%)$ precision, competitive with the theoretical uncertainty. The approach towards this goal will continue, and build upon, the successful strategy of the NA62 experiment.
The keystones of the measurement, required to suppress backgrounds at the $\mathcal{O}(10^{11})$ level, are the following:
\begin{itemize}
\item High efficiency and high-precision tracking of both the $K^{+}$ upstream and $\pi^{+}$ downstream. This, coupled with a careful choice of signal regions, will allow kinematic suppression of backgrounds by a factor of $\mathcal{O}(10^{3})$.
\item High precision time measurements, allowing time-matching between upstream and downstream detectors with $\mathcal{O}(20~\text{ps})$ precision. Simulations show that with this time-matching performance, the effect of the higher intensity can be totally compensated in the matching between upstream and downstream tracks, without losing efficiency for the signal selection and without increasing the relative contamination of background coming from upstream decays and interactions with respect to NA62.
\item Comprehensive and hermetic veto systems:
\begin{itemize}
\item Photon veto detectors with hermetic coverage of the 0--50~mrad range for photons from $\pi^0\to\gamma\gamma$ decays, allowing photon suppression by a factor of $\mathcal{O}(10^{8})$.
\item Veto detectors to cover downstream regions not covered by the principal detectors. 
\item Rejection of upstream decays and interactions, for suppressing possible upstream backgrounds by an additional factor of $\mathcal{O}(10)$, with changes to the beam line setup and new systems dedicated to the detection of the upstream background events.
\end{itemize}
\item High-performance particle identification system (for $\pi$/$\mu$ discrimination), suppressing backgrounds with muons by a factor $\mathcal{O}(10^{7})$.
\end{itemize}

\begin{figure}
\centering
\includegraphics[scale=0.5]{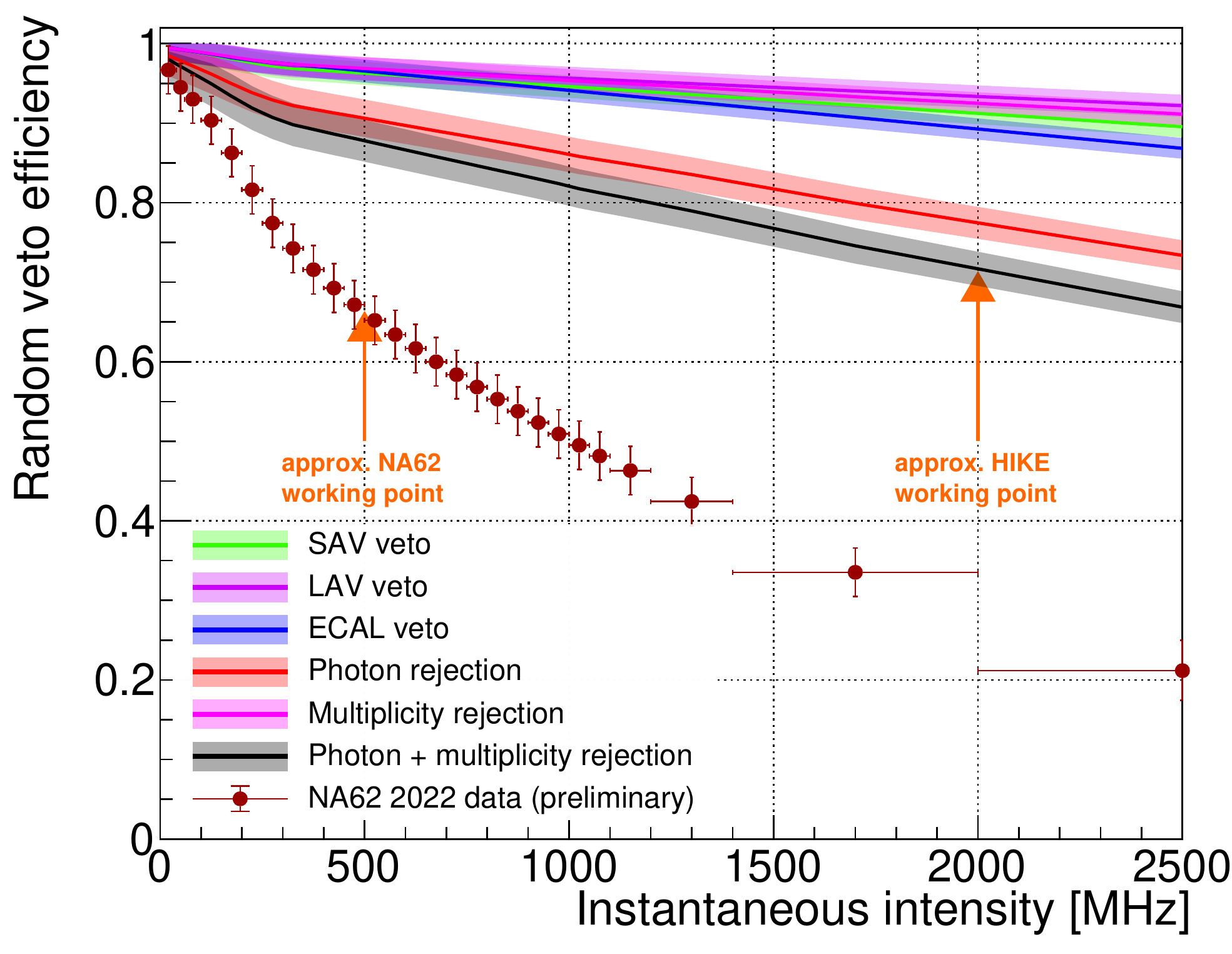}
\vspace{-3mm}
\caption{Expected HIKE Phase~1 random veto efficiency $\varepsilon_{\rm RV}$ for the $K^+\to\pi^+\nu\bar\nu$ analysis as a function of instantaneous beam intensity: the total effect (grey band) and the contributions from the individual photon/multiplicity veto conditions. The NA62 graph is also shown for comparison: the lower $\varepsilon_{\rm RV}$ values with respect to HIKE Phase~1 are due to the worse time resolution.}
\label{fig:RVprojections}
\end{figure}

Crucial to the study of the $K^+\to\pi^+\nu\bar\nu$ decay at HIKE will be the exploitation of the high-intensity environment, taking advantage of the high $K^+$ flux while mitigating detector pileup effects. 
A critical performance indicator is the ``random veto efficiency'', i.e., the fraction of events passing signal selection criteria that are sensitive to accidental activity in the detector systems. The random veto efficiency measured within the NA62 selection for the $K^+\to\pi^+\nu\bar\nu$ decay at the current intensity working-point
is $\varepsilon_{\rm RV}\approx 65\%$. This is dependent on selection criteria that are limited by the timing precision of the detectors. The random veto efficiency is approximately linear as a function of the instantaneous beam intensity.
At HIKE, this random veto efficiency must be maintained or improved, requiring an improvement in time resolution by the same factor as the intensity increase. 
With this performance, a projection for the random veto efficiency as a function of intensity is shown in Fig.~\ref{fig:RVprojections}. Individual contributions from photon veto subsystems (SAV, LAV, ECAL) are indicated which in total give the `photon rejection' curve. The curve labelled `multiplicity rejection' relates to the selection criteria required to reject additional activity in an event (outside the photon veto detectors). Finally, the total random veto efficiency, combining photon and multiplicity rejection, is shown. At a secondary beam intensity 4 times higher than NA62, with the detector updates and corresponding selection changes shrinking veto windows by a factor of 4, the random veto efficiency is maintained at the same level as currently achievable at NA62.

The HIKE sensitivity for the ${\cal B}(K^+\to\pi^+\nu\bar\nu)$ measurement is estimated as follows.
\begin{itemize}
\item The effective number of collected $K^+$ decays per year is $N_K/{\rm year} \simeq 2 \times 10^{13}$. This can be extrapolated from the data already collected by the NA62 experiment, scaling for an intensity four times larger than the NA62 nominal intensity and 200 days of data-taking per year. 
This automatically includes effects already present in the current setup, such as duty cycle, SPS beam availability, secondary beamline downtime, and detector and DAQ efficiencies.
%
\item Acceptance of the signal selection: $\varepsilon_A \simeq 0.1$. The improvements on the time-matching performance and on the resolution of the charged-particle identification system will lead to an increase of about 50\% relative to that at the current NA62 working point. 
\item Random veto efficiency ($\varepsilon_{\rm RV}$) and trigger efficiency ($\varepsilon_{\rm trig}$), which according to the projections already described should remain the same as for the current NA62 working point even at higher intensity: $\varepsilon_{\rm RV} \simeq 0.7$, $\varepsilon_{\rm trig} \simeq 0.9$.
\end{itemize}
Under these conditions, the single event sensitivity attainable in one year of data-taking is found to be ${\cal B}_{\rm SES}(1 \; {\rm year}) = (N_K/{\rm year} \cdot \varepsilon_A \cdot \varepsilon_{\rm RV} \cdot \varepsilon_{\rm trig})^{-1} \simeq 8\times 10^{-13}$, leading to the number of expected SM signal events per year:
\begin{displaymath}
N_{\pi\nu\bar{\nu}} / {\rm year} \simeq 100.
\end{displaymath}

With an $O(10\%)$ relative background contamination in the signal sample and systematic uncertainty sources under control to $O(1\%)$ or better, a measurement of the branching ratio $\mathcal{B}(K^{+}\to\pi^{+}\nu\bar{\nu})$ to $\mathcal{O}(5\%)$ precision can be reached by HIKE in 4~years of data-taking. Assuming that the number of observed events is equal to the SM expectation, values of ${\cal B}(K^+\to\pi^+\nu\bar\nu)$ higher than 10\% with respect to the SM prediction will be excluded at 95\% CL.


\subsubsection{Test for scalar amplitudes in the $K^+\to\pi^+\nu\bar\nu$ decay}

\begin{figure}[tb]
\centering
\includegraphics[width=0.5\columnwidth]{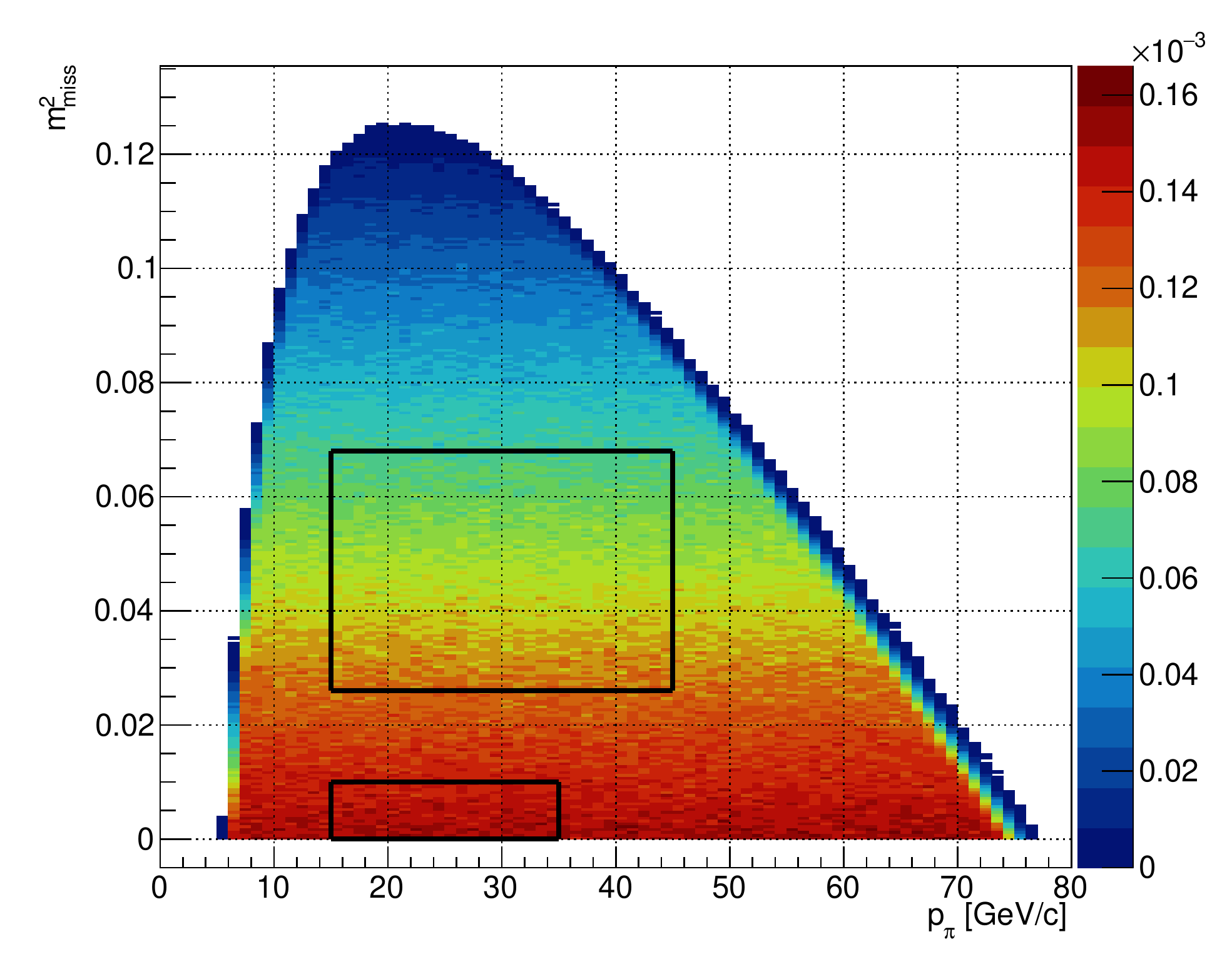}%
\includegraphics[width=0.5\columnwidth]{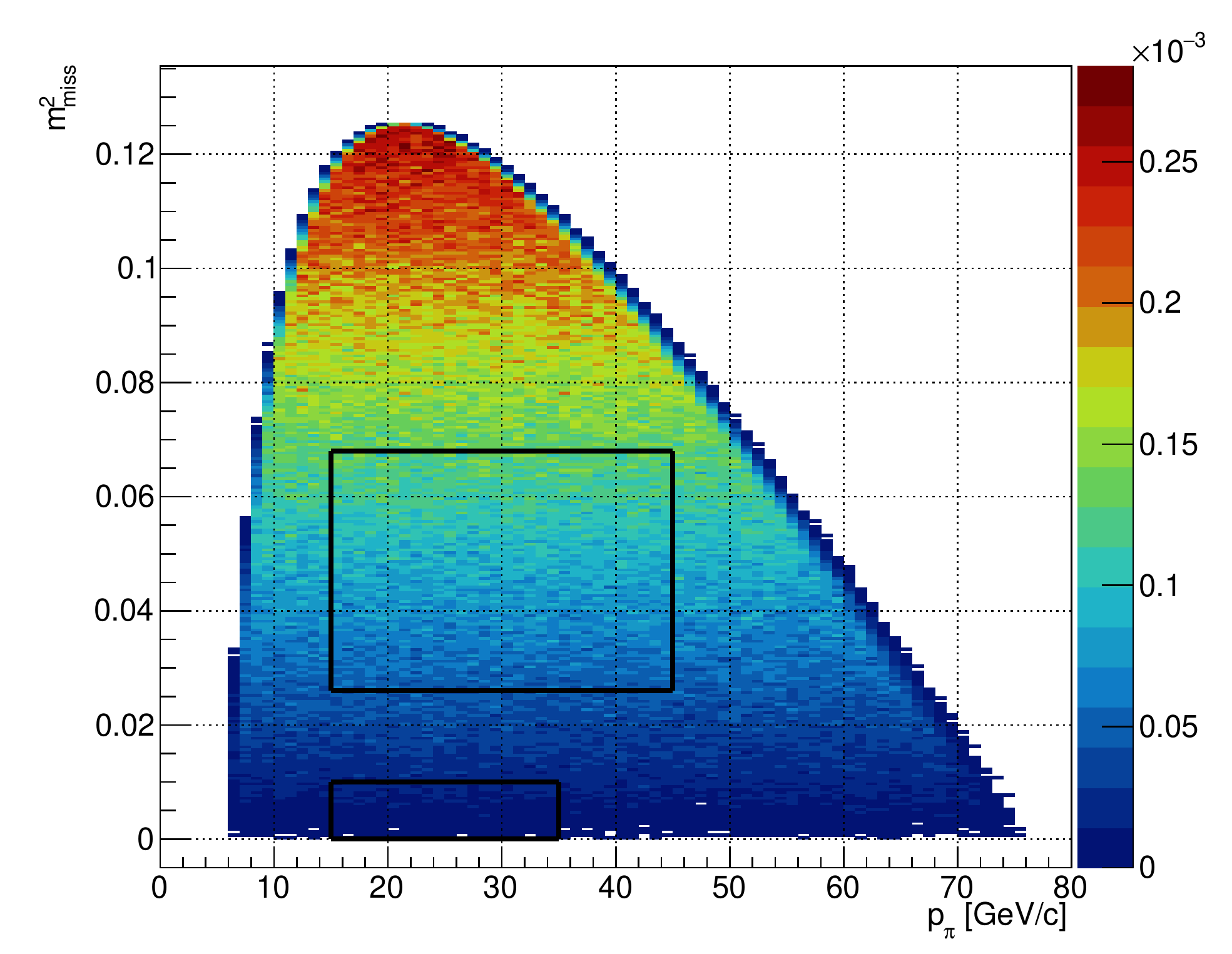}
\vspace{-5mm}
\caption{Simulated distributions of the squared missing mass $m_{\rm miss}^{2}=(P_{K^+}-P_{\pi^+})^2$ vs pion momentum in the laboratory frame for the SM $K^+\to\pi^+\nu\bar\nu$ decay with a vector nature (left) and the LNV $K^+\to\pi^+\nu\nu$ decay with a scalar nature (right). Black boxes indicate the signal regions for which experimental sensitivity is highest.}
\label{fig:SMvsLNVdistributions}
\end{figure}

As introduced in Section~\ref{sec:PhysicsKpinunu}, in BSM scenarios the measurable branching ratio of $K^+\to\pi^+\nu\bar\nu$ is formed from two components: the SM process $K^+\to\pi^+\nu\bar\nu$ with a purely vector nature and a possible 
BSM process $K^+\to\pi^+\nu\nu$ with a scalar nature~\cite{Deppisch:2020oyx,Crosas:2022quq}.
Therefore the experimentally measured branching ratio is given by~\cite{Aebischer:2022vky,Deppisch:2020oyx}:
\vspace{-1mm}
\begin{equation}
\mathcal{B}(K^{+}\to\pi^+\nu\bar\nu) = \mathcal{B}_{\rm SM}(K^+\to\pi^+\nu\bar\nu) + \sum_{i \leq j}^{3} \mathcal{B}_{\rm LNV}(K^+\to\pi^+\nu_i\nu_j).
\end{equation}
The SM and BSM contributions are predicted to have different kinematic distributions (Fig.~\ref{fig:SMvsLNVdistributions}). To identify the nature of the decay, an investigation of the shape of the distribution of selected signal candidates as a function of kinematic variables will be performed. In absolute terms, within the signal regions (as defined in the current NA62 analysis), the acceptance for the BSM mode is approximately 75\% of the acceptance for the SM mode. With a sample of several hundred $K^+\to\pi^+\nu\bar\nu$ candidates, HIKE will be able to study the shape of the kinematic distributions as well as compare sets of sub-categories of phase-space to distinguish between, or constrain the relative contributions of, the possible vector and scalar natures of the process.

\subsubsection{Feebly-interacting particle production in $K^+\to\pi^+X_{\rm inv}$ decays}
\label{sec:KpiXSensitivity}

As described in Section~\ref{sec:fips}, searches for the $K^+\to\pi^+X_{\rm inv}$ decay, where $X_{\rm inv}$ is a FIP (either a dark scalar or ALP), can be performed within the analysis framework developed for the study of the $K^+\to\pi^+\nu\bar\nu$ decay. This is possible because the signature of these two decays is identical except that the $K^{+}\to\pi^{+}X_{\rm inv}$ decay produces a peak in the $m_{\rm miss}^2=(P_{K^{+}}-P_{\pi^{+}})^{2}$ variable at $m_X^2$, where $m_X$ is the mass of the FIP. Searches are therefore performed by looking for evidence of a peaking signal on top of the background dominated by the SM $K^+\to\pi^+\nu\bar\nu$ decay. The search strategy is described in Ref.~\cite{NA62:2020xlg}, which furthermore presents results based on the analysis of 2017 data, which has been updated for the full Run~1 (2016--2018) dataset in Ref.~\cite{NA62:2021zjw}.
Based on this experience, a sensitivity projection for HIKE Phase~1 has been performed assuming a 40-fold increase in the size of the data sample with respect to NA62 Run~1.

\begin{figure}[p]
\begin{center}
\resizebox{0.5\textwidth}{!}{\includegraphics{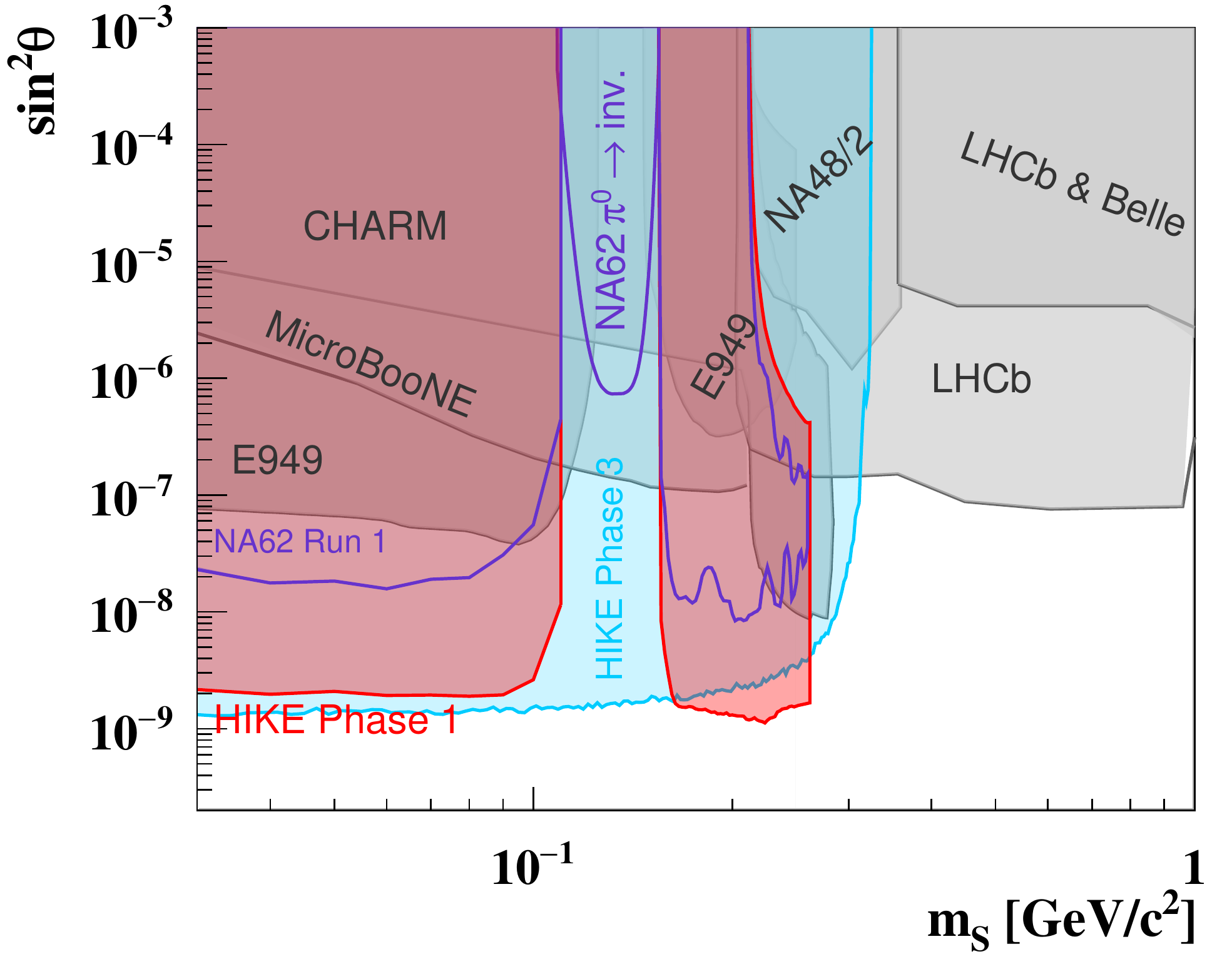}}%
\resizebox{0.5\textwidth}{!}{\includegraphics{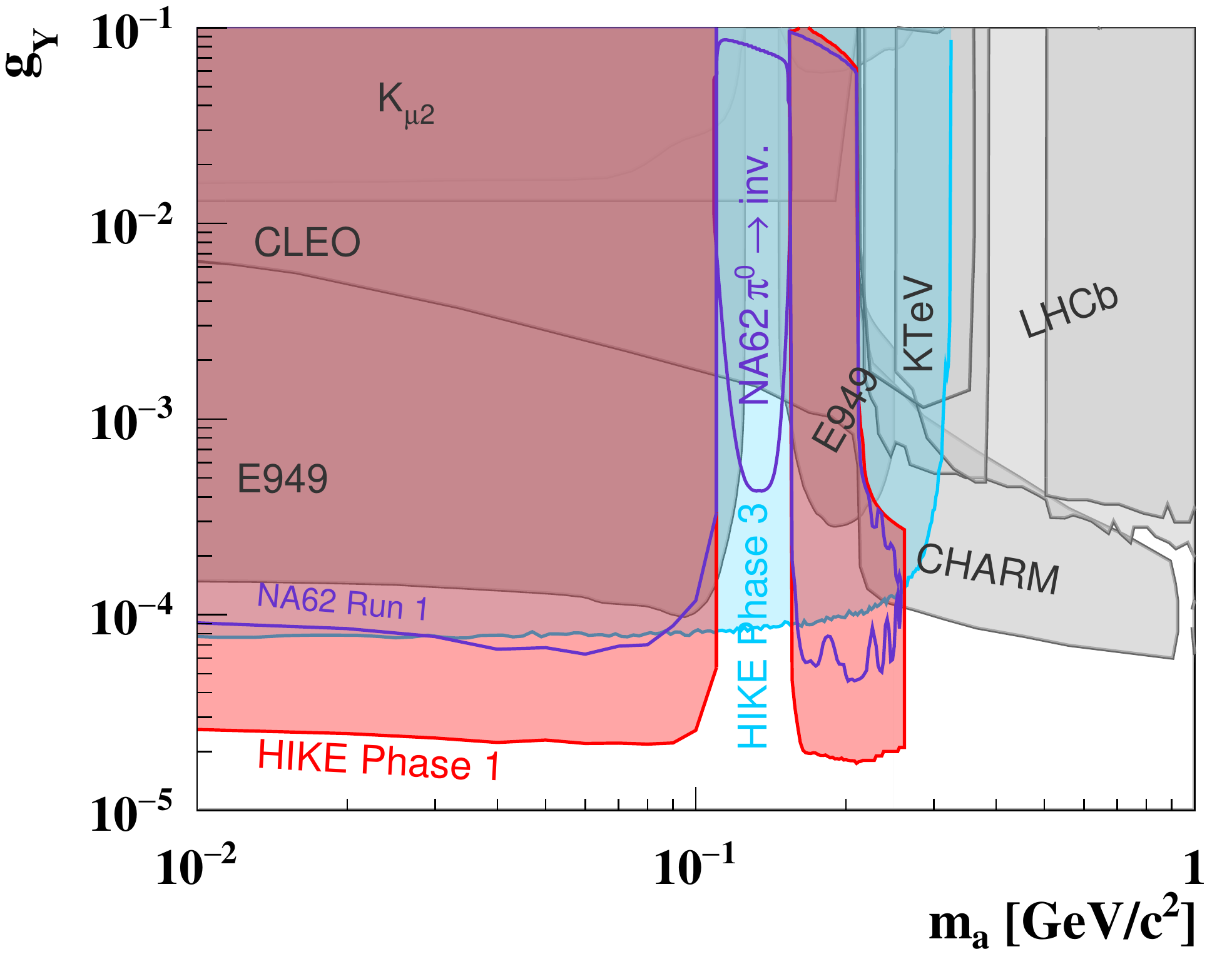}}%
\vspace{-6mm}
\end{center}
\caption{Left: excluded regions at 90\% CL of the $(m_S,\sin^2\theta)$ parameter space for a dark scalar, $S$, of the BC4 model~\cite{Beacham:2019nyx}. 
Right: excluded regions of the parameter space $(m_{a},g_{Y})$ for an ALP, $a$, of the BC10 model~\cite{Beacham:2019nyx}. 
The exclusion bound from analysis of NA62 Run~1 data~\cite{NA62:2020xlg,NA62:2020pwi,NA62:2021zjw} is shown in purple; the projected exclusion for HIKE Phase~1 and Phase~3 (KLEVER, assuming invisible $S$ or $a$ and updated from Ref.~\cite{Beacham:2019nyx}) is shown in red and light blue, respectively.
Other exclusion bounds are shown in grey, including results from E949~\cite{BNL-E949:2009dza}, CHARM~\cite{Winkler:2018qyg,CHARM:1985anb}, NA48/2~\cite{NA482:2016sfh}, LHCb~\cite{LHCb:2016awg,LHCb:2015nkv}, Belle~\cite{Belle:2009zue}, $K_{\mu2}$ experiment~\cite{Yamazaki:1984vg}, CLEO~\cite{CLEO:2001acz}, KTEV~\cite{KTEV:2000ngj}, MicroBooNE~\cite{MicroBooNE:2021usw}.}
\label{fig:KpiX_Projections}
\end{figure}

The principal background comes from the SM $K^+\to\pi^+\nu\bar\nu$ decay, and other backgrounds are negligible in comparison. The selection acceptance and efficiency are assumed to be consistent with the results obtained for the NA62 2018 data~\cite{NA62:2021zjw}. In this way realistic projections are performed, accounting for all first-order effects. The results are shown in Fig.~\ref{fig:KpiX_Projections} for the excluded regions in the interpretations of $X_{\rm inv}$ as either a dark scalar mixing with the Higgs boson or an ALP with fermionic couplings. Additional specialisation of the selection used for the $K^+\to\pi^+X_{\rm inv}$ decays might be possible, further enhancing sensitivity. HIKE Phase~3 sensitivity projections for the search for the $K_L\to\pi^0X_{\rm inv}$ decay, refined with respect to Ref.~\cite{Beacham:2019nyx}, are also shown in Fig.~\ref{fig:KpiX_Projections}. HIKE Phases~1 and~3 provide complementary sensitivities, allowing for exclusion of, or discovery within, new phase-space regions which are otherwise difficult to cover.



\newpage

\subsubsection{Heavy neutral lepton production in $K^+\to\ell^+N$ decays}
\label{sec:K-HNL-Sensitivity}

Searches for heavy neutral lepton (HNL) production in $K^+\to\ell^+N$ decays have been well established by the NA62 experiment using the main $K^+\to\pi^+\nu\bar\nu$ trigger chain for the $K^+\to e^+N$ case, and a downscaled control trigger chain for the $K^+\to\mu^+N$ case. NA62 has published world-leading exclusion on the HNL mixing parameters $|U_{\ell 4}|^2$ over much of the accessible mass range of 144--462~MeV/$c^2$ with the Run~1 dataset~\cite{NA62:2020mcv,NA62:2021bji}. Both searches are limited by background. In particular, the $K^+\to\mu^+\nu$ decay followed by $\mu^+\to e^+\nu\bar\nu$ decay in flight, and the $\pi^+\to e^+\nu$ decay of the pions in the unseparated beam, represent irreducible backgrounds to the $K^+\to e^+N$ process. The peaking nature of the $K^+\to\ell^+N$ signal in terms of the reconstructed missing mass allows for data-driven background evaluation, reducing the systematic uncertainties in the background estimates.

\begin{figure}[p]
\begin{center}
\resizebox{0.6\textwidth}{!}{\includegraphics{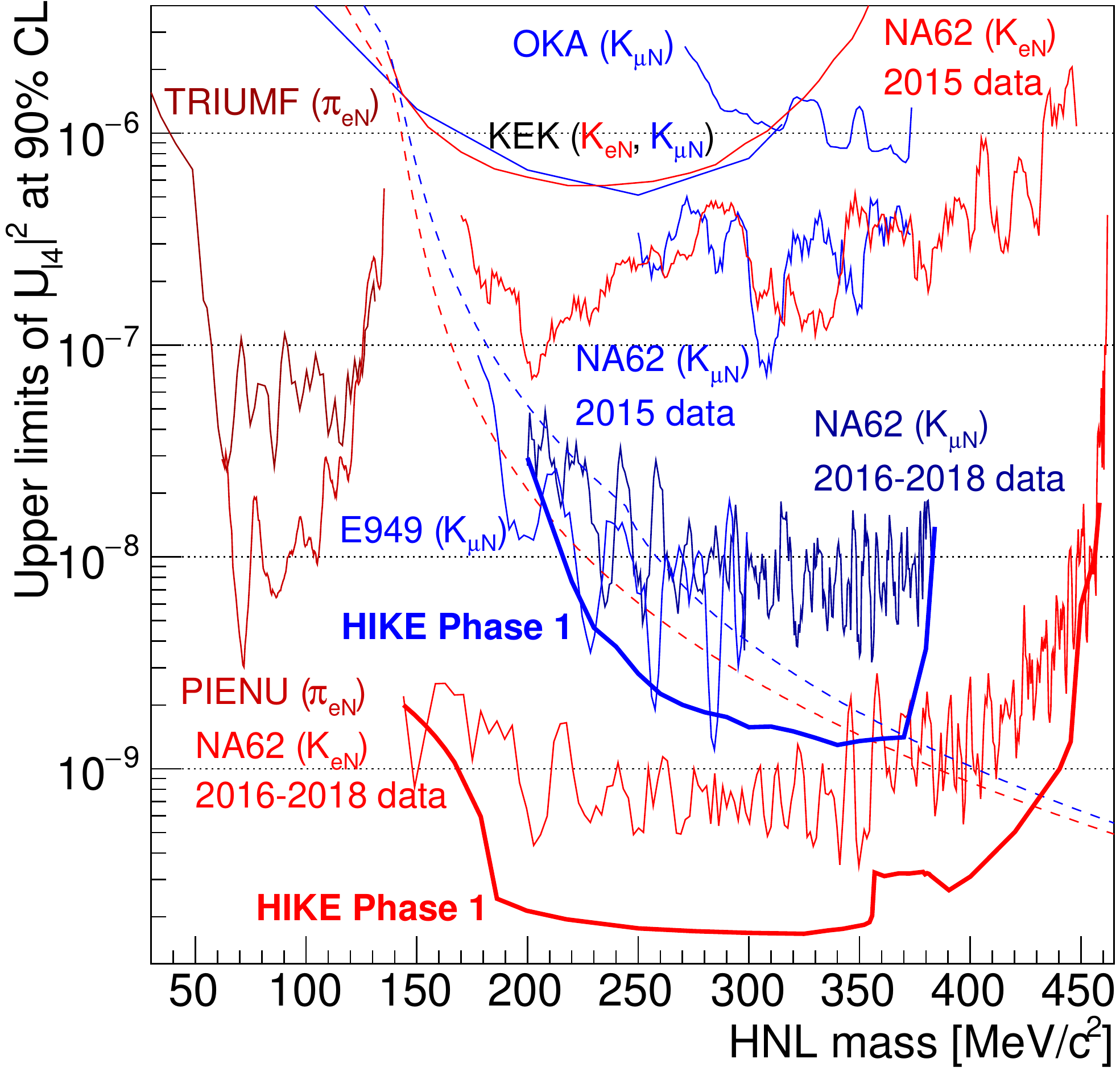}}
\vspace{-6mm}
\end{center}
\caption{Summary of the upper limits at 90\% CL of the HNL mixing parameters $|U_{e4}|^2$ and $|U_{\mu4}|^2$ obtained from production searches, and HIKE Phase~1 sensitivity with $K^+\to\ell^+N$ decays. HIKE projection for $|U_{\mu4}|^2$ assumes data collection with a highly-downscaled control trigger, and may improve by an order of magnitude in case a software trigger is employed.}
\label{fig:HNL}
\end{figure}

HIKE Phase~1 sensitivity projections for HNL production searches in $K^+\to\ell^+N$ decays, obtained from a detailed analysis assuming NA62-like trigger chains and conditions, are shown in Fig.~\ref{fig:HNL}. HIKE Phase~1 offers world-leading sensitivity to $|U_{e4}|^2$ in the region $m_N>140~{\rm MeV}/c^2$, approaching the seesaw bound and complementing future long-baseline neutrino experiments~\cite{Abdullahi:2022jlv}, and is expected to improve on the state-of-the-art for lower HNL masses via searches for the $K^+\to\pi^0e^+N$~\cite{Tastet:2020tzh} and $\pi^+\to e^+N$ decays. HIKE Phase~1 also provides competitive sensitivity to $|U_{\mu4}|^2$. The projection for $|U_{\mu4}|^2$ assumes data collection with a highly-downscaled control trigger, and may improve by an order of magnitude in case a software trigger based on the streaming readout (Section~\ref{sec:streaming-readout}) is employed.


%
%
%
%
%

\subsubsection{Other measurements}
\label{sec:phase1-other-measurements}

HIKE Phase~1 is expected to increase the world samples of many rare $K^+$ decays by an order of magnitude, developing upon the trigger strategies developed for the NA62 experiment. This would bring many rare decay measurements to a new level of precision. In particular, we expect to collect background-free samples of several times $10^5$ events of both $K^+\to\pi^+e^+e^-$ and $K^+\to\pi^+\mu^+\mu^-$ decays, leading to a powerful test of lepton universality (LFU) by comparing the form-factor parameters ($a_+$, $b_+$) of the two decay modes. This is of particular interest in the framework of the global NP fit~\cite{DAmbrosio:2022kvb} to the kaon decay data (Section~\ref{sec:global_fit}, Fig.~\ref{fig:cfitk}). Considering that the precision of the recent NA62 measurement in the muon channel~\cite{NA62:2022qes} is limited by the size of the dataset, HIKE Phase~1 is expected to measure the difference of the form-factor parameter $a_+$ between the two decay modes, $\Delta a_+^{e\mu}$, to a precision of $\pm0.007$, and the corresponding difference $\Delta b_+^{e\mu}$ to a precision of $\pm0.015$. Exploitation of the correlations between the measured $a_+$ and $b_+$ parameters in each mode will provide additional power in terms of LFU tests.

In terms of searches for $K^+$ and $\pi^0$ decays violating lepton number or flavour conservation, including $K^+\to\pi^-(\pi^0)\ell_1^+\ell_2^+$, $K^+\to\pi^+\mu^\pm e^\mp$, $K^+\to\ell_1^-\nu\ell_2^+\ell_2^+$ and $\pi^0\to\mu^\pm e^\mp$, the experimental technique based on dedicated di-lepton trigger chains has been firmly established by the NA62 experiment, leading to world-leading upper limits of ${\cal O}(10^{-11})$ on the branching ratios of a number of processes with the NA62 Run~1 dataset~\cite{NA62:2019eax,NA62:2021zxl,NA62:2022tte}. These searches are not limited by background, and HIKE Phase~1 sensitivity is expected to improve in the future almost linearly with the size of the dataset to the ${\cal O}(10^{-12})$ level.

\newpage

%% file: phase2.tex
\section{Phase 2: a multi-purpose $K_L$ decay experiment}
\label{sec:phase2}

\subsection{Experimental layout}

The baseline design of the multi-purpose $K_L$ decay experiment includes a 120~m long neutral beamline with the secondary beam opening angle of 0.4~mrad, proposed and studied extensively for the KLEVER phase of the HIKE programme. The beamline involves four stages of collimation, and provides adequate suppression of the short-lived $K_S$ and $\Lambda$ components. A detailed description of the beamline, including the expected $K_L$ yields and momentum spectra, is provided in Section~\ref{sec:neutral_beam} Unlike the KLEVER phase, HIKE Phase~2 detector will be equipped with a tracker, which offers the opportunity to perform characterisation of the KLEVER beam. For the forthcoming proposal, alternative beamline configurations can be studied.

The proposed $K_L$ production angle is 2.4~mrad (to be compared to 8~mrad for the KLEVER phase), which improves the $K_L$ yield and increases the mean $K_L$ momentum (Figs.~\ref{fig:beam_mom_ang}, \ref{fig:beam_mom}) therefore improving the acceptances for  $K_L\to(\pi^0)\ell^+\ell^-$ decays. Thanks to the relatively compact detector, HIKE Phase~2 allows for a 90~m long fiducial decay volume to be accommodated in the present ECN3 experimental hall, and no major civil engineering work is required in preparation for this phase of the programme.

\begin{figure}[h]
\begin{center}
\vspace{3mm}
\resizebox{\textwidth}{!}{\includegraphics{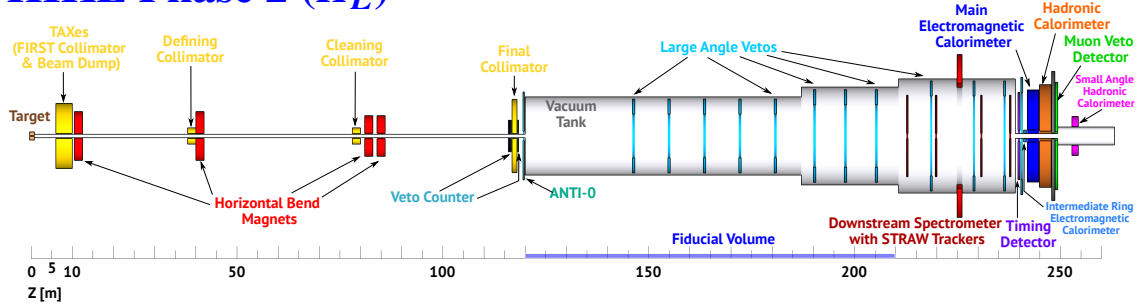}}
\put(-425,113){\color{blue}{\bf\LARGE HIKE Phase~2 ($K_L$)}}
\end{center}
\vspace{-6mm}
\caption{HIKE Phase~2 layout, with an aspect ratio of 1:10.}
\vspace{-3mm}
\label{fig:phase2-layout}
\end{figure}

It is proposed to use the HIKE Phase~1 experimental setup with minimal modifications (Fig.~\ref{fig:phase2-layout}). The GTK, KTAG, RICH and SAC detectors will be removed, the STRAW spectrometer will be shortened to a total length of 25~m, and central holes of the STRAW chambers will be realigned on the neutral beam axis (Section~\ref{sec:straw}). Reduction of the magnetic field of the spectrometer dipole magnet by about 20\%, leading to a momentum kick of 210~MeV/$c$, is possible without significant degradation of the mass resolution.

\subsection{Physics sensitivity}
\label{sec:phase2:sensitivity}

\begin{figure}[tb]
\begin{center}
\resizebox{0.5\textwidth}{!}{\includegraphics{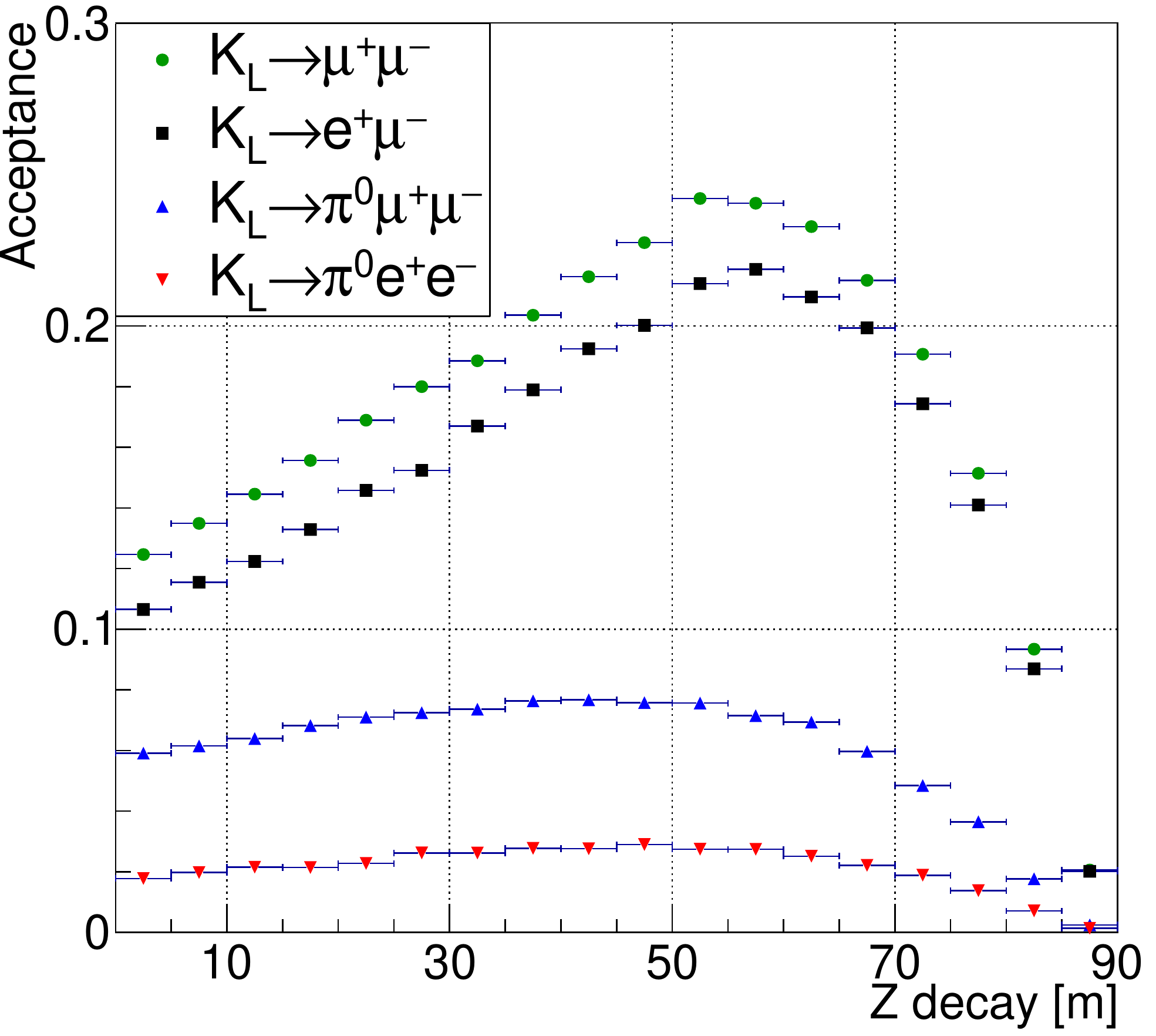}}%
\resizebox{0.5\textwidth}{!}{\includegraphics{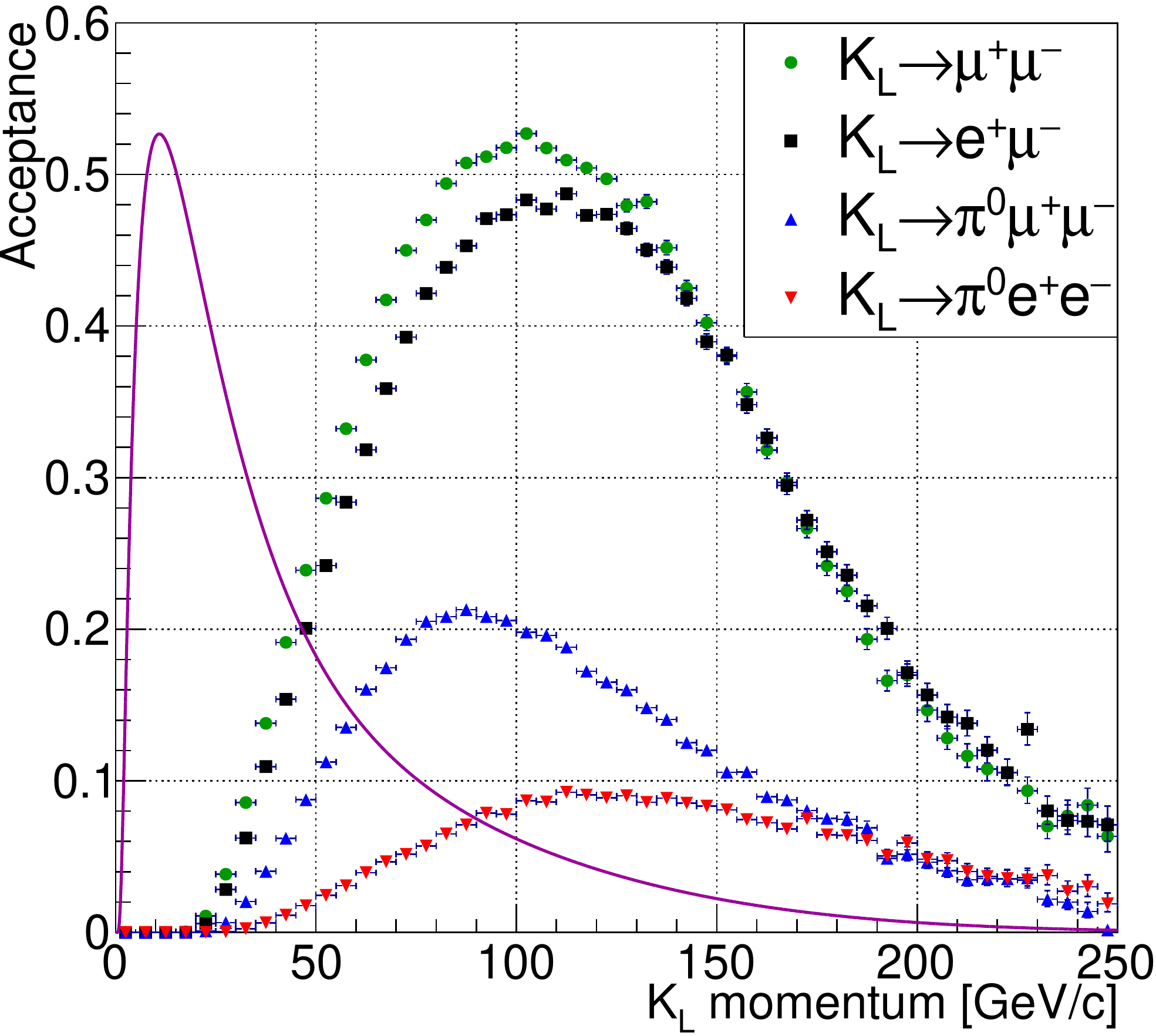}}
\vspace{-7mm}
\end{center}
\caption{Acceptances for rare $K_L$ decays for HIKE Phase~2, in bins of the decay vertex longitudinal coordinate (left) and $K_L$ momentum (right). Momentum distribution of the $K_L$ mesons decaying in the decay volume is shown by a continuous curve in the right panel.}
\label{fig:kl-acceptance}
\end{figure}

The expected $K_L$ yield in the beam is $5.4\times 10^{-5}$ per proton on target, and the assumed integrated proton flux of $1.2\times 10^{19}$~pot/year leads to the number of $K_L$ decays in the decay volume of $3.8\times 10^{13}$/year. The mean momentum of $K_L$ mesons entering the decay volume is 79~GeV/$c$, while the mean momentum of decaying $K_L$ mesons is 46~GeV/$c$ (Table~\ref{tab:beam_prod}, Fig.~\ref{fig:beam_mom_ang}). The expected signal yields for several principal rare and forbidden $K_L$ decays have been evaluated using a full Geant4-based simulation, reconstruction and analysis chain within the flexible HIKE offline software platform being developed on the basis of the NA62 software (Section~\ref{sec:computing}). The results are summarised in Table~\ref{tab:kl-hybrid}, and acceptances in bins of the longitudinal coordinate of the decay vertex and $K_L$ momentum are shown in Fig.~\ref{fig:kl-acceptance}.

\newpage


\begin{table}[tb]
\caption{HIKE Phase~2 sensitivity estimate for the principal rare and forbidden $K_L$ decays: assumed branching ratios (see Section~\ref{sec:KLpi0ll} for details), acceptances and expected signal yields in five years of data taking.}
\begin{center}
\vspace{-3mm}
\begin{tabular}{lccc}
\hline
Mode & Assumed branching ratio & Acceptance & Signal yield in five years \\
\hline
$K_L\to\pi^0 e^+e^-$ & $3.5\times 10^{-11}$ & 2.1\% & 140 \\
$K_L\to\pi^0 \mu^+\mu^-$ & $1.4\times 10^{-11}$ & 6.0\% & 160 \\
$K_L\to\mu^+\mu^-$ & $7\times 10^{-9}$ & 17\% & $2.3\times 10^5$ \\
$K_L\to\mu^\pm e^\mp$ & -- & 16\% & -- \\
\hline
\end{tabular}
\end{center}
\label{tab:kl-hybrid}
\end{table}

For the $K_L\to\pi^0\ell^+\ell^-$ decays, the expected SM Dalitz plot distributions~\cite{Isidori:2004rb} are used in the simulation, and a selection condition $m_{ee}>140~{\rm MeV}/c^2$ is applied in the $K_L\to\pi^0 e^+e^-$ case as required to suppress backgrounds from major $K_L$ decays into neutral pions followed by $\pi^0\to\gamma e^+e^-$ decays. The resulting single-event sensitivities for $K_L\to\pi^0\ell^+\ell^-$ decays (Table~\ref{tab:kl-hybrid}) improve by more than two orders of magnitude on the previous searches at the kTeV experiment~\cite{KTeV:2003sls,KTEV:2000ngj}. Suppression of the $K_L\to\gamma\gamma\ell^+\ell^-$ background, which represents a significant limitation especially in the $K_L\to\pi^0 e^+e^-$ case~\cite{Greenlee:1990qy}, relies on the good photon energy resolution provided by the EM calorimeter. The current NA62 LKr calorimeter~\cite{NA62:2017rwk} would provide a 2.2~MeV di-photon mass resolution for the proposed HIKE Phase~2 setup. Further potential backgrounds to $K_L\to\pi^0\ell^+\ell^-$ decays come from major $K_L$ decays such as $K_L\to\pi^+\pi^-\pi^0$ with $\pi^\pm$ misidentification or decay in flight, and from pileup effects.

Background estimation for the $K_L\to\pi^0\ell^+\ell^-$ decays is currently in progress, however the extensive experience of the NA62 lepton flavour/number violation programme in terms of background reduction~\cite{NA62:2019eax,NA62:2021zxl,NA62:2022tte} suggests that the proposed experimental technique allows for background suppression to the level of ${\cal O}(10^{-11})$ for rare and forbidden decays into multiple tracks and photons, for a well-collimated beam. Considering the expected signal yields in excess of 100~events (Table~\ref{tab:kl-hybrid}), we conclude that the first observation of the $K_L\to\pi^0\ell^+\ell^-$ decays is likely, and a sensitivity at the SM branching ratio level of ${\cal O}(10^{-11})$ can certainly be reached. Detailed HIKE Phase~2 sensitivity studies are in progress for the forthcoming proposal.

The expected $K_L\to\mu^+\mu^-$ signal yield suggests a 0.2\% statistical precision on the measurement of the decay branching ratio. We expect that overall a precision better than 1\% is attainable when normalising to the $K_L\to\pi^+\pi^-$ decay, which improves on the previous measurement by the BNL-E871 experiment~\cite{E871:2000wvm}.

HIKE Phase~2 offers sensitivities of ${\cal O}(10^{-12})$ for branching ratios of a broad range of rare and forbidden $K_L$ decay modes. As an example, the acceptance for the lepton flavour violating $K_L\to\mu^\pm e^\mp$ decays (Table~\ref{tab:kl-hybrid}) leads to expected upper limits of both decay branching ratios of $8\times 10^{-14}$ at 90\% CL in five years of operation (the null hypothesis and assuming low background), improving the limit by a factor of 60 on the previous search by the BNL-E871 experiment~\cite{BNL:1998apv}.

%% file: klever.tex
\section{Phase 3 (KLEVER): measurement of the $K_L\to\pi^0\nu\bar\nu$ decay}
\label{sec:klever} 

Exploratory work on an experiment to measure ${\cal B}(K_L\to\pi^0\nu\bar\nu)$ in a future stage of the ECN3 high-intensity kaon physics programme began even before the start of NA62 data taking. The basic design work for the KLEVER experiment ($K_L$ Experiment for VEry Rare events) was carried out as a part of the first Physics Beyond Colliders initiative at CERN, in preparation for the 2020 update of the European Strategy for Particle Physics~\cite{Ambrosino:2019qvz}.

As in the other phases of the programme, KLEVER makes use of the 400-GeV primary proton beam from the SPS, slow extracted and transported to the T10 target (\Sec{sec:proton_beam}) at an intensity of $2\times10^{13}$ protons per pulse (ppp), corresponding to about six times the nominal NA62 intensity. The neutral secondary beam for KLEVER is derived at an angle of 8~mrad; the beamline design is described in \Sec{sec:neutral_beam}. 
In practice, the choice of production angle has been optimised together with the limits of the fiducial volume (FV) to maximise the signal sensitivity and minimise $K_L\to\pi^0\pi^0$ and $\Lambda\to n\pi^0$ backgrounds.
For a production angle of 8~mrad, the neutral beam has a mean $K_L$ momentum of 40~GeV, so that 5\% of $K_L$ decay inside an FV extending from 130~m to 190~m downstream of the target. The $K_L$ mesons decaying inside the FV have a mean momentum of 27~GeV/$c$. The boost from the high-energy beam facilitates the rejection of background channels such as $K_L\to\pi^0\pi^0$ by detection of the additional photons in the final state. On the other hand, the layout poses particular challenges for the design of the small-angle vetoes, which must reject photons from $K_L$ decays escaping through the beam exit amidst an intense background from soft photons and neutrons in the beam.
Background from $\Lambda \to n\pi^0$ decays in the beam must also be kept under control.

The KLEVER goal is to achieve a sensitivity of about 60~events for the $K_L\to\pi^0\nu\bar\nu$ decay at the SM BR with an $S/B=1$. At the SM BR, this would correspond to a relative uncertainty of about 20\%. We would expect to be able to observe a discrepancy with SM predictions with $5\sigma$ significance if the true BR is a bit more than twice or less than one-quarter of the SM BR, or with $3\sigma$ significance if the true BR is less than half of the SM rate. 
As noted in \Sec{sec:neutral_beam}, with an opening angle of 0.4~mrad, the $K_L$ yield in the beam is $2.1\times10^{-5}$ per proton on target (pot).
With a fiducial-volume acceptance of 5\% and a selection efficiency of 3\%, collection of 60 SM events would require a total primary flux of $6\times10^{19}$ pot, corresponding to five years of running at an intensity of $2\times10^{13}$ ppp under NA62-like slow-extraction conditions with 3000 spills per day for 200 days per year.

The scheduling of the KLEVER phase is influenced by two considerations. Firstly, to provide maximum protection from $\Lambda\to n\pi^0$ background, the beamline would need to be lengthened by 150~m (Section~\ref{sec:neutral_beam_ext}). Secondly, despite the efforts to benchmark the KLEVER beam simulation by comparison with existing data on inclusive particle production by protons on lightweight targets at SPS energies (Section~\ref{sec:kl-beamline-sim}), confidence in the simulations and background estimates would be greatly increased by the possibility to acquire particle production data and measure inclusive rates in the KLEVER beamline setup with a tracking experiment in place. This would be possible in HIKE Phase~2 (Section~\ref{sec:phase2}).
Therefore, although in principle KLEVER could aim to start data taking after LS4 (for which injector physics is currently foreseen to begin in early 2034), the most natural timescale might envisage KLEVER as running after Phase~2. We are studying scheduling options to allow KLEVER data taking to begin before LS5 (foreseen to start in 2038).  


\subsection{Experimental layout}

\begin{figure}[h]
\centering
\vspace{2mm}
\includegraphics[width=\textwidth]{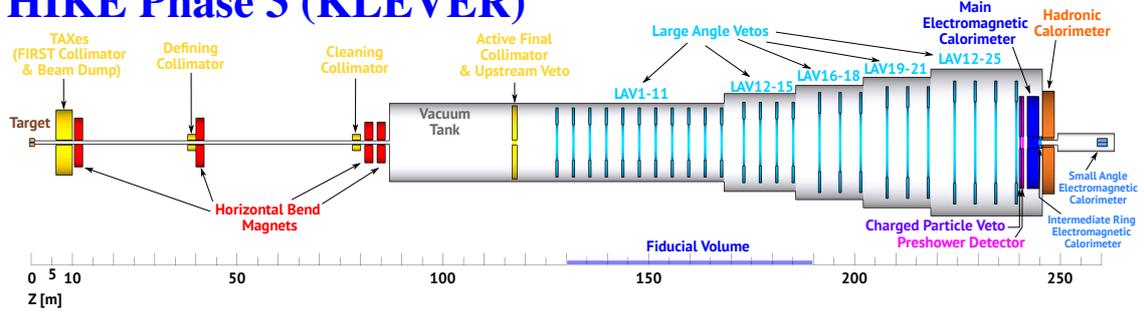}
\put(-425,110){\color{blue}{\bf\LARGE HIKE Phase~3 (KLEVER)}}
\vspace{-2mm}
\caption{HIKE Phase~3 (KLEVER) layout, with an aspect ratio of 1:10. The baseline configuration, without the extension of the beamline by 150 m is shown.}
\label{fig:klever_exp}
\end{figure}

The KLEVER setup largely consists of a collection of high-efficiency photon detectors arranged around a 160-m-long vacuum volume to guarantee hermetic coverage for photons from $K_L$ decays emitted at polar angles out to 100 mrad and to provide a nearly free path through vacuum up to the main electromagnetic calorimeter (MEC) for photons emitted into a cone of at least 7.5~mrad.
The fiducial volume (FV) spans about 60~m just downstream of the active final collimator (AFC), but the photon veto coverage extends along the entire length up to the MEC. The layout of the detector elements is schematically illustrated in Fig.~\ref{fig:klever_exp}.

The largest elements are about 3~m in diameter. The beginning of the vacuum volume is immediately downstream of the cleaning collimator at $z=80$~m, i.e., 80~m downstream of the T10 target. A 40-m vacuum decay region allows the upstream veto calorimeter (UV) surrounding the AFC to have an unobstructed view for the rejection of $K_L$ decays occurring upstream of the detector volume. The UV and AFC at $z=120$~m define the start of the detector volume, which is lined with 25 large-angle photon and charged-particle veto stations (LAV) in five different sizes, placed at intervals of 4 to 6~m to guarantee hermeticity for decay particles with polar angles out to 100~mrad. The MEC, at the downstream end of the vacuum volume,
replaces the NA48 LKr calorimeter used in NA62: it reconstructs the $\pi^0$ vertex for signal events and helps to reject events with extra photons.
A charged-particle veto detector (CPV) in front of the MEC rejects $K_L\to\pi^\pm e^\mp\nu$ and $K_L\to\pi^+\pi^-\pi^0$ backgrounds, and a preshower detector (PSD) allows reconstruction of the angles of incidence for photons, providing additional constraints on signal candidates even if only one photon converts. 
The small-angle vetoes on the downstream side of the MEC intercept photons from $K_L$ decays that pass through the beam pipe. The small-angle calorimeter (SAC) intercepts the neutral beam; its angular coverage as seen from the downstream end of the FV extends to $\pm2$~mrad. Because of the high neutron and photon rate in the beam, the SAC design is one of the most challenging aspects of the experiment. The intermediate-ring calorimeter (IRC) is a ring-shaped detector between the SAC and MEC and intercepts photons from downstream decays that make it through the calorimeter bore at slightly larger angles.
In addition to the photon vetoes, the experiment makes use of hadronic calorimeters downstream of the MEC, to help reject background from hadron interactions and the copious $K_L$ decays into charged particles such as muons and pions.

Because of the experimental challenges involved in the $K_L\to\pi^0\nu\bar\nu$ measurement, and in particular, the very high efficiency required for the photon veto systems, the performance requirements for KLEVER drive the specifications for many of the new detectors for HIKE. In this case, the LAVs, MEC, IRC, SAC, and hadronic calorimeters built for Phases 1 and 2 would all be able to be used for KLEVER in Phase 3. The UV, CPV, and PSD would have to be built new for KLEVER, possibly with recycled components from the analogous detectors for the earlier phases. 

In the baseline design with the original 120 m beamline, the FV covers the region $130~{\rm m}<z<190$~m. The positioning of the FV significantly upstream of the calorimeter, together with the relatively high $K_L$ momentum, is key to obtaining sufficient $K_L\to\pi^0\pi^0$ background rejection with the ring-shaped LAV geometry covering polar angles out to 100~mrad. This comes at a cost in acceptance for signal decays, which increases significantly as the FV is moved closer to the calorimeter. Extending the downstream limit of the FV would increase the sensitivity of the experiment but will require improvements in background rejection; further optimisation along these lines is under study.

Fig.~\ref{fig:klever_exp} shows the experimental configuration without a lengthened beamline to suppress background from $\Lambda$ decays. 
Lengthening the beamline would change the distance from the cleaning collimator to the AFC, so that the cleaning collimator would be upstream of the start of the vacuum tank
instead of right at its entrance. The detector configuration would be otherwise unchanged.


\subsection{Rates and timing performance}
\label{sec:klever_rates}

\begin{table}[tb]
\centering
\caption{Rates for events with hits on KLEVER detectors, by detector system (left) and for certain event classes (right), evaluated for the 120 m beamline configuration.}
\vspace{-2mm}
\begin{tabular}{lc|lc}
\hline
Detector & Event rate (MHz) & Event class & Rate (MHz) \\ \hline
AFC & 2.3 & Exactly 2 hits on MEC & 4.8  \\
UV & 7.1 & Exactly 2 photons on MEC & 1.0 \\
LAV & 14 & 2 hits on MEC with UV, LAV veto & 3.1 \\
MEC & 18 & 2 hits on MEC, no other hits & 0.007 \\
IRC & 22 & & \\
SAC & 95 & & \\
\hline
\end{tabular}
\label{tab:klever_rates}
\end{table}

The FLUKA simulation of the beamline performed by the Conventional Beams Working Group (\Sec{sec:neutral_beam}) contains an idealised representation of the experimental setup for the purposes of evaluating rates on the detectors, both from the decays of $K_L$ mesons and from the beam halo. The estimated rates by detector system
are listed in \Tab{tab:klever_rates}, assuming a primary intensity of $2\times10^{13}$ ppp and a (pessimistic) 3~s effective spill. The simulation was performed for the standard, 120 m beamline (without extension).
The rates are obtained with a geometrical representation of the detector volumes and assumed probabilities for 
hadrons to register signals. 
For the AFC, UV, LAV, and MEC, the probability for registering signals is assumed to be 25\% for neutrons and $K_L$ mesons, and 100\% for all other particles. For the IRC and SAC, the assumed probability is 10\% for neutral hadrons, and 100\% for all other particles. The particle tracking threshold is 1~GeV.
Since a particle detected in any of the detectors would effectively constitute a veto (the rates of signal candidates being negligible for these purposes), this provides an estimate of the total veto rate: 134~MHz, of which 104 MHz from events with hits on a single detector and 30 MHz from events with hits on multiple detectors.
The total rate is dominated by the rate of interactions of beam particles in the SAC, consisting of 13 MHz of hits from $K_L$ mesons, 44~MHz of hits from neutrons, and 40~MHz of hits from beamline photons with $E>5$~GeV.
Hadronic interactions can be efficiently recognised offline and we assume only 10\% of the corresponding rate contributes to the inefficiency from accidental coincidence. The remaining veto rate from accidentals is then at most 100 MHz, which we take as a figure of merit. The event time is obtained from the $\pi^0$ candidate reconstructed in the MEC, and the accidental
coincidence rate is dominated by events on the SAC. Assuming a $\pm5\sigma_t$ coincidence window, to limit the inefficiency from accidental coincidence to below 25\%, the time resolution $\sigma_t$ on the coincidence must be better than 250~ps, and the detector design must minimise any contributions from non-gaussian fluctuations to the timing distribution. These considerations lead to the specification that the nominal time resolution for the most critical detectors (the MEC and the SAC) must be better than 100~ps.

As noted above, the full FLUKA simulation was performed with the original, 120 m beamline configuration. Many of the rates in \Tab{tab:klever_rates} would be expected to decrease by the ratio of beam solid angle, i.e., by a factor of 0.41, when the beamline is extended. This scaling may not be exact for the upstream detectors (AFC, UV, LAV), as the rates on those detectors are determined in part by the details of the collimation and sweeping scheme, but is is certainly true for the downstream detectors, and in particular, for the SAC, which drives the random veto rate for the experiment. Thus, the downstream extension not only provides a significant general decrease in the rates on the detectors, but would lower the estimated random veto rate from 25\% to 10\%.

The rates for certain classes of events to be acquired are also listed in \Tab{tab:klever_rates}: for events with exactly two hits and with exactly two photons on the calorimeter, independently of the other detectors; for events with two hits on the calorimeter without hits on the UV or LAV (presumed to be in online veto); and for events with two hits on the calorimeter and no hits on any of the other detectors. These are estimates of the rates for the physical events---there is no simulation of the detector response apart from the efficiency assumptions outlined above---but these results do suggest that the total rates for event classes that would approximately correspond to level-0 trigger conditions in NA62 are in fact of about the same order of magnitude as in NA62.

\begin{figure}
\centering
\includegraphics[width=0.5\textwidth]{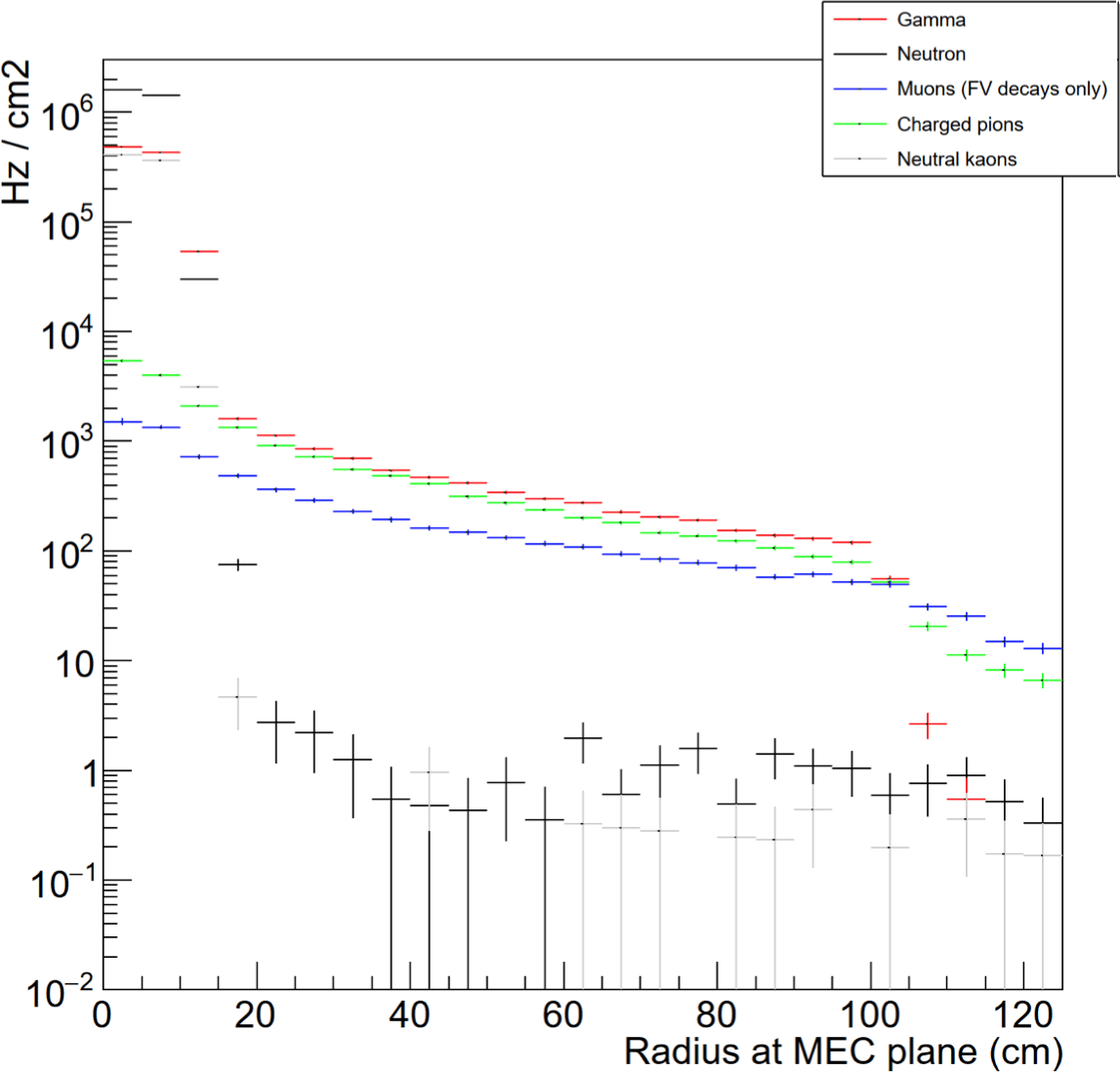}
\vspace{-2mm}
\caption{Radial distribution of particle fluxes at the MEC for KLEVER, from FLUKA beamline simulation.}
\label{fig:klever_halo}
\end{figure}

Fig.~\ref{fig:klever_halo} shows the radial distribution of the particle fluxes at the front face of the MEC, as estimated with the FLUKA simulation of the neutral beam. The beam halo from photons and neutral hadrons drops off rapidly for
$10 < r < 20$~cm, with $r$ the radial distance from the beam axis. For neutrons, the halo-to-core ratio decreases from $2.5\times10^{-2}$ to $8.8\times10^{-5}$ over this interval, corresponding to a decrease in the total neutron rate in the halo from 12~MHz to 43~kHz. For $r>20$~cm, the neutral hadron fluxes are on the order of a few Hz/cm$^2$. At $r = 20$~cm, there are 
1~kHz/cm$^2$ of photons and charged pions and a few hundred Hz/cm$^2$ of muons, dropping off more or less linearly with radius. 

Once the extended beamline layout is finalised, it will be necessary to re-run the FLUKA-based simulation to obtain definitive estimates of the beam flux, halo, and veto rates in the detectors.  

\subsection{Expected performance for $K_L\to\pi^0\nu\bar{\nu}$}
\label{sec:klever_sens}

Simulations of the experiment in the standard configuration carried out with fast-simulation techniques (idealised geometry, parameterised detector response, etc.) suggest that the target sensitivity (60 SM events with $S/B = 1$) is achievable. However, due to the the reduction in the beam solid angle, the $K_L$ flux is decreased by 60\% when the beamline is extended by 150~m. We are exploring options to increase the acceptance for signal decays. The acceptance for $K_L\to\pi^0\nu\bar{\nu}$ decays in the fiducial volume is only a few percent due to the cuts used to reject background, primarily from $K_L\to\pi^0\pi^0$ decays, so increasing the signal acceptance is closely related to optimising the background rejection. 

\subsubsection{Signal selection and rejection of $K_L$ background}

The basic criteria for signal selection are as follows:
\begin{itemize}
    \item Events with exactly two photon clusters on the MEC and no other activity in the detector, including in any of the photon vetoes, are selected.
    \item Assuming that the two photons have invariant mass equal to $m_{\pi^0}$ and that the $K_L$ decays on the nominal beam axis ($r=0$), the position of the decay vertex $z_{\rm rec}$ is calculated from the distance between the photon clusters on the MEC. The vertex is required to lie inside the FV.
    \item The requirement that both photons must have radius $r > r_{\rm min} = 35$~cm from the beam axis helps to increase the rejection for events with overlapping clusters, provides a safeguard against background from $K_L\to\pi^0\pi^0$ decays upstream of the AFC, and helps to reject residual $\Lambda\to n\pi^0$ decays.
    \item A specific background topology with two photons at large angle from an odd-paired $K_L\to\pi^0\pi^0$ decaying just upstream of the MEC but falsely reconstructing in the FV is rejected by the requirement that the photon of lower energy have $E_{\rm min} > 2~{\rm GeV}/r_{E_{\rm \min}}$, where $r_{E_{\rm min}}$ refers to the photon with $E = E_{\rm min}$.
    \item Signal events are required to have $p_{\perp\,{\rm min}} > 140$~MeV, where $p_\perp$ refers to the $\pi^0$ as reconstructed by the MEC. 
    \item At least one photon is required to convert in the PSD, allowing reconstruction of the vertex position and transverse momentum with relaxed assumptions on the radial position of the decaying $K_L$. The vertex position $z_{\rm PSD}$ must be upstream of the downstream edge of the FV and the transverse momentum $p_{\perp \,{\rm PSD}}$ must be greater than $p_{\perp\,{\rm min}}$.     
\end{itemize}

\begin{figure}
\centering
\includegraphics[width=\textwidth]{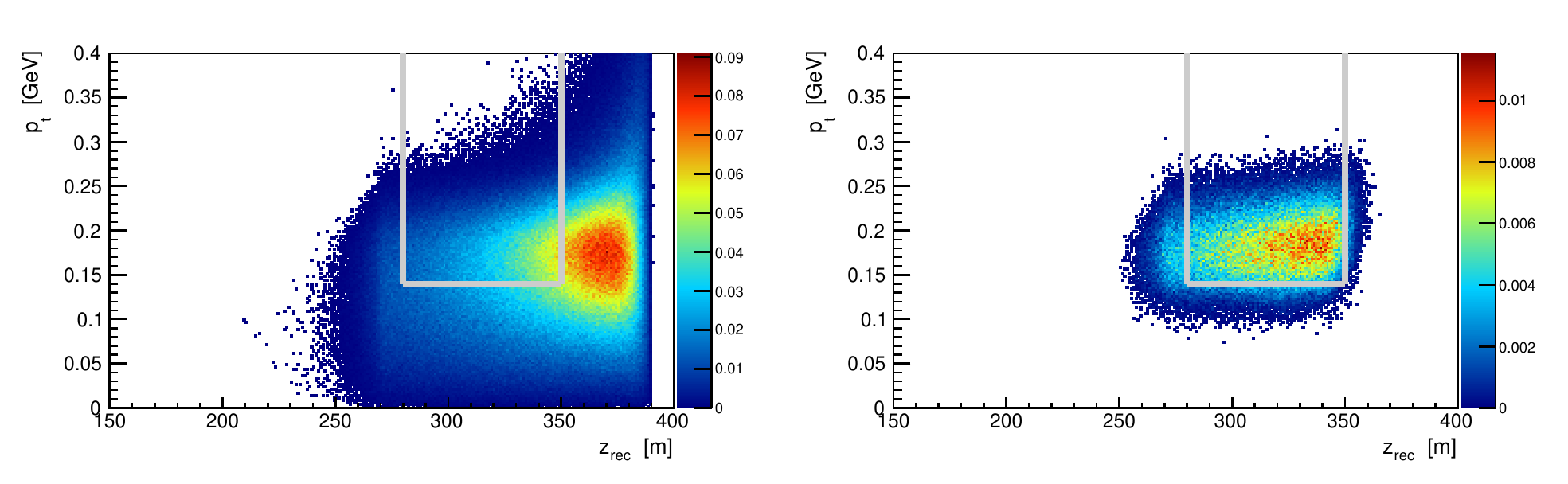}
\caption{Distribution of $K_L\to\pi^0\nu\bar{\nu}$ events in the $(z_{\rm rec}, p_\perp)$ plane, from fast MC simulation of the configuration with beamline extension, for all events with two photons on the MEC (left), and after $r_{\rm min}$, $E_{\rm min}$, and preshower cuts (right).}
\label{fig:sens_sig}
\end{figure}

\Tab{tab:sens_sig} shows the signal acceptance from the fast simulation. The numbers of events at each stage of selection when the above cuts are sequentially applied are listed, both for the original 120~m and extended 270~m beamline configurations, assuming ${\cal B}(K_L \to \pi^0\nu\bar{\nu}) = 3\times10^{-11}$. Note that for the extended beamline configuration, the FV extends 10 m further downstream than it does for the original configuration.
Plots of the distributions of the events satisfying these minimal criteria in the
plane of $p_\perp$ vs $z_{\rm rec}$ are shown in Fig.~\ref{fig:sens_sig}, for the extended beamline configuration only. The two panels show the distributions for all events with two photons on the MEC, and with the addition of the $r_{\rm min}$, $E_{\rm min}$, and PSD requirements. Two features are apparent. Firstly, the acceptance for signal events increases dramatically in proximity to the calorimeter: the majority of the events with two photons on the MEC actually lie downstream of the FV. Secondly, the $p_\perp$ resolution degrades rapidly downstream of the FV, due to the error in $p_\perp$ reconstruction resulting from ignoring the radial displacement of the decaying $K_L$. Note finally that the PSD cuts partially enforce the definition of the signal box, which explains the shape of the distribution in Fig.~\ref{fig:sens_sig} (right). For events passing the FV and cluster reconstruction cuts, the $p_\perp < 140$~MeV selection retains 60\% of signal events.

\begin{table}
\centering
\caption{Estimated sensitivity for $K_L\to\pi^0\nu\bar\nu$.}
\vspace{-2mm}
\begin{tabular}{llccc}
\hline
\rule{0pt}{2.5ex}\rule[-1.5ex]{0pt}{0pt} & Stage & Acceptance & Cumulative acceptance & Total events\\ \hline
\multicolumn{5}{l}{\rule{0pt}{2.5ex}\rule[-1.5ex]{0pt}{0pt}\bf 120 m beamline, 60 m fiducial volume (130--190 m)}\\
& Produced $K_L\to\pi^0\nu\bar{\nu}$ & 1 & 1 & 37800 \\
& \hspace{1cm}Decay in FV & 5.93\% & 5.93\% & 2240 \\
& \rule{0pt}{3ex}$2\gamma$ in MEC & 2.34\% & 2.34\% & 884 \\
& + reconstructed in FV & 0.205 & $7.13\times10^{-3}$ & 269 \\
& + $r_{\rm min} > 35$~cm & 0.504 & $3.60\times10^{-3}$ & 136 \\
& + $E_{\rm min}$ cut & 0.911 & $3.28\times10^{-3}$ & 124 \\
& + $p_\perp > 140$~MeV & 0.568 & $1.86\times10^{-3}$ & $70.4$ \\
& \rule{0pt}{3ex}+ $(z, p_\perp)_{\rm PSD}$ cuts with $0.5X_0$ & 0.469 & $0.874\times10^{-3}$ & $33.0$ \\
& \hspace{1cm}or with $1.0X_0$ & 0.684 & $1.28\times10^{-3}$ & $48.2$ \\ \hline
\multicolumn{5}{l}{\rule{0pt}{2.5ex}\rule[-1.5ex]{0pt}{0pt}\bf 270 m beamline, 70 m fiducial volume (280--350 m)} \\
& Produced $K_L\to\pi^0\nu\bar{\nu}$ & 1 & 1 & 15483 \\
& \hspace{1cm}Decay in FV & 5.07\% & 5.07\% & 785 \\
& \rule{0pt}{3ex}$2\gamma$ in MEC & 1.95\% & 1.95\% & 302 \\
& + reconstructed in FV & 0.429 & $8.38\times10^{-3}$ & 130 \\
& + $r_{\rm min} > 35$~cm & 0.474 & $3.97\times10^{-3}$ & 61.5 \\
& + $E_{\rm min}$ cut & 0.867 & $3.44\times10^{-3}$ & 53.3 \\
& + $p_\perp > 140$~MeV & 0.595 & $2.05\times10^{-3}$ & 31.7 \\
& \rule{0pt}{3ex}+ $(z, p_\perp)_{\rm PSD}$ cuts with $0.5X_0$ & 0.479 & $0.982\times10^{-3}$ & 15.2 \\
& \hspace{1cm}or with $1.0X_0$ & 0.699 & $1.43\times10^{-3}$ & 22.2 \\ \hline
\end{tabular}
\label{tab:sens_sig}
\end{table}

\begin{figure}
\centering
\includegraphics[width=\textwidth]{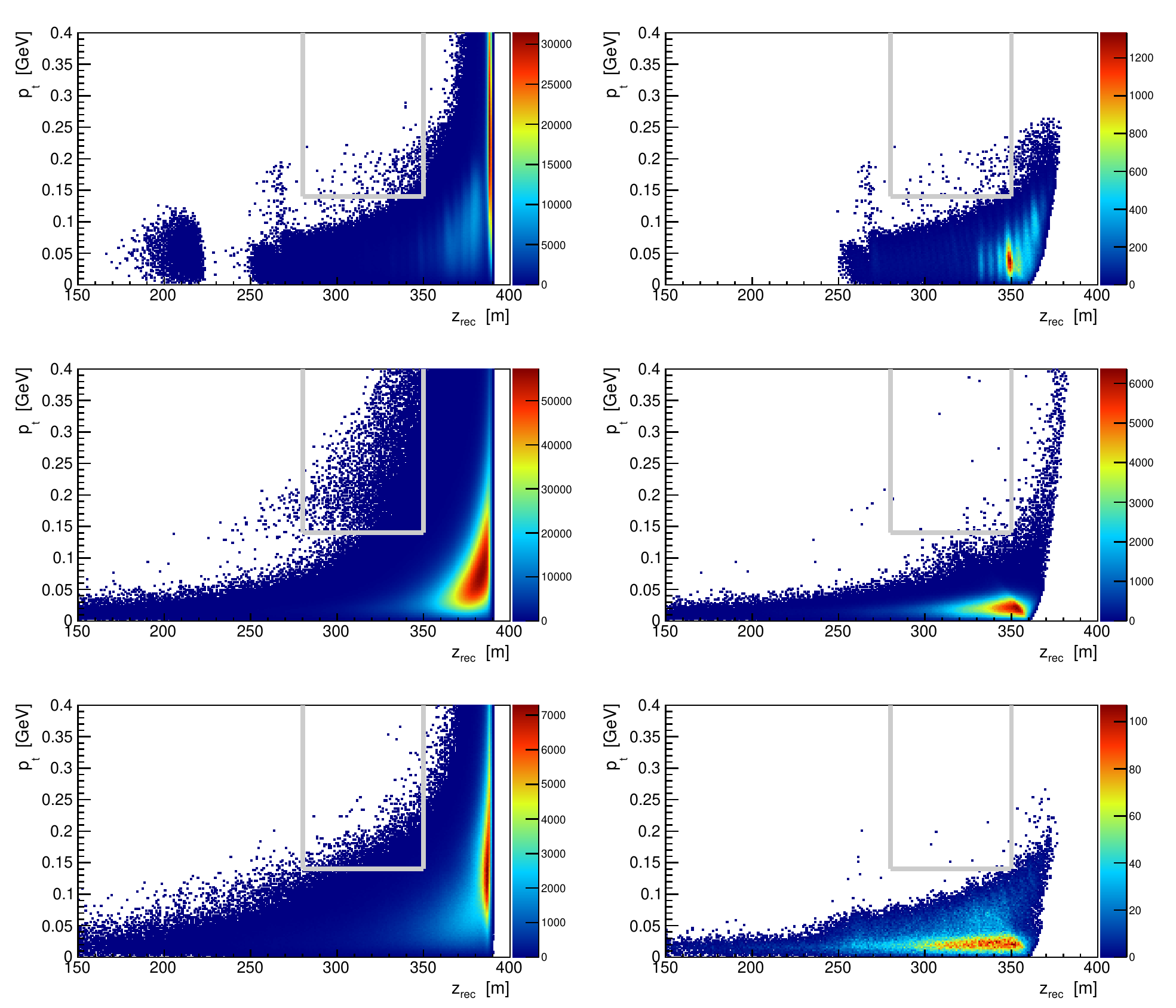}
\caption{Distribution of $K_L\to\pi^0\pi^0$ events in the $(z_{\rm rec}, p_\perp)$ plane, from fast MC simulation of the configuration with beamline extension, for all events with two photons on the MEC (left column), and after $r_{\rm min}$ and $E_{\rm min}$ cuts (right column). The preshower cuts are not applied. From top to bottom, rows show events with the two photons from the same $\pi^0$ (even), with the two photos from different $\pi^0$'s (odd), and with at least one cluster containing overlapping photons (fused).}
\label{fig:sens_2p0}
\end{figure}

In analogy to \Tab{tab:sens_sig}, \Tab{tab:sens_2p0} shows the number of $K_L\to\pi^0\pi^0$ events surviving the various signal selection cuts. Here, it is useful to distinguish between events with the two detected photons from the same $\pi^0$ (even pairings), with the two photons from different $\pi^0$'s (odd pairings), and with at least one cluster containing overlapping photons (fused clusters). As is readily seen from the left panels of Fig.~\ref{fig:sens_2p0}, the background from $\pi^0\pi^0$ events from all three categories is concentrated in the proximity of the MEC, even more so than for signal events. This arises from veto inefficiencies for high-angle, low-energy photons from $K_L$ decays close to the MEC, in addition to the geometrical effect also present for signal events. The background for odd events extends further upstream into the FV than for the other categories, because of the incorrect assumption that the two photon clusters come from a single $\pi^0$. These events are rejected effectively by the PSD.
Differently to Fig.~\ref{fig:sens_sig}, the right panels of Fig.~\ref{fig:sens_2p0} show the distributions without the application of the preshower cuts. This is because, in the case of the background, the cuts on $(z, p_\perp)_{\rm PSD}$ are almost as effective as those on $(z_{\rm rec}, p_\perp)$ from the calorimeter.  
\begin{table}
\centering
\caption{Estimated background from $K_L\to\pi^0\pi^0$.}
\vspace{-2mm}
\begin{tabular}{llcccc}
\hline
\rule{0pt}{2.5ex}\rule[-1.5ex]{0pt}{0pt} & Stage & Total & Even & Odd & Fused\\ \hline
\multicolumn{6}{l}{\rule{0pt}{2.5ex}\rule[-1.5ex]{0pt}{0pt}\bf 120 m beamline, 60 m fiducial volume (130--190 m)}\\
& Produced $K_L\to\pi^0\pi^0$ & \multicolumn{4}{c}{$1.089\times10^{12}$} \\
& \hspace{1cm}Decay in FV & \multicolumn{4}{c}{$5.93\% = 64.5\times10^9$} \\
& \rule{0pt}{3ex}$2\gamma$ in MEC & $3.83\times10^8$ & $0.94\times10^8$ & $2.64\times10^8$ & $0.26\times10^8$ \\
& + reconstructed in FV & $8.40\times10^6$ & $0.70\times10^6$ & $6.70\times10^6$ & $1.00\times10^6$ \\
& + $r_{\rm min}, E_{\rm min}$ cuts & $4.34\times10^6$ & $0.33\times10^6$ & $3.84\times10^6$ & $0.18\times10^6$ \\
& + $p_\perp > 140$~MeV & $261\pm16$ & $154\pm12$ & $86\pm9$ & $21\pm5$ \\
& \rule{0pt}{3ex}+ $(z, p_\perp)_{\rm PSD}$ cuts with $0.5X_0$ & $80\pm9$ & $61\pm8$ & $11\pm3$ & $8\pm3$ \\
& \hspace{1cm}or with $1.0X_0$ & $117\pm13$ & $89\pm11$ & $16\pm5$ & $12\pm4$ \\
\hline
\multicolumn{6}{l}{\rule{0pt}{2.5ex}\rule[-1.5ex]{0pt}{0pt}\bf 270 m beamline, 70 m fiducial volume (280--350 m)}\\
& Produced $K_L\to\pi^0\pi^0$ & \multicolumn{4}{c}{$0.446\times10^{12}$} \\
& \hspace{1cm}Decay in FV & \multicolumn{4}{c}{$5.07\% = 22.6\times10^9$} \\
& \rule{0pt}{3ex}$2\gamma$ in MEC & $9.45\times10^7$ & $2.02\times10^7$ & $6.71\times10^7$ & $0.72\times10^7$ \\
& + reconstructed in FV & $5.12\times10^6$ & $0.48\times10^6$ & $4.07\times10^6$ & $0.58\times10^6$ \\
& + $r_{\rm min}, E_{\rm min}$ cuts & $21.3\times10^5$ & $2.31\times10^5$ & $18.3\times10^5$ & $0.71\times10^5$ \\
& + $p_\perp > 140$~MeV & $234\pm15$ & $175\pm13$ & $48\pm7$ & $11\pm3$ \\
& \rule{0pt}{3ex}+ $(z, p_\perp)_{\rm PSD}$ cuts with $0.5X_0$ & $70\pm8$ & $59\pm8$ & $7\pm3$ & $4\pm2$ \\
& \hspace{1cm}or with $1.0X_0$ & $102\pm12$ & $86\pm11$ & $10\pm4$ & $6\pm3$ \\
\hline
\end{tabular}
\label{tab:sens_2p0}
\end{table}

To ensure that it does not interfere with the detection of low-energy photons by the MEC, the PSD radiator is thin, and the PSD cuts, although effective at suppressing background, penalize the signal acceptance because of the requirement for at least one photon to convert. With a $1X_0$ radiator, the acceptance is increased by about 50\% with respect to that obtained with a $0.5X_0$ radiator, but the optimum thickness will require more detailed simulations to determine. Tables~\ref{tab:sens_sig} and~\ref{tab:sens_2p0} thus give the final expected numbers of signal and background events in both hypotheses. 

For the original 120 m beamline configuration, assuming a $1X_0$ radiator, Tables~\ref{tab:sens_sig} and~\ref{tab:sens_2p0} give 48.2 expected signal events and $117\pm13$ background events from $K_L\to\pi^0\pi^0$.
For the case of the extended beamline, 22.2 signal events are expected after accounting for the reduction of the $K_L$ flux, and $102\pm12$ background events, where the increased background results mainly from extending the fiducial volume by an additional 10~m in the downstream direction to partially make up for the reduced $K_L$ flux.
This scheme is far from final. 
We are continuing to investigate potential improvements to both the experiment and the analysis to increase the sensitivity, reduce the background, and increase redundancy, including the following:
\begin{itemize}
    \item Recovering sensitivity for signal events by further extending the FV in the downstream direction. As noted above, this will require improved background rejection.
    \item Making better use of information from the PSD. In particular, a kinematic fit to combine information from calorimeter and PSD shows promise for improving the $(z, p_\perp)$ reconstruction in the downstream region, where the angles of incidence are greatest and the $p_\perp$ reconstruction is poorest, but where the potential for recovery of signal events is highest.
    \item Continuing to improve the effectiveness of the photon veto systems, especially in the downstream region. At the moment, we assume that the MEC gives no response for photons with $E<100$~MeV (\Sec{sec:klever_mec}). If this threshold can be lowered, which is likely with the design proposed for the MEC, the populations of events in the downstream regions seen in Fig.~\ref{fig:sens_2p0} will be significantly decreased.
    \item Use of multivariate analysis techniques to efficiently combine the event selection criteria. While efforts have been made in this direction, current attention is on development of the full simulation, which is prerequisite to establishing a definitive search strategy. 
\end{itemize}

\subsubsection{$\Lambda\to n\pi^0$ background and beamline extension}
\label{sec:klever_lambda}
As noted in \Sec{sec:neutral_beam_ext}, additional measures must be taken to ensure sufficient suppression of background from $\Lambda\to n\pi^0$ decays.
The most attractive option is to maintain the 8~mrad production angle and increase the length of the beamline from target to AFC by 150 m.%
\begin{table}
\centering
\caption{Estimated background from $\Lambda\to n\pi^0$.}
\vspace{-2mm}
\begin{tabular}{llccc}
\hline
\rule{0pt}{2.5ex}\rule[-1.5ex]{0pt}{0pt} & Stage & Acceptance & Cumulative acceptance & Total events\\ \hline
\multicolumn{5}{l}{\rule{0pt}{2.5ex}\rule[-1.5ex]{0pt}{0pt}\bf 120 m beamline, 60 m fiducial volume (130--190 m)}\\
& Produced $\Lambda\to\ n\pi^0$ & 1 & 1 & $2.00\times10^{14}$ \\
& \hspace{1cm}Decay in FV & $4.70\times10^{-6}$ & $4.70\times10^{-6}$ & $9.42\times10^8$ \\
& \rule{0pt}{3ex}$2\gamma$ in MEC & $7.86\times10^{-6}$ & $7.86\times10^{-6}$ & $1.58\times10^9$ \\
& + reconstructed in FV & 0.351 & $2.76\times10^{-6}$ & $5.53\times10^8$ \\
& + $r_{\rm min}, E_{\rm min}$ cuts & 0.316 & $8.71\times10^{-7}$ & $1.75\times10^8$ \\
& + $p_\perp > 140$~MeV & $6.80\times10^{-3}$ & $5.92\times10^{-9}$ & ($1.19\pm0.04)\times10^6$ \\
& \rule{0pt}{3ex}+ $(z, p_\perp)_{\rm PSD}$ cuts with $0.5X_0$ & 0.132 & $7.81\times10^{-10}$ & $(1.57\pm0.14)\times10^5$ \\
& \hspace{1cm}or with $1.0X_0$ & 0.193 & $1.14\times10^{-9}$ & $(2.29\pm0.21)\times10^5$ \\ \hline
\multicolumn{5}{l}{\rule{0pt}{2.5ex}\rule[-1.5ex]{0pt}{0pt}\bf 270 m beamline, 70 m fiducial volume (280--350 m)}\\
& Produced $\Lambda\to n\pi^0$ & 1 & 1 & $8.22\times10^{13}$ \\
& \hspace{1cm}Decay in FV & $2.07\times10^{-9}$ & $2.07\times10^{-9}$ & $1.70\times10^6$ \\
& \rule{0pt}{3ex}$2\gamma$ in MEC & $2.79\times10^{-9}$ & $2.79\times10^{-9}$ & $2.29\times10^5$ \\
& + reconstructed in FV & 0.445 & $1.24\times10^{-9}$ & $1.02\times10^5$ \\
& + $r_{\rm min}, E_{\min}$ cuts & 0.222 & $2.75\times10^{-10}$ & $2.26\times10^4$ \\
& + $p_\perp > 140$~MeV & 0.0207 & $5.70\times10^{-12}$ & $468\pm54$ \\
& \rule{0pt}{3ex}+ $(z, p_\perp)_{\rm PSD}$ cuts with $0.5X_0$ & 0.0760 & $4.33\times10^{-13}$ & $35.6\pm14.0$ \\
& \hspace{1cm}or with $1.0X_0$ & 0.111 & $6.32\times10^{-13}$ & $52.0\pm20.4$ \\ \hline
\end{tabular}
\vspace{-2mm}
\label{tab:sens_lambda}
\end{table}

\begin{figure}
\centering
\includegraphics[width=\textwidth]{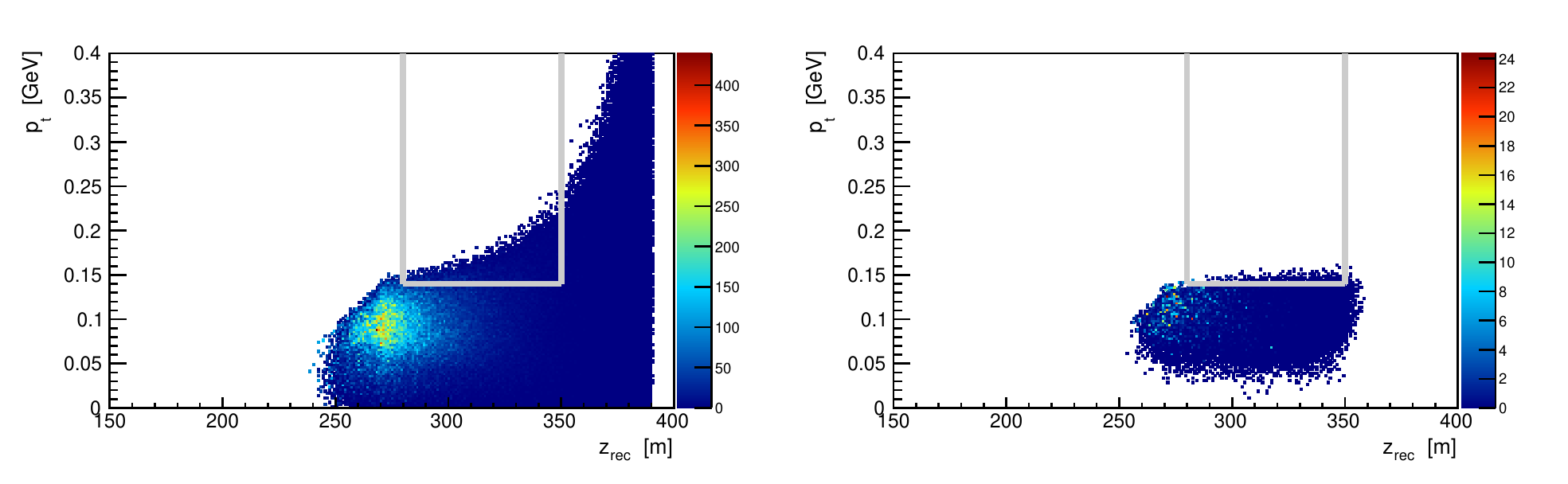}
\caption{Distribution of $\Lambda\to n\pi^0$ events in the $(z_{\rm rec}, p_\perp)$ plane from fast MC simulation of the configuration with beamline extension, for all events with two photons on the MEC (left), and after $r_{\rm min}$, $E_{\rm min}$, and preshower cuts (right).}
    \label{fig:sens_lambda}
\end{figure}
The effectiveness of this solution is demonstrated by \Tab{tab:sens_lambda}, from which it is seen that the beamline extension decreases the number of $\Lambda$ decays with two photons on the MEC by a factor of nearly 7000 and the number passing all cuts by a factor of more than 4000, from 229\,000 to $52\pm14$. 
The centre-of-mass decay momentum of the $\pi^0$ in $\Lambda\to n\pi^0$ decay is only 104~MeV, and even with the contribution from the neutral beam divergence, the true value of $p_\perp$ is always less than 120~MeV, so the $p_\perp$ and preshower cuts are particularly effective.
The distributions in $(z_{\rm rec}, p_\perp)$ are shown in Fig.~\ref{fig:sens_lambda}. Before the $r_{\rm min}$, $E_{\rm min}$, and preshower cuts, most of the $\Lambda$ decays occur at the most upstream end of the FV: the 70~m FV spans more than 3 average decay lengths. The distribution for the 52 surviving events is seen in the right panel. Most of these events can be eliminated using a hadronic calorimeter at small angle backing up the SAC to veto the high-energy neutron from the $\Lambda\to n\pi^0$ decay. Because of the mass asymmetry in the decay, the neutrons from decays in the FV that pass analysis cuts are {\em always} emitted into the SAC acceptance and have an energy distribution with a mean of 220~GeV; 98\% of these neutrons have $E>150$~GeV. Relatively few of the beam neutrons have energies this high (Fig.~\ref{fig:beam_flux}); a hadronic calorimeter with a threshold at $E=150$~GeV would add less than 10~MHz to the total ${\cal O}$(100~MHz) veto rate for the extended configuration. The concept requires full simulation, but it seems reasonable to assume that the residual $\Lambda$ background could be reduced by an order of magnitude or more.

\subsubsection{Other backgrounds}
In addition to the backgrounds from $K_L\to\pi^0\pi^0$ and $\Lambda\to n\pi^0$, there are a number other background channels that require further advances in the simulation to properly study, and in particular, the full incorporation of the detailed FLUKA beamline simulation into the HIKE Monte Carlo. 
While mitigation of potential background contributions from these sources might ultimately require specific modifications to the experimental setup, we expect this task to be less complicated than dealing with the primary challenges from $K_L\to\pi^0\pi^0$ and $\Lambda\to n\pi^0$.

\paragraph*{$K_L\to 3\pi^0$}
Background from $K_L\to 3\pi^0$ is not expected to be significant due to the relative improbability of losing four photons. A preliminary study was performed with the fast simulation for an earlier configuration of the experiment.
Essentially all of the selected events have the $K_L$ vertex very near the calorimeter; the detected photons have low energy and are concentrated at very small radius. For combinatoric reasons, there are significantly more events reconstructed in the FV with one or more photons lost to cluster fusion than in the case of $K_L\to\pi^0\pi^0$.
The fiducial volume cut and $r_{\rm min}$ cuts are highly effective at eliminating these events, and the residual background has very low $p_\perp$. This study will be repeated with the updated experimental configuration at a level of statistics sufficient to rule out or evaluate residual contributions to the background. 

\paragraph*{$K_L\to\gamma\gamma$}
This channel has also been studied by fast simulation. Because the transverse momentum of the $\gamma\gamma$ pair is zero in the rest frame, the $p_\perp$ cut used to reject $K_L\to\pi^0\pi^0$ background is extremely effective, and no events are expected from $K_L\to\gamma\gamma$ decays in the core of the neutral beam. Any background from this channel must come from the decays of $K_L$'s scattering from the collimator surfaces. 
The KOTO experiment has reported background from $K_L\to\gamma\gamma$ at a level corresponding to about 6 events at SM sensitivity~\cite{KOTO:2020prk}. Cluster shape cuts to constrain the angles of photon incidence were helpful in reducing this background. KLEVER has numerous advantages, including a long beamline with four stages of collimation, resulting in beam with an excellent core-to-halo ratio, and reconstruction constraints from the PSD.

\paragraph*{Decays to charged particles}
The role of the CPV in rejecting decays with charged particles in the final state is discussed in \Sec{sec:klever_detectors}. An important decay against which the CPV not completely effective is $K_L\to\pi^+\pi^-\pi^0$: because of the low decay momentum, it is possible for the $\pi^+$ and/or $\pi^-$ to escape down the beam pipe. KOTO has resolved this problem by installing additional charged particle vetoes along the walls of the downstream beam pipe, and it is straightforward to adopt the same solution for KLEVER.
Radiative semileptonic decays and $K_{e4}$ decays will also require simulation with the full HIKE MC.

\paragraph*{Beam interactions on collimator surfaces}
There are a number of potential backgrounds from the inelastic scattering of beam particles from the collimator surfaces.
Hyperons and $K_S$ mesons can produced in these interactions. Compared to the $\Lambda$ and $K_S$ components of the neutral beam, these secondaries have much lower momentum and should decay away if produced on the upstream collimators. Scattering in the AFC is more dangerous, so further beamline optimisation, including the possibility of introducing additional stages of collimation, may be necessary.
Based on KOTO's experience, $K^0p\to K^+n$ charge exchange is possibly the most important process~\cite{KOTO:2020prk}, because the $K^+$ can undergo $K_{e3}$ decay inside the FV with the $\pi^0$ reconstructed and the $e^+$ emitted at very low energy and at very high angle. KOTO is exploring the use of very thin upstream charged particle detectors to reduce this background. In KLEVER, the first three stages of collimation are followed by magnetic sweepers, so this background can only arise from scattering in the AFC, which for this process leaves a signal to veto.

\paragraph*{Neutron interactions in the MEC}
Neutron interactions in the calorimeter can create background due to the possibility for hadronic shower fragments to travel a substantial distance in the calorimeter before reinteracting, creating two clusters. The principal strategy for abatement of this background is careful beamline design to reduce the neutral beam halo. This is also the logic behind widening the calorimeter bore so that the beam penumbra intercepts the IRC downstream of the MEC. In addition, the particle-identification capabilties of the calorimeter play an important role. KOTO, which uses a CsI calorimeter, identifies neutron interactions via cluster shape analysis and double-sided light readout. The MEC design permits optimisation of the transverse segmentation to allow cluster shape analysis, while longitudinal shower information is obtained from the use of spy tiles (\Sec{sec:klever_mec}). Since the halo is concentrated on the inner regions of the calorimeter, the $r_{\rm min}$ cut is expected to be effective. KLEVER will also make use of information from the HIKE hadron calorimeters to reject hadronic showers.   
\paragraph*{Beam-gas interactions}
The background from single $\pi^0$ production in interactions of beam neutrons on residual gas has been estimated with the FLUKA-based simulation (\Sec{sec:neutral_beam}) to be at most a few percent of the expected signal for a residual gas pressure of $10^{-7}$~mbar.

\subsubsection{Outlook} 
As discussed in the previous sections, the current simulations give 22.2 expected signal events, $102\pm12$ expected background events from $K_L\to\pi^0\pi^0$, and $52\pm20$ expected background events from $\Lambda\to n\pi^0$, which can potentially be reduced to the low single digits or eliminated by the introduction of a hadronic calorimeter to back up the SAC. We have identified in-principle solutions for the mitigation of the remaining backgrounds that require further study. The designs of the beamline and experiment may require relatively minor modifications, such as the introduction of additional collimation stages or of small detectors to accomplish specific tasks such as upstream or downstream charged-particle veto, but these are well-defined problems amenable to solution.
We also consider the following factors in interpreting the signal and background estimates:
\begin{itemize}
    \item Random veto will reduce both signal and background by about 10\% (\Sec{sec:klever_rates}).
    \item The use of an oriented crystal metal photon converter in the TAX collimator will reduce the scattering of the neutral beam by about 20\%, leading to an effective increase in $K_L$ flux of 10\% (\Sec{sec:kl-beamline-sim}). Since this involves no significant technical difficulty, we assume that this will cancel the loss from random veto. 
    \item Alternatively, the use of a high-$Z$ target could increase the effective $K_L$ flux by up to 30\% due to the dramatic reduction in the thickness required for the photon converter. However, we do not take this into account for the moment, as the associated costs and technical difficulties have not been evaluated. 
\end{itemize}
With these factors taken into consideration, the signal and background estimates from the simulation remain unchanged.

Relative to the sensitivity target of 60 SM events, the apparent shortfall arises principally because of the reduction of the the $K_L$ flux a factor of 2.4 with the extended beamline. With the performance gains expected from reoptimisation of the analysis, with particular attention to extension of the FV, better use of the information from the PSD, and continued improvements to the effectiveness of the photon veto systems with emphasis on the calorimeter design, this loss can largely be recovered, so that the original sensitivity goal remains within reach.

An effort is underway to develop a comprehensive simulation and to use it to validate the results obtained so far. This work is being carried out within the framework of the flexible HIKE Monte Carlo platform, allowing new configurations of the existing and proposed detectors to be simulated on the basis of the NA62 software.

%% file: dump.tex
\section{Operation in beam-dump mode}
\label{sec:dump}


\subsection{Experimental layout}
The HIKE beam dump operation will build upon the experience accumulated in NA62 with beam-dump data taking.
In 2021, $1.4\times 10^{17}$ protons were collected in 10~days of data taking at NA62 in beam-dump mode, with the T10 target used to generate the standard NA62 secondary hadron beam removed from the beamline. Proton beam was made to interact in the NA62 movable collimators, called TAX, located 23~m downstream of T10 within two pair of dipoles, and 80~m upstream of the decay volume. An ad-hoc setting of the dipoles allows a substantial reduction of the rate of muons emitted by pion decays in the proton-induced hadronic showers in the TAX (the so called ``halo'' muons). Secondary interactions of halo muons 
can induce background to the searches proposed, and optimisation of the beamline reduces this background by at least one order of magnitude~\cite{Gatignon:2650989}. On-going and completed studies have shown that the residual background is negligible, in particular when searching for two-body decays of new-physics mediators. In this case, the background rejection is improved by requiring that the trajectory of the mediator points to the beam dump. Based on the NA62 experience, it is assumed that the background will remain negligible at HIKE after an increase of 10 to 50 times in statistics with respect to the NA62 beam-dump operation. Therefore sensitivity projections are performed assuming no background limitation.

Sensitivity projections are produced assuming $10^{19}$ and $5\times10^{19}$ POT accumulated in beam-dump mode.
Operation at $4\times10^{13}$ POT for 4.8~s spills is assumed. When compared to the NA62 2021 beam-dump period, this corresponds to an intensity increase by a factor of 8. Assuming the same efficiencies as for the 2021 NA62 data taking, HIKE will be able to collect $10^{19}$ POT in three to four months. The trigger rate foreseen during beam-dump operation is modest, amounting to less than 200~kHz overall. 

Using a movable dump
allows for a quick switch between a kaon-beam and a beam-dump operation, making it possible to perform specific online calibration procedures in kaon-beam mode during beam-dump operation periods. The importance of this aspect was demonstrated in 2021, significantly improving the overall data quality. A careful consideration concerning the cooling of the dump is needed: the intensity foreseen demands a cooling power exceeding the maximum presently available for the NA62 TAX collimators by a factor of four. Cooling and radio-protection considerations are discussed in Section~\ref{sec:beam}.

As demonstrated by the 2021 NA62 beam-dump data, frequent and accurate calibration of the beam secondary-emission intensity monitors (BSI) is necessary, to ensure an overall uncertainty in the integrated proton flux at the level of 5\%.  
Monitors based on the present BSI-technology are 
adequate to withstand the projected beam intensity. Exploratory experimental studies by the CERN beam-group experts show that the goal of achieving a 5\% uncertainty in the proton flux measurement is within reach. 


\subsection{Physics sensitivity}
\label{sec:DumpSensitivity}

\begin{figure}[p]
\begin{center}
\includegraphics[width=0.6\textwidth]{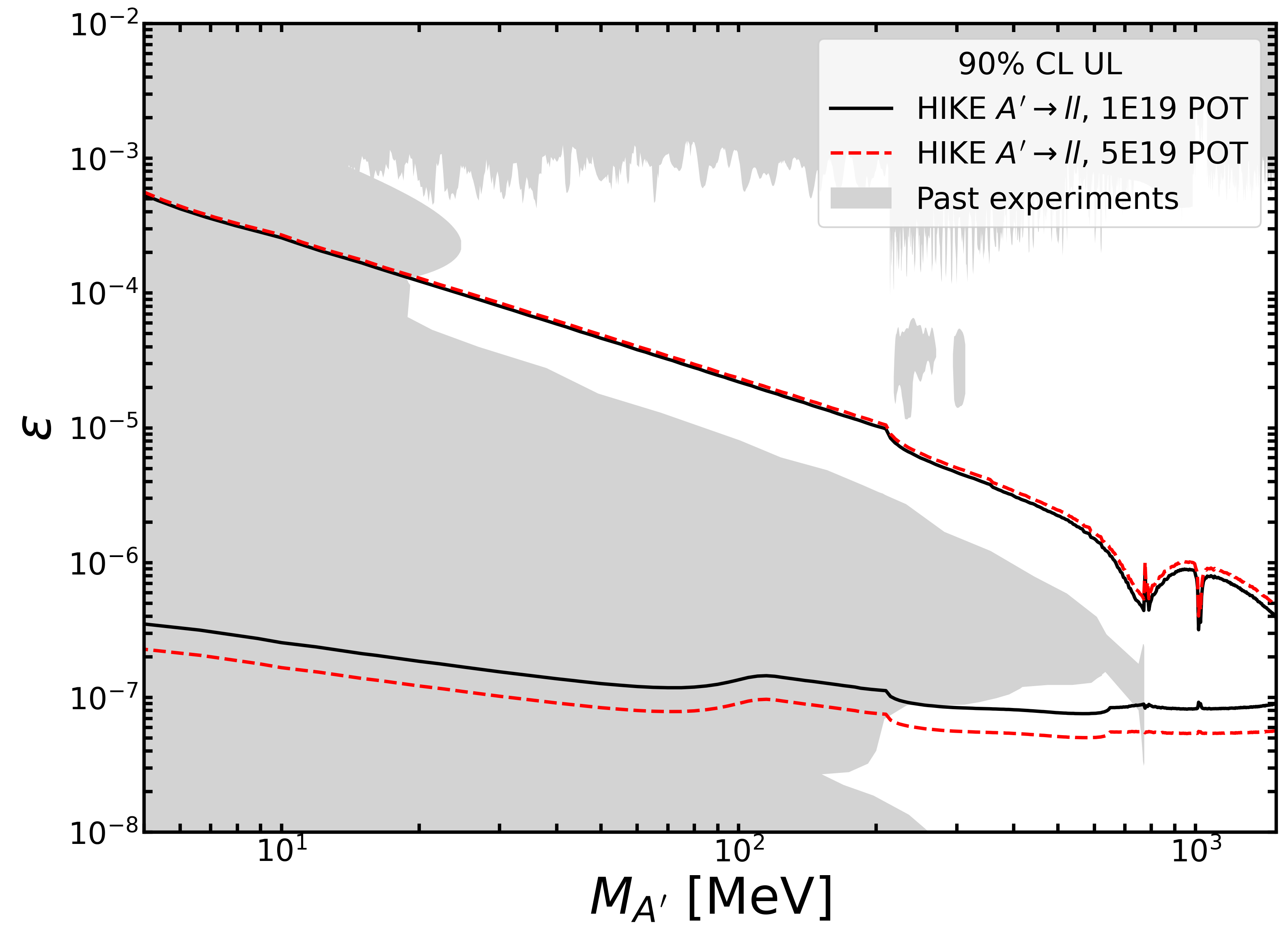}
\caption{\label{fig:excl_BC1} Expected HIKE 90\% CL exclusion limits for the BC1 scenario ($A^\prime\to\ell^+\ell^-$). The currently excluded areas of parameter space are shown in grey.}
\end{center}
\end{figure}

\begin{figure}[p]
\begin{center}
\includegraphics[width=0.48\textwidth]{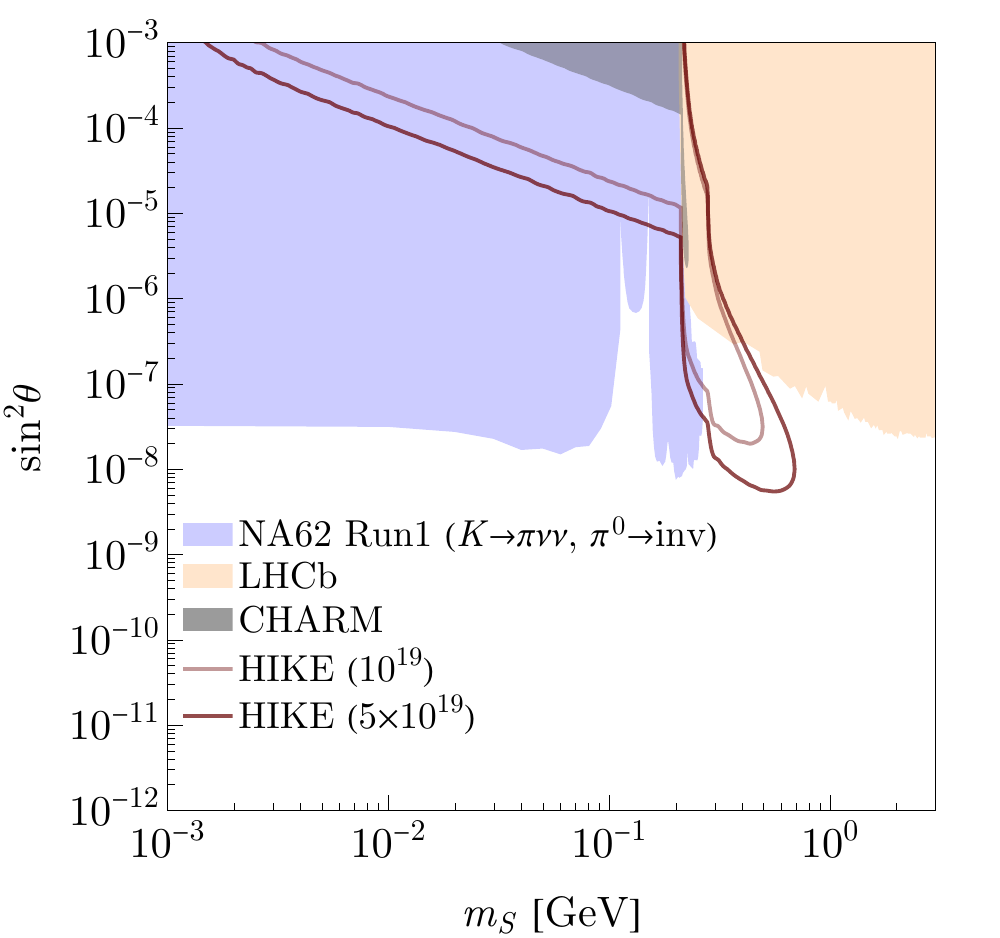}
\includegraphics[width=0.48\textwidth]{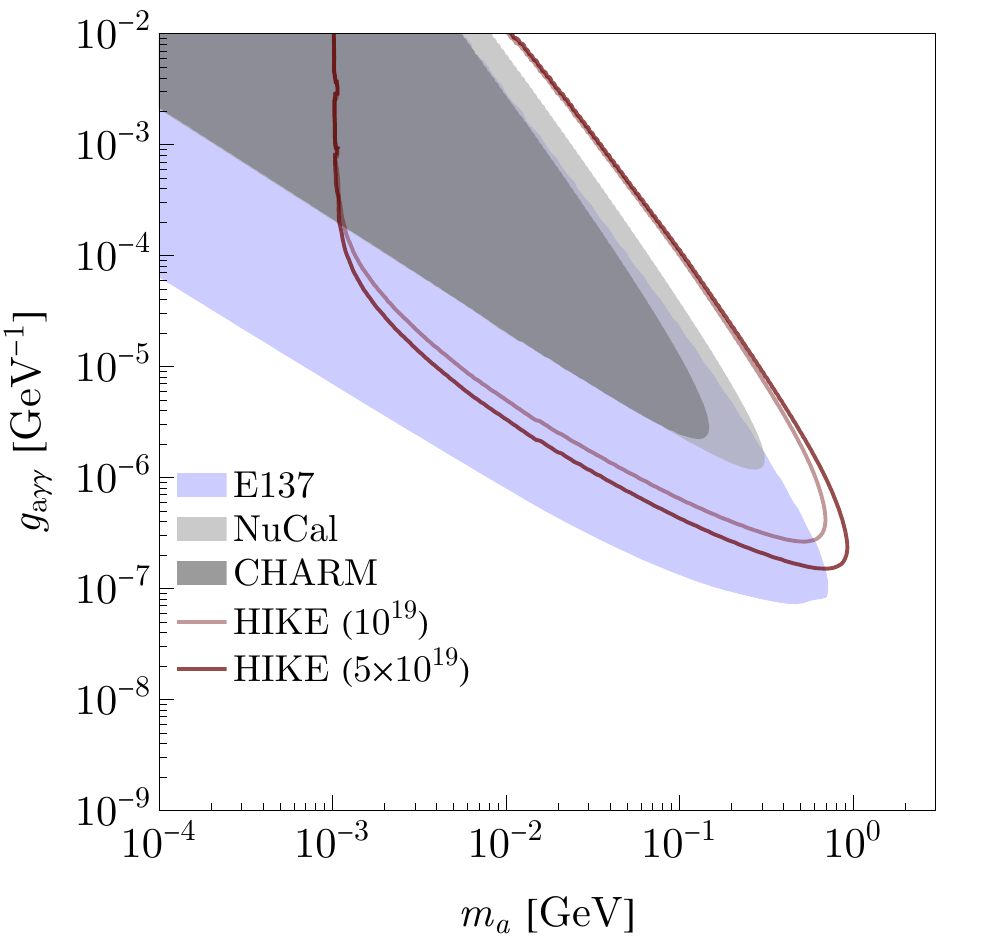}
\caption{\label{fig:excl_BC4BC9} 90\% CL exclusion limits obtained in simulations using the \texttt{ALPINIST} framework with the selection criteria described in~\cite{Jerhot:2022chi}. Current exclusion limits are shown with filled and future projections with empty contours. Left: Search for a dark scalar $S$ in the BC4 scenario. For HIKE, production from $B \to K^{(\ast)} S$ decays and sensitivity to $S\to ee,\mu\mu, \pi\pi, KK$ decays are considered. The $S$ hadronic decay widths derived in~\cite{Winkler:2018qyg} have been used. Exclusions from LHCb~\cite{LHCb:2016awg,LHCb:2015nkv}, NA62 Run 1~\cite{NA62:2021zjw,NA62:2020pwi}, E137, and CHARM measurements are separately shown. The E137 exclusion is obtained using the \texttt{ALPINIST} framework with data provided in~\cite{Dolan:2017osp}.
Right: search for an axion-like particle $a$ in the BC9 scenario. HIKE sensitivity to $a\to \gamma\gamma$ decays is considered.}
\end{center}
\end{figure}

\begin{figure}[p]
\begin{center}
\includegraphics[width=0.41\textwidth]{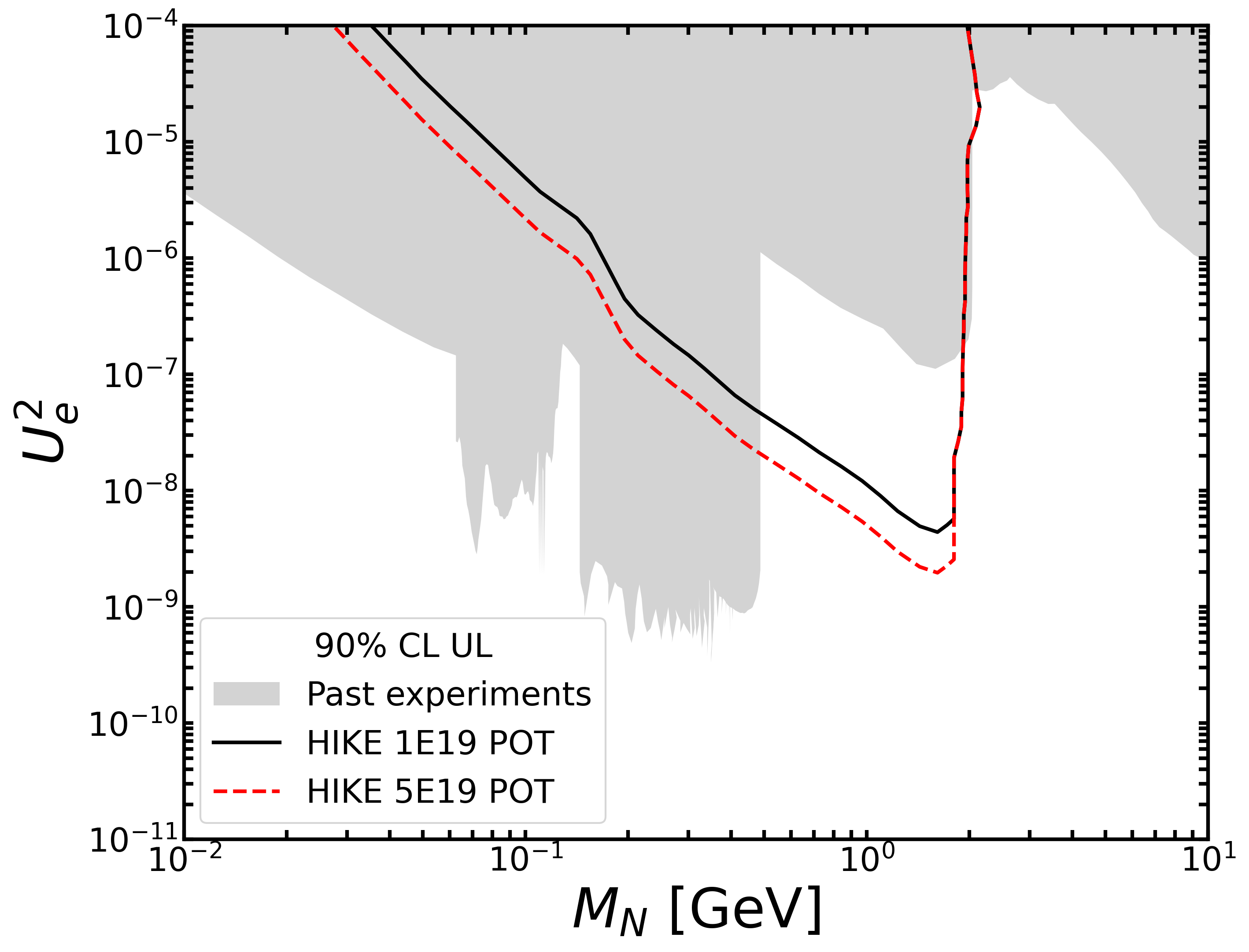}%
\includegraphics[width=0.41\textwidth]{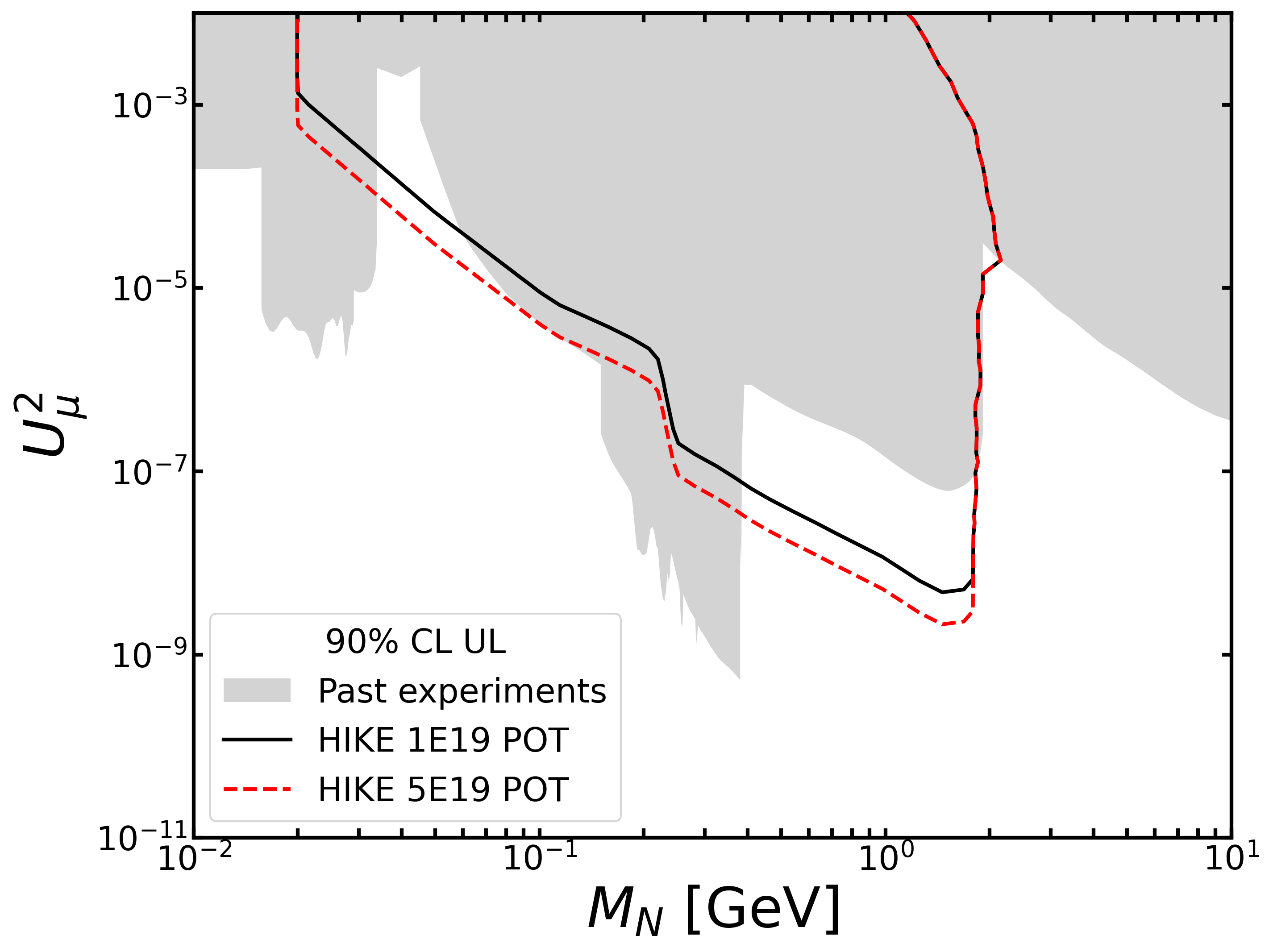}\\
\includegraphics[width=0.41\textwidth]{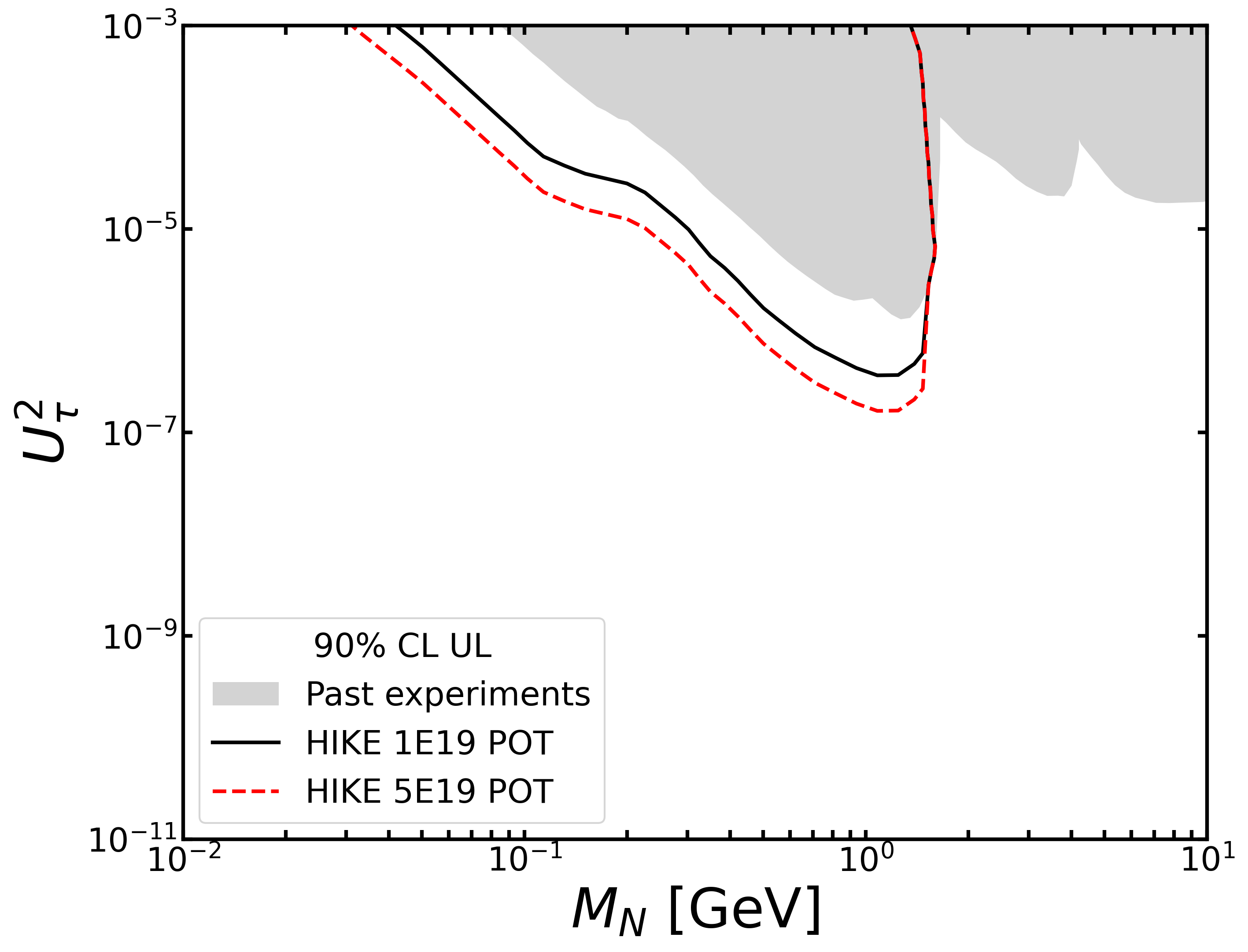}
\vspace{-2mm}
\caption{\label{fig:excl_BC6BC7BC8} HIKE 90\% CL exclusion limits for BC6 (top-left), BC7 (top-centre) and BC8 (bottom) scenarios in the beam-dump mode, assuming sensitivity to two-track final states and NA62 geometrical acceptance. The currently excluded areas of parameter space are shown in grey.}
\end{center}
\end{figure}

\begin{figure}[p]
\begin{center} 
\includegraphics[width=0.5\textwidth]{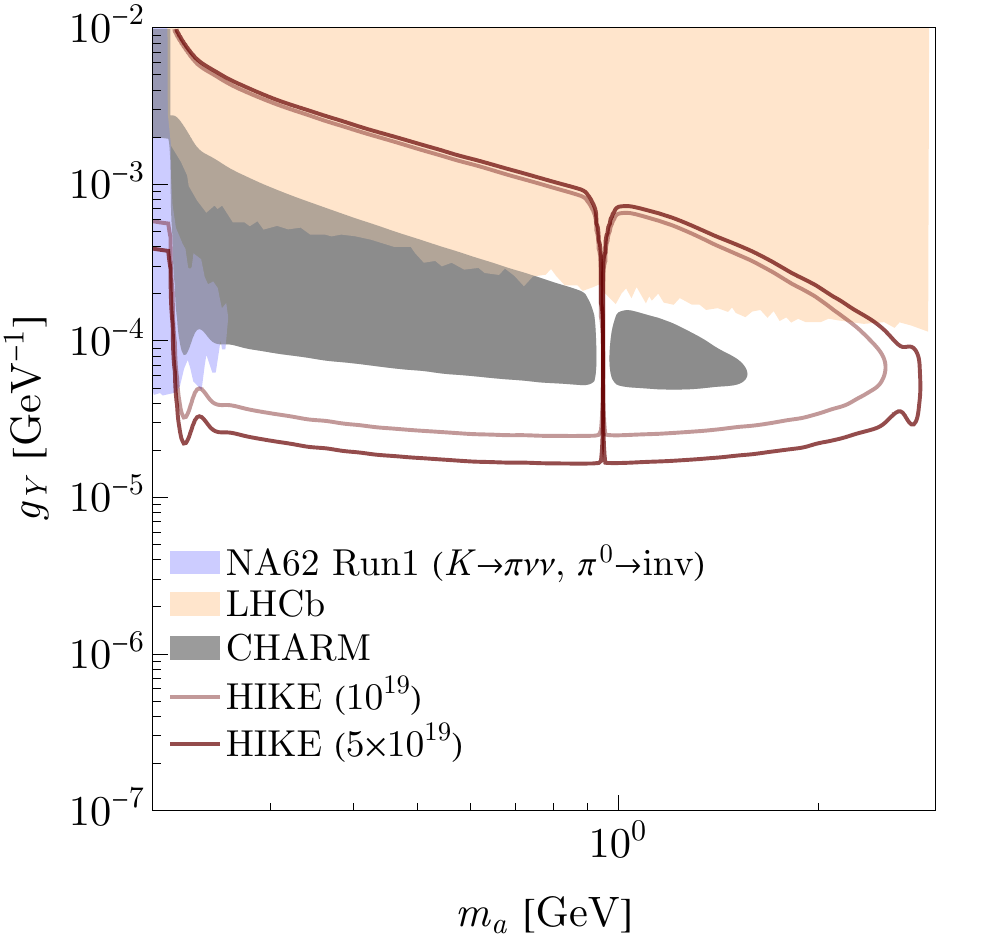}%
\includegraphics[width=0.5\textwidth]{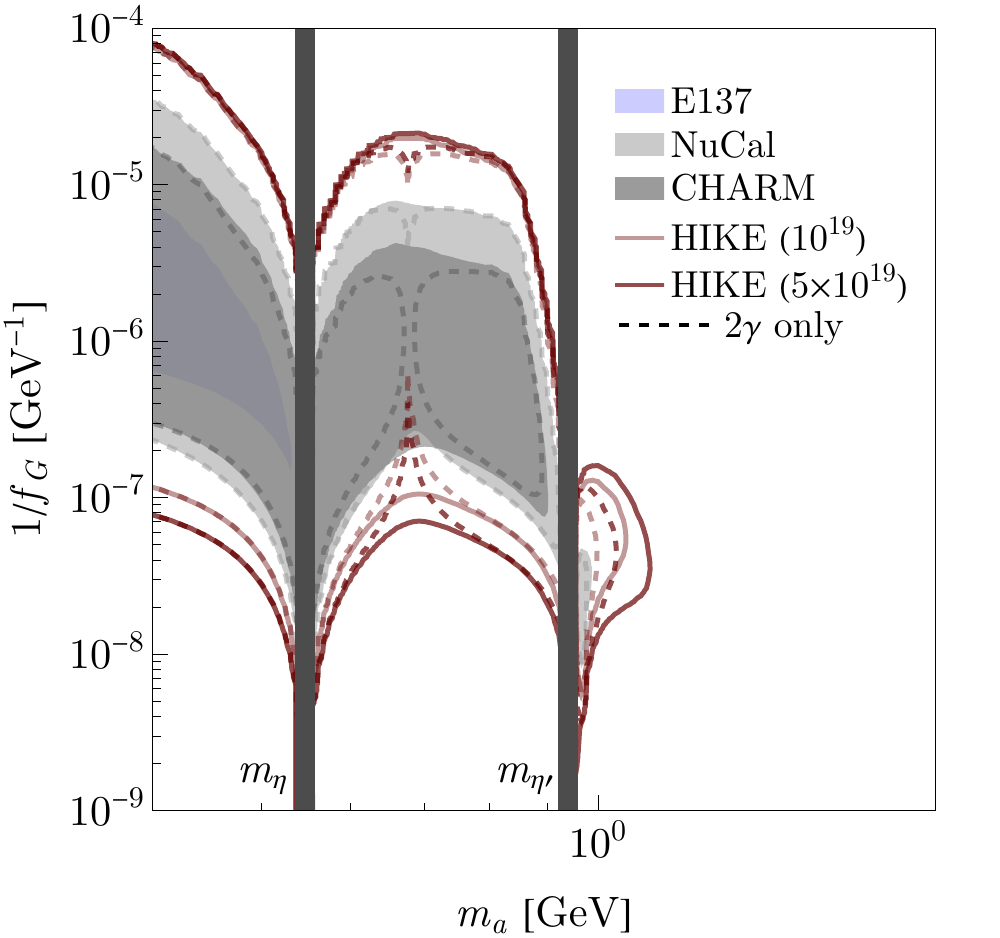}
\vspace{-6mm}
\caption{\label{fig:excl_BC10BC11} Exclusion limits at 90\% CL obtained with the \texttt{ALPINIST} framework for an axion-like particle $a$. Current exclusion limits are shown with filled and future projections with empty contours. In the BC10 scenario (left), we assume HIKE sensitivity to ALP production from $B$ meson decays and to the ALP decays $a \to ee$ and $a \to \mu\mu$. In the BC11 scenario (right), we assume HIKE sensitivity to $a \to \gamma\gamma$ decays (dashed contour) and also to the hadronic decays $a \to \pi^+\pi^-\gamma$, $3\pi$, $2\pi\eta$, etc. (full contour).}
\end{center}
\end{figure}

The projected sensitivity curves of the HIKE experiments in beam-dump mode are evaluated as exclusion plots at 90\% CL, obtained in the assumption of zero observed events and negligible expected background. Note that some of the scenarios are also accessible by HIKE operating in kaon mode (Section~\ref{sec:fips}). With reference to the portal classification discussed in Ref.~\cite{Beacham:2019nyx}, the models relevant to the HIKE sensitivity projections are:
\begin{itemize}
\item[BC1] A new single vector state $A^\prime$, that interacts with the electromagnetic current through a small coupling constant, $\varepsilon$. In the NA62 beam-dump setup, such dark photon can be produced by proton strahlung and through the decays of secondary mesons ($\pi^0$, $\eta^{(\prime)}$, etc.). The dark photon decays to SM particles, as final states with dark matter fields are assumed to be kinematically forbidden. Below 700~MeV, di-lepton decays dominate the $A^\prime$ width. The free parameters of the model are $M_{A^\prime}$ and $\varepsilon$. The expected sensitivity of HIKE to the dark photon production and decay is shown in Fig.~\ref{fig:excl_BC1}. 
\item[BC4] A new scalar state $S$, that interacts with the SM Higgs through a small mixing constant, $\theta$. In the beam-dump setup, such dark scalar is mostly produced through flavor-changing neutral-current decays of $B$ mesons ($B\to K^{(\ast)}S$). The effective coupling constant is proportional to the fermion mass. Hence, when scanning the $S$ mass from MeV to GeV and above, each new decay mode kinematically accessible ($e^+e^-$, $\mu^+\mu^-$, $\pi^+\pi^-$, etc.) tends to either saturate or significantly increase the $S$ width. The free parameters of the model are $m_{S}$ and $\theta$. The expected HIKE sensitivity is shown in Fig.~\ref{fig:excl_BC4BC9}~(left), where we have used the scalar hadronic decay widths from~\cite{Winkler:2018qyg}.
\item[BC6, 7, 8] New fermions $N$, that interact with the SM lepton doublets. These heavy-neutral leptons (HNL) mix with the SM neutrinos through a matrix called $U$. Models BC6, BC7 and BC8 assume the presence of a single HNL of Majorana type and the dominance of electron, $\mu$, or $\tau$ neutrino couplings, respectively. In the beam-dump setup, the production of HNL is dominated by leptonic and semileptonic decays of charmed mesons or $\tau$ leptons, each model favouring the corresponding lepton flavour. The decay width is dominated by meson-lepton or meson-neutrino two-body final states. The free parameters of the BC6,7,8 models are $m_N$ and $U_{e,\mu,\tau}$, respectively. The expected HIKE sensitivity curves are obtained with the same assumptions of~\cite{Drewes:2018gkc} and are shown in Fig.~\ref{fig:excl_BC6BC7BC8}.
\item[BC9, 10, 11] A new pseudoscalar state $a$ (axion-like particle, ALP), that interacts with the SM fields. Models BC9, BC10 and BC11 assume a dominant coupling of an ALP with SM photons, fermions, and gluons, respectively. In BC9, the production in a proton beam dump is dominated by the interaction of photons from the decays of secondary mesons ($\pi^0$, $\eta^{(\prime)}$) with the EM fields of the TAX nuclei. The decay width is dominated by the ALP decay to two photons in the whole $m_a$ range. The free parameters of the BC9 model are $m_{a}$ and the coupling $C_{\gamma\gamma}$. The expected HIKE sensitivity is shown in Fig.~\ref{fig:excl_BC4BC9}~(right). In BC10, the production is dominated by decays of secondary $B$ mesons ($B\to K^{(\ast)} a$). For $m_a$ below 700~MeV, the decay width is dominated by decays to di-lepton final states. The free parameters of the BC10 model are $m_{a}$ and the Yukawa coupling $g_Y$ to the SM fermionic fields. The expected HIKE sensitivity is shown in Fig.~\ref{fig:excl_BC10BC11}~(left). The broad phenomenology of BC11 can reproduce either of the two former models, depending on the ALP mass: one-loop corrections lead to an effective coupling to photons, and an effective coupling to quarks is generated. The former case leads to the ALP di-photon decay; the latter to hadronic decays, e.g. $a\to\pi^+\pi^-\gamma$, but not to leptonic decays. The expected HIKE sensitivity is shown in Fig.~\ref{fig:excl_BC10BC11}~(right).
\end{itemize}


\clearpage
\newpage

%% file: detectors.tex
\section{The HIKE detectors}

The aim is to achieve the same detector performances of NA62 at a factor 4--5 higher beam intensity, with suitable time resolution and rate capability. Describe which detector modifications and readout we need.


\subsection{Detectors upstream of the fiducial decay volume}

\input{detectors/KTAG.tex}
\input{detectors/GTK.tex}

\input{detectors/VC.tex}

\input{detectors/anti0.tex}

\subsection{Fiducial decay volume and its detectors}

\input{detectors/CHANTI.tex}

\input{detectors/LAV.tex}

\input{detectors/STRAW.tex}

\subsection{Detectors downstream of the fiducial decay volume} 

\input{detectors/RICH.tex}

\input{detectors/Timingplane.tex}

\input{detectors/EMCalo.tex}

\input{detectors/IRC}

\input{detectors/HCAL.tex} 

\input{detectors/SAC.tex}

\input{detectors/HASC.tex}

\input{detectors/klever_detectors.tex}

%% file: detectors/KTAG.tex
\subsubsection{Cherenkov kaon tagger (KTAG)}
\label{sec:ktag}

A differential ring-focusing Cherenkov detector (KTAG), used for $K^+$ tagging in the NA62 experiment at a kaon rate of about 40~MHz
provides tagging efficiency above 95$\%$~\cite{NA62:2017rwk,Goudzovski:2015xaa}. The KTAG uses nitrogen radiator gas (to be replaced with hydrogen before LS3) and ring-imaging optics to focus Cherenkov photons onto eight spherical mirrors reflecting them onto eight photodetector arrays. A diaphragm is used so that only Cherenkov light emitted by kaons reaches the photodetectors. Each array is equipped with a matrix of 48 Hamamatsu R7400-U03 and R9880-210 single-anode phototubes (PMTs) with peak quantum efficiency (QE) of 20--40$\%$. The average rate of detected photons per PMT depends on the QE and PMT position and varies between 3--5~MHz, for an effective area of 2.5~cm$^2$. A single-photon time resolution of 300~ps and the mean number of detected photons per kaon of about 20 lead to a 70~ps overall kaon time resolution.

For HIKE Phase~1, the KTAG is expected to provide a time resolution of 15--20~ps and a tagging efficiency above 95\%. Simulations of the KTAG with hydrogen radiator gas show that a mean number of detected photons per kaon of about 30 is expected; the improvement in the light yield with respect to nitrogen radiator is due to newly built ring-imaging optical components optimised for hydrogen.
The expected $K^+$ rate in the HIKE beam is about 200~MHz, corresponding to a 10~MHz/cm$^2$ maximum rate of detected photons (including a safety margin) assuming the NA62 values of the QE and acceptance of the photodetection system. The KTAG should be upgraded with new photodetection, front-end and readout systems to satisfy these requirements. Photodetection devices under consideration for the HIKE KTAG detector are micro-channel plate photomultipliers (MCP-PMTs) which are compact devices providing a single-photon time resolution of 50--70~ps with a low dark count rate (below 1~kHz/cm$^2$), able to operate at the expected maximum photon rate of 10~MHz/cm$^2$. Assuming a gain of $10^6$, 300~days of runtime and a 50\% duty cycle, the above photon rate leads to an integrated anode charge (IAC) of 16~C/cm$^2$.


\begin{figure}[p]
\centering
\includegraphics[width=0.5\linewidth]{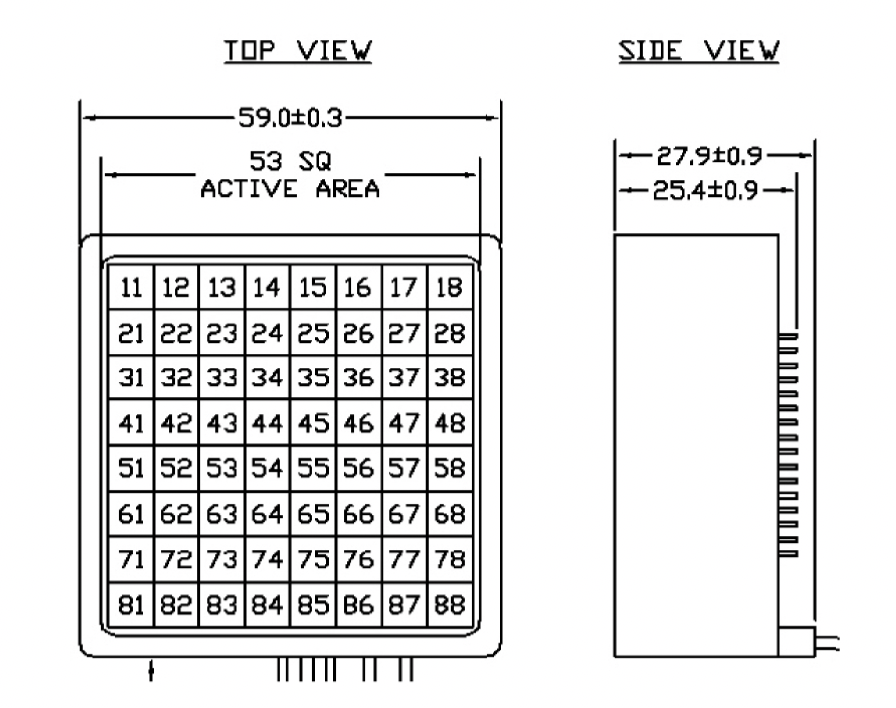}
\caption{Schematics of a Photonis Planacon square-shaped MCP-PMT array under investigation.}
\label{fig:mcp-pmt}
\end{figure}

\begin{figure}[p]
\centering
\includegraphics[width=0.4\textwidth]{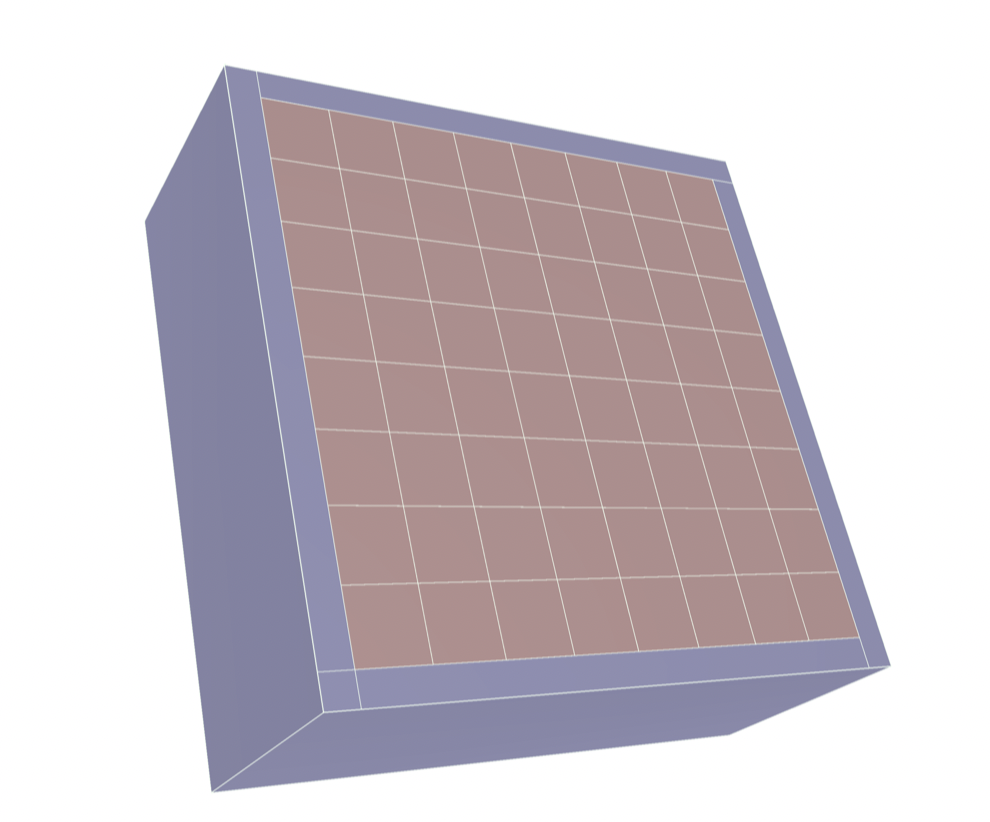}%
\includegraphics[width=0.4\textwidth]{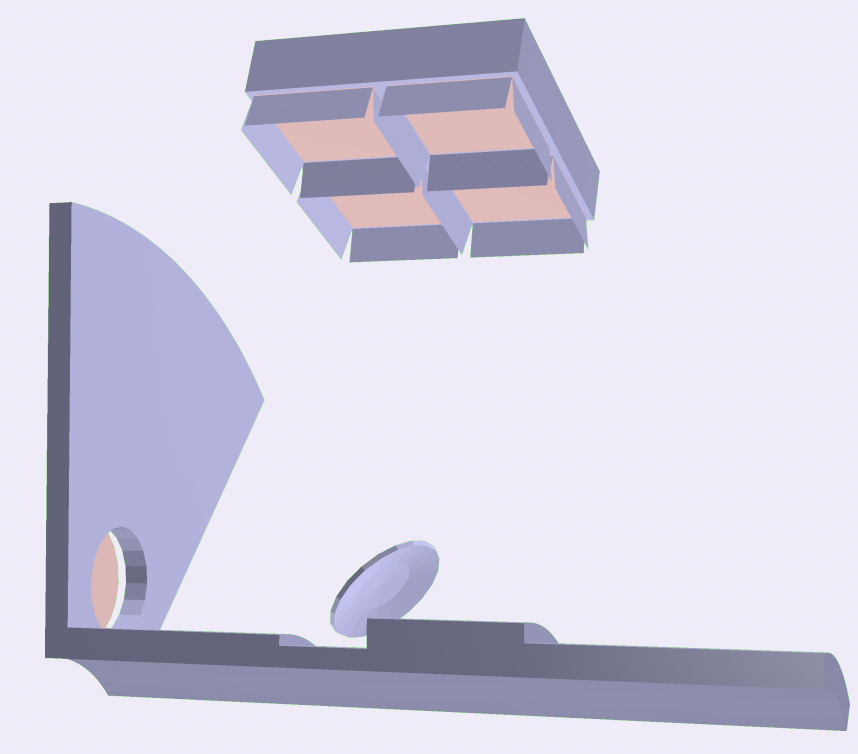}
\caption{Sketches of a single MCP-PMT array (left) and a matrix of four MCP-PMTs (right) to be used to detect photons in each of the eight octants of the HIKE KTAG detector.}
\label{fig:octant}
\end{figure}

\begin{figure}[p]
\centering
\includegraphics[width=0.5\linewidth]{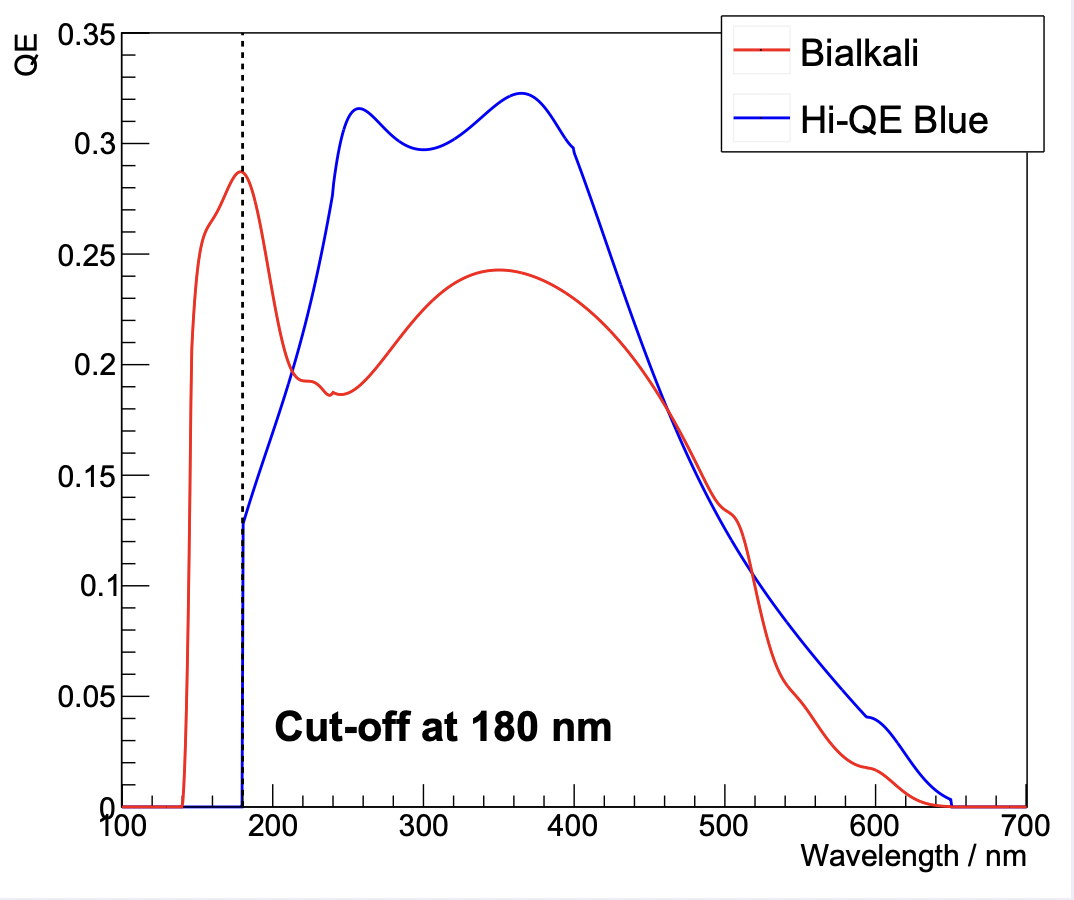}
\caption{QE of the Photonis Planacon MCP-PMTs as functions of wavelength for the two photocathode types considered (standard bialkali and Hi-QE Blue). A cutoff at 180~nm corresponds to the minimum wavelength of the Cherenkov photons reaching the photodetection system.}
\label{fig:QE}
\end{figure}


A limitation of standard MCP-PMTs is their limited lifetime, caused by a rapid decrease in the QE of the photocathode (PC) with increasing IAC. This ageing of the PC is caused by heavy feedback ions from the residual gas. For standard MCP-PMTs the QE was observed to drop by 50\% for IAC of 0.2~C/cm$^2$~\cite{Britting:2011zz}. 
MCP-PMT treatment with atomic layer deposition (ALD) coating increases the lifetime dramatically~\cite{Pfaffinger:2018huh}. The Planacon XP85112 from Photonis, a square-shaped MCP-PMT treated with an improved ALD technique involving two layers, has been demonstrated to reach a lifetime of 33~C/cm$^2$ without any sign of ageing~\cite{LehmanTBC}. This represents an improvement in lifetime by a factor of more than~150 with respect to standard MCP-PMTs, and a factor of at least~10 compared to earlier lifetime optimisation techniques.

A Photonis Planacon MCP-PMT of 2''$\times$2'' size consisting of $8\times 8$ pixel sensors (Fig.~\ref{fig:mcp-pmt}), treated with two-layer ALD coating, is a viable solution for the photodetection system of the KTAG in high intensity applications. 
The Cherenkov photons at each photodetector plane can be detected with a matrix of four MCP-PMTs (Fig.~\ref{fig:octant}). The MCP-PMT is available with standard bialkali and Hi-QE Blue photocathode types; the two QE parameterisations provided by the manufacturer are shown in Fig.~\ref{fig:QE}. Simulation results obtained with a geometrical filling factor of 75\% and a collection efficiency of 60\% show that the 15--20~ps kaon time resolution is achievable. The linearity of MCP-PMTs with rate at high gain is under investigation. However the rate stability can be adjusted by modifying the resistance and capacitance of the MCP coating inside the MCP-PMT.

The above results for the kaon time resolution do not include contributions from front-end and readout systems. The fastIC~\cite{Gomez_2022} and picoTDC~\cite{picoTDC2015Talk} electronics (Section~\ref{sec:ro-boards}) being currently developed for other applications are viable options. The FastIC technology provides a 20~ps time resolution and better linearity for signals from SiPM and MCP-PMTs. PicoTDC chips with 64~channels and 12~ps binning can provide a 4~ps time resolution. Both time contributions will only have a marginal effect on the KTAG timing capability.


%% file: detectors/GTK.tex
\subsubsection{Beam tracker} 
The GigaTracker~\cite{Rinella_2019} is the beam tracker used by the NA62 experiment to measure the momentum and time of the beam particles with high momentum, angular and time resolutions.
The detector is made by planes of hybrid pixel silicon detectors, with a dipole magnets between the planes to allow the momentum measurement. For NA62 Run~1 (2016--2018) three silicon planes were used, then for NA62 Run~2 (2021--LS3) a fourth plane was added.
Each plane is made of a n-in-p planar sensor, with a thickness of \SI{200}{\micro \meter} and a sensitive area of $60 \times 27 ~\rm{mm^2}$, bump-bonded to 10 readout ASICs (TDCPix). The ASICs need to operate with a cooling system: the NA62 GigaTracker used a silicon microchannel cooling plate that represented the first use of this technology in high-energy experiments.
In order to minimise the probability of interactions between the beam particles and the detector, the material budget must be kept as low as possible: the total thickness of a single plane is around \SI{500}{\micro\meter}, namely $0.5\%X_0$.
The NA62 GigaTracker has achieved very good performances, namely a track time resolution of
${\cal O}(100~{\rm ps})$, an angular resolution of 16~$\mu$rad and a momentum resolution of 0.2\%.

Among all the above characteristics of the detector, the time resolution represents a limiting parameter for its operations in the context of HIKE. With more than four-fold instantaneous intensity increase, the time resolution should scale accordingly, namely from less than 200~ps to below 50~ps for a single hit. The radiation hardness of the new detector needs to be improved by at least a factor four if a replacement approach is to be preserved, or more in case of a single installation for the entire HIKE data taking campaign.
Table~\ref{tab:gtk_comparison} summarises the comparison between the current NA62 GigaTracker and the new detector needed for HIKE.

\begin{table}[tb]
\centering
\caption{Comparison between the NA62 Gigatracker main characteristics and the requirements for the HIKE beam tracker. One year corresponds to 200 days of beam.}
\vspace{-2mm}
\begin{tabular}{l|c|c}
\hline
&NA62 GigaTracker& New beam tracker \\
\hline
Single hit time resolution & < 200 ps & < 50 ps \\
\hline
Track time resolution & < 100 ps & < 25 ps \\
\hline
Peak hit rate & $2 ~\rm{MHz/mm^2}$ & $8 ~\rm{MHz/mm^2}$\\
\hline
Pixel efficiency & > 99 \% & > 99 \% \\
\hline
Peak fluence / 1 year [$10^{14} ~ 1~\rm{MeV~n_{eq}/cm^2}$] & 4 & 16 \\
\hline

\end{tabular}
\label{tab:gtk_comparison}
\end{table}

The interest for silicon detectors with fast timing information capable to operate in a high-radiation environment is shared among different particle-physics collaborations, including the LHC experiments
for the high luminosity phase of the collider. 
Following the 2020 Update to the European Strategy for Particle Physics, 4D tracking at high fluences was identified as one of the most urgently needed technologies for future detectors (2021 European Collaboration for Future Accelerators; Detector Research and Development Roadmap DRDT 3.2 and DRDT 3.3). 
Therefore several promising R\&D projects are ongoing and their outcome could be adopted for HIKE.

\begin{figure}[tb]
\centering
\includegraphics[width=0.75\linewidth]{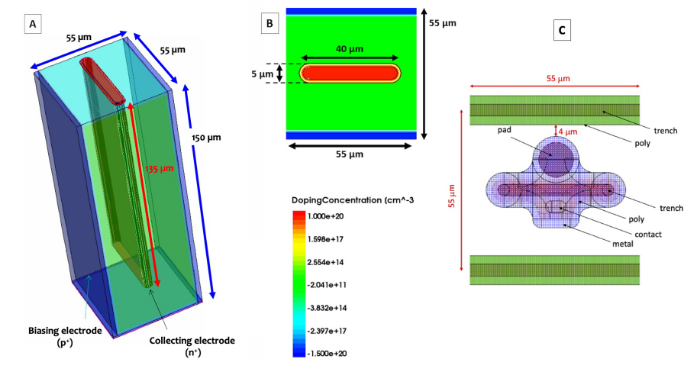}
\caption{Elementary cell of a 3D-trench silicon detector, as designed by the TIMESPOT project~\cite{Anderlini_2020}. Doping profiles are shown (n++ in red, p-- in green, p++ in blue). A) 3D rendering; B) pixel section; C) pixel layout.}
\label{fig:3dtrench}
\end{figure}

A strong option is the timeSPOT project~\cite{hep-ph_Lai_2018,hep-ph_ignite_ALai_2022}: 
which develops a technology called hybrid 3D-trenched pixels in which the pixel electrode geometry is optimised for timing performance.  A representation of one cell of a 3D-trench sensor is reported in Fig.~\ref{fig:3dtrench}. As for standard 3D-pixels, the sensors are able to withstand very large irradiation. The project has experimentally demonstrated that sensors with a pitch of \SI{55}{\micro\meter} provide a time resolution of \SI{10}{ps} up to fluence of \SI{2.5e16}{MeV~n_{eq}/cm^2}~\cite{hep-ph_Lampis_2022}. The timeSPOT collaboration is planning to extend these tests up to fluences of \SI{1e17}{MeV~n_{eq}/cm^2}, that has never been achieved so far. Excellent detection efficiencies were also measured~\cite{hep-ph_Lampis_2022} by operating the sensor inclined by an angle of \SI{20}{\degree} with respect to the beam incidence. As the electrode itself is not sensitive, the inclination ensures that particle always cross some active pixel region. Sensor with a size of $2\times2$~cm$^2$ can be produced and technical solution like stitching are being explored to produce larger devices.

A first prototype of the ASIC, TIMESPOT1, has been realised~\cite{hep-ph_CadedduEtAl_2022}. The chip contains a pixel matrix of $32\times 32$ elements, with a pixel pitch of $55 ~\rm{\mu m}$. Each pixel in the matrix contains an analogue and a digital circuit, with its TDC to measure the time of the arrival of the signals from the sensor. Digitised signals are then sent to the periphery of the matrix where 8~multiplexed output links send the data out at 1.28~Gb/s speed. Efforts are on-going to scale the size of the ASIC.
The above considerations make timeSPOT a viable option for the HIKE beam tracker.

Other projects involving different technologies, such as monolithic pixel sensors~\cite{hep-ph_IacobucciEtAl_2022} or LGADs~\cite{hep-ph_PellegriniEtAl_2015,hep-ph_PaternosterEtAl_2020}, are also taken in consideration. Monolithic detectors can be made with very small thickness and excellent time resolutions, however they have a lower radiation tolerance compared to the 3D-trench sensors. 


%% file: detectors/VC.tex
\subsubsection{Veto counter} 
\label{sssec:vetocounter}

The introduction of the veto counter detector (VC) surrounding the beamline has helped reduce the upstream background for the $K^+\to\pi^+\nu\bar\nu$ analysis. However, a part of the background remain undetected, passing at a very small angle inside the holes accommodating the beam pipe. The random veto due to the usage of the veto counter is less than 2\% at NA62 nominal intensity, but is expected to increase with the beam intensity. When more than one particle is present in the event, the granularity of the current setup does not allow to reliably separate them and distinguish between accidental activity due to halo muons and genuine background candidates.

\begin{figure}[tb]
\centering
\includegraphics[width=0.75\linewidth]{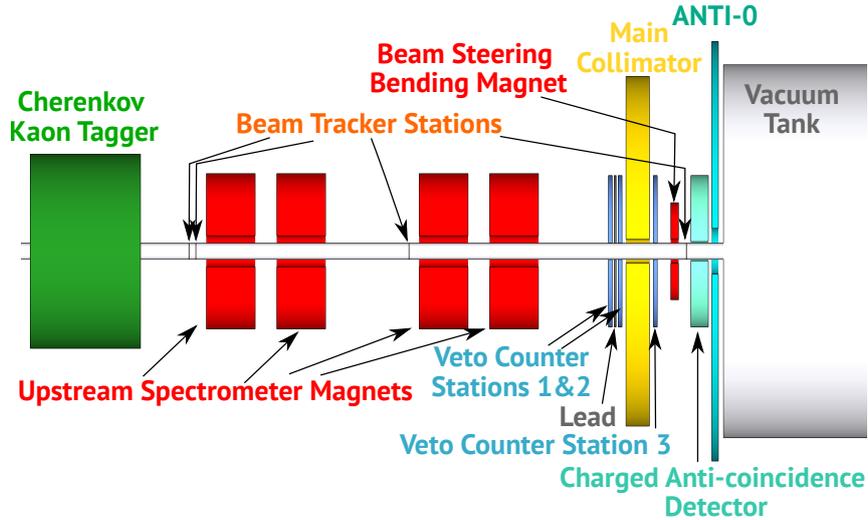}
\caption{Zoomed view of upstream region of HIKE Phase~1 (see Fig.~\ref{fig:phase1-layout}), highlighting the location of the veto counter stations.}
\label{fig:UpstreamRegionZoom}
\end{figure}

An improved design based on scintillating fibre (SciFi) technology, developed for the SciFi tracking detector at LHCb~\cite{Joram2015LHCbSF}, is proposed to tackle these limitations. The detector will consist of three detection stations: two located in front of the main collimator, separated by a layer of lead acting as a photon converter, and a third one located immediately behind the collimator, see figure~\ref{fig:UpstreamRegionZoom}. Each station needs to extend at least up to \SI{10}{\cm} above the beamline and \SI{30}{\cm} below the beamline to cover the full range where upstream decays are expected to cross the detection planes. For the same reason, the horizontal coverage needs to extend up to \SI{6}{\cm} on either side of the beam centre. A central hole will accommodate the reduced-size beam pipe, allowing it to reach only \SI{2}{\cm} from the beam centre. The thin SciFi modules would then be attached to the beam pipe allowing for an increased detection efficiency for interactions of photons and charged particles with the beam pipe. Following these specifications, the active area of the new veto counter will cover a rectangular surface area of $26\times $\SI{78}{\cm^2} with a hole at the centre to accommodate the passing beam. With the present state of the technology as used in LHCb, each station is made of scintillating fibres of diameter \SI{250}{\um}, read out by silicon photo-multipliers (SiPMs). The fibres are arranged in mats of multiple fibre layers forming a detection plane. Two orthogonal SciFi planes form a single veto counter station. The described detector scheme will provide X-Y reading of charged particles crossing a detection station with a spatial resolution $\sigma_{x,y}\sim\SI{200}{\um}$, a time resolution below \SI{200}{\ps}, and detection efficiency larger than 99$\%$. The number of readout channels for the above design will be about 4000 channels per station. If needed, several fibres could be coupled together into a single SiPM to reduce the number of channels at the expense of some spatial resolution. The thickness of the detection planes can also be modified to increase detection efficiency and improve time resolution. The small granularity and high spatial resolution of each detection plane will provide tracking capabilities by combining information from consecutive stations, further improving the time resolution. 

The expected hit rate in the detector is extrapolated from the rates at the nominal NA62 beam intensity, assuming it grows linearly with the beam intensity. The hit rate at four times the NA62 beam intensity is expected to be 30~MHz per detection plane, which is too high for the existing readout electronics of the SciFi detector but can be addressed with the TDC-Felix electronics commissioned in NA62 (Section~\ref{sec:readout}). The front-end electronics must also be adapted for HIKE. The present front-end ASIC used for the LHCb SciFi receives signals from the SiPMs, which are then shaped and digitised by a 2-bit ADC with a sampling rate of 40~MHz. Due to the time resolution requirements for HIKE, new front-end electronics will be used, based on ToT or FADC, similar to the solutions for other detector systems (see for example the KTAG and LAV sections). 


A different technology option is an update on the design currently used in NA62 using tiles of plastic scintillator. In this case, the use of tiles with smaller height (4~cm in NA62) and the usage of a double plane for each station (a plane with tiles placed horizontally and a plane with tiles placed vertically) will allow to achieve a smaller granularity and separate signals left by multiple particles. An increased acceptance coverage will be achieved by adding tiles also on the left and right of the beam. A rectangular hole of $42\times 78$~mm$^2$ size will allow the beam pipe to pass through the detector. In addition, some tiles could be placed 
along the beam pipe to increase the detection of particles escaping the beam pipe between stations.

Further improvements in upstream background rejection power could be achieved by rearranging the beamline elements in the veto counter region. Combined with the proposed hardware improvements, the new veto counter detector will reduce the upstream background at least three times more than the present system.

%% file: detectors/anti0.tex
\subsubsection{ANTI-0}

The baseline option for the charged particle veto upstream of the decay volume (very upstream ANTI station called ANTI-0) is a cell structure hodoscope covering area of $\oslash$2.2~m around the beampipe at the entrance of the fiducial decay volume. The NA62 ANTI-0 hodoscope design is based on rectangular scintillating tiles with transverse size of $120\times120$~mm$^2$ read by SiPMs through short lightguides arranged in a chessboard style at both sides of the central foil. The general design of ANTI-0 for HIKE follows that of the NA62 ANTI-0 hodoscope~\cite{Danielsson:2020opy} with the two major changes: finer granularity to sustain up to x6 intensity, and two active layers to ensure higher efficiency.

An additional improvement, currently under investigation, is the possibility to use reabsorption-free fast nanocrystal-based plastic scintillators~\cite{Gandini:2020aaa}. The new scintillator already now can sustain a higher radiation dose, having an emission peak shifted to $\lambda>550$~nm, tolerating transparency losses in UV and blue regions. Some perovskite-based scintillators show very fast decay time, with the first time component of the order of $\sim 0.3$~ns~\cite{JANA2022110}, which could be crucial for building fast future detectors. These technological possibilities will be further investigated for the HIKE Proposal. More details can be found in the description of the MEC.


%% file: detectors/CHANTI.tex
\subsubsection{Charged anti-coincidence detector}

The Charged Anti-coincidence detector (CHANTI) provides veto for events with particles scattered inelastically off the last station of the 
beam tracker
and halo particles originating from upstream decays entering the fiducial decay region close to the beam. 

The NA62 CHANTI comprises six stations, each station covering an area of $30\times30$~cm$^{2}$ with a $9\times\SI{5}{\cm^2}$ inner opening for the beam. The CHANTI covers hermetically the angular region between \SI{34}{\milli rad} and \SI{1.38}{rad} with respect to the beam axis. The CHANTI, with its time resolution of about \SI{800}{\pico s}, induces a signal loss due to accidental activity of about 5$\%$ at the nominal intensity but is expected to become more significant with increasing intensity.

\begin{figure}[tb]
\centering
\includegraphics[width=0.45\textwidth]{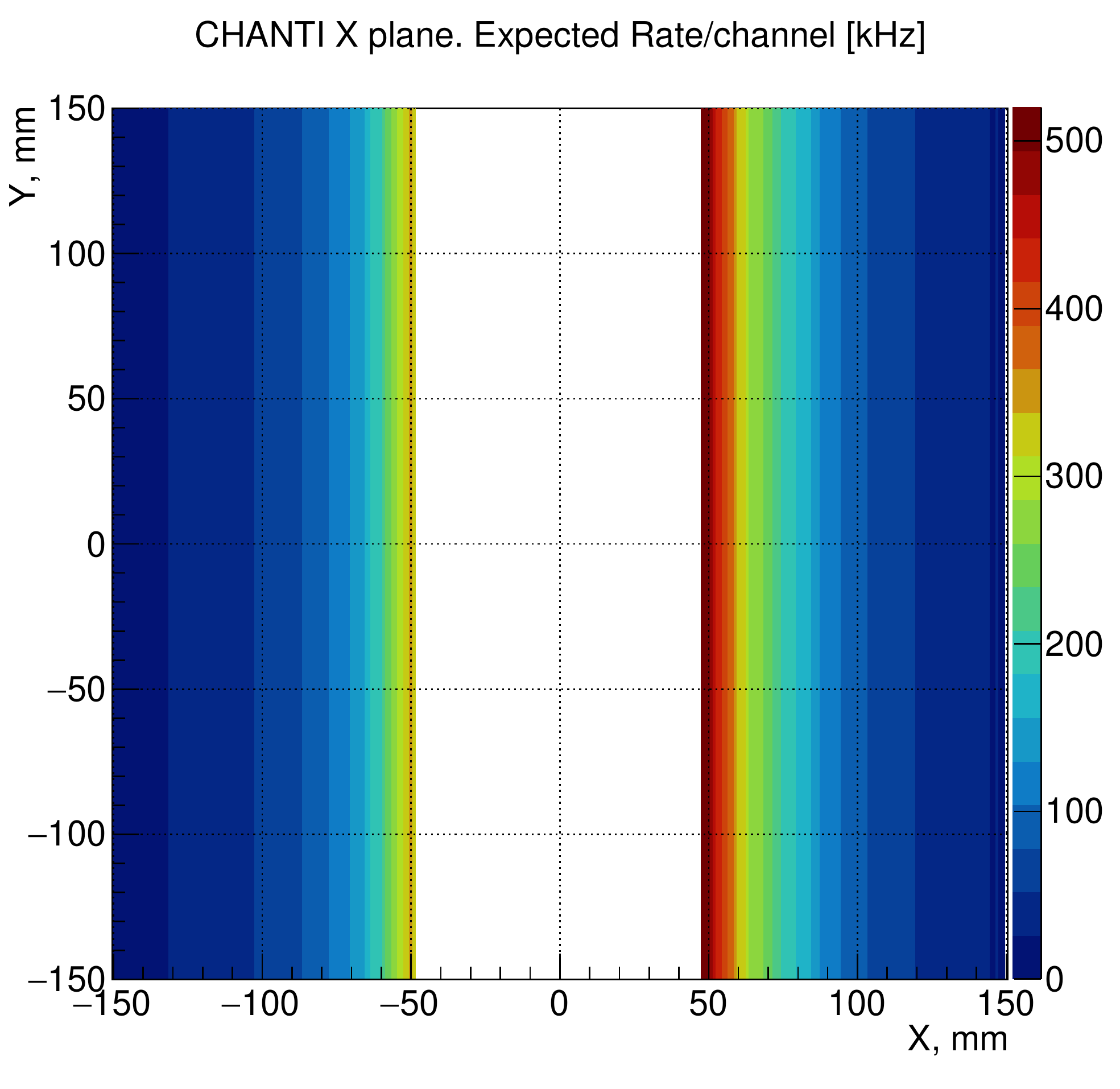}%
\includegraphics[width=0.45\textwidth]{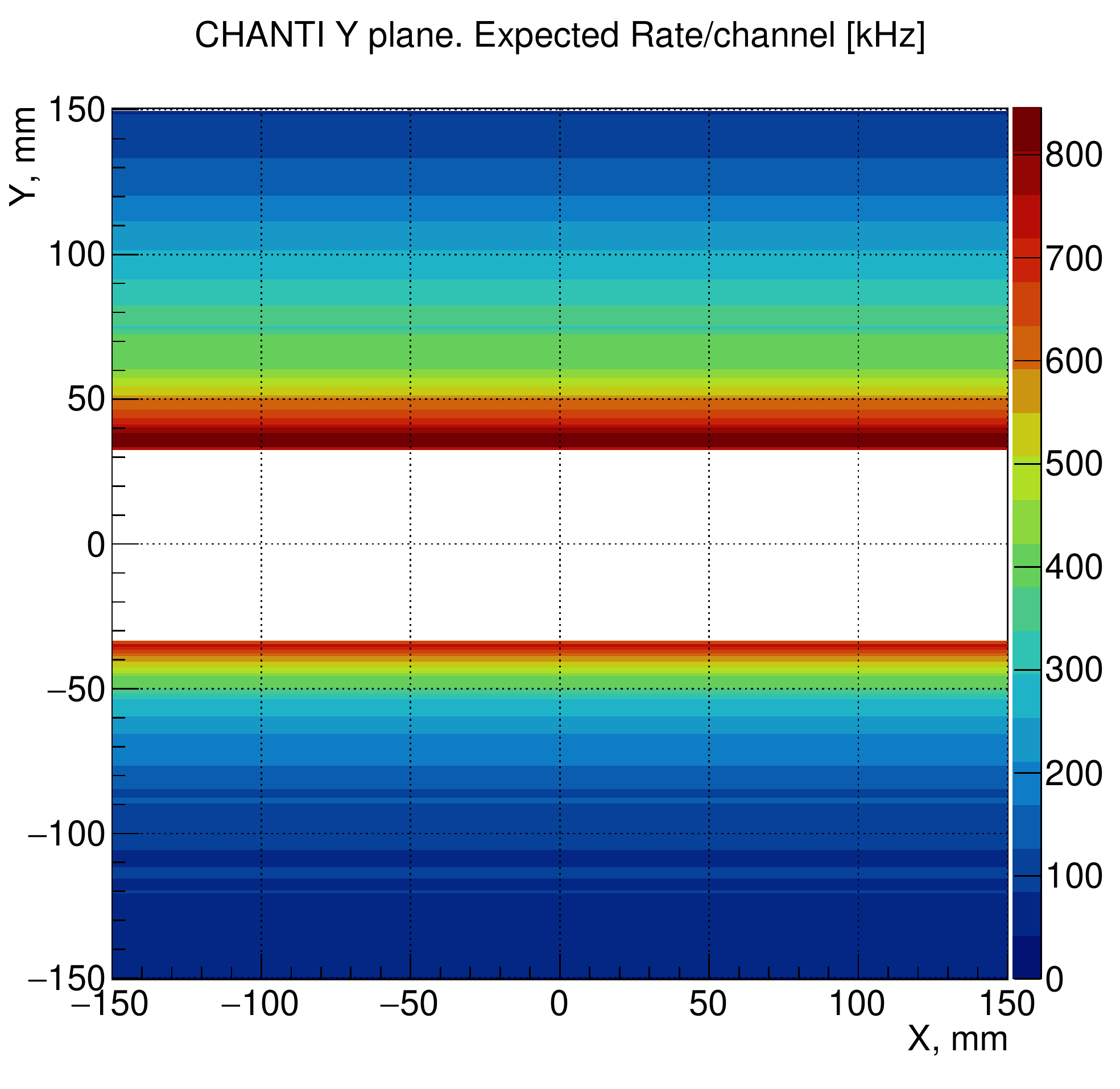}
\vspace{-4mm}
\caption{CHANTI expected rate per fibre for the vertical and horizontal planes.}
\label{fig:CHANTI_Exp_Rate}
\end{figure}

The design of the HIKE CHANTI stations is based on the usage of thick plastic scintillating fibres (from $\oslash 2$ up to $\oslash 3$~mm) read by SiPMs from both ends.
The expected particle rate has been estimated using the NA62 beamline simulation, assuming an hadron beam intensity six times higher than the nominal 750MHz of NA62. The expected rate per channel in kHz is shown in Fig.~\ref{fig:CHANTI_Exp_Rate} for the option with $\oslash 2$~mm thick fibres overlapping by 0.5~mm. The option with $\oslash 2$~mm fibres, that is the maximum diameter currently widely available on the market, assumes 464 fibres per station (208 forvertically oriented X station, and 256 for horizontally oriented Y station) and about 1000 SiPMs/station. The required number of electronic channels could be reduced by connecting signals from the peripheral fibres into analogue OR before digitization.

\paragraph{Alternative detector design using SciFi technology}

A more ambitious option would be to pursue the SciFi detector technology rather than scintillator technology. The new detector would consists of six stations with $39\times $\SI{39}{\cm^2} transverse dimensions, positioned to hermetically cover the required angular region with respect to the beam axis. Each station is made of scintillating fibres ($\oslash \SI{250}{\um}$) arranged in two planes that provide X-Y information about the incoming charged particles. Each plane is made of mats of several fibre layers, which are read out by SiPMs. The layout allows a time resolution below \SI{200}{\pico s} per station and a spatial resolution $\sigma_{x,y}\sim\SI{200}{\um}$. The time resolution would be at least a four-fold improvement, and the exceptional spatial resolution would provide tracking capabilities. The stations would be thicker than the ones currently used at LHCb~\cite{Joram2015LHCbSF}, bringing the detection efficiency to above 99$\%$. 

The new CHANTI detector could also be used in conjunction with the new VetoCounter
(Section~\ref{sssec:vetocounter}), both made of SciFi technology, to separate halo muons from charged particles produced in interactions in the beam-tracker. Owing to their identical spatial and time resolutions and largely overlapping sensitive regions, a combination of the information between the two detectors can reduce significantly the accidental veto due to halo-muon activity.

Using a SciFi detector for the new CHANTI presents a technological challenge. At present detectors using SciFi technology are operated in air and a dedicated R$\&$D program is required to enable their operation in vacuum. A solution to this challenge will allow cooling of the SiPM to liquid Nitrogen temperatures, eliminating the noise completely and improving radiation tolerance. This is an important improvement over the present CHANTI, which uses SiPMs without cooling. The SiPMs currently in operation at NA62 are heavily impacted by the effect of accumulated radiation dose; such issue would not be present for a SciFi detector with cooled SiPMs in vacuum. 

The number of readout channels for the above design would be about 3000 channels per station, but it can be reduced if several fibres are coupled together and fed into a single SiPM. This would be at the expense of some spatial resolution, and would not be an issue. The readout is planned to be based on the TDC-Felix electronics~\ref{sec:readout}, that is able to sustain the hit rate presented in Fig.~\ref{fig:CHANTI_Exp_Rate}.



%% file: detectors/LAV.tex
\subsubsection{Large angle vetos} 
\label{sec:lav}

As demonstrated by the successful $K^+\to\pi^+\nu\bar\nu$ measurements at NA62, the efficiency of the existing photon veto systems, including the existing Large Angle Vetoes (LAV)~\cite{NA62:2017rwk} based on the OPAL lead glass~\cite{OPAL:1990yff}, is adequate for the HIKE $K^+\to\pi^+\nu\bar\nu$ measurement. In the analysis of NA62 Run~1 data, the overall $\pi^0$ detection inefficiency was about $1\times10^{-8}$~\cite{NA62:2020fhy}, and the expected number of background events from $K^+\to\pi^+\pi^0$ (the only significant background channel with photons) was about 7\% of the expected number of signal events~\cite{NA62:2021zjw}. The $\pi^0$ veto efficiency estimate was validated by studies of the single-photon detection efficiency for each veto subsystem~\cite{NA62:2020pwi}.
The inefficiency of the NA62 LAVs for single photons was found to rapidly decrease with photon energy to a value of about $3\times10^{-3}$ for photons with $E=300$~MeV, and thereafter to slowly decrease with energy to values of a few $10^{-4}$ for photons of $4$--$6$~GeV.\footnote{This refers to the LAV system in operation, including the effects of the non-hermeticity of the retrofitted arrangement of lead-glass blocks. The actual detection efficiency for the lead-glass detectors themselves was found with electrons to be an order of magnitude higher~\cite{Ambrosino:2007ss}.}
In the decay-in-flight technique for the $K^+\to\pi^+\nu\bar\nu
$ measurement, the $K^+$/$\pi^+$ vertex is accurately reconstructed and required to be in the fiducial volume, and the fiducial momentum cuts on the secondary $\pi^+$ guarantee that the $\pi^0$ in $K^+\to\pi^+\pi^0$ decays has at least 40~GeV of energy. As a result, the LAV system only has to cover out to 50~mrad in the polar angle as seen from the fiducial volume. Moreover, the two photons from $K^+\to\pi^+\pi^0$ decays have an anticorrelation such that for events with one low-energy photon heading into the LAVs, the second photon has high energy and is seen in the LKr. These conditions allow the very low detection inefficiency for $\pi^0$'s to be obtained even with LAVs with single-photon inefficiencies on the order of $10^{-3}$.

On the other hand, the time resolution obtained with the existing LAVs is problematic for HIKE. The Cherenkov light produced in the large lead-glass blocks is highly directional, and the light propagates to the PMT photocathodes via complicated paths with multiple reflections. There is considerable time spread arising from the angle of particle incidence, and perhaps for different particle species, the Cherenkov characteristics of mips and electromagnetic showers being quite different. Above all, the light yield of the lead-glass blocks for electromagnetic showers is just 0.2 p.e./MeV. As a result, the LAVs have a time resolution of about 1 ns with significant extra-gaussian tails~\cite{NA62:2017rwk}. In order to maintain the random veto rate constant in HIKE, the time resolution needs to be improved by at least a factor of 4, and the tails substantially reduced or eliminated. 

The LAVs are potentially the most 
resource-intensive subsystem to build for HIKE, so it is essential that they be reusable for all phases of the program. 
The design of the LAVs for HIKE is therefore driven by the efficiency requirements for the $K_L$ program, and in particular, for the measurement of $K_L\to\pi^0\nu\bar{\nu}$.
The time resolution requirements for the $K_L$ phase are similar to those for the $K^+$ phase. The preliminary estimate of the total hit rate on the LAVs in KLEVER is 14~MHz (\Tab{tab:klever_rates}), implying a 3.5\% random veto rate for a $\pm5\sigma$ coincidence window of 2.5~ns.
The efficiency requirements for the $K_L$ phase, on the other hand, are much more stringent than for the $K^+$ phase.
Unlike for the case of $K^+\to\pi^+\nu\bar{\nu}$, for $K_L\to\pi^0\nu\bar{\nu}$ decays, the reconstruction of the $K_L$ decay vertex is uncertain, so the photon vetoes need to cover the entire length of the experiment from the upstream end of the fiducial volume to the calorimeter. Moreover, the kinematic anticorrelation between the energies (or angles) of missed photons from $K_L\to\pi^0\pi^0$ decays is largely washed out. Nevertheless, because of the boost from the high-energy beam, the KLEVER simulations show that it is sufficient for the large-angle photon vetoes (LAVs) to cover polar angles out to 100~mrad, as long as the fiducial volume is located well upstream of the calorimeter and the LAV detectors themselves satisfy the efficiency requirements discussed below: indicatively, the photon detection inefficiency must be less than 5\% at 10~MeV, less than $2.5\times10^{-4}$ at 100~MeV, and less than $2.5\times10^{-6}$ for energies above 2.5~GeV.


\paragraph{Layout for the $K^+$ phase}

\begin{table}[tb]
\centering
\caption{Parameters of the existing NA62 Large Angle Vetoes.}
\vspace{-2mm}
\begin{tabular}{@{}cccccc@{}}
\hline
Stations & Diameter [mm] & \multicolumn{2} {c}{Block radius [mm]} & Layers & Blocks\\
& Outer wall & Inner & Outer & & \\
\hline
LAV1--LAV5  & 2168  &   537 &  907 & 5 & 160\\
LAV6--LAV8  & 2662  &   767 & 1137 & 5 & 240\\
LAV9--LAV11 & 3060  &   980 & 1350 & 4 & 240\\
LAV12       & 3320  &  1070 & 1440 & 4 & 256\\
\hline
\end{tabular}
\label{tab:na62_lav}
\end{table}
The positions and radii of the LAV stations for the HIKE $K^+$ phase are expected to be quite similar to those for NA62. In NA62, there are 12 LAV stations: 11 incorporated in the vacuum tank, and one in air just upstream of the calorimeter. Parameters of the NA62 LAV stations are listed in \Tab{tab:na62_lav}.


\paragraph{Layout for KLEVER}

For KLEVER, a total of 25 LAV stations
in five different sizes, operated in vacuum and placed at intervals of 4 to 6~m, are needed to guarantee coverage out to $\theta = 100$~mrad.
The most downstream LAV station (LAV25) leaves as much of the front face of the Main Electromagnetic Calorimeter (MEC) uncovered as possible, and the dimensions and positions of the other LAVs have been chosen so that LAV25 defines the most restrictive aperture
for photons from the fiducial volume. The dimensions and segmentation of the 25 LAV stations are specified in \Tab{tab:klever_lav}.


\begin{table}[tb]
\caption{Dimensions, segmentation, number of readout channels, and total quantity of plastic scintillator for the five different types of LAV stations.}
\vspace{-2mm}
\centering
\begin{tabular}{cccccc}
\hline
LAVs & $r_{\rm int}$ (m) & $r_{\rm ext}$ (m) & Sectors & Total channels & Tot. scint. (kg)\\ \hline
1--11  & 0.44 & 0.85 & 40 & 3520 & 9690 \\
12--15 & 0.58 & 0.99 & 48 & 1536 & 4290 \\
16--18 & 0.72 & 1.23 & 56 & 1344 & 4970 \\
19--21 & 0.86 & 1.37 & 64 & 1536 & 5680 \\
22--25 & 1.00 & 1.51 & 72 & 2304 & 8525 \\
\hline
\end{tabular}
\label{tab:klever_lav}
\end{table}

As noted above, 11 of the LAV stations will be constructed for the HIKE Phase~1. The 12th station, operated in air rather than in vacuum, might be used for Phase~1 only. The remaining 14~stations in vacuum will be constructed for the HIKE Phase~2, as needed, and for KLEVER. The sizes and positions of the LAVs were implemented in the KLEVER simulation before the HIKE $K^+$ phase was proposed. Because the polar angle coverage was extended by increasing the number of LAV stations rather than their dimensions, the dimensional parameters of the KLEVER LAVs, though not the same as for NA62, are quite similar. Although the LAVs must be positioned carefully in KLEVER to guarantee the needed coverage, with some careful engineering work, it should be possible to reconcile the LAV designs for the HIKE $K^+$ and $K_L$ phases. For example, from the comparison of Tables \ref{tab:na62_lav} and \ref{tab:klever_lav}, it seems likely that LAVs 1--5 for the $K^+$ program could be recycled as LAVs 11--15 for $K_L$, LAVs 6--8 for $K^+$ would become LAVs 16--18 for $K_L$, and LAVs 9--11 for $K^+$ would become LAVs 19--21 for $K_L$. Various sections of the NA62 vacuum tank would have to be modified or remade.

\paragraph{Design and expected performance}

\begin{figure}[tb]
\centering
\includegraphics[width=0.5\textwidth]{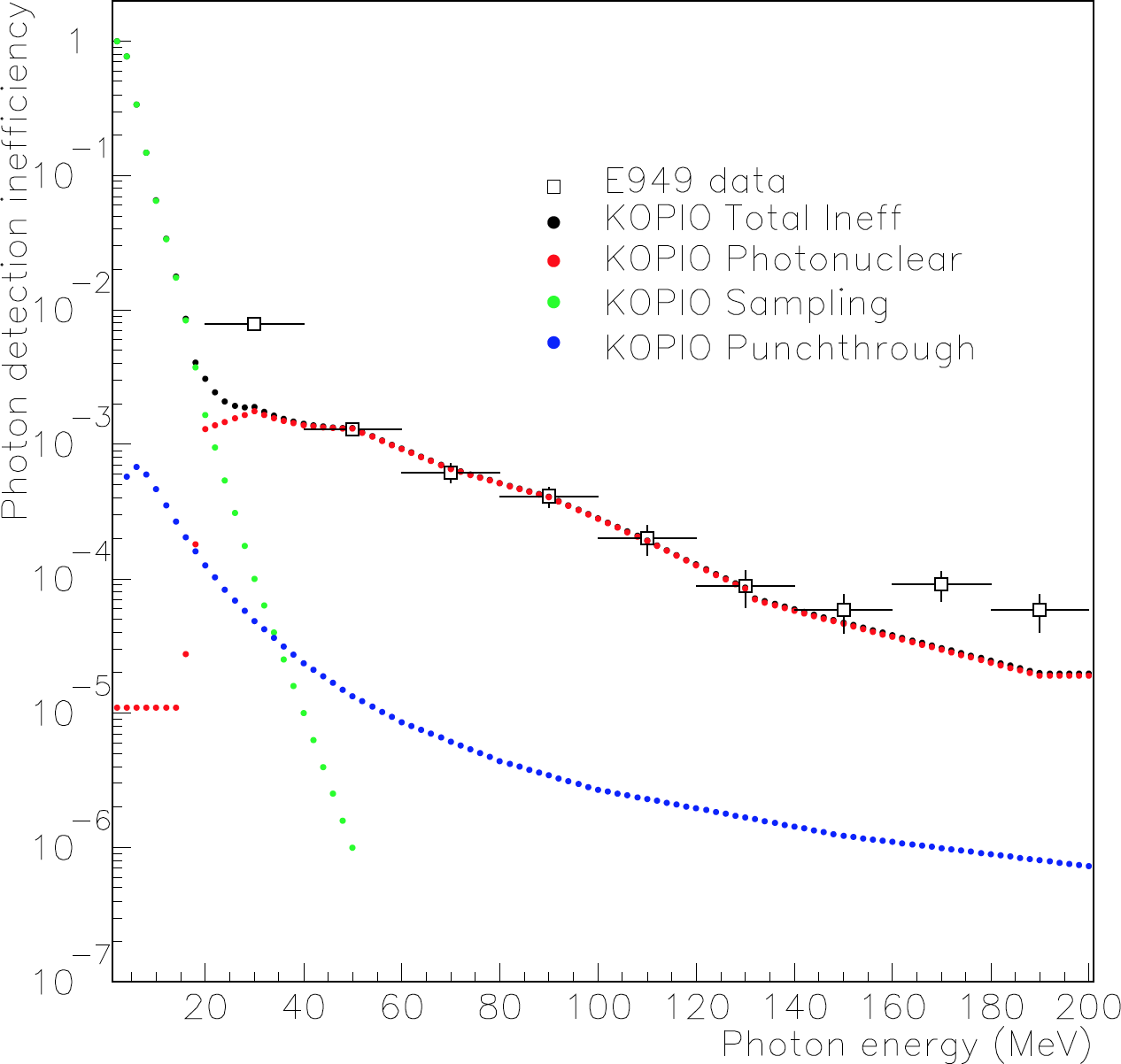}
\caption{Photon detection inefficiency parameterization from KOPIO~\cite{KOPIO+05:CDR}, broken down by source. Measured
inefficiencies for the E949 barrel veto~\cite{Atiya:1992vh} are also plotted.}
\label{fig:kopio_eff}
\end{figure}

As a reference for what low-energy photon detection efficiencies can be
achieved for the new LAVs, Fig.~\ref{fig:kopio_eff} shows the inefficiency
parameterization used for the KOPIO proposal~\cite{KOPIO+05:CDR}.
For the energy range
$50~{\rm MeV} < E < 170~{\rm MeV}$,
the overall parameterization (black circles) is based on detection
inefficiencies measured for the
E787/949 barrel photon veto~\cite{Atiya:1992vh} using $K^+\to\pi^+\pi^0$
events; these data are also shown in the figure.
Outside of this range, the parameterization
is guided by FLUKA simulations with different detector designs, and
the overall result is (slightly) adjusted to reflect the segmentation
of the KOPIO shashlyk calorimeter, but for most of the interval
$E < 200$~MeV, the results do not differ much from the E949 measurements.
The contributions to the inefficiency from photonuclear interactions,
sampling fluctuations, and punch through were estimated from
known cross sections, statistical considerations, and mass attenuation 
coefficients.

\begin{figure}
\centering
\includegraphics[width=0.5\textwidth]{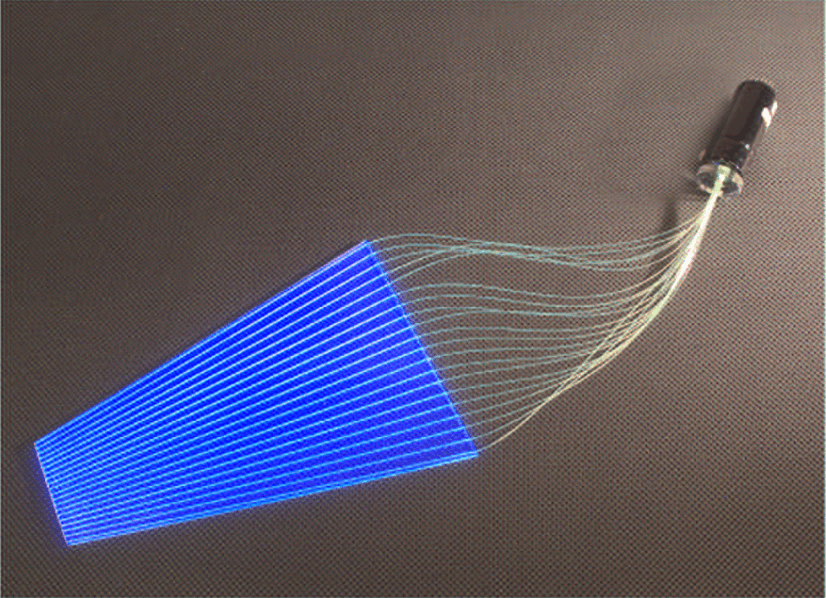}
\caption{Prototype tile with WLS fiber readout for the CKM VVS detector~\cite{Frank:2001aa}.}
\label{fig:ckm_tile}
\end{figure}

One possible design for the HIKE LAVs would
be similar to the Vacuum Veto System (VVS) detectors planned for the CKM
experiment at Fermilab~\cite{Frank:2001aa}. The CKM VVS is a lead/scintillator-tile detector with a segmentation of 1~mm Pb + 5~mm scintillator, for an electromagnetic sampling fraction of 36\%. This segmentation is the same as for the
E787/949 barrel photon veto, so the same low-energy efficiencies
might be expected. The wedge-shaped tiles are stacked into modules and
arranged to form a ring-shaped detector. The scintillation light is
collected and transported by 1-mm-diameter WLS fibers in radial grooves, as seen in Fig.~\ref{fig:ckm_tile}.
In the approximate geometry for HIKE, the LAV modules
would consist of 96 layers, for a total thickness of $\sim$60~cm, corresponding to
$\sim$18 $X_0$. There are 40--72 modules per detector, each with 20 fibers per tile. 
In the
original VVS design, the fibers brought the light to optical windows for
readout by PMTs outside of the vacuum. Readout by SiPMs
inside the vacuum would make for shorter fibers and would facilitate the
mechanical design---the availability of economical
SiPM arrays with large effective area makes this an attractive option. In the HIKE geometry, the fibers in a module would be bundled for readout by eight SiPM arrays. As in the VVS design, alternating fibers from each tile would be read by different SiPMs to provide redundancy, and there would be four readout layers in depth.

\begin{figure}[tb]
\centering
\includegraphics[width=0.6\textwidth]{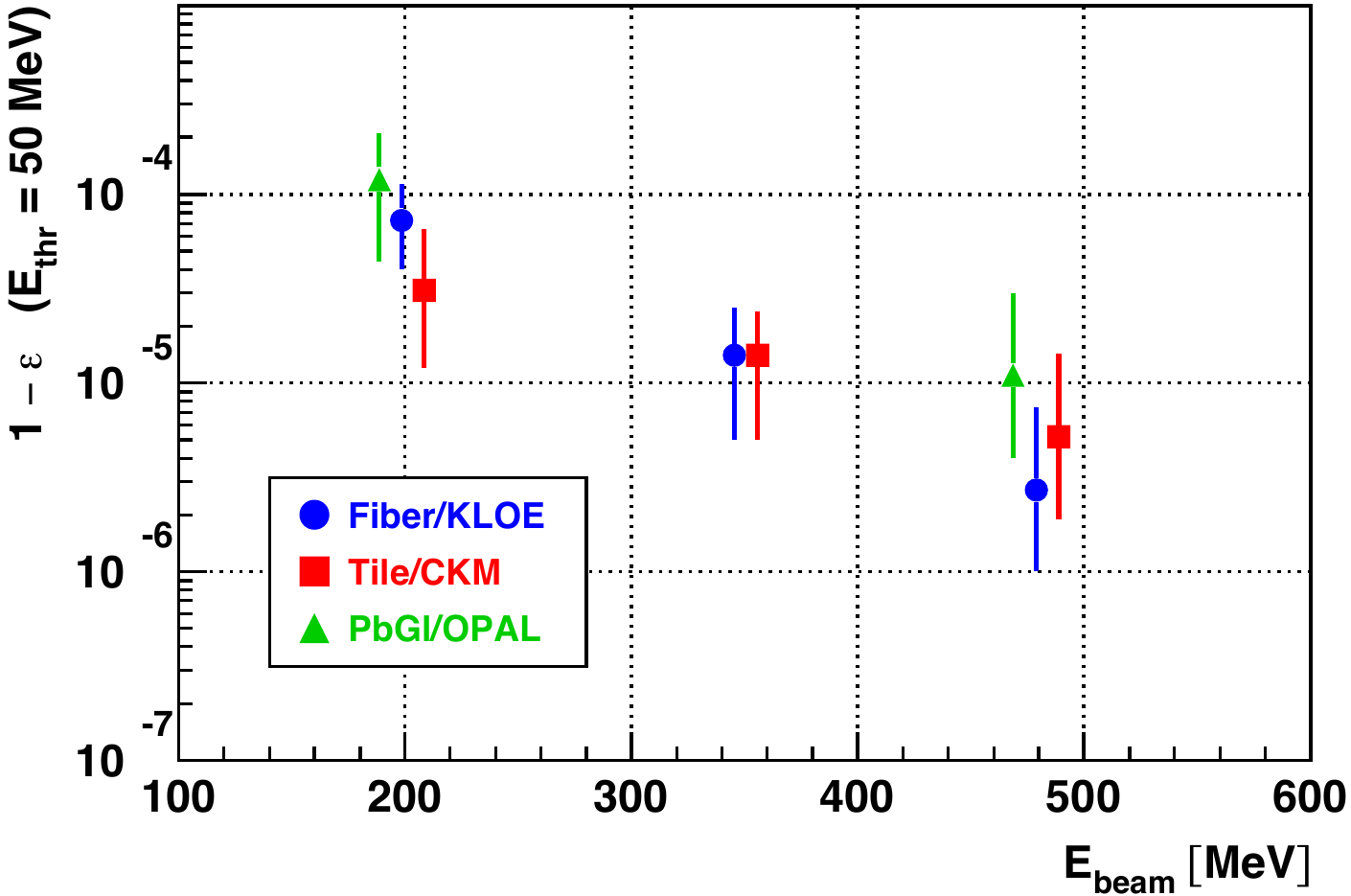}
\vspace{-2mm}
\caption{Measurements of detection inefficiency for tagged electrons with 203, 350, 483~MeV energies for three veto prototypes, made at the Frascati BTF in 2007. The red squares are for the CKM VVS prototype.}
\label{fig:hawaii}
\end{figure}

In the original NA62 proposal~\cite{Anelli:2005xxx}, before the OPAL lead glass became available,
very similar detectors were the baseline solution for the existing LAVs,
and in 2007, the efficiency of the CKM VVS prototype in the energy range 200--500~MeV was measured using a tagged electron beam at the Frascati
Beam-Test Facility (BTF) \cite{Ambrosino:2007ss}.
The results are shown in Fig.~\ref{fig:hawaii}. Earlier, the efficiency of the same prototype was measured at the
Jefferson National Laboratory with tagged electrons in the interval 500--1200~MeV. An inefficiency of $3\times10^{-6}$ was found at 1200~MeV,
even with a high threshold (80~MeV, or 1 mip)~\cite{Ramberg:2004en}.
The efficiency parameterization for the LAVs used in our simulations is based on the
KOPIO efficiencies up to the point at 129~MeV, and then extrapolated
through the three points measured at the BTF, to $2.5\times10^{-6}$ for photons
with $E > 2.5$~GeV. This is not unreasonable,
considering that the Jefferson Lab measurement shows that nearly this
inefficiency is already obtained at 1.2~GeV.
The tests of this prototype at the BTF were not optimized for the measurement of the time resolution, but indicated a time resolution of better than 250~ps for 500 MeV electrons, which should be sufficient for HIKE, including KLEVER.


\paragraph{Readout}

A leading contribution to the random veto inefficiency for the LAVs in NA62 is from halo muons. If the interactions of halo muons could be reliably distinguished from photon showers, the random veto inefficiency for the LAVs would be decreased, potentially relaxing the requirements on the LAV time resolution. In NA62, the LAV signals are discriminated against two thresholds, and signal amplitudes are obtained using the time over threshold technique. 
While some use of this information has been made in NA62 to partially recover the LAV random veto inefficiency and efforts are continuing in this direction, in practice, the separation between mips and low-energy photon showers has not so far proved to be reliable. To improve upon this situation, we are investigating the gains to be had with a fully digitizing FADC readout for the LAVs, as is foreseen for all of the photon veto detectors in KLEVER.


%% file: detectors/STRAW.tex
\subsubsection{Spectrometer }
\label{sec:straw}

\begin{figure}[h]
\centering
\includegraphics[width=0.75\textwidth]{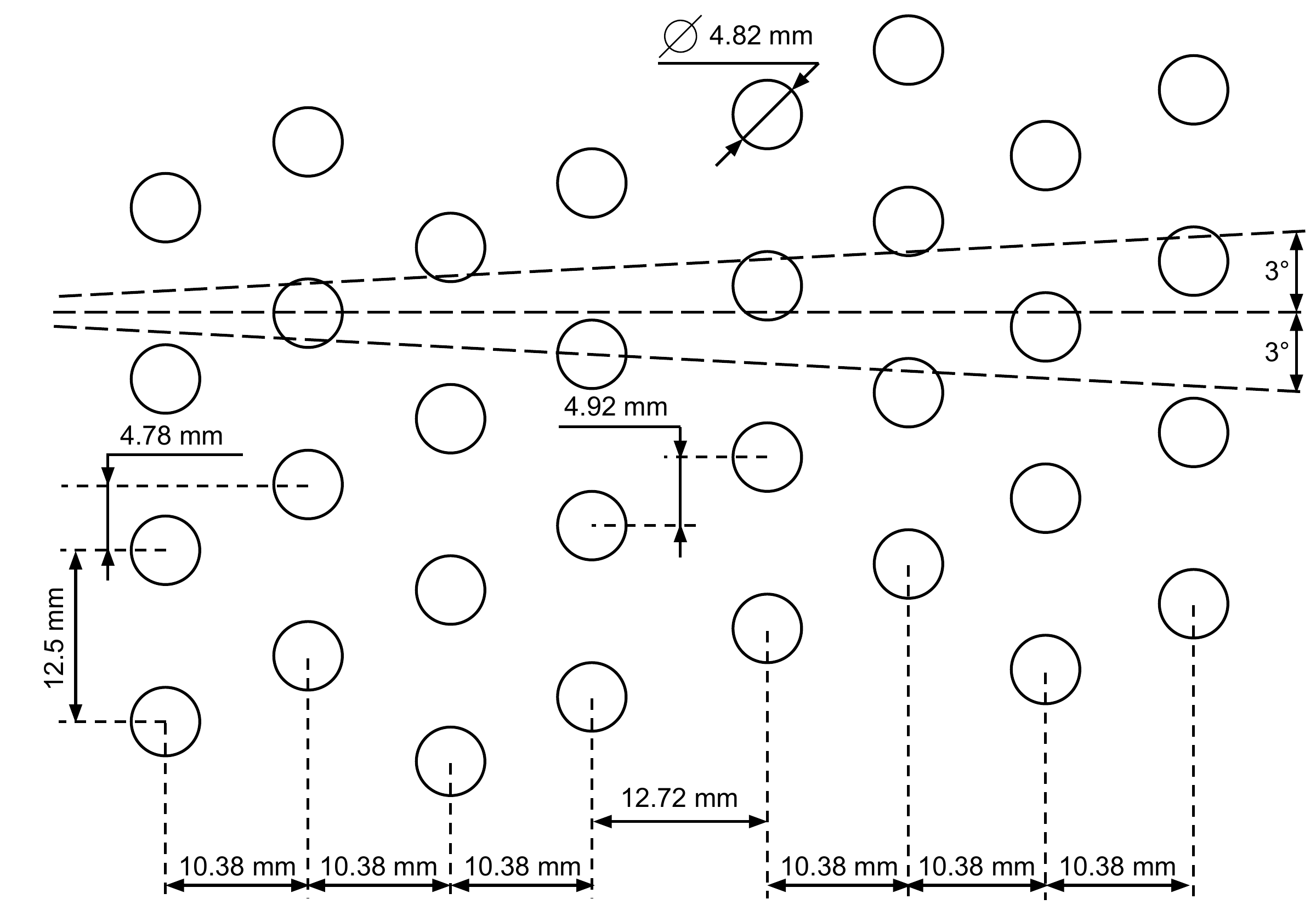}
\vspace{-2mm}
\caption{Optimised layout of straw tubes with a diameter of 4.82~mm in a view.}
\label{fig:straw_new_layout}
\end{figure}

A spectrometer similar to the NA62 straw tracker~\cite{NA62:2017rwk}, comprising four straw chambers and a dipole magnet, is planned to reconstruct the momentum and direction of charged particles in the final state.
The straw tubes are expected to remain in the vacuum tank containing the decay region, profiting from the successful technology developed for NA62.
Straws with diameter less than 5~mm are necessary to handle the expected particle rates at the higher beam intensity.
Straw diameter reduction by a factor $\sim 2$ with respect to the NA62 straws will lead to shorter drift times and an improvement in the resolution of the trailing edge time from the current ${\sim}30$~ns to ${\sim}6$~ns.
A smaller diameter of the straw also requires a change in the geometric placement of the straws in a single view.
Design work based on Monte Carlo simulations was performed, and the straw layout was optimised taking into account realistic spacing and dimension requirements, resulting in a choice of eight straw layers per view, shown in Fig.~\ref{fig:straw_new_layout}.

The material used to make new straws using the ultrasonic welding technique will be the same as in the current spectrometer, namely Mylar coated with 50~nm of copper and 20~nm of gold on the inside.
To reduce the detector material budget, the thickness of the mylar will be reduced from $36~\mu$m to either $12~\mu$m or $19~\mu$m.
The diameter of the gold-plated tungsten anode wires might be reduced from $30~\mu$m to $20~\mu$m.
The final decision on the mylar thickness and the wire diameter will be made based on mechanical stability tests.
The development of small-diameter thin-walled straws has synergies with R\&D work for COMET phase II at J-PARC~\cite{Nishiguchi:2017gei}, and is included in the ECFA detector R\&D roadmap~\cite{ecfareport:2021}.

\begin{figure}[p]
\centering
\includegraphics[width=0.83\textwidth]{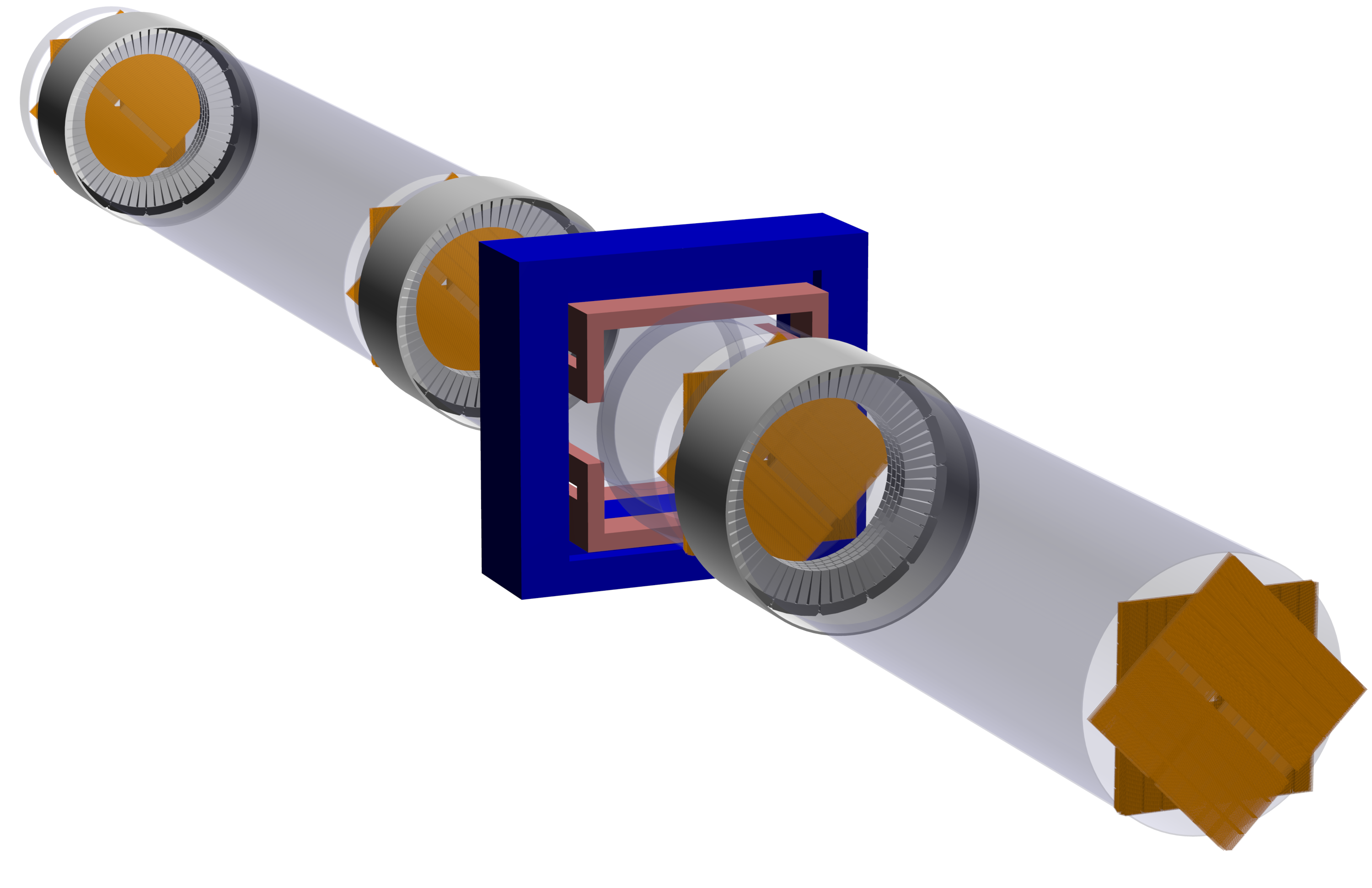}
\vspace{-6mm}
\caption{
\label{fig:straw_visualization}
Geant4 visualisation of the new straw spectrometer.}
\end{figure}

Based on the results of the design study, a Geant4-based simulation of the new spectrometer was developed using the same dimensions and positions of the straw chambers, the number and orientation of views in the chamber, the gas composition (Ar+CO$_2$ with 70:30 ratio) and the properties of the dipole magnets as in the current NA62 layout. 
A comparison between the two straw detectors is given in Table~\ref{tab:straw_comparison}, and a Geant4 visualisation of the new spectrometer is shown in Fig.~\ref{fig:straw_visualization}.
The new spectrometer is planned to have the capability of aligning the central holes of the straw chambers on the beam axis for both $K^+$ (Section~\ref{sec:phase1}) and $K_L$ (Section~\ref{sec:phase2}) modes of operation.

The NA62 track reconstruction algorithm was adapted for the new detector, and a preliminary resolution comparison indicates that the new spectrometer could improve the resolution for the reconstructed track angles and momenta by 10--20\% with respect to the existing NA62 spectrometer while maintaining the high track reconstruction efficiency.

\begin{figure}[p]
\centering
\includegraphics[width=0.45\textwidth]{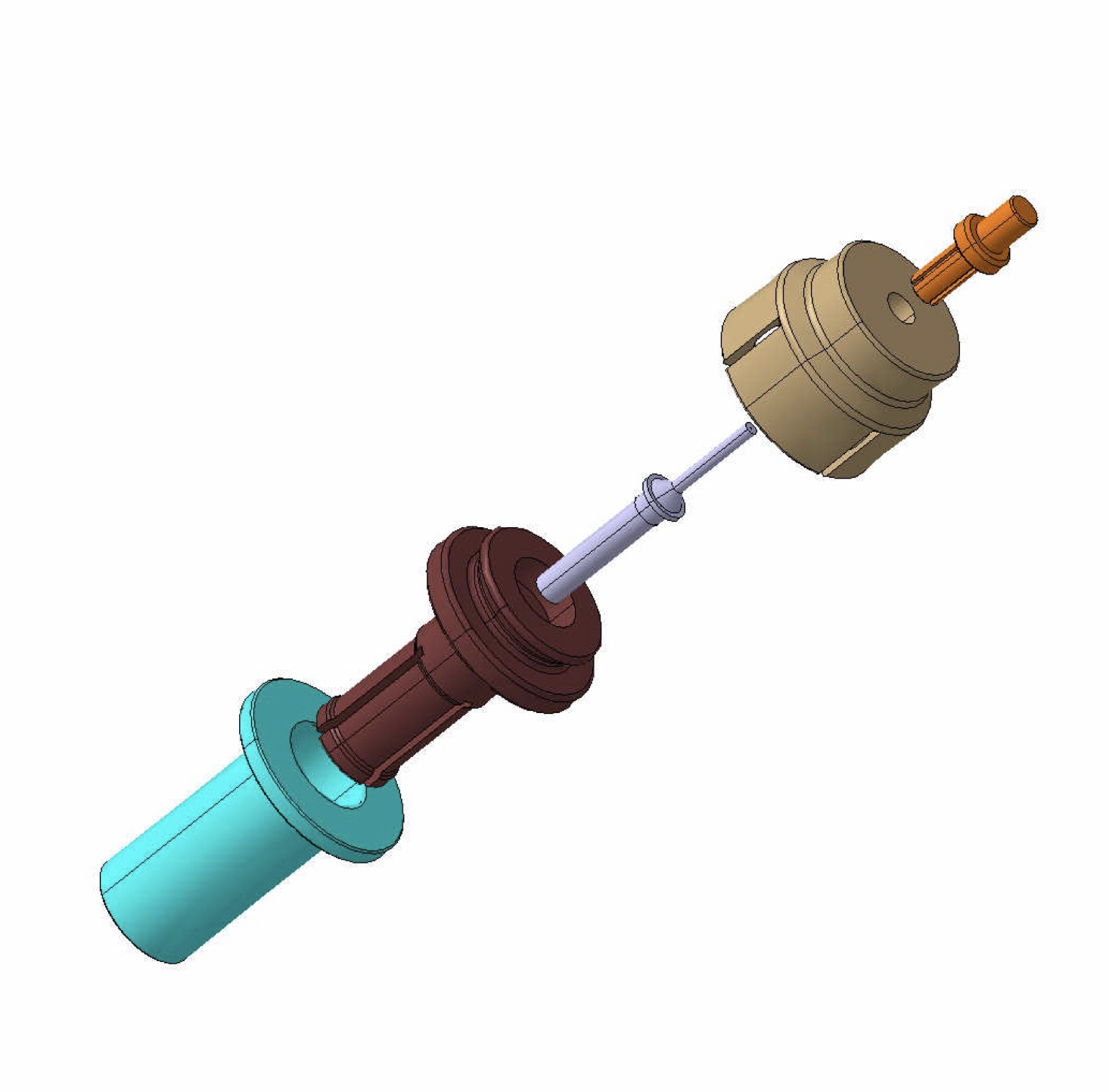}
\includegraphics[width=0.45\textwidth]{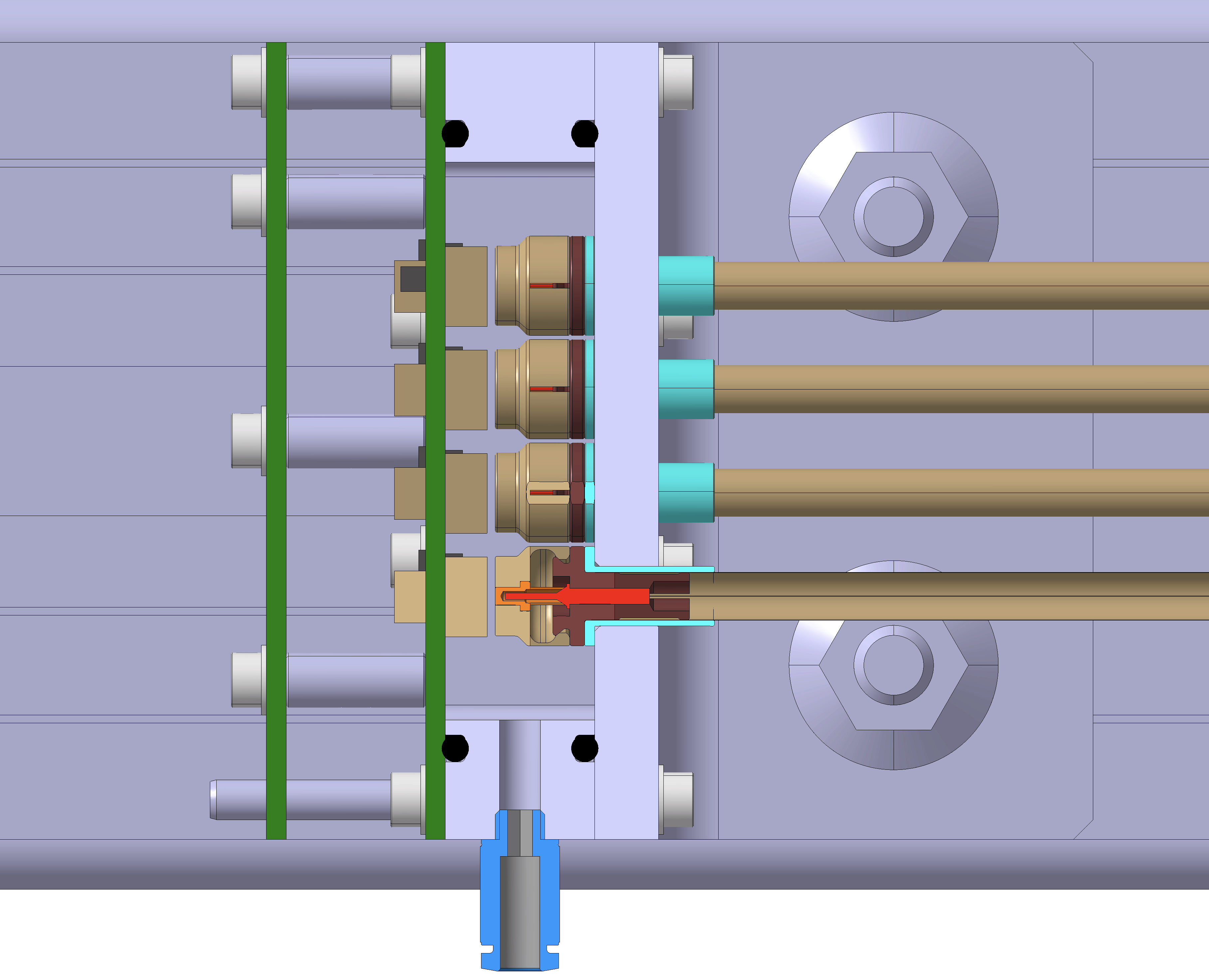}
\vspace{-2mm}
\caption{Left: detail of the connectivity elements (coaxial) in order to minimise the signal path to the front-end electronics.
Right: cross-section of the prototype to validate connectivity and basic performance of a 4.82~mm diameter straw using new front-end electronics with the capability to measure the trailing edge.}
\label{fig:spectrometer:connectivity}
\end{figure}

\begin{figure}[p]
\centering
\includegraphics[width=0.46\textwidth]{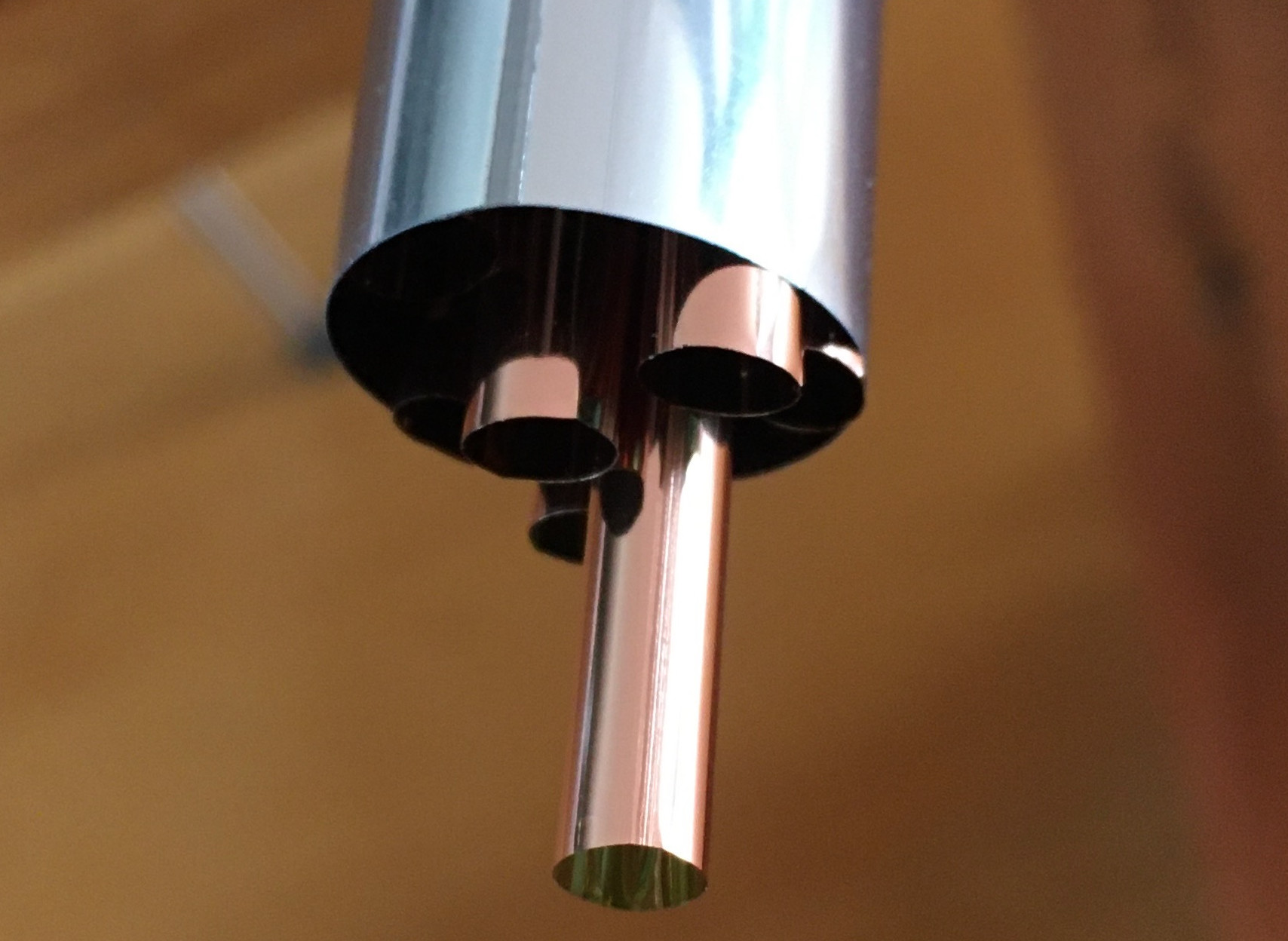}~
\includegraphics[width=0.45\textwidth]{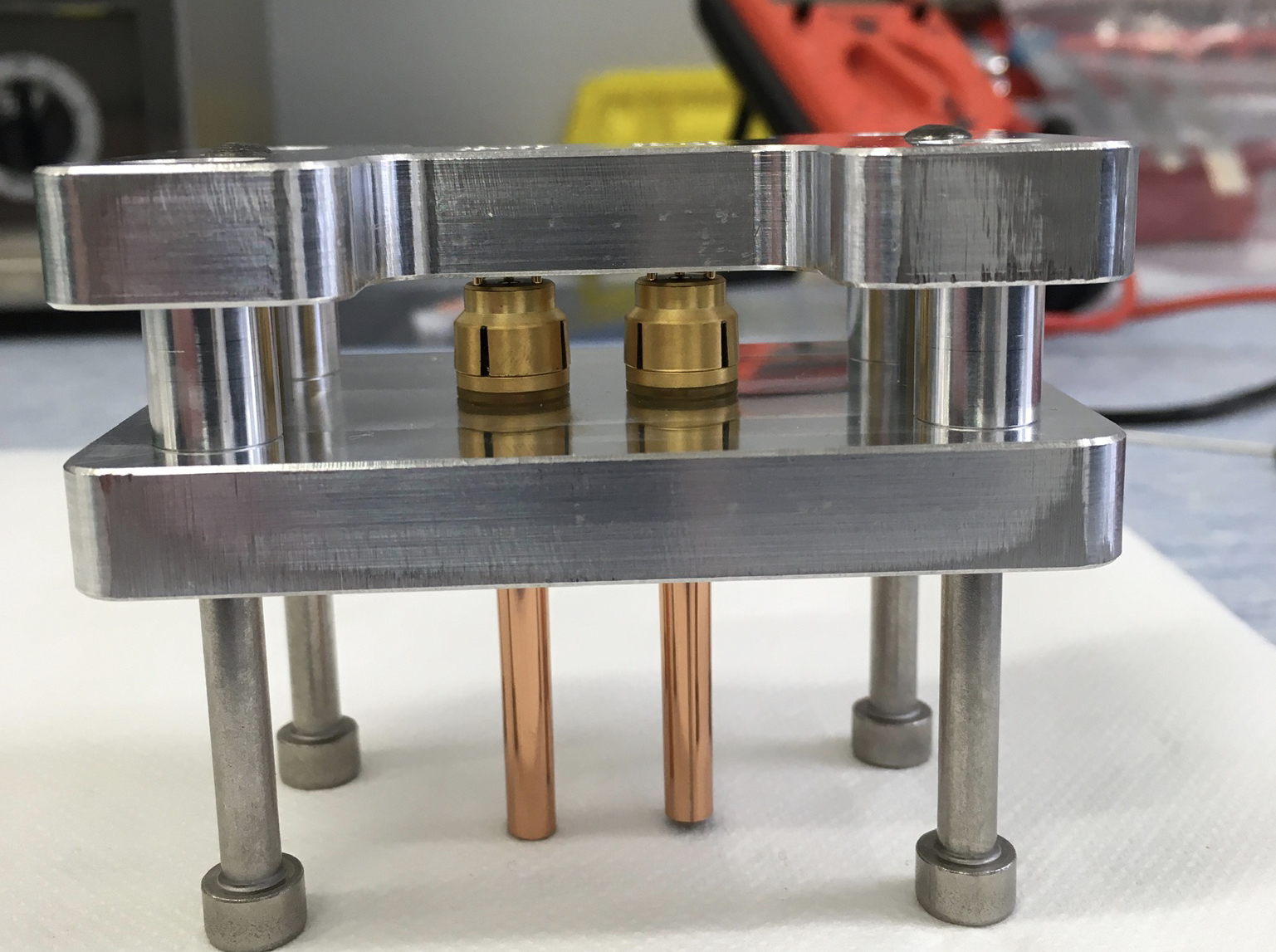}
\caption{Left: pre-production of straws with a diameter of 4.82~mm and a wall thickness of 19~$\mu$m. Right: test of signal connectivity and high-voltage stability of individual components. }    \label{fig:spectrometer:prototest}
\end{figure}

\begin{figure}[p]
\centering
\includegraphics[width=0.7\textwidth]{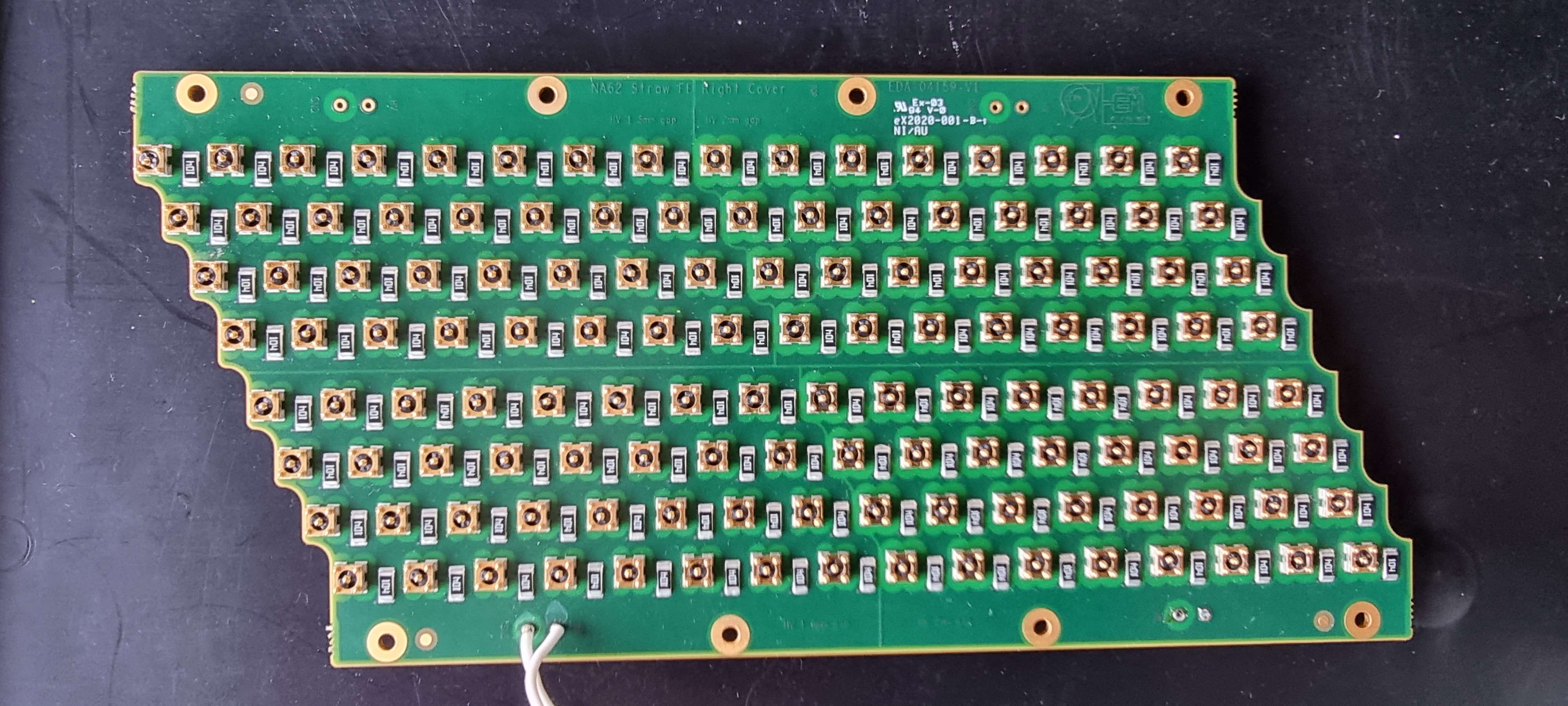}
\caption{Prototype of a HV board for the new straw spectrometer.}
\label{fig:spectrometer:hv_board}
\end{figure}

Investigations of possible technological solutions for the straw connectivity, design of a new high-voltage board, and a pre-production of straw tubes with a diameter of 4.82~mm have already been started (Figs.~\ref{fig:spectrometer:connectivity}, \ref{fig:spectrometer:prototest}, \ref{fig:spectrometer:hv_board}).


\begin{table}[h]
\centering
\caption{Comparison of the NA62 spectrometer and the new straw spectrometer.}
\vspace{-2mm}
\begin{tabular}{l|r|r}
\hline
&Current NA62 spectrometer& New straw spectrometer\\
\hline
Straw diameter & 9.82~mm & 4.82~mm\\
Straw length & 2100~mm  & 2100~mm \\
Planes per view & 4 & 8 \\
Straws per plane & 112 & ${\sim}160$\\
Straws per chamber & 1792 & ${\sim}5200$\\
\hline
Mylar thickness & $36~\upmu$m  &  (12 or 19)~$\upmu$m \\ 
Anode wire diameter & 30~$\upmu$m & (20 or 30)~$\upmu$m \\
Total material budget & 1.7\% $X_0$ & (1.0 -- 1.5)\% $X_0$ \\
\hline
Maximum drift time &  ${\sim}150~$ns &  ${\sim}80~$ns \\
Hit leading time resolution & (3 -- 4)~ns & (1 -- 4)~ns\\
Hit trailing time resolution & ${\sim}30~$ns & ${\sim}6~$ns \\
Average number of hits hits per view & 2.2 & 3.1\\
\hline
\end{tabular}
\label{tab:straw_comparison}
\end{table}


\begin{figure}[p]
\centering
\includegraphics[width=0.37\textwidth]{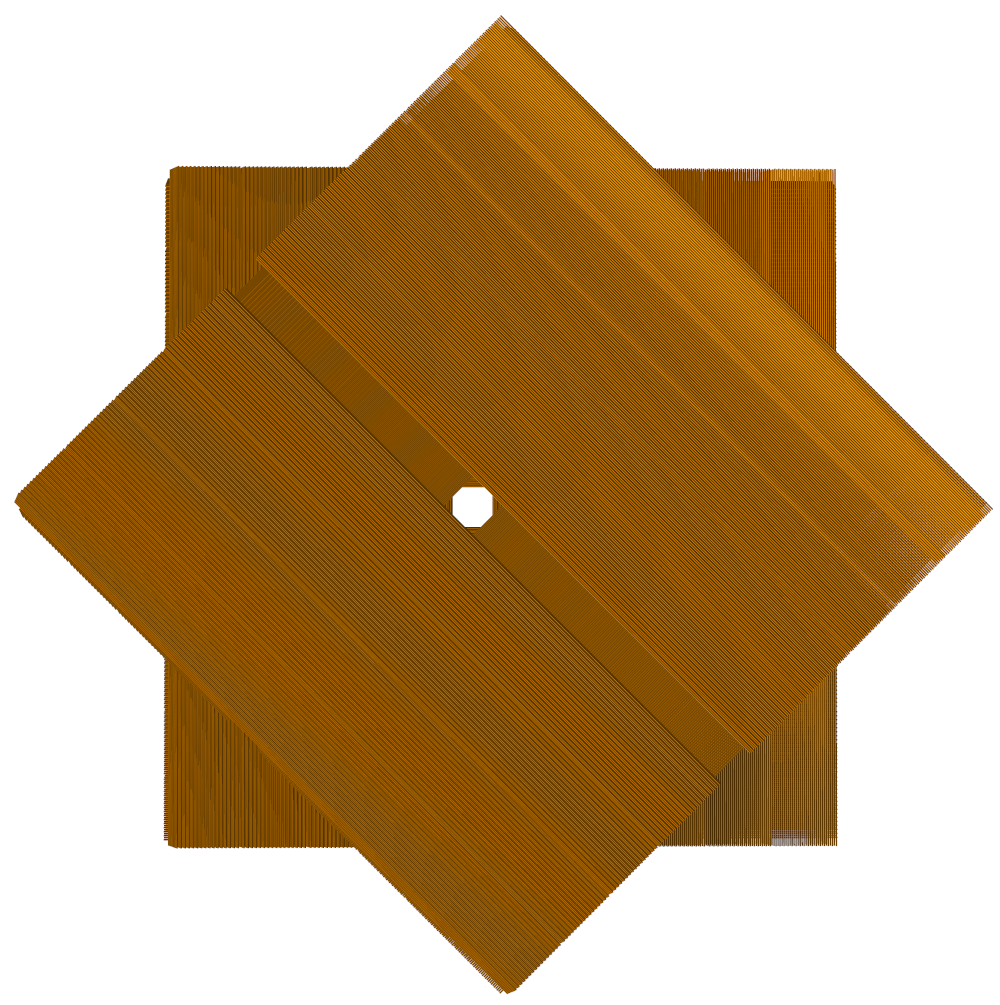}%
\includegraphics[width=0.37\textwidth]{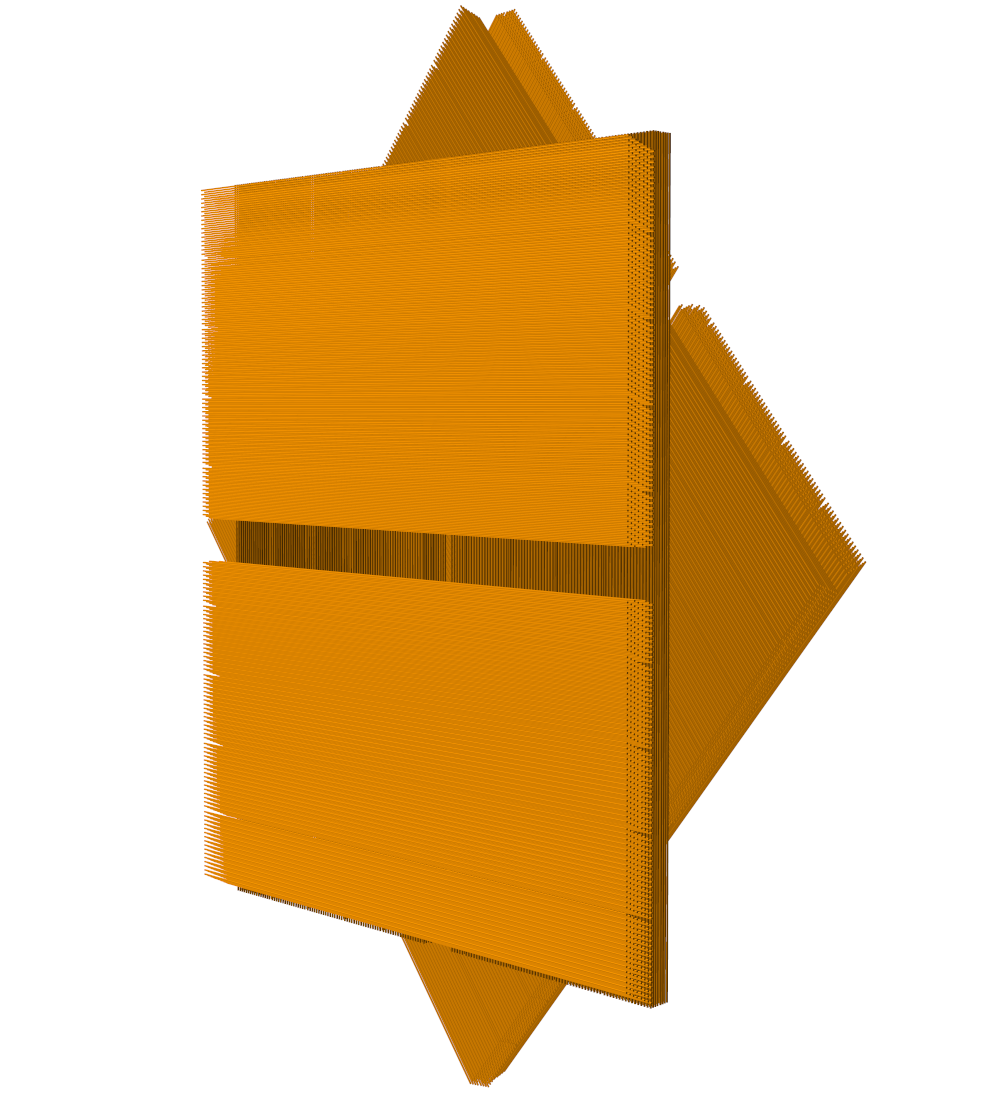}
\vspace{-2mm}
\caption{
\label{fig:straw_visualization2}
Geant4 visualisation of a new straw chamber: (left) front view; (right) tilted back view.}
\end{figure}

\begin{figure}[p]
\centering
\includegraphics[width=0.5\textwidth, page=1, trim={0.0cm 0.0cm 1.8cm 0.0cm}, clip]{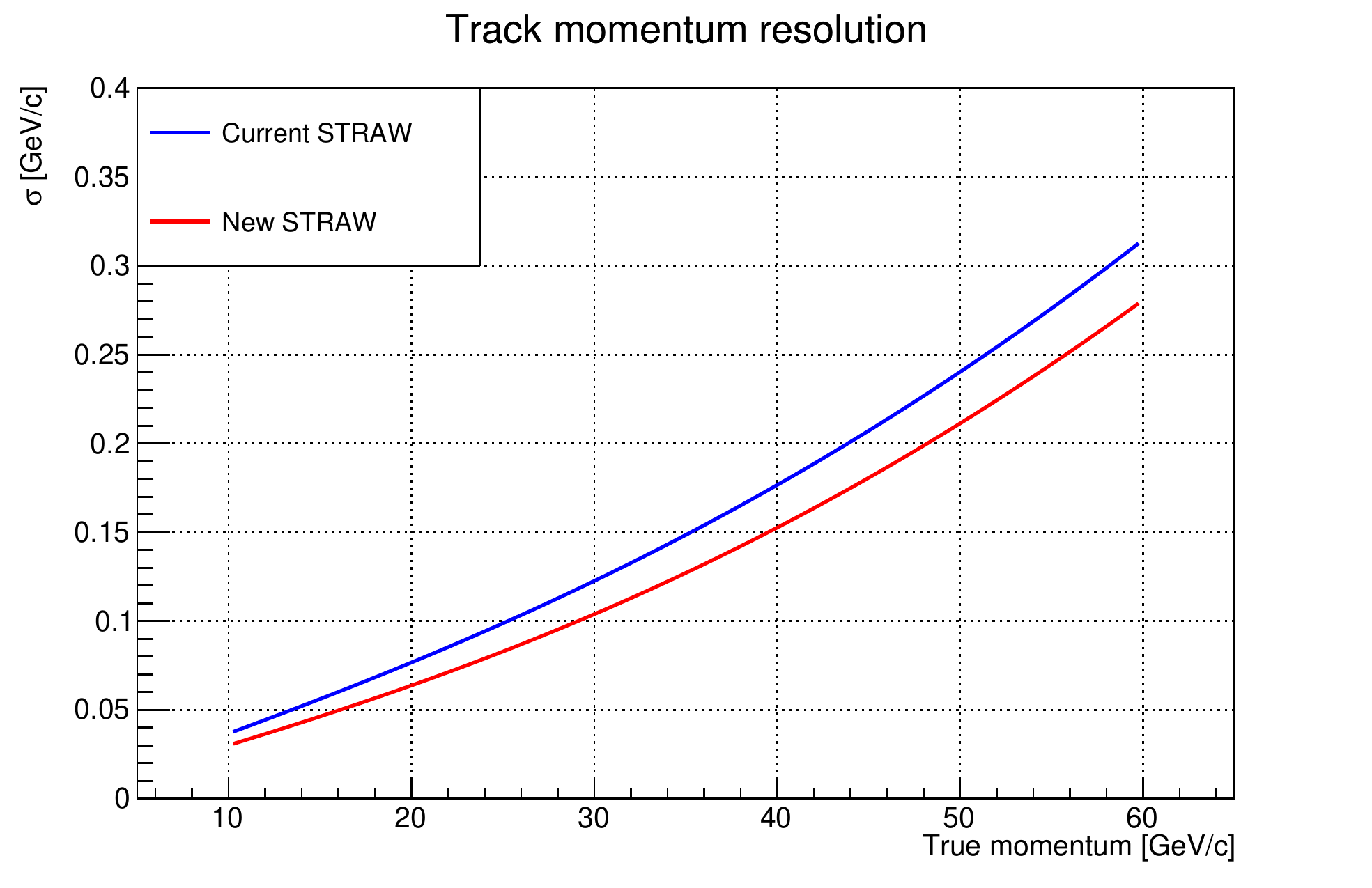}%
\includegraphics[width=0.5\textwidth, page=3, trim={0.0cm 0.0cm 1.8cm 0.0cm}, clip]{figs/straw_resolutions.pdf}
\vspace{-8mm}
\caption{
\label{fig:straw_resolution}
Preliminary comparison of resolutions of track momentum (left) and track $\theta_X$ angle (right) between the existing NA62 spectrometer (blue) and the new spectrometer with $12~\mu$m mylar thickness (red). A similar improvement is observed in the reconstructed $\theta_Y$ angle.}
\end{figure}

\clearpage




%% file: detectors/RICH.tex
\subsubsection{The RICH detector}

The NA62 RICH detector, which uses neon at atmospheric pressure as the radiator, is well suited for operation within the HIKE programme in terms of the mechanical structure (vessel, mirror support, end-caps). Major changes should only concern the Cherenkov light sensors and the two flanges hosting them. Improvement of the geometrical acceptance for negative particles is also being considered.

\paragraph{New photodetectors}

The present NA62 RICH is equipped with Hamamatsu R7400-U03 phototubes with a time resolution of 240~ps for single photons, quantum efficiency (QE) of $\sim$20$\%$ and with a distance between the centres of adjacent sensors of 18~mm \cite{Anzivino_2020}. This distance, constrained by the sensor size, gives the main contribution to the resolution on the single hit position ($\sim$4.7~mm) and consequently to the overall resolution on the ring radius ($\sim$1.5~mm), the main parameter driving the performance on the particle identification (PID) of the RICH detector \cite{Anzivino_2018}. The single hit time resolution together with the number of hits associated to each ring (which depends on the QE) determines the overall time resolution of the RICH for positive tracks of 80~ps on average.

The main requirement for RICH operation at HIKE Phase~1 is a ring time resolution of 20--30~ps, in order to reduce the coincidence window with the upstream detectors (beam tracker and KTAG). The RICH is the only downstream detector that could reach such resolution for the kaon decay products. Besides, a reduction of the sensor size would improve the particle identification performance.
Silicon photomultiplier (SiPMs)
meet these requirements: SiPMs with 100~ps time resolution and QE of 40\% or above are already available from the main manufacturers.

Considering a similar active area geometrical filling factor (80\%) and a similar average number of hits per ring to the NA62 case, and taking into account a factor~2 improvement in the QE, we have evaluated the number of hits per ring and the track time resolution with the new configuration. The time resolution for pion momenta of 15 and 45~GeV/$c$
(the limits of the RICH working region) for the NA62 RICH, and those expected for the future RICH instrumented with SiPMs, are listed in Table~\ref{tab:RICH1}. The latter meets the HIKE requirements.

\begin{table}[tb]
\begin{center}
\caption{Comparison of the time resolution of the NA62 RICH and the new RICH.}
\vspace{-2mm}
\begin{tabular}{l|c|c}
\hline
& NA62 RICH & Future RICH \\ 
\hline
Sensor type & PMT & SiPM \\
Sensor time resolution & 240~ps & 100~ps \\
Sensor quantum efficiency & 20\% & 40\% \\
Number of hit for $\pi^+$ at 15~GeV/$c$ & 7 & 14 \\
Number of hit for $\pi^+$ at 45~GeV/$c$ & 12 & 24 \\
Time resolution for $\pi^+$ at 15~GeV/$c$ & 90~ps & 27~ps \\
Time resolution for $\pi^+$ at 45~GeV/$c$ & 70~ps & 20~ps \\
\hline
\end{tabular}
\label{tab:RICH1}
\end{center}
\vspace{-4mm}
\end{table}

\begin{figure}[tb]
\centering
\includegraphics[width=0.8\linewidth]{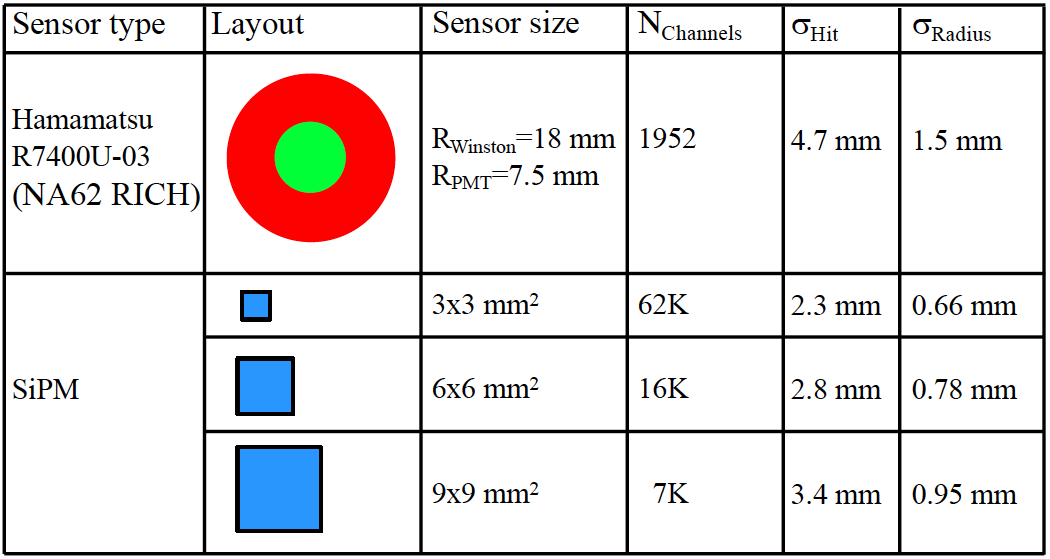}
\caption{Layout of the present RICH NA62 sensors compared to SiPM candidates (in scale) and the corresponding performance in terms of the ring radius resolution.}
\label{fig:RICH1}
\end{figure}

Replacement of the light sensors, necessary to improve the time resolution, also represents an opportunity to improve the RICH performance in terms of particle identification. A smaller sensor size and an improved QE will allow to establish a more optimal RICH working point, improving both the muon rejection and pion identification efficiency in the $K^+\to\pi^+\nu\bar\nu$ analysis. Fig.~\ref{fig:RICH1} illustrates the improvement in the ring radius resolution $\sigma_{\rm Radius}$ achieved by sensor size reduction and QE improvement. For a small SiPM size, the contributions to the single hit resolution coming from misalignment of the reflecting mirrors and neon (radiator) dispersion, 0.6~mm and 2.1~mm respectively, become dominant.

In considering SiPMs as photodetector candidates for the HIKE RICH, the effects of the main drawbacks related to these sensors must be evaluated:
\begin{itemize}
\item[-] Dark count: for a 800~ps coincidence window, and an annulus area of $7\times 10^{4}~{\rm mm}^2$ considered for evaluation of the possible particle identification hypotheses, a dark count rate of several kHz/mm$^2$ would produce a non-negligible number of spurious hits. To lower the contamination to the level of few percent, the SiPM should be operated at low temperature. A possible layout of the flanges housing the SiPM and the cooling system is discussed in Section~\ref{RICH-mechanics}.
\item[-] Cross-talk is strongly dependent on the SiPM type, can be reduced by cooling the SiPM, and will be considered for the choice of the final sensor. Cross-talk leads to an extra contribution (for a small fraction of events) to the hit space resolution, and consequently to the ring radius resolution.
\item[-] Ageing/radiation hardness: the RICH flanges are located at a distance of 1.5~m from the beam pipe and are not traversed by the bulk of the particle flux. Nevertheless the radiation level in the high intensity environment should be investigated. Cooling of the SiPMs would reduce the effects of radiation damage.
\end{itemize}
Alternative photodetectors are the MCP-PMTs considered for the HIKE KTAG detector and described in Section~\ref{sec:ktag}.

\paragraph{New mechanics}
\label{RICH-mechanics}

The NA62 RICH vessel, end-caps and the mirror support panel will not be changed in the HIKE configuration, apart for minor modifications. 
Concerning the mirrors, a replacement or a re-aluminisation should be
considered to recover from the deterioration of the coating layer already observed in 2014 during their installation.
The new reflecting surface must match, in terms of reflectivity as a function of light wavelength, the quantum efficiency of the photodetectors, so both the reflective and coating layers will be chosen accordingly.

\begin{figure}[tb]
\centering
\includegraphics[width=0.4\linewidth]{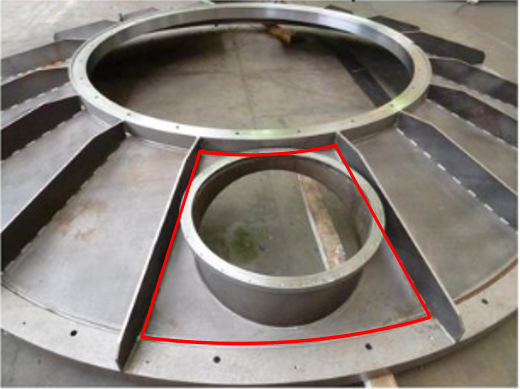}
\caption{The upstream RICH end-cap after its construction, in 2013. The circular hole hosts the sensor flange in the NA62 RICH. The red contour delimits the region that can be instrumented with new photo-sensors.}
\label{fig:RICH2}
\end{figure}

The change of sensors will request modification of the two
flanges hosting them. This represents an opportunity to increase the RICH acceptance for negatively-charged particles (note that the NA62 RICH is optimised
for the $K^+\to\pi^+\nu\bar\nu$ decay).
Simulations show that increasing the instrumented area from 5700 cm$^2$ to 7600~cm$^2$ would lead to a good acceptance for negative tracks. The end-cap region that can be instrumented without compromising the vacuum-proof mechanics is shown in Fig.~\ref{fig:RICH2}. In the redesign of the new flanges, a photosensor cooling system will be introduced, which would avoid
inducing a temperature gradient in the RICH radiator gas. An adequate system to guarantee thermal insulation between the sensor
flanges and the vessel must be considered.  

%% file: detectors/Timingplane.tex
\subsubsection{The timing detector}

The timing detector will allow to have a redundant and complementary measurement of the transit time of charged particles in the downstream detector region with respect to the RICH. The information of the timing plane can be used for efficiency studies, trigger purposes and to veto accidental activity. The timing detector can also be used to have more than one time measurement for pions and muons with momentum below the Cherenkov thresholds in the RICH of 12 and 8~GeV/$c$, respectively. The requested time resolution is about 100~ps.

\begin{figure}[tb]
\centering
\includegraphics[width=0.75\linewidth]{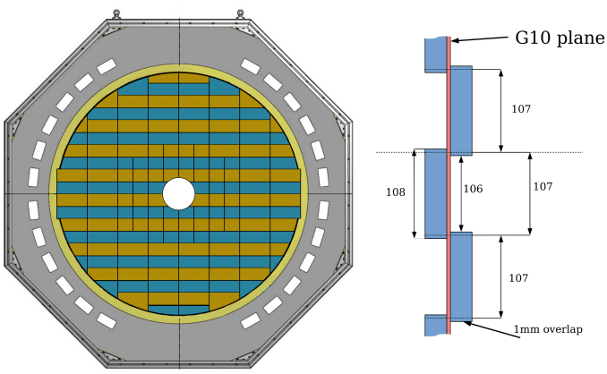}
\caption{Layout of the present NA62 CHOD in the $xy$ and $xz$ planes.}
\label{fig:CHOD1}
\end{figure}

For the construction of this detector HIKE will profit from the expertise gained by NA62 with the NA48-CHOD and NA62-CHOD detectors. The NA48-CHOD, consisting of two planes of scintillator
slabs of approximately $100\times6$~cm$^2$ size, cannot be used in the HIKE environment; the high particle rate (and the high probability of more than one track hitting the same counter) will not allow to 
correct the measured time for the light propagation time in the scintillator to the photo-sensors depending on the track impact point.
On the contrary, the NA62-CHOD layout (Fig.~\ref{fig:CHOD1}),
consisting of tiles of approximately $10\times 13$~cm$^2$ (in the central region) will be suitable for HIKE. The expected maximum rate per tile of 2.8 MHz (inferred from experience with NA62 data taking) is affordable and can be lowered further by reducing the dimension of the tiles near the beam pipe, where the particle rate is higher. 

The NA62-CHOD time resolution (500~ps) is limited by the light transportation outside the detector acceptance via wavelength shifting fibers. In the HIKE setup, the photosensors will be attached
directly to the tiles to avoid the degradation. In this layout the sensors will be irradiated by the downstream particle flux. The photosensor type will be chosen profiting from the studies foreseen for other detectors (KTAG, RICH, HCAL, MUV).  To further improve the time resolution, an additional detector plane can be considered.


%% file: detectors/EMCalo.tex
\subsubsection{The HIKE electromagnetic calorimeter}
\label{sec:calorimeter}

The forward electromagnetic calorimeter is crucial for any kaon program. In particular, in the HIKE $K^+$ phase this detector serves as the principal photon veto for the measurement of $K^+\to\pi^+\nu\bar\nu$ and is essential for the reconstruction of the final states for decays like $K_{\ell2}$, $K^+\to\pi^+\ell^+\ell^-$, and $K^+\to\pi^+\gamma\gamma$. For the neutral beam configurations, the electromagnetic calorimeter reconstructs the $\pi^0$ vertex for $K_L\to\pi^0\nu\bar{\nu}$ events and helps to reject events with extra photons. Many of the same issues arise in the design of the electromagnetic calorimeter for the $K^+$ and $K_L$ phases. 
We therefore seek a design for a fast calorimeter with excellent photon detection efficiency and energy resolution to be used in all phases of the HIKE program. 
While the new calorimeter is investigated and built, 
the existing NA62 LKr calorimeter is presented as a possible solution for the $K^+$ program since it could meet the performance requirements.

\paragraph*{Performance requirements}

The principal performance requirements for the electromagnetic calorimeter are excellent energy resolution and intrinsic detection efficiency for high-energy photons, good two-cluster separation for photons, and excellent time resolution.

It is natural to inquire as to whether the liquid-krypton calorimeter (LKr) \cite{Fanti:2007vi}
currently used in NA62 can be reused for HIKE.
The energy, position, and time resolution of the LKr calorimeter were measured in NA48 to be
\begin{align}
\frac{\sigma_E}{E} &=0.0042\oplus\frac{0.032}{\sqrt{E {\rm (GeV)}}}\oplus\frac{0.09}{E {\rm (GeV)}}, \\
\sigma_{x, y} & = 0.06~{\rm cm} \oplus \frac{0.42~{\rm cm}}{\sqrt{E {\rm (GeV)}}}, \\
\sigma_t & = \frac{2.5~{\rm ns}}{\sqrt{E {\rm(GeV)}}}.
\end{align}
Indeed, the efficiency
and energy resolution of the
LKr appear to be satisfactory for all phases of HIKE.
Studies of $K^+\to\pi^+\pi^0$ decays in NA48 data and tests conducted in 2006 with tagged photons from an electron beam confirmed that the LKr has an inefficiency of less than $10^{-5}$ for photons with $E > 10$~GeV, providing the needed rejection for forward photons \cite{NA62+07:M760}.
These studies were fully confirmed in NA62. Notwithstanding the presence of a much larger amount of material upstream of the LKr calorimeter in NA62 than in NA48, a study of single-photon efficiency underpinning the NA62 measurement of ${\rm BR}(K^+\to\pi^+\nu\bar{\nu})$ (and used to obtain a limit on ${\rm BR}(\pi^0\to{\rm invisible})$)
found an inefficiency of about $10^{-5}$ at $E=20$~GeV, slightly decreasing at higher energies \cite{NA62:2020pwi}.
The LKr time resolution, however, is a significant issue. The necessary upgrade for the LKr calorimeter to be used in the $K^+$ phase is discussed below (\Sec{sec:LKr_calo}). For the $K_L$ phases, and for KLEVER in particular, the calorimeter provides the measurement of the event time, and must have a time resolution of 100~ps or better for the reconstruction of $\pi^0$'s with energies of a few GeV. Additionally, the size of the LKr inner bore would limit the beam solid angle and hence the kaon flux during the $K_L$ phases. A new calorimeter that would meet these requirements is described in \Sec{sec:klever_mec}.

\paragraph{The NA62 LKr calorimeter}
\label{sec:LKr_calo}
The LKr energy resolution meets the HIKE requirements, while the time resolution must be substantially improved. This can be achieved by performing a major upgrade of the LKr readout electronics, as discussed below.

\paragraph*{Major upgrade of the LKr readout electronics}

The readout of the NA62 liquid krypton calorimeter is based on the measurement of the initial current induced by the charge generated by the showers, to have a fast response and to separate pile-up pulses. The electronics chain for each cell is composed of three elements:
\begin{itemize}
\item a preamplifier, sitting in the cold liquid,  to collect the charge with an integration time of about 150~ns;
\item a so-called transceiver module, mounted immediately after the cryostat feed-throughs, whose function is to restore the fast signal, removing the pole of the integration in the preamplifier, and to prepare a differential signal to drive long cables to the next stage;
\item a shaper, to have a narrow signal (70~ns), followed by an FADC to digitise the signal at a rate of 40~MHz.
\end{itemize}

The increase in intensity by a factor 4 will produce more pile-up events, whose detection could be difficult with the existing chain. To cope with this, the baseline for a new readout structure will be the reduction of the shaping time to the minimum possible of about 28~ns (the transceiver output has a risetime of about 20~ns which adds to the time constants of the shaper) with a reduction of the amplitude of about 40\%, and subsequent digitization at 160~MHz, which is above the Nyquist limit, but which could help in identifying superposition of pulses looking at the width of the pulse.

\paragraph*{Space charge}
The increase of intensity will also worsen the effect of the space charge: already in the current operation we observe a loss of response in the hottest cells, due to the build up of the space charge during the burst. This effect is inversely proportional to the square of the value of the high voltage across the cell and directly proportional to the energy density created by ions. The effect of the space charge is, at low densities, to reduce the electric field at the anode and to increase it at the cathode. There is then a threshold value after which the field at the anode is zero and above this value an increasing fraction of the cell is unresponsive.

With the current NA62 beam intensity we see already that in the hottest cells we are near or just above the critical value, clearly seen both from the decrease in the response across the burst and from a modulation of the response across the $x$ coordinate of the cells. An increase in intensity will worsen this behaviour for the hottest cells, but it will also extend the area affected by this problem. To mitigate this issue, an increase of the high voltage is possible: going from the actual 3.5 kV to 5 kV will reduce by two the value of the critical parameter and the performance will be the same as now, which is not a problem for veto operations. An additional improvement for all analyses that do not use the calorimeter as a veto will be to increase the size of the actual Intermediate Ring Counter (IRC, \Sec{sec:irc}), to cover the area most affected by the space charge.




\paragraph{The new HIKE electromagnetic calorimeter} 
\label{sec:klever_mec}
We are investigating the possibility of replacing the
LKr with a shashlyk calorimeter patterned on the PANDA FS calorimeter, in turn
based on the calorimeter designed for the KOPIO experiment~\cite{Atoian:2007up}. This design featured modules $110\times110$~mm$^2$ in cross section made of 
alternating layers of 0.275-mm-thick lead absorber and 1.5-mm-thick injection-moulded polystyrene scintillator.
This composition has a radiation length of 3.80 cm and a sampling fraction of 39\%. 
The scintillator layers were optically divided into four $55\times55$~mm$^2$ segments; the scintillation light was collected by WLS fibres traversing the stack longitudinally and read out at the back by avalanche photodiodes (APDs). KOPIO was able to obtain an energy and time resolution of 3.3\% and 73~ps at 1~GeV with this design, establishing that it is capable of providing the same energy resolution as the LKr while meeting the time resolution requirements for HIKE.

For HIKE, the design would be updated to use silicon photomultipliers (SiPMs) instead of APDs. The final choice of module size and readout granularity has yet to be determined, but on the basis of KLEVER simulations that assume that clusters are resolved if more than 6~cm apart, readout cells of $5\times5$~cm$^2$ seem reasonable. The Moli\`ere radius for the fine-sampling shashlyk design described above is 3.39~cm,
so smaller cells can be used if needed. For comparison, 
the Moli\`ere radius of liquid krypton is 5.86~cm, and the NA48 LKr calorimeter features $2\times2$~cm$^2$ cells.
\begin{figure}
    \centering
    \includegraphics[width=0.52\textwidth]{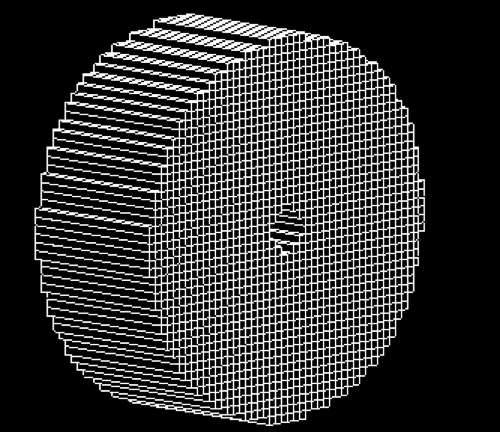}
    \caption{Wireframe drawing of the MEC as implemented in Geant4 for the NA62 Monte Carlo.}
    \label{fig:mec}
\end{figure}
The new calorimeter, referred to as the main electromagnetic calorimeter (MEC) in KLEVER, would have an inner bore of at least 12~cm in radius to allow the passage of the neutral beam. The bore could be widened to as much as 15~cm to allow the penumbra of beam photons and neutral hadrons to pass through and be intercepted by the small-angle vetoes. For the $K^+$ phase, a smaller inner aperture would be required; in NA62, this angular region is covered by the IRC (\Sec{sec:irc}), and a new IRC based on the same shashlyk technology would be used together with the MEC during the $K^+$ phase. The sensitive area has an outer radius of 125~cm. \Fig{fig:mec} shows a wireframe drawing of the calorimeter as implemented in the NA62 Geant4-based Monte Carlo for HIKE.

\begin{figure}
\centering
\includegraphics[width=0.7\textwidth]{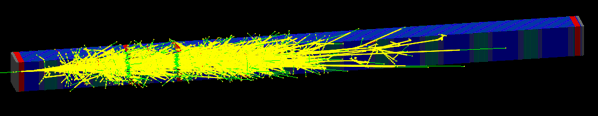}
\caption{Geant4 simulation of a 5~GeV photon showering in a MEC module.}
\label{fig:shashlyk_shower}
\end{figure}
\begin{figure}
\centering
\includegraphics[width=0.6\textwidth]{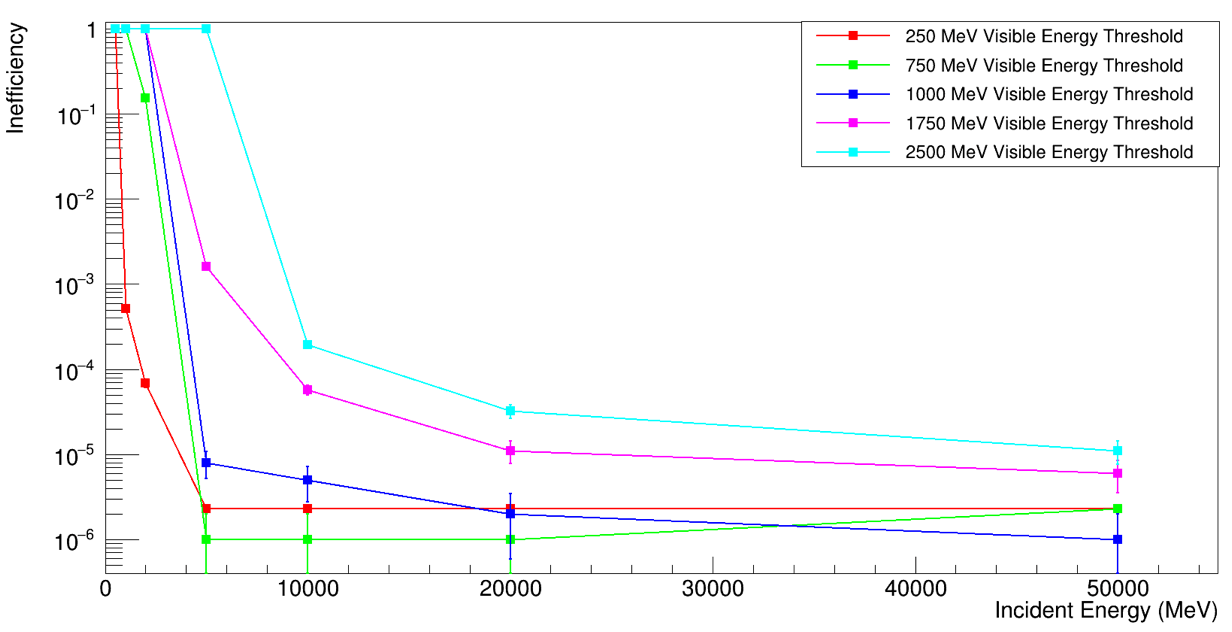}
\caption{Detection efficiency as a function of incident particle energy, for various thresholds on visible energy, from the Geant4 simulation.}
\label{fig:shashlyk_efficiency}
\end{figure}

For the KLEVER sensitivity studies of \Sec{sec:klever_sens}, the inefficiency for photon detection assumed for the MEC was assumed to be 1 for energies below 100~MeV, falling exponentially to $10^{-3}$ at 1~GeV, $10^{-4}$ at 5~GeV, and thence to $10^{-5}$ at 15~GeV, in accordance with experience with the LKr as discussed above. The model of the shashlyk calorimeter developed for the HIKE Monte Carlo does not include the light and signal readout, which are still being defined, but does contain a detailed representation of the module structure, allowing studies of the efficiency as a function of visible energy deposition. \Fig{fig:shashlyk_shower} shows the simulated interaction of a 5~GeV photon in a module of the calorimeter. \Fig{fig:shashlyk_efficiency} shows results obtained from the simulation for the photon detection inefficiency as a function of incident energy for various thresholds on visible energy. The simulation, which confirms the visible energy fraction of 39\%, demonstrates that the fine-samplng shashlyk design satisfies the efficiency requirements for KLEVER, as expected. 

The radiation resistance of the scintillator is a potential concern, especially for the $K_L$ phase. In KLEVER, the dose rate is dominated by photons from $K_L$ decays. The rate is most intense on the innermost layers of the calorimeter, for which it is about 2~kHz/cm$^2$.
Precise dose rate calculations have yet to be performed, but an estimate suggests a dose of 4~kGy/yr to the scintillator for the innermost layers. This estimate would suggest that radiation damage, while a concern, is likely manageable. Radiation robustness may be a factor in the final choice of scintillator.

\begin{figure}
    \centering
    \includegraphics[width=0.4\textwidth]{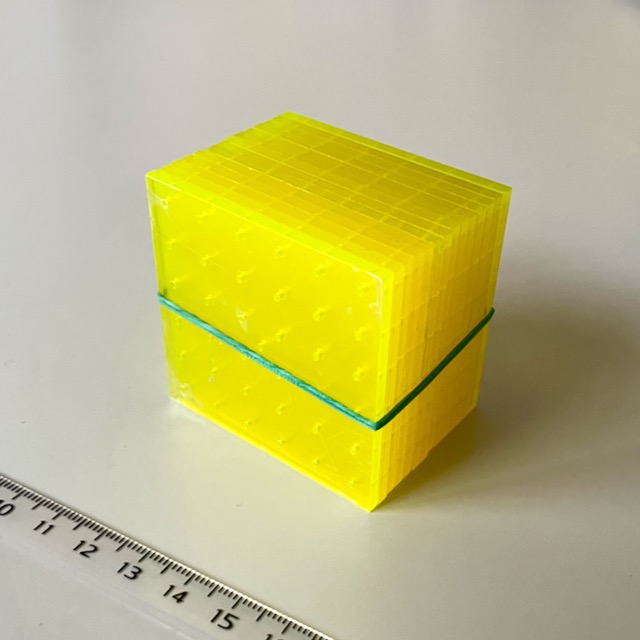}\hspace{0.075\textwidth}
    \includegraphics[width=0.4\textwidth]{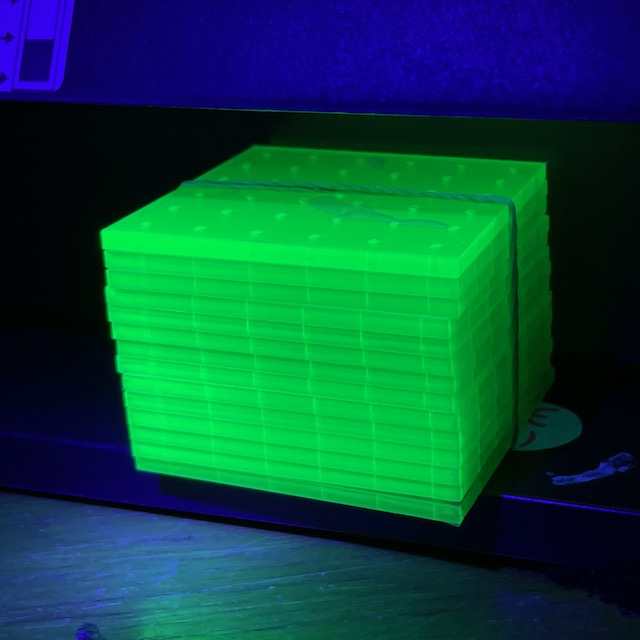}
    \caption{Shashlyk tiles made of perovskite nanocomposite scintillator tested as an alternative to conventional scintillator, in ambient light (left) and under ultraviolet light (right).}
    \label{fig:nc}
\end{figure}
Although current information suggests that optimized formulations of conventional polystyrene scintillator are sufficiently luminous, fast, and radiation resistant, in synergy with the AIDAinnova project NanoCal, we are evaluating the advantages that can be obtained with less conventional choices for the light emitter (e.g., perovskite \cite{Gandini:2020aaa, Dec:2022aaa} or chalcogenide \cite{Liu:2017aaa} quantum dots) or matrix material (e.g., polysiloxane \cite{Acerbi:2020itd}). In particular, we have participated in a recent head-to-head experimental comparison of small shashlyk prototypes made from conventional scintillator, specifically, the extrusion-moulded polystyrene scintillator formulated at IHEP Protvino for KOPIO \cite{Atoian:2007up}, with 1.5\% PTP and 0.04\% POPOP, and a nanocomposite scintillator consisting of 0.2\% lead halide bromide (CsPbBr$_3$) nanocrystals in PMMA\footnote{Glass To Power SpA, Rovereto TV, Italy.}. The latter, which emits at 520~nm (\Fig{fig:nc}), is expected to be a very fast and bright alternative to conventional scintillators; its comparatively long wavelength emission and use of PMMA as a matrix material is expected to confer good radiation hardness. 
\begin{figure}
    \centering
    \includegraphics[height=0.3\textheight]{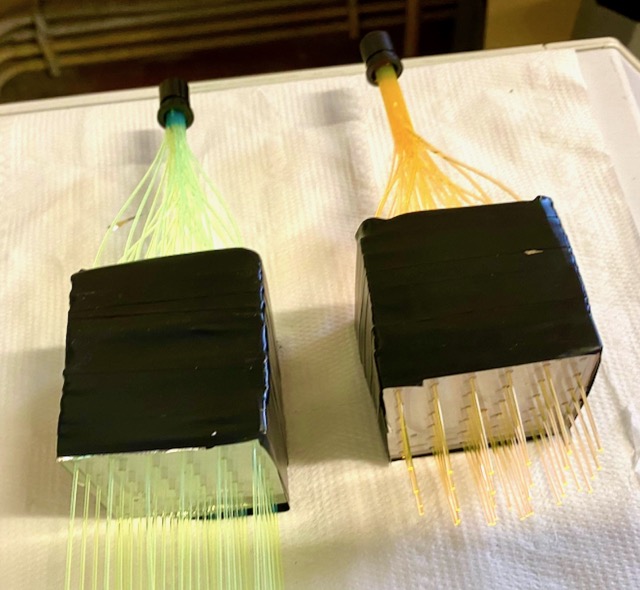}\hspace{0.075\textwidth}
    \includegraphics[height=0.3\textheight]{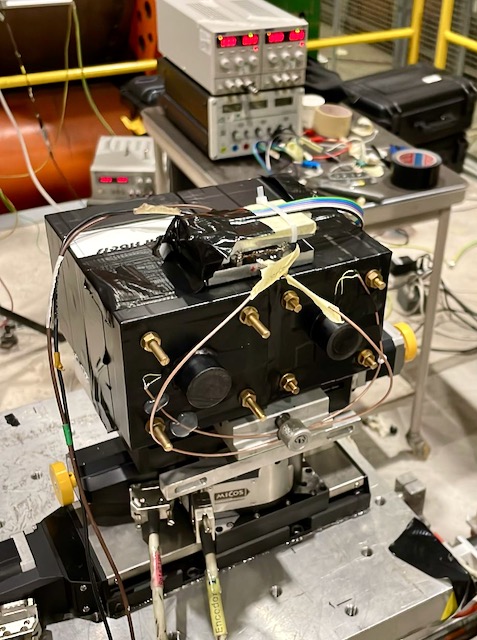}
    \caption{Small shashlyk prototypes tested in fall 2022, during construction (left) and on the H2 beam line for testing (right).}
    \label{fig:nc_test}
\end{figure}
Identical prototypes consisting of 12 layers of 0.6-mm lead and 3-mm scintillator were constructed out of components from the PANDA/KOPIO calorimeter (\Fig{fig:nc_test}).
For the conventional prototype, the original scintillating tiles were used with Kuraray Y-11(200) green-emitting WLS fibre. The nanocomposite prototype used Kuraray O-2(100) orange-emitting WLS fibre. Each was read out with a single Hamamatsu 13360-6050 SiPM ($6\times6$~mm$^2$, 50~$\mu$m pixel size) and fast amplifier with a gain of 4. Both protoypes worked well; the conventional prototype showed significantly more light output than the nanocomposite prototype, but we have reason to believe that the luminosity and attenuation length of the orange fibres plays a significant role. A series of tests are planned to explore different fibre-scintillator pairings (including the use of custom-produced fibres) as well as successively optimised nanocomposite scintillator formulations.     

In addition to the basic criteria on energy resolution, efficiency, time resolution, and two-cluster separation, ideally, the MEC would provide information useful for particle identification. For example, in KLEVER, identification of pion interactions would provide additional suppression of background for decays with charged particles in the final state, and, as the experience with KOTO suggests, it is crucial to have as much information as possible to assist with $\gamma/n$ discrimination. The fine transverse segmentation of the MEC will play an important role: a simple cut on cluster RMS with the existing LKr can suppress up to 95\% of pion interactions in NA62 data. Fast digitisation of the signals from the MEC is expected to provide additional $\gamma/n$ discrimination. For both the $K^+$ and $K_L$ phases, the MEC will be backed up with hadronic veto calorimeters as discussed in \Sec{sec:hcal}.

\begin{figure}
    \centering
    \includegraphics[width=0.8\textwidth]{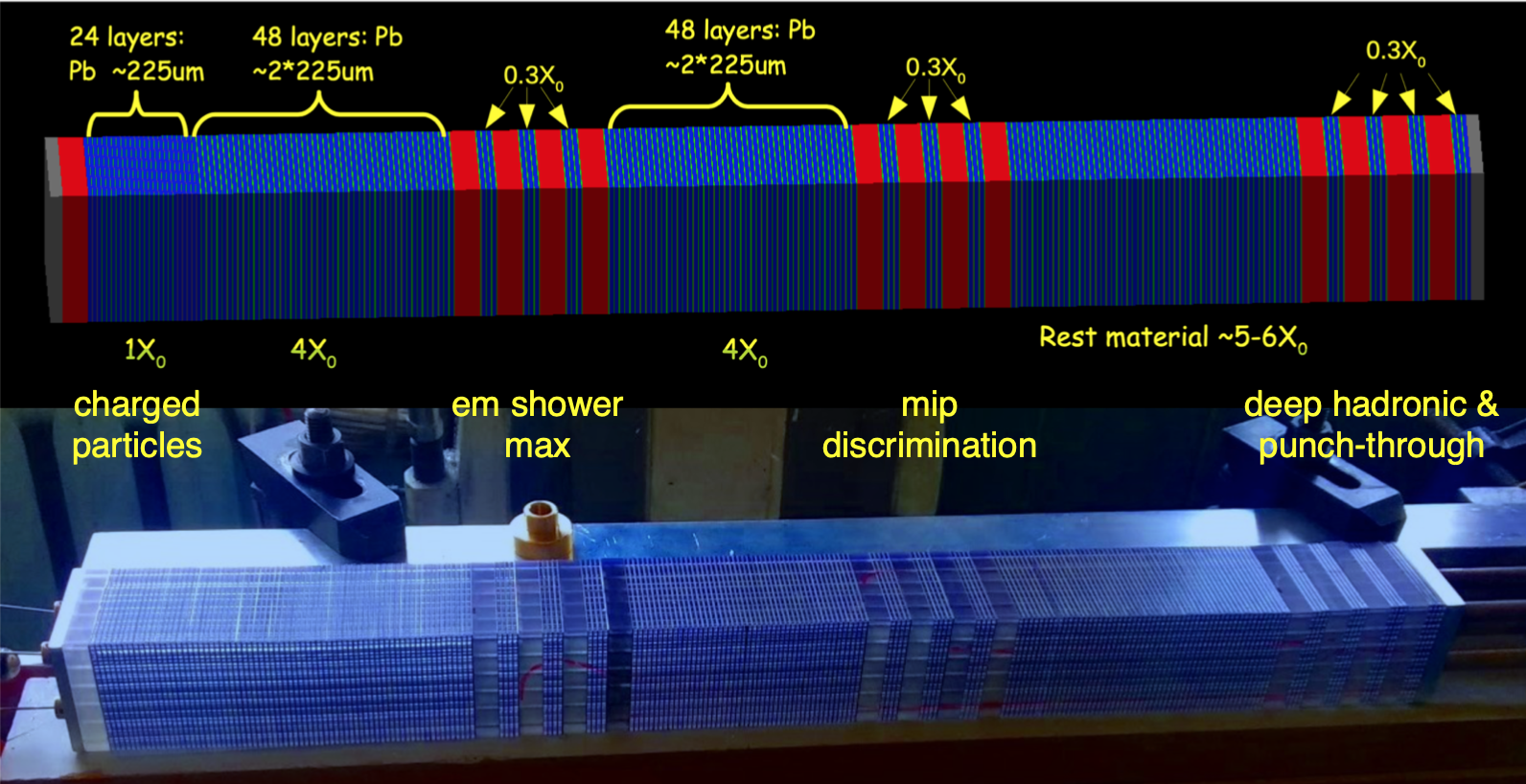}\\ \vspace*{5mm}
    \includegraphics[width=0.8\textwidth]{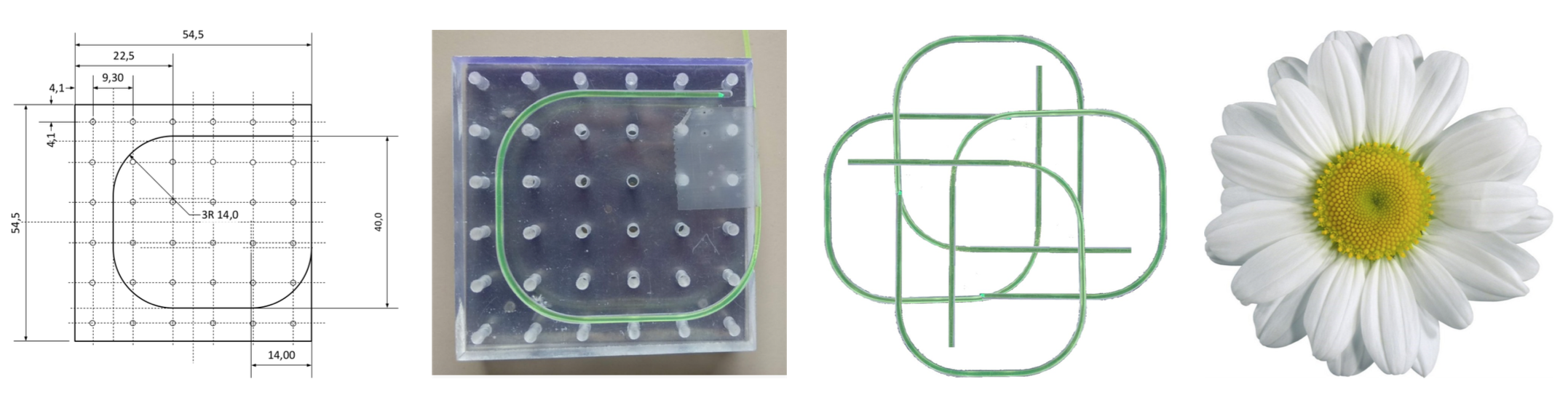}\\
    \caption{Top: Geant4 model of small prototype for {\it romashka} calorimeter, featuring spy tiles placed at key points in the shashlyk stack, together with a photograph of the module constructed at Protvino. Bottom: Fiber routing scheme for independent readout of spy tiles, giving rise to the name {\it romashka} (chamomile).}
    \label{fig:klever_spy_tiles}
\end{figure}
We are also experimenting with concepts to obtain information on the longitudinal shower development from the shashlyk design. One possible concept makes use of ``spy tiles'', 10-mm thick scintillator bricks
incorporated into the shashlyk stack but optically isolated from it and
read out by separate WLS fibres. The spy tiles are located at key points
in the longitudinal shower development: near the front of the stack,
near shower maximum, and in the shower tail, as illustrated in 
Fig.~\ref{fig:klever_spy_tiles}. This provides longitudinal sampling of the shower
development, resulting in additional information for $\gamma/n$ separation.
The prototype shown in  Fig.~\ref{fig:klever_spy_tiles} was constructed at Protvino in early 2018. Its basic functionality was tested in the OKA beamline in April 2018, and more comprehensive tests were carried out in September 2019 at DESY in collaboration with LHCb. Simulations suggest (and preliminary test beam data validate, to a certain extent) that the {\it romashka} design with spy tiles can give at least an order of magnitude of additional neutron rejection relative to what can be obtained from the transverse segmentation of the calorimeter alone, providing an overall suppression of up to 99.9\% for neutron interactions. 
The small prototype has a cross sectional area of only $55\times55$~mm$^2$ (one readout cell) and a depth of $14 X_0$ (60\% of the design depth), and both transverse and longitudinal leakage significantly complicated attempts to measure the time resolution. For electrons with $1 < E < 5$~GeV, the measured time resolution was about 200~ps and virtually constant; we expect that this can be significantly improved. The time resolution for hits on the spy tiles (independently of information from the shashlyk stack) was on the order of 500--600~ps, which may be difficult to improve. This is not expected to be a problem, however: the main shashlyk signal establishes the event time and the association of the PID information from the shashlyk tiles is based on the segmentation, with occupancies per cell of at most a few tens of kHz on the innermost layers.  

Although promising, the {\it romashka} design is not the only solution under investigation for obtaining information on the longitudinal shower development. Other concepts being explored, besides variations on the spy tile readout such as on-board SiPMs, include alternatives such as two-sided front/back readout and explicit segmentation into two or more modules in depth. 

We plan to continue R\&D work on improvements to the shashlyk design such as from the use of innovative scintillators and longitudinal segmentation schemes through 2023, while at the same time, we are designing a full-scale prototype of the baseline solution with conventional scintillator and a uniform shashlyk stack for beam testing next year. 


%% file: detectors/IRC.tex
\subsubsection{Intermediate-ring calorimeter (IRC)}
\label{sec:irc}

In NA62, the IRC is a small, ring-shaped calorimeter that sits just upstream of the LKr to provide photon veto coverage for the angular interval corresponding to the inactive region near the inner surface of the LKr cryostat. 
The NA62 IRC has an inner radius of 60~mm and an outer radius of 145~mm. The inner and outer radii are not concentric---the inner bore is offset in the horizontal direction by 10 mm, so that it is centered on the beam axis.
In NA62, the IRC is a shashlyk calorimeter with layers consisting of 1.5~mm of lead and 1.5~mm of scintillator. It is segmented into quadrants and read out with PMTs. 

In HIKE, as discussed in \Sec{sec:calorimeter}, the IRC will be used to extend the coverage of the calorimeter to small radii, and performance specifications similar to those for the calorimeter will be necessary. 
If the LKr used is during the $K^+$ phase, the IRC will cover the region of highest rates on the LKr, where space-charge effects are important. 
The new MEC, not needing a cryostat, will have minimum dead space at its inner radius. However, it could be convenient to have a bore of 15 cm or more to allow passage of the neutral beam halo during the $K_L$ phase of the experiment. In that case, during the $K^+$ phase, the IRC would be used in the same way with the MEC as it would with the LKr. 
For the HIKE $K_L$ phases, the IRC will then be moved downstream of the calorimeter and used to veto photons at the small radii occupied by the penumbra of the neutral beam. 

Ideally, the same instrument would be used for both phases, but due to the horizontal displacement of the $K^+$ beam at the location of the IRC, changes to the geometry may be necessary. 
If two different IRCs are needed, they can certainly be built with identical technology, namely, as shashlyk calorimeters with geometry similar to that for the NA62 IRC and sampling and readout granularity as for the HIKE MEC, providing more light for better time resolution and higher readout granularity for better rate resistance.

%% file: detectors/HCAL.tex
\subsubsection{Hadron Calorimeter (HCAL) and Muon Veto Detector (MUV) }
\label{sec:hcal}

\paragraph{Hadron calorimeter requirements}

As in NA62, the hadron calorimeter (HCAL) is the main detector for 
$\pi/\mu$ identification and separation. 
While minimum ionizing muons can be distinguished from pion showers without difficulty,
so-called catastrophically interacting muons, which deposit all or a large fraction of 
their energy in the calorimeter, are of more concern.
The $\pi/\mu$ separation is therefore achieved both by longitudinal tracking 
through the whole calorimeter and by examining the transversal shower shapes 
of electromagnetic muon versus hadronic pion showers. 
Similar to the NA62 MUV1-3 detectors, an average muon misidentification probability of 
${\cal O}(10^{-6})$ over the momentum range from 10 to about 50~GeV is necessary 
(see Section~\ref{sec:phase1}). The HCAL therefore needs segmentation in all three dimensions.  

The 4--6 times higher intensity in HIKE with respect to NA62 leads to a total rate of 
charged tracks from $K^+$ decays of ${\cal O}(10~\text{MHz})$ on the calorimeters.
This converts into a sub-nanosecond time resolution which is required 
to avoid dead-time from random veto.
Such a time resolution cannot be achieved with scintillating strips as in the 
NA62 MUV1/MUV2, since the long light paths cause a resolution of 1~ns or worse. 
In addition, a strip read-out would suffer from double hits on the same strip at high rates.

\begin{figure}
    \centering
    \includegraphics[width=0.6\linewidth]{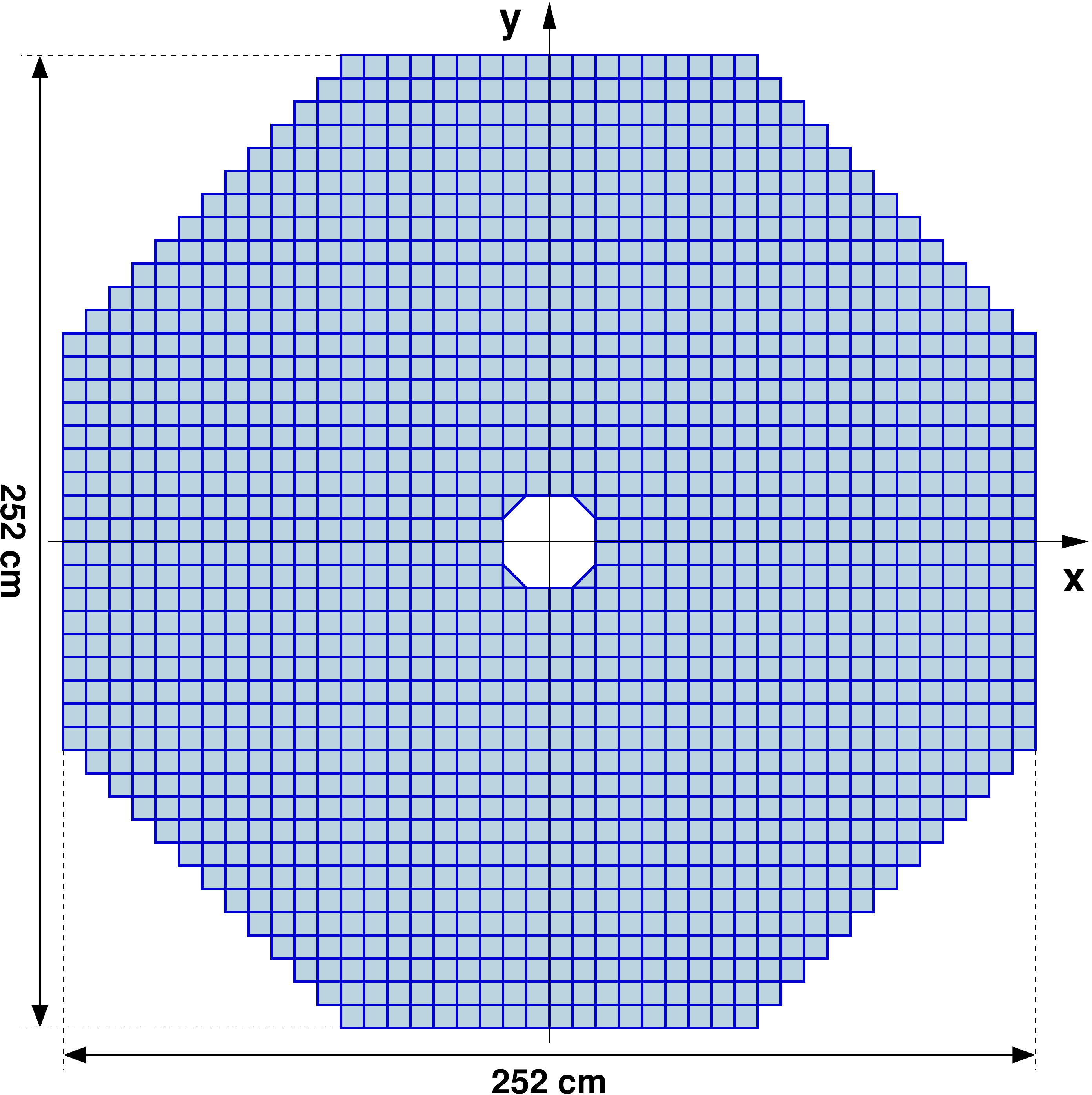}
    \caption{Transverse HCAL layout with an inner radius of 126~cm and readout cells of $6\times6$~cm$^2$ cross section. The plane comprises in total 1440 cells.}
    \label{fig:HCAL_Layout}
\end{figure}

The HIKE HCAL therefore needs to have a cellular layout to reduce both the rate on each channel and the time resolution.
With an iron absorber with a Moli\`ere radius of 1.72~cm, a cell size of 
$6\times 6$~cm$^2$ allows a sufficient distinction between hadronic pion showers
and electromagnetic showers from catastrophically interacting muons.
A corresponding layout of an HCAL with an octagonal shape with an inner radius 
of 126~cm is shown in Fig.~\ref{fig:HCAL_Layout}. 
It consists of a grid of $42 \times 42$~cells.
The subtraction of the beam pipe region and the corners to obtain an octagon results in 1440 cells in the transverse plane.

We consider two options for an iron-scintillator based calorimeter, 
which are described in the following.
The decision between the two options will be taken in the near future, after more detailed simulation of the particle ID capabilities of the two designs.

\paragraph{High-granularity hadron calorimeter}

For the ILD detector at the proposed International Linear Collider (ILC) 
the CALICE Collaboration has developed a highly granular analogue hadron calorimeter 
(AHCAL)~\cite{Sefkow:2018rhp} and a similar detector is being constructed for the 
CMS High Granularity Calorimeter Upgrade (HGCAL)~\cite{CMS:2017jpq}. 
In both these cases the high granularity is needed to improve 
jet energy resolution with particle-flow methods. 
While jet reconstruction is not necessary in HIKE, high granularity is nevertheless 
required for optimum $\pi/\mu$ separation and sub-nanosecond time resolution in a high-
rate environment as described above. 

\begin{figure}
\centering
\raisebox{1pt}{%
\includegraphics[width=0.31\linewidth]{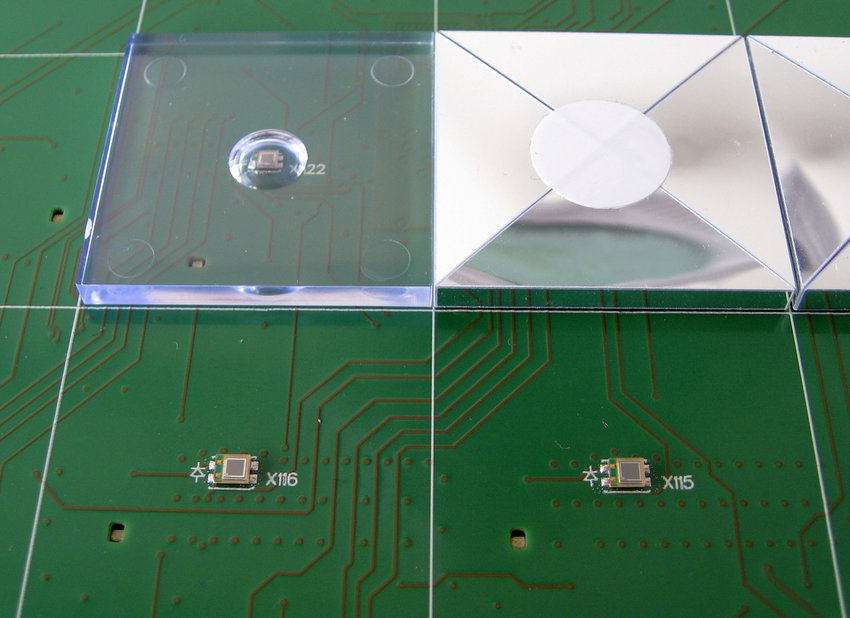}}
\hfill
\includegraphics[width=0.65\linewidth]{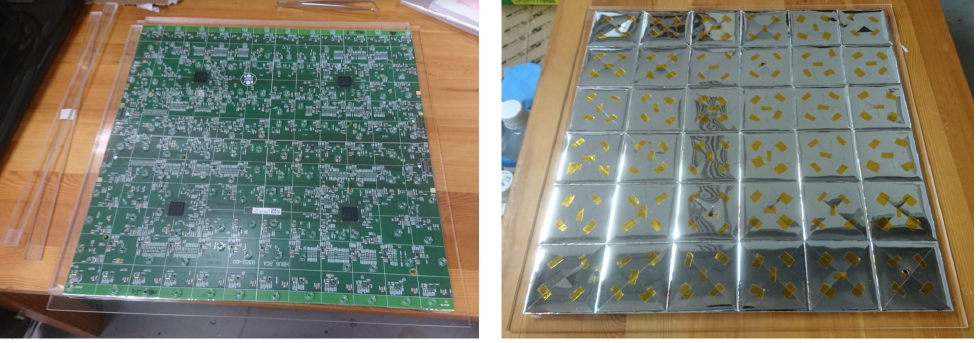}
\caption{Left: $3\times3$~cm$^2$ scintillator tiles of the CALICE AHCAL, wrapped and unwrapped, mounted on a common readout board with
SiPMs~\cite{Sefkow:2018rhp}.
Centre and right: back and front sides of a readout board with $6\times6$~cm$^2$ scintillator tiles~\cite{Tsuji:2019zuj}.}
\label{fig:CaliceTiles}
\end{figure}

The basic design of a high-granularity HCAL is an iron-scintillator sampling calorimeter
with 3~cm iron absorber plates interleaved with 40~layers of scintillating tiles of 
$6 \times 6$~cm$^2$ cross section and 1~cm thickness, which gives enough light yield 
to detect minimum ionizing particles. The total HCAL therefore comprises 
7.2~nuclear interaction lengths and has a length of about 2~m,
when considering additional space for electronics and air.
Each scintillating tile is readout separately by a single SiPM, which is connected 
to a common readout board (see Fig.~\ref{fig:CaliceTiles} for the Calice AHCAL layout). 
The SiPM needs a relatively large dynamic range, i.e. a large number of pixels,
for a linear energy measurement for both minimum ionizing particles 
and electromagnetic and hadronic showers. 
Candidates for such SiPMs are e.g. the existing Hamamatsu types S13360 or S14160 with
more than 14000 pixels and a sensitive area of $3 \times 3$ or $6 \times 6$~mm$^2$. 

Each readout board of $36\times36$~cm$^2$ size carries $6\times6=36$ scintillating
tiles, wrapped in reflective foil and glued to the PCB.
The SiPMs are surface-mounted on the PCB and are housed in small cavities 
on the bottom side of the tiles (Fig.~\ref{fig:CaliceTiles}). 
The front-end electronics together with a digitizing ASIC occupy the 
back side of the readout board. The digitization therefore takes place directly 
next to the photo detectors, ensuring an excellent time resolution.

Assuming the general design shown in Fig.~\ref{fig:HCAL_Layout} with 1440 cells 
in each layer and in total 40 scintillating layers, 
the number of readout boards is 1800 (not all fully equipped) and 
the total number of channels is 57600.
The possibility of a reduced channel count by e.g. using larger tiles in the 
outer region or less active layers is being studied at the moment.

As described above, the digitization of the SiPM signals is done with an on-board ASIC. 
A suitable such chip is the HGCROC~\cite{Bouyjou:2022wii}, 
which has been developed for the high-rate environment of the CMS HGCAL.

\paragraph{Shashlik hadron calorimeter}

\begin{figure}
    \centering
    \includegraphics[width=0.7\linewidth]{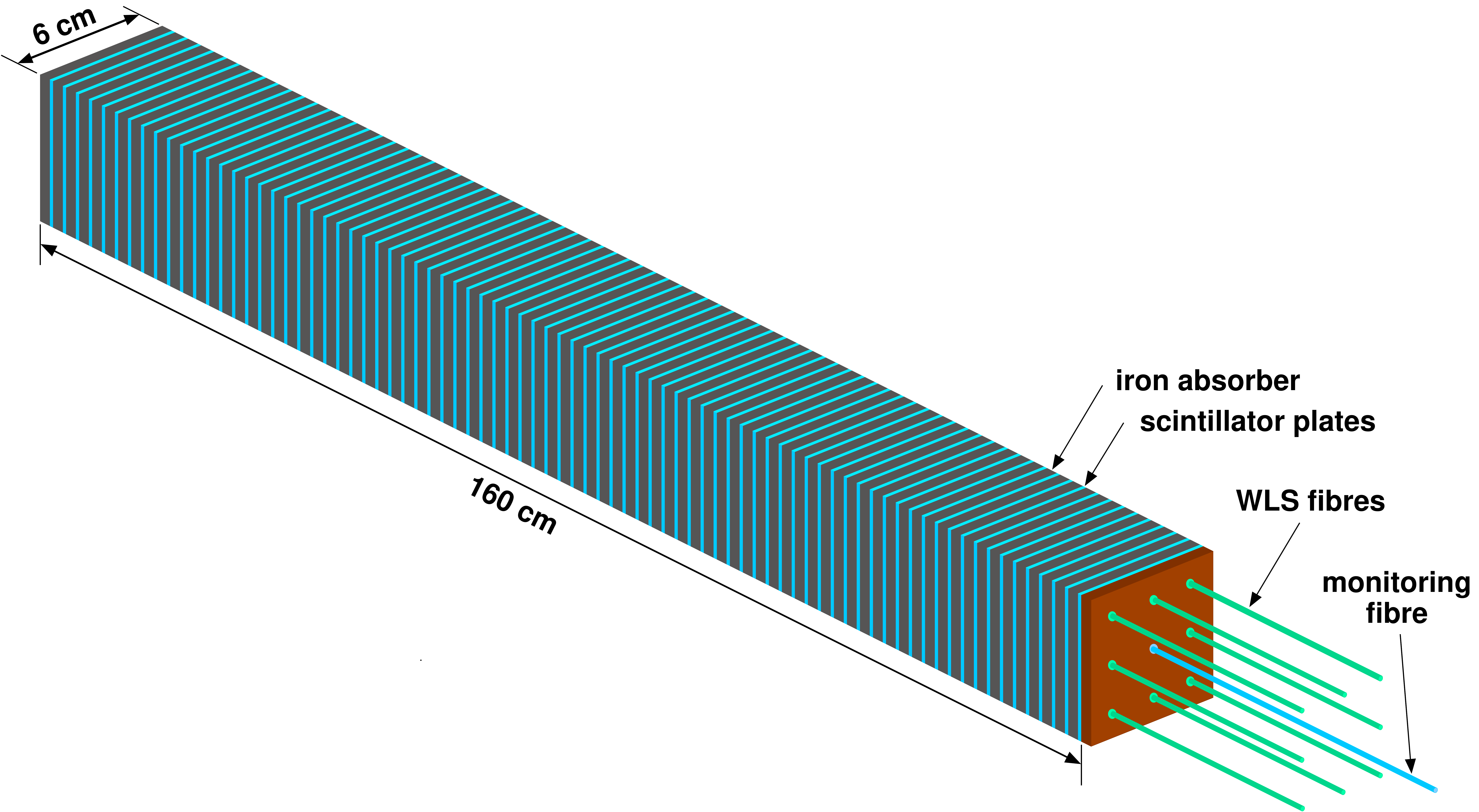}
    \caption{A shashlik HCAL module. The beam comes from the left.
            The central fibre serves for injection of monitoring light.}
    \label{fig:HCAL_Shashlik}
\end{figure}

As second option we consider a shashlik calorimeter with the same $6 \times 6$~cm$^2$ cell size as described above. Each module (cell) consists of  80~absorber layers of 1.5~cm iron and 80~active layers of 0.5~cm plastic scintillator
(Fig.~\ref{fig:HCAL_Shashlik}).
The scintillator light is read out with wavelength-shifting (WLS) fibres to 
either photomultiplier tubes or large SiPMs or SiPM arrays. 
For calibration, the scintillators can be activated by light input through an 
additional clear fibre. 

In the simplest set-up the number of channels is just 1440, 
greatly reduced with respect to the Calice-like design. However, such a design would not 
provide any information about the longitudinal shower development. 
In NA62, the two separate modules MUV1 and MUV2 are of great importance for 
$\pi/\mu$ separation, therefore a shashlik design should also at least contain two 
modules with separate readout or, alternatively, having  light readout 
both on the front and the back side of the detector.

\paragraph{Muon veto detector (MUV)}

The muon veto detector (MUV) vetoes muons 
and contributes to the muon identification and suppression 
at a event-building-farm or the analysis level. It is placed following the 
HCAL behind a 80~cm thick iron wall 
(corresponding to about 5 additional nuclear interaction lengths).
The MUV needs a time resolution of the order of 100--150~ps, to keep the loss of signal due to random activity at few percent level, as it is presently achieved by NA62 with a time resolution of about 500~ps.

The MUV will be built from scintillating tiles with direct photo detector readout. 
Two feasible options exist. The first scenario is applicable with a 
high-granularity HCAL, where the MUV detector would be just another scintillating layer,
but behind the iron wall and with scintillating tiles of 5~cm thickness to ensure 
high light yields for an optimum time resolution.
The second MUV option is a copy of the NA62 MUV3 detector, 
also consisting of 5~cm thick scintillating tiles, but read out 
by PMTs from the back side. In contrast to NA62, where the tile front faces measure 
$22 \times 22$~cm$^2$, the tile size would need to be reduced to about 
$8 \times 8$~cm$^2$ to be able to cope with the higher rate in HIKE.


%% file: detectors/SAC.tex
\subsubsection{Small-angle electromagnetic calorimeter (SAC)}

For the measurements of $K^+\to\pi^+\nu\bar\nu$ and $K_L\to\pi^0\nu\bar\nu$, as well as other measurements of rare decays requiring hermetic photon vetoes, the coverage of the veto system must extend down to zero in the polar angle, to intercept photons that would otherwise escape the detector through the downstream beam pipe.
Vetoing photons at small angle is easier for the $K^+$ phase than for the $K_L$ phases, because the $K^+$ beam can be diverted and dumped outside the acceptance of the small-angle calorimeter. For the $K_L$ phase, instead, the SAC sits directly in the neutral beam; it must reject photons from $K_L$ decays that would otherwise escape via the downstream beam exit while remaining relatively insensitive to the very high flux of neutral hadrons in the beam, so that the experiment is not blinded by the random vetoes from these hadrons. The design and construction of the SAC for the $K_L$ phase is thus a unique challenge. To the extent that R\&D for the KLEVER SAC is well underway, its construction should be complete before LS3, allowing it to be validated for the $K^+$ phase before the end of the present run. The operational experience gained during the $K^+$ phase will then be used to validate the SAC with respect to the more stringent requirements for KLEVER, which form the basis for the following discussion. Note that for much of HIKE Phase~2 (specifically, for $K_L\to\pi^0\ell^+\ell^-$ measurements) a production angle of 2--4~mrad is preferred. For running at the smallest production angles, the SAC might not be installed because of the extremely high total neutral beam rate.

\begin{figure}
\centering
\includegraphics[width=0.6\textwidth]{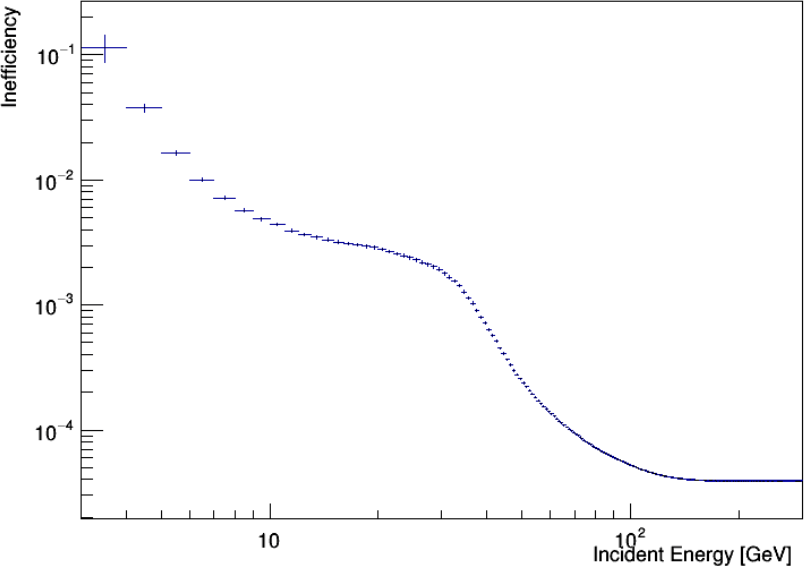}
\caption{Maximum tolerable inefficiency of the KLEVER SAC vs incident photon energy, for photons from $K_L\to\pi^0\pi^0$ events, from simulation.}
\label{fig:sac_eff_req}
\end{figure}

The maximum tolerable inefficiency for photon detection for the KLEVER SAC is illustrated as a function of incident photon energy in Fig.~\ref{fig:sac_eff_req}. This plot was obtained from the fast simulation described in \Sec{sec:klever_sens} by cumulating the energy distribution in the SAC for $K_L\to\pi^0\pi^0$ events that evade all other vetoes and pass all other cuts used to select $K_L\to\pi^0\nu\bar\nu$ events. 
The following, indicative efficiency requirements can be identified:
\begin{itemize}
\item For $E < 5~{\rm GeV}$, the SAC can be blind. For the $K\to\pi\nu\bar{\nu}$ studies in both the $K^+$ and $K_L$ experiments, background processes that are not otherwise efficiently vetoed do not have photons on the SAC.
\item For photons with $5~{\rm GeV} < E < 30~{\rm GeV}$, the SAC inefficiency must be less than 1\%.
Most photons on the SAC with energies in this range
from $K_L$ decays that are otherwise accepted as signal candidates
are from events in which there are two photons on the SAC, so that the efficiency requirement is relatively relaxed. 
\item Only for photons with $E > 30~{\rm GeV}$ must the inefficiency be very low ($<10^{-4}$).
For $K_L$ decays passing analysis cuts with a photon on the SAC in this energy range, the other photon is emitted at large angle and has low energy, so the SAC veto is important.
\end{itemize}
Although these SAC efficiency requirements are not intrinsically challenging, from the simulations of the beamline without extension (\Sec{sec:neutral_beam}), there are about 130 MHz of $K_L$ mesons, 440~MHz of neutrons, and 40~MHz of high-energy ($E>5$~GeV) beam photons incident on the SAC, and the required efficiencies must be attained while maintaining insensitivity to the nearly 600~MHz
of neutral hadrons in the beam. 
In order to keep the false veto rate to an acceptable level, the hadronic component must be reduced to at most a few tens of MHz, so that the total 
accidental rate is dominated by the beam photons and in any case significantly less than the 100~MHz target cited in \Sec{sec:klever_rates}.
These requirements lead to the following considerations:

\begin{itemize}
\item The SAC must be as transparent as possible to the interactions of neutral hadrons. In practice, this means that the nuclear interaction length $\lambda_{\rm int}$ of the SAC must be much greater than its radiation length $X_0$.
\item The SAC must have good transverse segmentation to provide $\gamma/n$
discrimination. 
\item It would be desirable for the SAC to provide additional information useful for offline $\gamma/n$ discrimination, for example, from longitudinal segmentation, from pulse-shape analysis, or both.
\item The SAC must have a time resolution of 100~ps or less.
\item The SAC must have double-pulse resolution capability at the level of a few ns.
\end{itemize}
In addition, in five years of operation, the SAC will be exposed to a neutral hadron fluence of about $10^{14}$~$n$/cm$^{2}$, as well as a dose of up to several MGy from photons.

\begin{figure}
\centering
\includegraphics[width=0.5\textwidth]{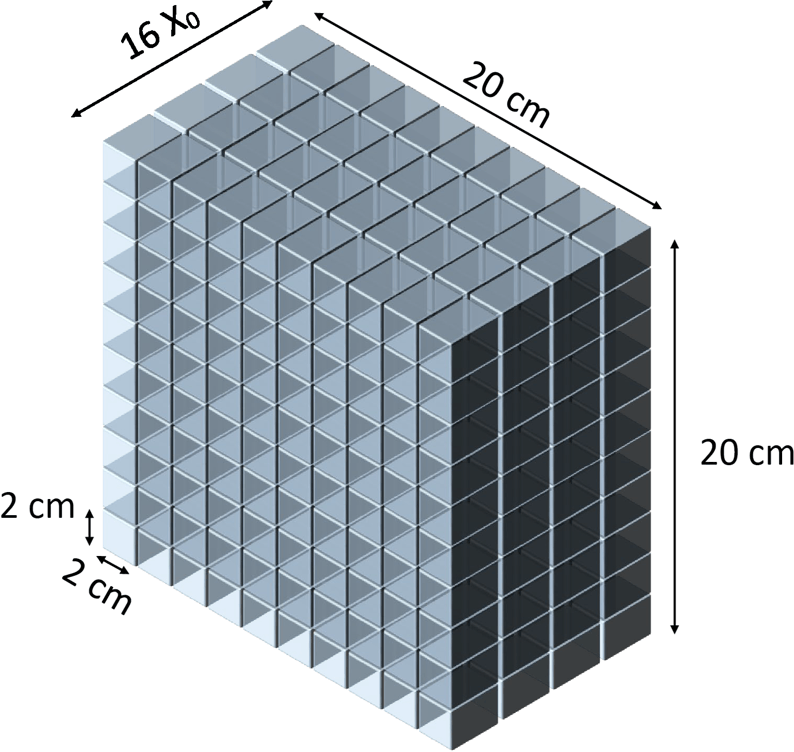}
\caption{Dimensional sketch of a SAC for KLEVER based on dense, high-$Z$ crystals with both transverse and longitudinal segmentation.}
\label{fig:klever_sac}
\end{figure}

One possible design that is well-matched to the KLEVER requirements would be to use a highly segmented, homogeneous calorimeter with dense, high-$Z$ crystals providing very fast light output. As an example, the small-angle calorimeter for the PADME experiment used an array of 25~lead fluoride (PbF$_2$) crystals of $30\times30\times140~{\rm mm}^3$ dimensions. PbF$_2$ is a Cherenkov radiator and provides very fast signals.
For single crystals read out with PMTs, a time resolution of 81~ps and double-pulse separation of 1.8~ns were obtained for 100--400~MeV electrons~\cite{Frankenthal:2018yvf}, satisfying the KLEVER timing performance requirements. The PADME performance was obtained with fully digitising waveform readout at 5 GS/s; waveform digitisation would also be required for KLEVER (Section~\ref{sec:readout}).

At the doses expected at KLEVER, radiation-induced loss of transparency to Cherenkov light could be significant for PbF$_2$, as suggested by existing studies with ionising doses of up to ${\cal O}(10~{\rm kGy})$~\cite{Cemmi:2021uum,Kozma:2002km,Anderson:1989uj}. However, these studies also found significant annealing and dose-rate effects in PbF$_2$, as well as the effectiveness of bleaching with UV light. If the effects of radiation damage to PbF$_2$ prove to be a significant problem, a good, radiation-hard alternative could be recently developed, optimised lead tungstate (PbWO$_4$, PWO) 
\cite{PANDA:2011hqx,Auffray:2016xtu,Follin:2021kgn}. In particular, ultrafast PWO (PWO-UF) with a decay time constant of 640 ps, good light yield, and high radiation tolerance has recently been developed~\cite{Korzhik:2022xln}.

For the KLEVER SAC, a design with longitudinal segmentation is under study, as shown in Fig.~\ref{fig:klever_sac}. This design would feature four layers of $20\times20\times40$~mm$^3$ PbF$_2$ or PWO-UF crystals (each $\sim$4$X_0$ in depth). To minimise leakage, the gaps between layers would be kept as small as possible. Compact PMTs such as Hamamatsu's R14755U-100 could fit into a gap of as little as 12~mm. 
Readout with SiPMs would facilitate a compact SAC design even further, but may require advances in SiPM radiation resistance and timing performance.

\begin{figure}
\centering
\includegraphics[width=0.5\textwidth]{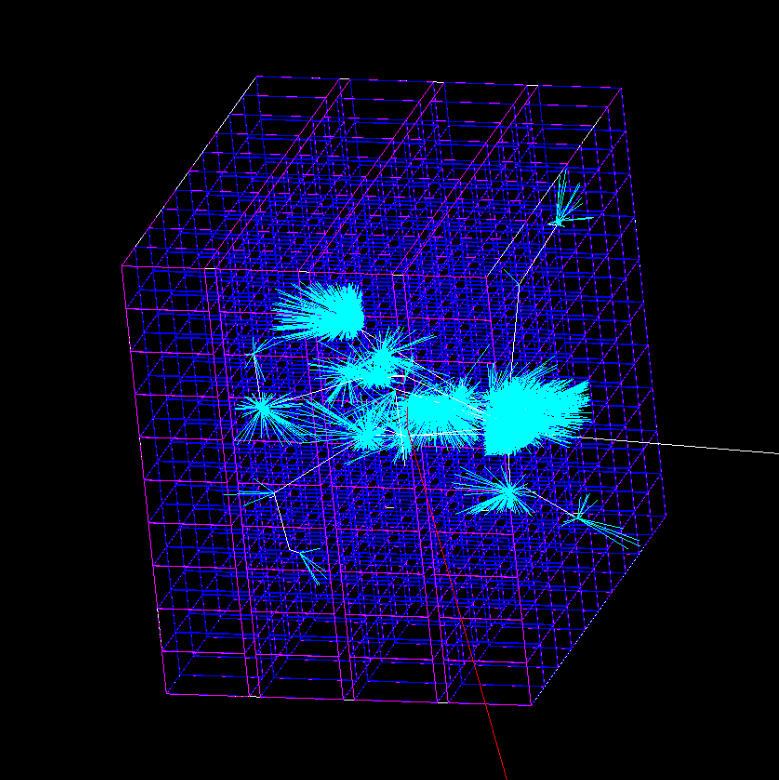}
\caption{Geant4 simulation of the KLEVER SAC showing a 1-GeV photon incident from the right. White tracks show photons, cyan tracks show Cherenkov photons, and the red track shows a negatively charged particle.}
\label{fig:sac_interaction}
\end{figure}

A Geant4 simulation of the KLEVER SAC has been developed for the HIKE Monte Carlo. This simulation can be used to study the performance with different crystals and geometries (the readout response is not yet simulated).
Fig.~\ref{fig:sac_interaction} shows the simulation of an interaction of a 1-GeV photon in the KLEVER SAC with PbF$_2$
crystals and a 5-mm gap width between layers for readout with SiPMs. The Cherenkov photon yield is 85,000 per GeV of incident photon energy. There is a $\sim$10\% loss of shower containment at the highest photon energies, which is not seen to depend on the size of the gaps between layers over the interval of 0--20~mm. This is considered to be an acceptable trade-off for the purpose of maintaining the detector relatively transparent to hadrons. The response 
to the main components of the neutral beam has been studied. 
\begin{figure}
\centering    \includegraphics[width=0.48\textwidth]{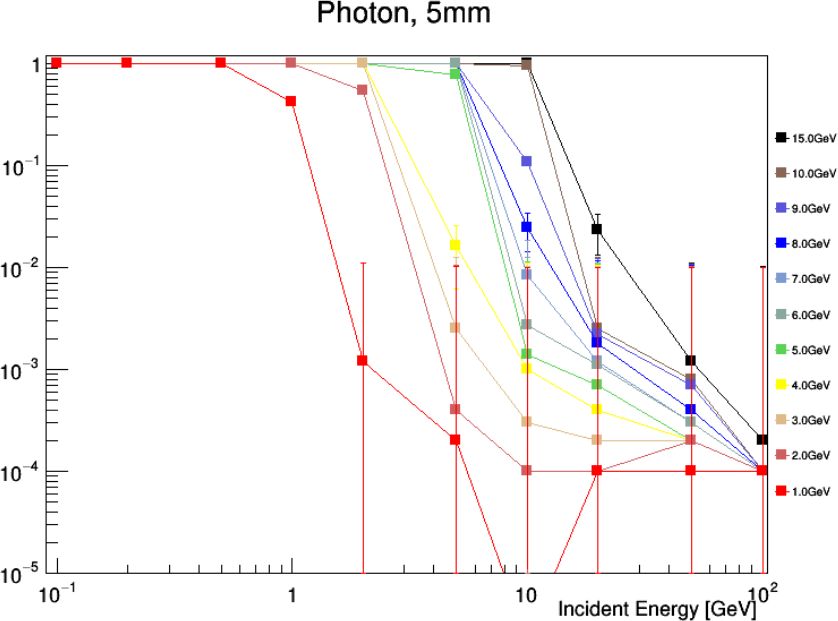}\\
\includegraphics[width=0.48\textwidth]{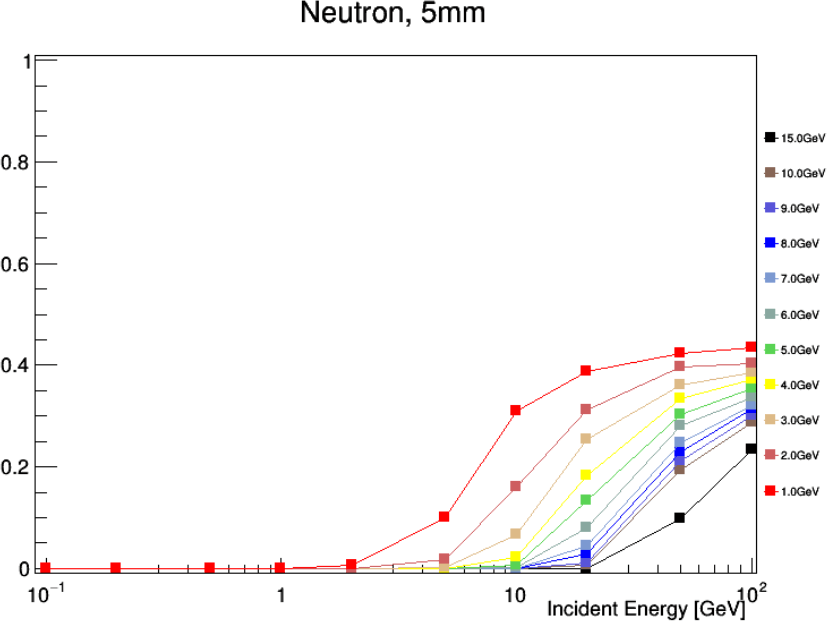}%
\hspace{0.02\textwidth}%
\includegraphics[width=0.48\textwidth]{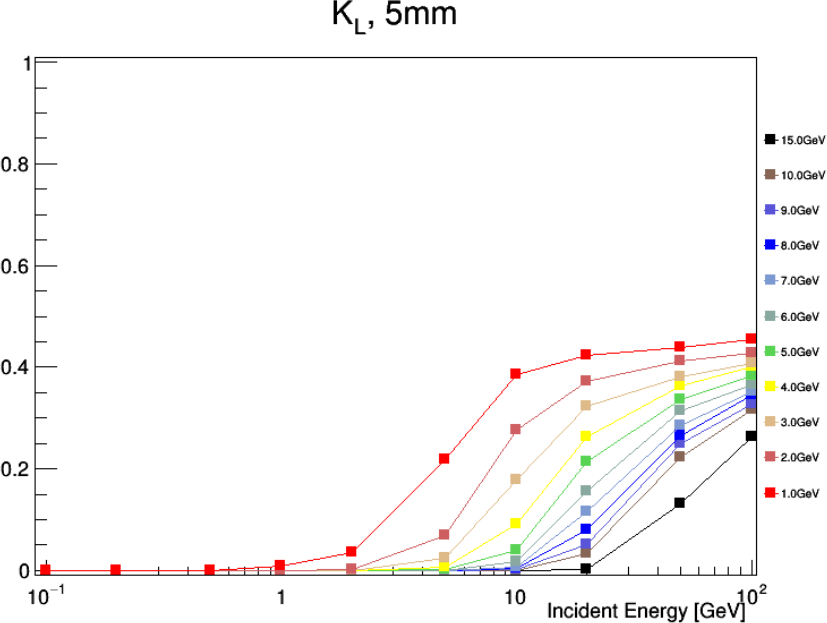}
\caption{SAC inefficiency for photons (top) and  efficiency for neutrons and $K_L$ mesons (bottom left and right) vs.\ incident particle energy, for various thresholds on observed incident particle energy as measured from the number of Cherenkov photons produced.}
\label{fig:sac_eff}
\end{figure}
Fig.~\ref{fig:sac_eff} shows the inefficiency of the SAC for photon detection and the efficiency of the SAC for neutron and $K_L$ detection as a function of incident particle energy, for various thresholds set on the measured incident particle energy from the number of Cherenkov photons produced. It is seen that with a threshold set at the number of Cherenkov photons produced by a photon of 4--5 GeV, the SAC efficiency requirements are satisfied, while about 30\% of the neutral hadrons leave signals above threshold, contributing to the random veto rate.   

\begin{figure}
\centering
\includegraphics[height=0.2\textheight]{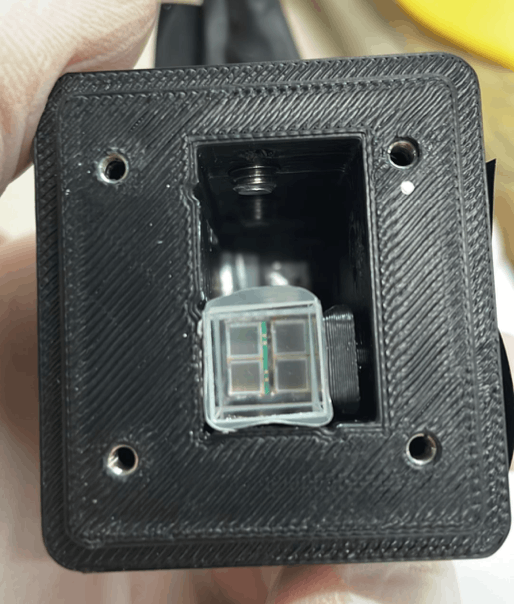}\hspace{0.1\textwidth}
\includegraphics[height=0.2\textheight]{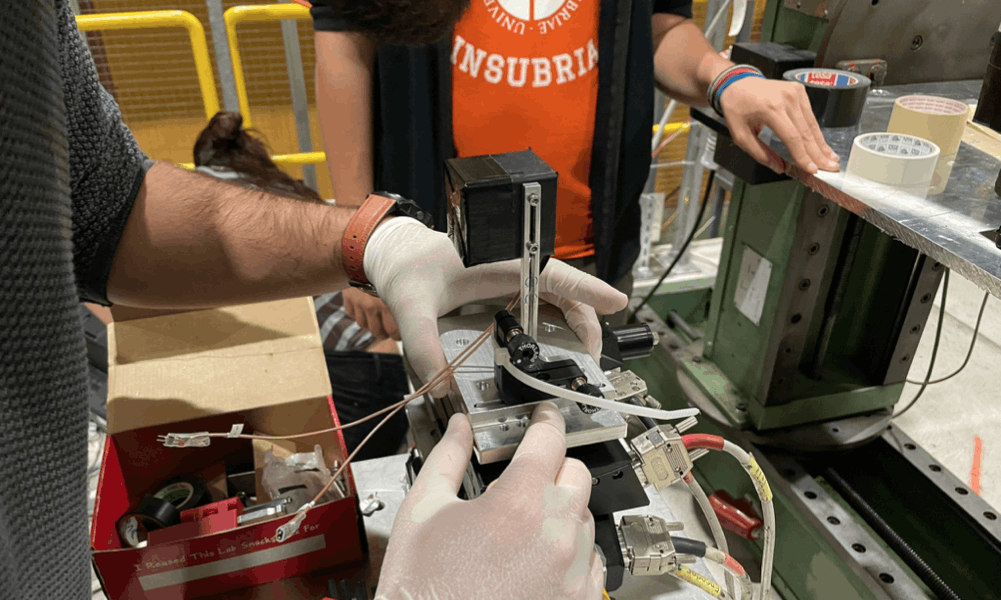}
\caption{Left: Single crystal read out with $2\times2$ matrix of four $4\times4$~mm$^2$ SiPMs. Right: Installation of crystal on H2 beam line during fall 2022 KLEVER/CRILIN test beam.}
\label{fig:crilin_tb}
\end{figure}

R\&D work on the KLEVER SAC is currently being carried out in synergy with other collaborations, facilitated in part by participation in the AIDAinnova research network. 
CRILIN, an electromagnetic calorimeter under development for the International Muon Collider Collaboration, is an independently proposed, highly granular, longitudinally segmented, fast crystal calorimeter with SiPM readout and performance requirements similar to those for the KLEVER SAC~\cite{Ceravolo:2022rag}, and much development work for KLEVER is being carried out in collaboration with CRILIN, with particular emphasis on the SiPMs, front-end electronics, and signal readout, as well as on solutions for detector mechanics and SiPM cooling. The first test beam measurements with individual PbF$_2$ and PWO crystals were performed in summer 2021 at the Frascati BTF and the SPS North Area, followed by additional tests in fall 2022 in which some of the first commercially available samples of PWO-UF were also tested.
These tests were focused on understanding the best possible time resolution that can be obtained, studying the systematics of light collection in the small crystals, and validating the CRILIN choices of SiPMs and the design of the readout amplifier. In autumn 2022, single $10\times10\times40$~mm$^3$ crystals of PbF$_2$ (4.3$X_0$) and PWO-UF (4.5$X_0$) were exposed to high-energy (60--120 GeV) electron beams at the SPS H2 beamline. 
Each crystal was viewed by a matrix of four Hamamatsu 14160-4010 SiPMs ($4\times4$~mm$^2$, 10~$\mu$m pixel size), which were read out in pairs, providing independent readout channels for the left and right sides of the crystal (Fig.~\ref{fig:crilin_tb}).
The SiPM was chosen to represent a plausible choice in the current state of the art for high-speed response, short pulse width, and good radiation resistance.
The signals from the SiPMs were amplified
with the CRILIN electronics and digitized at 5 GS/s. 
The data collected are currently being analyzed, but some preliminary conclusions are already apparent:
\begin{itemize}
    \item The time resolution obtainable from the combination of either crystal, PbF$_2$ or PWO-UF, with the chosen SiPM and CRILIN electronics is excellent, with the final time resolution expected to be significantly better than required for the KLEVER SAC.
    \item The light yield for PWO-UF is approximately twice that for PbF$_2$; the time resolution for PbF$_2$ is slightly better. 
    \item For either crystal, the light produced is highly localized on the rear face of the short crystal, requiring some care with the segmented readout.
    \item The CRILIN electronics performed very well; KLEVER should evaluate the possibility of faster shaping to obtain better double pulse discrimination.
\end{itemize}

\begin{figure}
    \centering
    \includegraphics[height=0.175\textheight]{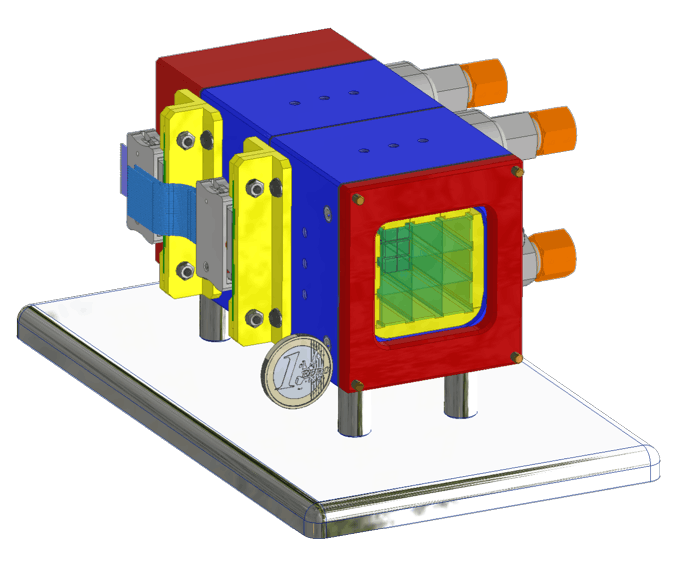}
    \includegraphics[height=0.175\textheight]{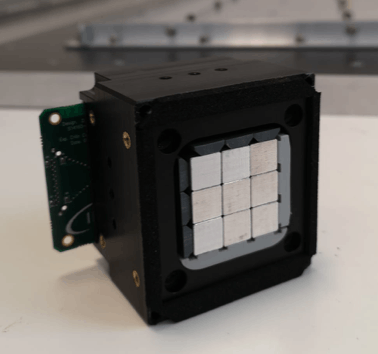}
    \includegraphics[height=0.175\textheight]{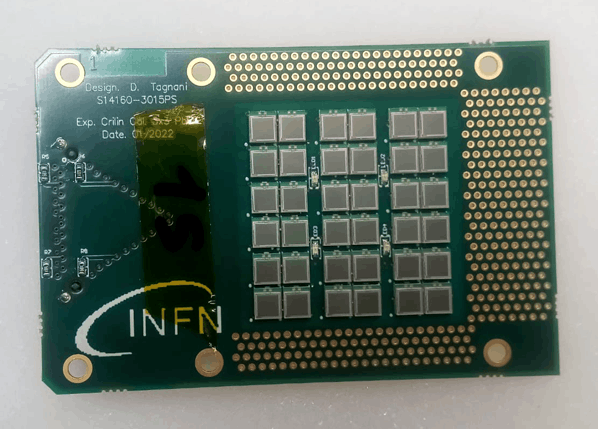}
    \caption{Two-layer $3\times3$ crystal CRILIN prototype: mechancial drawing (left), assembled front module (center), and PCB with SiPMs for one layer (right).}
    \label{fig:crilin}
\end{figure}
The next step of CRILIN development is to test a two-layer, $3\times3$ crystal array. The mechanical prototype (Fig.~\ref{fig:crilin}) features a scheme for SiPM cooling. The light is read out in the same manner as for the single crystals tested in 2021--2022, with two channels per crystal, for a total of 36~channels. Modular electronics for power distribution, SiPM signal processing, and control have also been developed. This prototype will be populated with both PbF$_2$ and PWO-UF crystals and tested in the coming months to validate the segmentation scheme, test the time resolution, study various optimisations such as the crystal surface preparation, and evaluate the performance of the engineering solutions adopted.

As noted above, the full SAC design with 4 layers of 4-cm crystals suffers slightly from shower leakage, while the interaction probability for hadrons in the beam is about 30\%. A possibility to make the SAC more hermetic for photon showers and less sensitive to hadrons is to exploit the effects of the coherent interactions of high-energy photons in oriented crystals to induce prompt electromagnetic showering. 
This is the same technique to be used for optimisation of the photon converter in the neutral beam (\Sec{sec:kl-beamline-sim}).
Because of the relative ease in producing high-$Z$ optical crystals of high quality, there are good prospects for using this technique to construct a compact
calorimeter with a very small radiation length, referred to the primary interaction~\cite{Bandiera:2019tmg}. A decrease by a factor of 5 in the effective radiation length for a 4-mm thick PWO crystal has been observed for 120~GeV electrons incident to within 1~mrad of the \hkl[001] axis~\cite{Bandiera:2018ymh}. Considering that the SAC acceptance for photons from $K_L\to\pi^0\pi^0$ decays in the fiducial volume extends at most to $\pm2$~mrad, it should be possible to orient the crystals to enhance the probability for photon conversion. This would allow the thickness of the SAC layers to be reduced, reducing the rate of neutral hadron interactions while maintaining shower containment and photon detection efficiency.

The potential gains from aligning 
the crystals,
as well as the procedures and mechanics needed, are under study.
In summer 2021, KLEVER and CRILIN, together with the STORM collaboration, used a tagged photon beam to measure the enhancement of shower processes in thicker PbF$_2$ and PWO crystals as a function of angular alignment, correlating variables such as the energy dispersed in the crystal and multiplicity of charged particles produced in the shower with the production of Cherenkov and scintillation light in the crystals. STORM collected similar data with electron beams in summer 2021 and 2022. The data are still under analysis, but as a preliminary generalisation, the effective radiation length for crystals of 2--4 cm thickness is reduced by 20--30\% by alignment, with an angular acceptance of about 1~mrad~\cite{Monti-Guarnieri:2022ezh}.
Further studies are planned with the CRILIN prototypes, and the OREO collaboration, continuing the work of STORM, has plans to build a prototype PWO-UF calorimeter
in which the \hkl[100] axes of the crystals on the front layer are aligned. In addition to allowing studies of the variation of the energy deposition, radiation length and Moli\`ere radius with beam energy and crystal alignment, the OREO prototype will serve as a platform for developing the procedures
needed to build a calorimeter with aligned crystals.

%% file: detectors/HASC.tex
\subsubsection{The hadron sampling calorimeter (HASC)}

The original NA62 setup
was improved by the addition of a hadron sampling calorimeter (HASC-1) adjacent to the beampipe, downstream of the muon detector. The primary purpose of this detector is to reduce the $K\to\pi^+\pi^+\pi^-$ background to the $K^+\to\pi^+\nu\bar\nu$ decay, by vetoing the topology in which the $\pi^-$ undergoes hadronic interaction in the first STRAW chamber, while an energetic $\pi^+$ travels in the beampipe undetected by the IRC and emerges downstream. Analysis of the NA62 Run~1 dataset has revealed that HASC-1 is also efficient as a photon veto, providing a 30\% reduction of the $K^+\to\pi^+\pi^0$ background to the $K^+\to\pi^+\nu\bar\nu$ decay. Consequently, a second calorimeter (HASC-2) was installed in 2021 at the HASC-1 longitudinal position, symmetrically with respect to the beam axis, which improves further the $K^+\to\pi^+\pi^0$ rejection.

Each of the HASC-1 and HASC-2 stations consists of 9~identical modules. 
Each module is a sandwich of 120~lead/scintillator alternating tiles, with a total volume of $10\times 10\times 120 ~\mathrm{cm}^3$ (W x H x L). The sampling ratio is 4:1, the scintillator tiles having a dimension of $100 \times 100 \times 4~{\rm mm}^3$ while the lead thickness is 16~mm. Each module is organised in 10~longitudinal readout sections, each scintillator tile of every section being optically coupled with a wavelength shifting (WLS) optical fiber of 1~mm$^2$ cross-section. At the rear side of each module there are 10~optical connectors, originally designed to be coupled with 3$\times$3~mm$^2$ green-sensitive micro-pixel avalanche photodiodes (MAPD) (currently the S12572-015C Hamamatsu SiPM sensors are used).

The FE electronics and SiPMs installed on the HASC-2 station are cooled down to 21ºC with a custom-made system  
consisting of three Peltier thermoelectric coolers / module and a water-air heat exchanger used to blow cold air in the modules end-cap cases. The temperature is maintained constant by a temperature controller – MCU based – which regulates the Peltier supply voltage via a PID routine. For HIKE, this scheme will be extended to HASC-1,
improving the quality of SiPM signals and reducing ageing due to radiation to an acceptable level.

Signal rates observed in the HASC during the NA62 data taking in 2022 at nominal beam intensity are summarised in Table~\ref{tab:hasc_rates}. The rates fall rapidly with the distance from the beampipe. The modules next to the beampipe contain $35\%$ and $25\%$ of the total activity in the HASC-1 and HASC-2 stations, respectively.

\begin{table}[tb]
\centering
\caption{Signal rates observed in the HASC during the NA62 data taking in 2022.}
\vspace{-2mm}
\begin{tabular}{l|c}
\hline
Detector part & Rates in NA62 setup at nominal intensity\\
\hline
HASC-1 & 1~MHz\\
HASC-1, most active module & 0.35~MHz\\
HASC-1, most active channel & 0.05~MHz\\
HASC-2 & 0.21~MHz  \\
HASC-2, most active module & 0.05~MHz\\
HASC-2, most active channel & 6~kHz\\
\hline
Total (HASC-1 and HASC-2) & 1.21~MHz \\
\hline
\end{tabular}
\label{tab:hasc_rates}
\end{table}

The NA62 HASC provides near-optimal geometric coverage for the relevant topologies of the $K^+\to\pi^+\pi^+\pi^-$ and $K^+\to\pi^+\pi^0$ backgrounds to the $K^+\to\pi^+\nu\bar\nu$ decay, and is suitable also for HIKE Phase~1. Assuming the HIKE beam profile to be similar to the NA62 one, we expect at most a rate of 200~kHz in the most active channel for the HIKE HASC. The main issue in the operation of the current HASC at the HIKE beam intensity is the random veto. It is determined chiefly by the time resolution of the detector, presently at the level of 370~ps FWHM for HASC-2 and 430~ps for HASC-1, in agreement with HAMAMATSU specifications~\cite{Hamamatsu:S1272-015C}. For HIKE, the resolution should be improved to $\mathcal{O}(100~\mathrm{ps})$. The HAMAMATSU S13362-3050DG SiPMs achieving this value are available on the market~\cite{Hamamatsu:S13362-3050DG}, and represent a promising candidate for the upgrade. Table~\ref{tab:HASC_SiPM_performances} shows a comparison between the current HASC SiPMs and the candidate for the upgrade. Another appealing feature of the new SiPMs is the embedded two-stage thermoelectric cooling which greatly simplifies the current cooling scheme.

\begin{table}[tb]
\centering
\caption{\label{tab:HASC_SiPM_performances} Summary of SIPM characteristics for the model currently employed in the NA62 HASC and the one proposed for the HIKE upgrade.}
\vspace{-2mm}
\begin{tabular}{l|c|c}
\hline
SIPM model & S12572-015C (current) & S13362-3050DG (upgrade)\\
\hline
Photon detection efficiency     & 25\% & 40\% \\
Typical dark count & 1000~kcps & 25 kcps \\
Gain & $2.3\times 10^5$ & $1.7\times 10^6$ \\
Time resolution FWHM & 400~ps & 110~ps~\cite{Kravchenko_2017}\\
\hline
\end{tabular}
\end{table}

An alternative option for the HASC upgrade, which is currently under study, involves PMTs. A particularly interesting example is HAMAMATSU H14220A, which could be used to read out individual scintillator plates (instead of groups of six, as done presently), eliminating pileup due to detecting a convoluted signal hence offering lower rise time ($\tau_\mathrm{r}$) and fall time ($\tau_{\mathrm{f}}$). The main drawback of reading single scintillators is the lower number of collected photons ($N_{\mathrm{ph}}$), compared with SiPM readout, which could have a negative impact on the time resolution which is known to be proportional with $\sqrt{\tau_\mathrm{r}\tau_\mathrm{f}/N_{\mathrm{ph}}}$~\cite{Gundacker_2013}.

%% file: detectors/klever_detectors.tex
\subsection{Detectors specific to KLEVER}
\label{sec:klever_detectors}

\subsubsection{Upstream veto (UV) and active final collimator (AFC)}

The upstream veto (UV) rejects $K_L\to\pi^0\pi^0$
decays in the 40~m upstream of the fiducial volume where there are
no large-angle photon vetoes.
The UV is a shashlyk calorimeter with the same basic structure as the MEC
(without the spy tiles). Because the UV does not participate in event reconstruction, the readout granularity can be somewhat coarser than for the MEC. The sensitive area of the UV has inner and outer radii of 10~cm and 100~cm.

The active final collimator (AFC) is inserted into the hole in centre of the UV. The AFC is a LYSO collar counter with angled inner surfaces to
provide the last stage of beam collimation while vetoing photons from $K_L$'s
that decay in transit through the collimator itself. The collar is made of 24 crystals of trapezoidal cross section, forming a detector with an
inner radius of 60 mm and an outer radius of 100 mm.
The UV and AFC are both 800 mm in depth. The maximum crystal length for
a practical AFC design is about 250 mm, so the detector consists of 3 or 4 longitudinal segments. Each crystal is read out on the downstream side with two avalanche photodiodes (APDs).
These devices couple well with LYSO and offer high quantum efficiency and
simple signal and HV management. Studies indicate that a light yield in excess of 4000~p.e./MeV should be easy to achieve.

\subsubsection{Charged-particle rejection}

For the rejection of charged particles, $K_L\to\pi^+e^-\nu$ is a benchmark channel
because of its large branching ratio and because the final state electron can be mistaken
for a photon. Simulations indicate that the needed rejection can be achieved
with two planes of charged-particle veto (CPV) each providing 99.5\%
detection efficiency, supplemented by the $\mu^\pm$ and $\pi^\pm$ recognition
capabilities of the MEC (assumed in this case to be equal to those of the LKr)
and the HIKE hadronic calorimeters, which could be
reused in KLEVER. The CPVs are positioned $\sim$3~m upstream of the MEC
and are assumed to be constructed out of thin scintillator tiles.
In thicker scintillation hodoscopes, the detection inefficiency arises
mainly from the gaps between scintillators. For KLEVER, the scintillators
will be only $\sim$5~mm thick ($1.2\%X_0$), and the design will be carefully
optimized to avoid insensitive gaps.

\subsubsection{Preshower detector}

The PSD measures the directions for photons incident on the MEC.
Without the PSD, the $z$-position of the $\pi^0$ decay vertex can only be
reconstructed by assuming that two clusters on the MEC are indeed
photons from the decay of a single $\pi^0$. With the PSD, a vertex can be
reconstructed by projecting the photon trajectories to the beamline.
The invariant mass is then an independent quantity, and
$K_L\to\pi^0\pi^0$ decays with mispaired photons can be efficiently
rejected.
The vertex can be reconstructed using a single photon and the constraint
from the nominal beam axis. 
Simulations show that with $0.5 X_0$ of converter
(corresponding to a probability of at least one conversion of 50\%)
and two tracking planes with a spatial resolution of 100~$\mu$m, placed 50~cm apart, the mass resolution is about 20~MeV and the vertex position resolution is about 10~m. The tracking detectors
must cover a surface of about 5 m$^2$ with minimal material.
Micropattern gas detector (MPGD) technology seems perfectly suited for the PSD. Information from the PSD will be used for bifurcation
studies of the background and for the selection of control samples,
as well as in signal selection.

%% file: DAQ.tex
\section{Data acquisition and high-level trigger}
\label{sec:readout}

Particle rate in HIKE Phase~1 is 3000~MHz in the beam tracker and 200~MHz for $K^+$ identification. Single-particle rates in the downstream detectors is about 50~MHz, including both $K^+$ decay products and the charged beam's muon halo. 
The neutral beam of HIKE Phases~2 and 3 is 50\% more intense that the Phase~1 charged beam. The small-angle calorimeter (SAC) in the neutral beam will encounter rates of about 100~MHz, and the other detectors encounter rates of about 20~MHz. 
The high particle rates will give rise to a harsh radiation environment in the HIKE experimental cavern.
To limit the impact of radiation on the DAQ system, we aim to minimise the amount of electronics located in the cavern, particularly electronics related to data processing and storage, and use radiation-tolerant optical links and error-correcting data formats such as (lp)GBT.

\subsection{Readout boards}
\label{sec:ro-boards}

The readout boards serve two purposes: to host devices that timestamp digital signals produced by the detectors, and to encode and transmit these data away. The readout boards are located in the high-radiation environment close to the detector, so radiation-hardened technologies will be used. 
The time of signals from the detectors will be recorded either with time-digital converters (TDC), analogue-to-digital converters (ADC), or dedicated ASIC devices. The 64-channel PicoTDC is under development at CERN and is anticipated to be ready before HIKE Phase~1. The PicoTDC can timestamp signals at 12~ps precision~\cite{PicoTDCDataSheet} and outputs data in the GBT format, allowing simple readout boards to be used that only convert electric signals from the PicoTDC into optical signals that are sent to the rest of the DAQ system.
Possible commercial solutions for fast FADC-based readout boards, to be adapted to the requirements of the DAQ in synergy with the manufacturer, are under investigation. The HIKE Phase~1 beam tracker data will be handled by dedicated ASICs.

The number of detector channels read by TDC or ADC in HIKE Phase~1 is estimated to be about~50000. Approximately half of the detector channels will be read by TDC devices, the other half by ADCs. Therefore an estimated 400 TDC devices with 64 channels each, and 1600 ADC devices with~16 channels each, will be required for HIKE Phase~1. For HIKE Phase~2, the 10240 channels of the LAV, $\mathcal{O}(2000)$ channels of the MEC, and 400 channels of the SAC will be read using ADCs, requiring about 800 ADC devices with 16 channels each. Spy tiles in the MEC would contribute an additional 8000 channels that can be read using TDCs, in which case about 125 TDC devices with 64 channels each will be needed.



\subsection{Streaming readout}
\label{sec:streaming-readout}

Moving the bulk of the DAQ hardware away from the high-radiation environment suggests implementing a streaming readout system, where detector signals are collected by radiation-tolerant readout boards and then transferred immediately, without an external trigger signal and without complicated processing or storage. The transferred data are received and buffered in the RAM of a set of servers, then transmitted to a computing farm for selection and filtering.

The first stage of data readout may be implemented using a FELIX-based system, developed at CERN and exploited in several HEP use cases.
The current version of the FELIX PCIe card has 24 optical links with a 9.6~Gb/s line rate and supports several data formats. When operating with the radiation-tolerant GBT format the effective data bandwidth is 3.2~Gb/s per link.
New versions of FELIX, currently under development and anticipated on the timescale of HIKE, can provide faster links (e.g.~25~Gb/s) or a larger number of links (e.g.~48).
The FELIX system also takes care of distributing clock, timing and commands to the front-ends, as well as receiving monitoring data for detector control purposes.
FELIX cards are hosted in servers, and all data received by the FELIX cards are transferred to the host and buffered in memory.
The servers then provide data to the computing farm as needed, over a high-speed ethernet network.
As modern servers can be equipped with 1~TB of RAM, the data recorded during a whole SPS spill (4.8~s) may be buffered, greatly relaxing any latency requirement on the data processing as the inter-spill period (at least 9.6~s) can be exploited. Other options for interfacing the detector data with the computing farm are presently available, such as the PCI40 board developed for the LHCb and ALICE experiments, and the NaNet board, used by the KM3Net experiment.

The main challenges of a streaming readout system are the transfer and processing of large quantities of data.
The highest data rates are expected during the KLEVER phase, in particular from the SAC, MEC, and LAV detectors. The raw data rates are summarised in Table~\ref{tab:phase2rates1} assuming the use of 14-bit ADCs. The largest rate is 72~Tb/s from the LAV. This can be considerably reduced by only reading channels that contain non-zero energy, by implementing a zero-suppression (ZS) algorithm in the readout boards. Assuming that reading data covering 16~ns is enough to contain the data from one event, the zero-suppressed data rates are given in Table~\ref{tab:phase2rates2}. In this case, the SAC dominates the data rate. Taking the FELIX system with GBT data format as the benchmark solution, the zero-suppressed SAC data could be collected by 9 FELIX cards.
Reading the non-suppressed data is unfeasible, and the streaming readout design depends on the successful implementation of zero-suppression in the ADC-based readout boards.

\begin{table}[H]
\centering
\caption{Raw data rates expected from the MEC, SAC, and LAV in HIKE Phase~3.}
\vspace{-2mm}
\begin{tabular}{lccc}
\hline
Detector & Channels & Sampling rate (GHz) & Raw data rate (Tb/s) \\ 
\hline
LAV & 10240 & 0.5 & 72  \\
MEC &  2000 & 1 & 26  \\
SAC &   400 & 1 & 6 \\
\hline
\end{tabular}
\label{tab:phase2rates1}
\end{table}

\begin{table}[H]
\centering
\caption{Zero-suppressed data rates expected from the MEC, SAC, and LAV in HIKE Phase~3.}
\vspace{-2mm}
\begin{tabular}{lcccc}
\hline
Detector & Event rate (MHz) & Average N channels hit per event & ZS data rate (Gb/s) \\ 
\hline
LAV & 14 &  5 & 8   \\
MEC & 18 & 10 & 40  \\
SAC & 95 & 30 & 640 \\
\hline
\end{tabular}
\label{tab:phase2rates2}
\end{table}

\subsection{Triggered readout}

In case zero suppression of the calorimeter data proves unfeasible, or data transmission at the required speed cannot be achieved on the calorimeter readout boards, a trigger may still be necessary to reduce the required data throughput.
The key parameters of the trigger are latency, which dictates how much data will have to be stored on the readout boards, and selectivity, which dictates how much data will have to be transferred to the computing farm.
The level of data reduction that the trigger can achieve is a compromise between the sophistication of the algorithms, with more sophisticated algorithms providing better selectivity, and the time taken to form the trigger decision.

We envisage a software trigger, with the trigger decision prepared using dedicated trigger algorithms running on the computing farm, most likely on a subset of the event data.
Recalling the anticipated LAV data rate is 72~Tb/s (9~TB/s), assuming a software-trigger latency of 100~ms and 100~LAV readout boards, each LAV readout board must be able to cache about 9\,GB of data. 
If the trigger can achieve a reduction factor of 100 -- the same as the combined hardware and software trigger of NA62 -- the transfer rate would be 720~Gb/s, which can be handled by 10 FELIX cards.




\subsection{Event-building farm and event filter}
\label{sec:ComputingFarm}

Regardless of whether a streaming or triggered DAQ is implemented, an event filter will be used to select candidate events, reducing the amount of data sent to permanent storage. The first stage of the event-filter is to combine fragments of events coming from the different detectors and reconstruct basic objects in each detector to identify physics events. Taking $K^{+} \to \pi^{+}\nu\bar{\nu}$ decays as an example, the event-building code will identify coincident signals in the RICH, STRAW, and KTAG as a potential $K^+\to \pi^+\nu\bar\nu$ event. This stage will be performed in an event-building computing farm located in the surface building of the experimental cavern.
The backbone of the event-building farm will be standard server computers running C++ algorithms. However, the event-building process centres around combinatorial algorithms that are ideally suited for hardware acceleration using FPGA or GPU devices. We envisage the use of heterogeneous computing architectures to handle the demanding processing task given to the event-building farm.
Subsequently, the event filter will examine signals in the events to identify the small fraction that should be kept for detailed analysis offline. This stage will be performed using dedicated servers in the Computing Centre on the CERN Pr\'evessin site.
A number of machine-learning-based algorithms have been developed at NA62 for particle identification and other tasks, and these can be the basis for evaluating events in the HIKE event filter.


%% file: Computing.tex
\section{Online and offline computing }
\label{sec:computing}


The baseline computing model is a scaled version of the current NA62 system. A flexible HIKE software platform is currently under development, on the basis of the NA62 software. HIKE will profit from the use of dedicated machines in Pr\'evessin for many, if not all, of the described functions below, provided enough band capability is present between the experiment site and the computing centre.


\subsection{Online monitoring}

The HIKE online monitoring system will be based on an updated model of the current one used in NA62.
At present in NA62, a subset of the data are sent from farm nodes running the event-building processes to dedicated four monitoring computers; 
each event is then processed by the standard reconstruction software.
The reconstructed events are then forwarded to a last machine where the different streams are recombined for displaying by dedicated processes. A set of low-level monitoring histograms produced by the reconstruction processes are also combined and displayed, for the purpose of quick data quality checks. 

For HIKE, a multi-level analysis pipeline, where each level performs more complex and time consuming tasks, should be implemented to shorten the latency between the acquisition of the data and the first feedback. Low level (data corruption) and high level (data quality) algorithms will perform error detection and alert the shift crew and the run control software. As often as possible automatic recovery procedures will be implemented to minimise data losses. A set of low-level calibration constants will also be computed during this first reconstruction and used both for monitoring purposes and to be fed back into the acquisition system. Depending on the details of the trigger scheme, the computing farm could also perform some of data filtering tasks. 

Reconstructed events will be processed with data quality algorithms, also used during full offline data processing (see next section), yielding output information collected over a whole run (about 1500~bursts) to be reviewed by the detector experts.


\subsection{Calibration and data quality}

Calibration and data quality monitoring strategy will be built on the NA62 systems. Calibration will be performed in several stages. First, geometrical detector alignments will be initially obtained from special ``muon'' runs, where TAXes are closed and the spectrometer magnet is switched off to obtain straight penetrating tracks. Later, these geometrical alignments will be refined and monitored using kaon decays in standard data-taking conditions. Raw timing alignments within individual detectors, like slewing corrections, will be obtained from data and assumed to be constant over a certain data taking period, although they are permanently monitored. 
A similar procedure to that of NA62 for relative timing alignment between different detectors (``T0s'') will be developed, to achieve an average precision at the picosecond level, with a width corresponding to the individual timing resolutions of each detector. This will be done on a per-burst basis. During processing information on detector-specific calibrations will also be obtained, on a per-burst basis, like energy conversions, dead/noisy channels etc, which will be fed back to the analysis software.

Data quality will be additionally monitored during full offline processing 
by software specific to each detector, but also including more common properties like efficiencies for different (online or offline) trigger algorithms. 
Graphical and numerical outputs will be produced for every detector, together with a database containing the monitored (and calibrated) quantities and alarms raised if certain thresholds are passed for one burst.


\subsection{Data processing model}

Based on experience, the NA62 processing model, with some improvements and changes, can be adapted for HIKE. In NA62, raw data processing and user analysis is performed only on CERN computing clusters, while Monte Carlo production is performed using the international GRID, including CERN Grid resources.

Processing of raw data for HIKE will be performed at CERN and using GRID resources (with a split still to be defined depending on the details of available computing and data sizes at the time of data taking), due to the increased amount of data. The splitting of the processed data according to trigger and/or physics analysis interests (``filtering'') must be refined to obtain stream sizes that are smaller and easier to handle. Stronger selection criteria will be applied, and the amount of information written for every selected event (slimming) will be reduced.

\subsection{Distributed computing model}

Large-scale generation of Monte Carlo simulations will be performed exclusively on the Grid. The existing NA62 Grid framework is a one-of-its-kind system, created specifically for NA62 and providing an easy-to-use and fully automated production system \cite{NA62:Grid}. With an uptime of 99.99\% over the last ten years, it is a proven system that we envisage to extend to HIKE. Besides its simulation and reconstruction capabilities, our Grid framework can be extended to handle user analysis, in case this becomes a necessity for HIKE.


\subsection{Data reduction model}

Due to the expected increase in the amount of data (due to the increased intensity) and in the size of data (due to the increase of signal channels), both filtering and slimming are foreseen for HIKE.
For filtering, overlap between different filtering streams must be minimised, while several filter outputs may be combined at analysis level.
For slimming, after reconstruction no low-level hit information will be stored, discarding as much as possible any information that can be:
\begin{enumerate}
\item Recomputed from stored information.
\item Likely not needed at analysis level.
\end{enumerate}
In specific cases of detector studies, or when an analysis requires access to low-level information, a selected list of events/bursts will be reconstructed again. The low-level information  
will be kept for those small dedicated samples for the duration of the study.

Generally, most analyses make use of standard high-level objects (downstream tracks, vertices) without the need to access most of the information stored in the underlying objects from which they were built (clusters, tracks, candidates). Those high level object could therefore contain only the subset of information commonly used (time and position at associated sub-detectors) and stored in high-level analysis files which would be the basis for most user analyses. Again, some dedicated studies occasionally will require access to the full reconstructed objects, but these are generally aiming at deeper understanding of detector effects and to develop procedures to deal with them. In most cases the final procedure will not rely on the full reconstruction information, but only on that already available in the high-level objects. In addition, such procedures can generally be standardised and applied to several analyses without further study. It will therefore be sufficient to reconstruct a subset of data with full information for the duration of the study.

%% file: infrastructure.tex
\section{Infrastructure and safety}
HIKE will be housed in TCC8 and ECN3 where NA62 is at present. An evaluation of the necessary modifications to the experimental area has been carried out in collaboration with the ECN3 Task Force~\cite{Brugger:2022}, the North Area Consolidation Project (NA-CONS)~\cite{Kadi:2018, Kadi:2019, Kadi:2021} and the Conventional Beams Working Group~\cite{Gatignon:2650989}, where topics such as requirements for beam infrastructure, vacuum, cooling and ventilation, electrical distribution, handling and transport of detectors, access to the cavern, IT infrastructure, gas distribution, cryogenic systems, civil engineering and radiation protection have been addressed. The full list of requirements is provided in Ref.~\cite{Charalambous:2022}. In order to ensure full compatibility with the North Area consolidation and as a central access point to all CERN service groups, discussion for the implementation of all requirements is steered through the NA-CONS Technical Coordination Committee (TCC) meetings, while integration aspects are handled in the Integration Committee for Experimental Areas (ICEA) as part of NA-CONS. 
The best match between the HIKE requirements and the infrastructure studies will be further investigated for the Proposal.
We would like to highlight that the experimental layout of HIKE does not require any civil engineering work before at least LS4.
The minimal changes and updates in the infrastructure will have a positive impact on the foreseen costs. 
Furthermore, the requirements for the electrical distribution and the detector cooling will be similar to that of NA62. The vacuum system with seven cryo-pumps went through thorough maintenance in 2021 and will be suitable to provide a good vacuum at the $10^{-6}$ mbar level for HIKE.

To keep the NA62 LKr calorimeter operational for HIKE Phase~1, continuous maintenance of the cryogenics is needed. In addition, besides the installation of a new readout back end, the legacy components of the readout chain should be reviewed.
Actions in this direction will be started already in the next YETS and procedures for regular review intervention will be defined.


The human, technical and financial resources needed for HIKE are being evaluated. Table~\ref{tab:costs} shows a tentative indication of the cost of the main detectors for HIKE. The total for Phase~1 is 22.3 MCHF. The Phase~2 will see the addition of the Main Electromagnetic Calorimeter (for an extra 5~MCHF). The remaining costs for Phase~3 (KLEVER) will be mainly the increase in the number of Large Angle Vetoes from 12 to 25 units (for an extra 9~MCHF). The numbers are only intended to give an idea of the financial extent of the project. A refined estimate will be prepared for the forthcoming proposal.

\begin{table}[H]
\centering
\caption{Tentative indication of detector costs for HIKE.}
\vspace{-2mm}
\begin{tabular}{r|c|l}
\hline
Detector & Cost (MCHF) & Comments \\ 
\hline
Kaon ID (KTAG) & 0.5 & Using MCP-PMTs   \\
Beam Tracker & 2.5 & Development, and production of 4 planes \\
Charged particle veto (CHANTI) & 0.4 & 6 stations, with SiPMs \\
VetoCounter (VC) & 0.2 & 3 stations (SciFi technology) \\
Anti0 & 0.1 & 1 plane, same technology as TimingPlanes\\
Large Angle Vetos (LAV) & 8 & 12 modules (Phase 1)\\
STRAW & 3.5 & 4 Straw chambers and associated readout \\
LKr upgrade OR MEC & 2.5 OR 5 & Readout upgrade OR new MEC \\
Small Angle Calorimeter (SAC) & 2 & High-Z crystals \\
Pion ID (RICH) & 0.8 & Using SiPMs  \\
Timing Detector & 0.2 & 2 planes, scint. tiles and SiPMs \\
Hadronic Calorimeter + Muon plane & 1.5 & Shashlyk technology and SiPMs \\
HASC & 0.1 & Using SiPMs \\
\hline
MEC (if not already in Phase 1) & 5 & Shashlyk technology and SiPMs (Phase 2)\\
\hline
Large Angle Vetoes (LAV) & 9 & Additional 13 modules (Phase 3)\\
\hline
\end{tabular}
\label{tab:costs}
\end{table}

%% file: conclusions.tex
\section{Summary and conclusions}


The HIKE project presented in this document constitutes a multi-phase experimental programme taking full advantage of the ECN3 experimental hall to provide a comprehensive study of flavour physics in the kaon sector. 
This physics programme is complementary to, and out of reach of, the LHC experiments. Importantly, HIKE will be sensitive to new physics up to the highest mass scales. 
The HIKE facility will, in addition, search for feebly-interacting particles with unprecedented sensitivity.
A summary of the HIKE phases and related intensity requirements is presented in Table~\ref{tab:conclusions}.


The experimental apparatus changes over time with a staged approach, which allows HIKE to evolve and adapt the physics scope in its three phases, an important feature
for a project that embraces a time scale of more than decade during which the physics landscape could change. Thanks to the successful experience of NA62 and its predecessor NA48, the experimental technique is well established and robust expectations of sensitivity can be provided driven by the analysis of the existing data. 
In this way the expected sensitivity for benchmark studies have been presented above.
The HIKE facility allows other users to run in parallel, even in the same area. This facilitates a high degree of diversity in the North-Area physics programme which, we believe, is crucial for the future of particle physics.


\begin{table}[htb]
\caption{Summary of intensity requirements for the HIKE Phases. The number of kaon decays refers to those inside a fiducial decay volume (see Sections about physics sensitivity). 
For the $K^+$ phase, the maximum required intensity is specified; possibilities of reducing the intensity and increasing the momentum bite will be studied for the full proposal. An intensity higher than $2\times 10^{13}$ can be used by HIKE when operating in dump mode.
}
\begin{center}
\vspace{-6mm}
\begin{tabular}{lccc}
\hline
Phase & Protons on target/spill & $K$ decays/year & Protons on TAX \\
\hline
Phase 1 ($K^+$) & $1.2\times 10^{13}$
& $2\times 10^{13}$ & - \\
Phase 2 ($K_L$+tracking) & $2\times 10^{13}$ & $3.8\times 10^{13}$ & - \\
Dump Mode & - & - & (2--4)$\times 10^{13}$ \\
Phase 3 ($K_L$, KLEVER) & $ 2\times 10^{13}$ & $1.3\times 10^{13}$ & - \\
\hline
\end{tabular}
\end{center}
\label{tab:conclusions}
\end{table}